\documentclass[10pt]{article}
\usepackage[bookmarks=true,bookmarksopen=false,bookmarksnumbered=true,colorlinks=true,linkcolor=Maroon]{hyperref}
\usepackage[T1]{fontenc}
\usepackage{graphicx,sectsty,latexsym,textcomp,enumitem,upgreek}
\usepackage[scr]{rsfso}
\DeclareMathAlphabet{\mathdutchcal}{U}{dutchcal}{m}{n}
\SetMathAlphabet{\mathdutchcal}{bold}{U}{dutchcal}{b}{n}
\DeclareMathAlphabet{\mathdutchbcal}{U}{dutchcal}{b}{n}
\usepackage[centertags]{amsmath}
\usepackage{float,amssymb,dsfont,xparse}
\usepackage{stackengine}
\setstackgap{S}{1pt}
\usepackage{amsrefs,amsthm}
\usepackage{fancyhdr}
\usepackage{xcolor,color}
\definecolor{mypink1}{rgb}{0.1, 0.00, 0.000}
\definecolor{mypink2}{RGB}{219, 48, 122}
\definecolor{mypink3}{cmyk}{0, 0.2, 0.4429, 0.0}
\definecolor{Maroon}{rgb}{0.2,0,0.8}
\makeatletter
\def\SK@@ref#1>#2\SK@{%
\leavevmode\vbox to\z@{{%
    \vss
    \SK@refcolor
    \rlap{\vrule\raise .75em%
\hbox{{\normalfont\footnotesize\ttfamily#2}}}}}}
\makeatother
\usepackage[depth=4]{bookmark}
\title{Hamiltonian quantization of solitons in the 
\(\phi^4_{1+1} \) quantum field theory}
\author{David Stuart
 \\{\it{\small Centre for Mathematical Sciences, Wilberforce Road, 
 Cambridge, CB3 OWA,
 England}}\\{\it\small{ email:dmas2@cam.ac.uk}}
}
\date{} 
\DeclareMathOperator{\Poly}{\mcP}
\DeclareMathOperator{\Pexp}{\mbox{${\mathdutchcal{PExp}}$}}
\DeclareMathOperator{\Exp}{\mbox{${\mathdutchcal{Exp}}$}}
\NewDocumentCommand{\oldnorm}{sO{}m}{%
  {\IfBooleanTF{#1}
    {\oldnormaux{\left|}{\right|}{#3}}
    {\oldnormaux{#2|}{#2|}{#3}}}
}

\makeatletter
\newcommand{\oldnormaux}[3]{\mathpalette\oldnormaux@i{{#1}{#2}{#3}}}
\newcommand{\oldnormaux@i}[2]{\oldnormaux@ii#1#2}
\newcommand{\oldnormaux@ii}[4]{%
  \sbox\z@{$\m@th#1#2#4#3$}%
  \sbox\tw@{$\m@th\|$}%
  \mathopen{\hbox to\wd\tw@{\hss\vrule height \ht\z@ depth \dp\z@ width .3\wd\tw@\hss}}%
  #4
  \mathclose{\hbox to\wd\tw@{\hss\vrule height \ht\z@ depth \dp\z@ width .3\wd\tw@\hss}}%
}
\makeatother
\begin{document}
\baselineskip=10pt

\newcommand{\dirop}{\partial\hspace{-1.25ex}\slash}
\newcommand{\Dirop}{{D}\hspace{-1.55ex}\slash}
\newcommand{\cav}{\cee_{0}}
\newcommand{\fav}{\eff_{0}}
\newcommand{\ttF}{\texttt{F}}
\newcommand{\ttea}{\texttt{e}_a}
\newcommand{\tteb}{\texttt{e}_b}
\newcommand{\tteo}{\texttt{e}_0}
\newcommand{\tteokap}{\texttt{e}_{0\kappa}}
\newcommand{\tteone}{\texttt{e}_{1}}
\newcommand{\tteonekap}{\texttt{e}_{1\kappa}}
\newcommand{\tteox}{\texttt{e}_{0\xi}}
\newcommand{\tteoxs}{\texttt{e}_{0\xi(s)}}
\newcommand{\tteonex}{\texttt{e}_{1\xi}}
\newcommand{\bfG}{\mathbf{G}}
\newcommand{\bfR}{\mathbf{R}}
\newcommand{\bfU}{\mathbf{U}}
\newcommand{\bfH}{\mathbf{H}}
\newcommand{\bfalpha}{\mbox{\boldmath$\alpha$}}
\newcommand{\bfPhi}{\mathbf{\Phi}}
\newcommand{\bfPi}{\mathbf{\Pi}}
\newcommand{\bfhatOm}{\mathbf{\hat\Omega}}
\newcommand{\bftildeOm}{\mathbf{\tilde\Omega}}
\newcommand{\bfOm}{\mathbf{\Omega}}
\newcommand{\bfp}{\mathbf{\Psi_0}}
\newtheorem{lemma}{Lemma}[section]
\newtheorem{prop}[lemma]{Proposition}
\newtheorem{theorem}[lemma]{Theorem}
\newtheorem{corollary}[lemma]{Corollary}
\newtheorem{definition}[lemma]{Definition}
\theoremstyle{remark}
\newtheorem{remark}[lemma]{Remark}
\newtheorem{example}[lemma]{Example}
\newtheorem{notation}[lemma]{Notation}
\newcommand{\sign}{\hbox{sign}\,}
\newcommand{\eps}{\epsilon}
\newcommand{\id}{\mathds{1}}
\newcommand{\be}{\begin{equation}}
\newcommand{\ee}{\end{equation}}
\newcommand{\ba}{\begin{eqnarray}}
\newcommand{\ea}{\end{eqnarray}}
\newcommand{\bes}{\[}
\newcommand{\ees}{\]}
\newcommand{\bas}{\begin{eqnarray*}}
\newcommand{\eas}{\end{eqnarray*}}
\newcommand{\hoa}{{H^{1}_{\mbA}}}
\newcommand{\hta}{{H^{2}_{\mbA}}}
\newcommand{\linf}{{L^\infty}}
\newcommand{\tp}{{\tilde P}}
\newcommand{\cf}{{\cal F}}
\newcommand{\bfP}{\mathbf{P}}
\newcommand{\ch}{{\cal H}}
\newcommand{\cfnd}{{\cal F}^{n}-{\cal F}^{n-1}}
\newcommand{\p}{\partial}
\newcommand{\abs}[1]{\vert #1 \vert}
\newcommand{\Balpha}{\mbox{$\hspace{0.12em}\shortmid\hspace{-0.62em}\alpha$}}
\newcommand{\scm}{\textsc m}
\newcommand{\spsi}{_{{{\Psi}}}}
\newcommand{\pt}{\frac{\partial}{\partial t}}
\newcommand{\pxk}{\frac{\partial}{\partial {x^k}}}
\renewcommand{\theequation}{\arabic{equation}}
\newcommand{\rgt}{\rightarrow}
\newcommand{\lngrgt}{\longrightarrow}
\newcommand{\intsT}{ \int_{0}^{T}\!\!\int_{\Sigma} }
\newcommand{\dxdt}{\;dx\,dt}
\newcommand{\sublt}{_{L^2}}
\newcommand{\sublf}{_{L^4}}
\newcommand{\sech}{{\rm sech}}
\newcommand{\tmx}{\tanh mx\,}
\newcommand{\smx}{\sech\, mx\,}
\newcommand{\tmy}{\tanh my\,}
\newcommand{\smy}{\sech\, my\,}
\newcommand{\ttmx}{\tanh{}^2 mx\,}
\newcommand{\tmxi}{\tanh{}m(x-\xi)\,}
\newcommand{\tmxip}{\tanh{}m(x-\xi')\,}
\newcommand{\tmxih}{\tanh{}m(x-\xi-h)\,}
\newcommand{\tmxiy}{\tanh{}m(y-\xi)\,}
\newcommand{\tmxiyh}{\tanh{}m(y-\xi-h)\,}
\newcommand{\ttmxi}{\tanh{}^2m(x-\xi)\,}
\newcommand{\ssm}{\sech{}^2 m}
\newcommand{\ssssm}{\sech{}^4 m}
\newcommand{\ssmx}{\sech{}^2 mx\,}
\newcommand{\ssmxi}{\sech{}^2 m(x-\xi)\,}
\newcommand{\ssmxih}{\sech{}^2 m(x-\xi-h)\,}
\newcommand{\ssmxiy}{\sech{}^2 m(y-\xi)\,}
\newcommand{\ssmxiyh}{\sech{}^2 m(y-\xi-h)\,}
\newcommand{\phxi}{}
\newcommand{\ttmy}{\tanh{}^2 my\,}
\newcommand{\ssmy}{\sech{}^2 my\,}
\newcommand{\naf}{\nabla_\mbA \Phi}
\newcommand{\covt}{(\pt -iA_0) }
\newcommand{\ano}{A^{n}_{0}}
\newcommand{\aNo}{A^{N}_{0}}
\newcommand{\Psino}{\Psi^{n}_{0}}
\newcommand{\PsiNo}{\Psi^{N}_{0}}
\newcommand{\Nmo}{{N\!-\!1}}
\newcommand{\nmo}{{n-1}}
\newcommand{\nmt}{{n-2}}
\newcommand{\Nmt}{{N\!-\!2}}
\newcommand{\gotm}{\frac{\gamma}{2\mu}}
\newcommand{\ootm}{\frac{1}{2\mu}}
\newcommand{\tloc}{T_{loc}}
\newcommand{\tmax}{T_{max}}
\newcommand{\beefb}{\pmb{b}}
\newcommand{\beed}{{\mathrm{b}}_d}
\newcommand{\beek}{{\mathrm{b}}_k}
\newcommand{\beekm}{{\mathrm{b}}_{-k}}
\font\msym=msbm10
\font\smallmsym=msbm7
\def\smr{{\mathop{\hbox{\smallmsym \char '122}}}}
\def\smc{{\mathop{\hbox{\smallmsym \char '103}}}}
\def\Complex{{\mathop{\hbox{\msym\char'103}}}}
\newcommand{\wkarr}{\; \rightharpoonup \;}
\def\Weak{\,\,\relbar\joinrel\rightharpoonup\,\,}
\newcommand{\To}{\longrightarrow}
\newcommand{\rp}{\hbox{Re\,}}
\newcommand{\pa}{\partial_A}
\newcommand{\pbfa}{\partial_{\mathbf A}}
\newcommand{\pao}{\partial_{A_1}}
\newcommand{\pat}{\partial_{A_2}}
\newcommand{\eire}{Err^0_{\mbox{\tiny IR}}}
\newcommand{\eirup}{{{\rm Err}^0_{\mbox{\tiny IR}}}}
\newcommand{\eirtd}{{\rm Err}_{\mbox{\tiny TD}}}
\newcommand{\barpa}{\bar{\partial}_{A}}
\newcommand{\barpbfa}{\bar{\partial}_{\mathbf A}}
\newcommand{\barpaphi}{\bar{\partial}_{A}\Phi}
\newcommand{\barpbfaphi}{\bar{\partial}_{\mathbf A}\Phi}
\newcommand{\myqed}{\hfill $\Box$}
\newcommand{\cd}{{\cal D}}
\newcommand{\dt}{\hbox{det}\,}
\newcommand{\sma}{_{{ A}}}
\newcommand{\bfpi}{{\mbox{\boldmath$\pi$}}}
\newcommand{\ce}{{\cal E}}
\newcommand{\ulh}{{\underline h}}
\newcommand{\ulg}{{\underline g}}
\newcommand{\ulX}{{\underline {\bf X}}}
\newcommand\la{\label}
\newcommand{\lamo}{\stackrel{\circ}{\lambda}}
\newcommand{\bfjo}{\underline{{\bf J}}}
\newcommand{\Vflato}{V^\flat_0}
\newcommand{\cm}{{\cal M}}
\newcommand{\dist}{{\mbox{dist}}}
\newcommand{\cs}{{\mathcal S}}
\newcommand{\mcH}{{\mathcal H}}
\newcommand{\mcD}{{\mathcal D}}
\newcommand{\mcK}{{\mathcal K}}
\newcommand{\mfrH}{{\mathfrak H}}
\newcommand{\mfrbH}{\mbox{\boldmath$\mathfrak H$}}
\newcommand{\mfrF}{{\mathfrak F}}
\newcommand{\mfrbF}{\mbox{\boldmath$\mathfrak F$}}
\newcommand{\mfrS}{{\mathfrak S}}
\newcommand{\mcF}{{\mathcal F}}
\newcommand{\mcbF}{\mbox{\boldmath$\cF$}}

\newcommand{\mcP}{\mbox{$\mathscr{P}$}}
\newcommand{\mcbP}{\mbox{\boldmath$\mcP$}}
\newcommand{\mcR}{{\mathcal R}}
\newcommand{\mcn}{{\mathcal  V}}
\newcommand{\mce}{{\mathcal  E}}
\newcommand{\mcb}{{\mathcal B}}
\newcommand{\mca}{{\mathcal A}}
\newcommand{\mcdpsi}{{\mathcal D}_{\psi}}
\newcommand{\mcd}{{\mathcal D}}
\newcommand{\mcl}{{\mathcal L}}
\newcommand{\mclaphi}{{\mathcal L}_{(\mbA,\Phi)}}
\newcommand{\mcdadjp}{\mcdadj_{\psi}}
\newcommand{\mcdadj}{{\mathcal D}^{\ast}}
\newcommand{\zl}{{Z_\Lambda}}
\newcommand{\thl}{{\Theta_\Lambda}}
\newcommand{\ca}{{\cal A}}
\newcommand{\cb}{{\cal B}}
\newcommand{\cg}{{\cal G}}
\newcommand{\cU}{{\cal U}}
\newcommand{\co}{{\cal O}}
\newcommand{\smA}{\small A}
\newcommand{\ttheta}{\tilde\theta}
\newcommand{\tn}{{\tilde\|}}
\newcommand{\rbar}{\overline{r}}
\newcommand{\oeps}{\overline{\varepsilon}}
\newcommand{\cgl}{\hbox{Lie\,}{\cal G}}
\newcommand{\Ker}{\hbox{\rm Ker\,}}
\newcommand{\Dom}{\hbox{\rm Dom\,}}
\newcommand{\Sym}{{\rm Sym\,}}
\newcommand{\Symn}{{\rm Sym}^n\,}
\newcommand{\Symhat}{\widehat{\rm Sym}\,}
\newcommand{\Symhatl}{\widehat{\rm Sym}{L}^2\,}
\newcommand{\fock}{\mfrH_0}
\newcommand{\AC}{\hbox{AC\,}}
\newcommand{\const}{\hbox{const.\,}}
\newcommand{\tr}{\hbox{tr\,}}
\newcommand{\grad}{\hbox{grad\,}}
\newcommand{\ttd}{{\tt d}}
\newcommand{\ttdel}{{\tt \delta}}
\newcommand{\ns}{\nabla_*}
\newcommand{\csl}{{\cal SL}}
\newcommand{\Fin}{\hbox{\rm Fin}}
\newcommand{\beq}{\begin{equation}}
\newcommand{\eeq}{\end{equation}}
\newcommand{\pr}{\hbox{proj\,}}
\newcommand{\proj}{{\mathbb P}}
\newcommand{\tproj}{\tilde{\mathbb P}}
\newcommand{\projq}{{\mathbb Q}}
\newcommand{\Q}{\projq}
\newcommand{\h}{{\mathds{h}}}
\newcommand{\eff}{{\mathrm{f}}}
\newcommand{\cee}{{\mathrm{c}}}
\newcommand{\tprojq}{\tilde{\mathbb Q}}
\newcommand{\bbvac}{{\Omega}}
\newcommand{\oN}{\overline N}
\newcommand{\cN}{{\mathcal N}}
\newcommand{\cF}{{\mathcal F}}
\newcommand{\cmet}{{{\hbox{${\mathcal Met}$}}}}
\newcommand{\met}{{\hbox{Met}}}
\newcommand{\bfga}{{\mbox{\boldmath$\gamma$}}}
\newcommand{\bfOmega}{\mbox{\boldmath$\Omega$}}
\newcommand{\bfTh}{\mbox{\boldmath$\Theta$}}
\newcommand{\bfmw}{{\bf m}_{\mbox{\boldmath$\smo$}}}
\newcommand{\bfmu}{{\mbox{\boldmath$\mu$}}}
\newcommand{\bfdel}{{\mbox{\boldmath$\delta$}}}
\newcommand{\bftau}{{\mbox{\boldmath$\tau$}}}
\newcommand{\upphiE}{\upphi_{\small E}}
\newcommand{\phisch}{{\mbox{\boldmath$\varphi$}}}
\newcommand{\pisch}{{\mbox{\boldmath$\pi$}}}
\newcommand{\phivs}{\bfupphi}
\newcommand{\pivs}{\bfuppi}
\newcommand{\bfupphi}{\mbox{\boldmath$\upphi$}}
\newcommand{\bfmcD}{\mbox{\boldmath$\mcD$}}
\newcommand{\bfupalpha}{\mbox{\boldmath$\upalpha$}}
\newcommand{\bfuppi}{\mbox{\boldmath$\uppi$}}
\newcommand{\frakm}{\mathfrak{m}}
\newcommand{\frakM}{\mathfrak{M}}
\newcommand{\bfulX}{{\mbox{\boldmath${X}$}}}
\newcommand{\bfultX}{\tilde{\mbox{\boldmath${X}$}}}
\newcommand{\bfmuw}{{\mbox{\boldmath$\mu$}}_{\mbox{\boldmath$\smo$}}}
\newcommand{\dmug}{d\mu_g}
\newcommand{\ltwo}{{L^{2}}}
\newcommand{\lfour}{{L^{4}}}
\newcommand{\hone}{H^{1}}
\newcommand{\honea}{H^{1}_{\smA}}
\newcommand{\er}{e^{-2\rho}}
\newcommand{\onetwo}{\frac{1}{2}}
\newcommand{\lra}{\longrightarrow}
\newcommand{\dv}{\hbox{div\,}}
\newcommand{\mbQ}{\mathbf Q}
\newcommand{\mbP}{\mathbf P}
\newcommand{\mbA}{\mathbf A}
\newcommand{\mba}{\mathbf a}
\newcommand{\mbm}{\mathbf m}
\newcommand{\mbn}{\mathbf n}
\newcommand{\mbx}{\mathbf x}
\newcommand{\mbPhi}{\mathbf \Phi}
\newcommand{\mbPi}{\mathbf \Pi}
\newcommand{\Fl}{\boldsymbol {\cal X}}
\newcommand{\fl}{\mbchi}
\newcommand{\mbg}{\boldsymbol \gamma}
\newcommand{\mbchi}{\boldsymbol \chi}
\newcommand{\mbChi}{\Fl}
\newcommand{\opsi}{\overline{\psi}}
\newcommand{\vac}{{\Omega^{}_0}}
\newcommand{\cvo}{{C_0^{\frac{1}{2}}}}
\newcommand{\cvot}{{({K_{0}^{\frac{1}{2}}}+\theta^{\frac{1}{2}}\P_0)^{-1}}}
\newcommand{\kq}{{K_0^{\frac{1}{4}}}}
\newcommand{\cv}{{(C^\theta)^{\frac{1}{2}}}}
\newcommand{\Dis}{\Updelta}

\newcommand{\doublecolon}{\setstackgap{S}{1.25pt}\raisebox{0.25pt}{\Shortstack{ . . }}}
\newcommand{\triplecolon}{\setstackgap{S}{.75pt}\raisebox{-0.25pt}{\Shortstack{. . . }}}
\numberwithin{equation}{section}
\numberwithin{lemma}{section}

\font\msym=msbm10
\def\Real{{\mathop{\hbox{\msym \char '122}}}}
\def\R{\Real}
\def\evot{\T}
\def\bevot{\mathbf{T}}
\def\T{\mathbb T}
\def\G{\mathbb G}
\def\D{\mathbb D}
\def\I{\mathbb I}
\def\U{\mathbb U}
\def\mbU{\mathbf U}
\def\Y{\mathbb Y}
\def\mbY{\mathbf Y}
\def\Z{\mathbb Z}
\def\E{\mathbb E}
\def\A{\mathbb A}
\newcommand{\elec}{\E}
\def\B{\mathbb B}
\def\P{\mathbb P}
\def\K{\mathbb K}
\def\k{\pmb k}
\def\J{\mathbb J}
\def\L{\mathbb L}
\def\D{\mathbb D}
\def\N{\mathbb N}
\def\M{\mathbb M}
\def\V{\mathbb V}
\def\mbV{\mathbf V}
\def\W{\mathbb W}
\def\mbW{\mathbf W}
\def\Op{\hbox {Op}}
\def\mbB{\mathds B}
\def\mbS{\mathbf{S}}
\def\bbS{\mathbb S}
\def\bbdotS{\dot {\mathbb S}}
\def\mbtS{\tilde{\bbS}}
\def\mbO{\mathbb O}
\def\T{\mathbb T}
\def\mbT{\mathbf{T}}
\def\Mink{{\mathop{\hbox{\msym \char '115}}}}
\def\Integers{{\mathop{\hbox{\msym \char '132}}}}
\def\Complex{{\mathop{\hbox{\msym\char'103}}}}
\def\C{\Complex}
\def\He{\hbox{He}}
\def\v{\hbox{v}}
\def\lb{D}
\def\upphihat{\hat\upphi}
\def\uppihat{\hat\uppi}
\def\varphihat{\hat\varphi}
\def\pihat{\hat\pi}
\def\upphiperp{\upphi^\perp}
\def\upphiperposc{\upphi^{\perp,osc}}
\def\uppiperp{\uppi^\perp}
\def\varphiperp{\varphi^\perp}
\def\piperp{\pi^\perp}
\def\bfupphiperp{\bfupphi^\perp}
\def\phischperp{\phisch^\perp}
\def\bfuppiperp{\bfuppi^\perp}
\def\bfvarphiperp{\bfvarphi^\perp}
\def\bfpiperp{\bfpi^\perp}
\font\smallmsym=msbm7
\def\smr{{\mathop{\hbox{\smallmsym \char '122}}}}
\newcommand{\define}{\stackrel{\text{\tiny def}}{=}}
\definecolor{refkey}{rgb}{0.75,0.25,0}
\definecolor{labelkey}{rgb}{0,0,1}
\maketitle
\thispagestyle{empty}
\vspace{-0.3in}
\begin{abstract}
  We first carry out the soliton sector quantization of the spatially cut-off
  $\phi^4_{1+1}$ theory with double well potential
  in the semiclassical limit, deriving the nonrelativistic
  Schr\"odinger equation as an equation describing the limiting soliton dynamics. In the 
  process we prove the semiclassical mass shift formula of Dashen, Hasslacher and Neveu,
  which is interpreted in terms of a unitary equivalence between normal ordered semiclassical quadratic Hamiltonians
  in two different representations of the Heisenberg relations. Secondly, we consider the $\phi^4_{1+1}$ theory
  coupled topologically to an external electromagnetic field and prove the main result, which is
  an approximation theorem reminiscent of the Born-Oppenheimer method, which describes
  the nonrelativistic dynamics of the soliton coupled to infinitely many transverse bosonic degrees of freedom,
  extending the techniques of soliton modulation theory from classical to quantum field theory.
\end{abstract}
\tableofcontents
\section{Introduction}
\setcounter{equation}{0}
\label{secint}
We study the interaction of a
scalar quantum field $\phi $ with a fixed (external)
electromagnetic field with potential $\A_\mu dx^\mu$ in two dimensional
space-time. The dynamics is determined
by the action
functional
\[
S_\lambda\,=\,\int\,\Bigl(\frac{1}{2}\partial_\mu\phi\partial^\mu\phi-
\frac{1}{2}\,g^2\bigl(\phi^2-\frac{m^2}{g^2}\bigr)^2
+\lambda\epsilon_{\mu\nu}\partial_\mu \A_\nu \phi\,\Bigr)\,dxdt\,.
\]
Observe that (modulo boundary terms) the electromagnetic coupling is via the topological current
\(J^\mu_{top}=\epsilon^{\mu\nu}\partial_\nu\phi
\) which is conserved via the identity \(\partial_\mu J^\mu_{top}=\epsilon^{\mu\nu}\partial_\nu\partial_\mu\phi\equiv 0\).
The associated topological charge \(\int_{-\infty}^{+\infty} J^0_{top}\,dx=\int_{-\infty}^{+\infty}\partial_x\phi\,dx\) ensures existence of
solitons for appropriate nonlinear potentials, as we recall below.
In particular the quartic interaction with a double well potential, under consideration in this paper, supports the existence
of solitons in the classical theory. The existence of the corresponding 
quantum theory can be proved by the Hamiltonian
methods of constructive quantum field theory, together with Kato's theory for
evolution operators generated by time-dependent Hamiltonians to incorporate the electromagnetic field. 
The aim is to analyze the dynamics of
the soliton in this quantized theory as ${g\downarrow 0} $, which
corresponds to a nonrelativistic limit for the soliton, which has mass
which diverges as $g^{-2} $ in this limit.  We develop the analytical 
framework for quantizing the theory, identify the appropriate 
degrees of freedom to
describe the soliton, allowing
\begin{itemize}
  \item a precise interpretation of the 
Dashen-Hasslacher-Neveu semiclassical mass correction formula (from \cite{Dashen}) in terms of
a unitary equivalence between the second quantized Hamiltonians obtained via two
different representations of the Heisenberg relations, see \eqref{suml}-\eqref{cba};
\item
the statement and proof of the main theorem, which gives an explicit description of
the \(g\to 0+\) limiting soliton dynamics induced by an external electromagnetic field, see Theorem \ref{main}. This represents
an extension to quantum field theory of modulation theory for solitary waves in classical field theory, as developed in \cite {MR0783974}
for nonlinear Schr\"odinger equations and \cite{MR1186038,MR1810509} for relativistic theories.
\end{itemize}
We refer to
\cite{MR0503137} and \cite{cole}*{Chapter 6} for a general physical discussion
of quantum solitons, and in particular in \S 4 of the latter reference draw attention to the discussion of the limit
\(g\to 0+\) which is both a semiclassical and a nonrelativistic limit for the soliton; also see
\cite{MR887102}*{\S 23.8} for a review of mathematical work on solitons in the
context of constructive field theory, and in particular to \cite{MR0496006} for
bounds on the soliton mass.
\paragraph{Classical Theory.} In this first section of the paper attention
is focused on the case of zero electric field (or $\lambda=0$). We work with the Euler-Lagrange equations for the action
\(S_\lambda|_{\lambda=0}\) in Hamiltonian form: the Hamiltonian is the functional
\begin{equation}\label{ham}
  H({\phi,\pi})\,=\,\int_{\smr}\,\mcH({\phi,\pi})\,dx\,,\quad
  \mcH({\phi,\pi})\,=\,\frac{1}{2}\,\bigl(
\pi^2\,+\,\partial_x\phi^2\bigr)\,+\,\cU(\phi)
\,.
\end{equation}
The potential function \(\cU \) is the double-well potential
\begin{equation}\label{dwell}
\cU(\phi)\,=\,\frac{m^4}{2g^2}\,\bigl(1-\frac{g^2\phi^2}{m^2}\bigr)^2\,=\,
\frac{1}{2}\,g^2\,\bigl(\phi^2-\Phi_0^2\bigr)^2\,.
\end{equation}
The parameters \(m,g \) are assumed to be positive numbers.
The functional is well-defined as a non-negative number, possibly equal to \(+\infty\),
on pairs \(({\phi,\pi})\in H^1_{loc}\times L^2_{loc}\); the pairs for which \(H({\phi,\pi})<\infty\)
are the finite energy configurations.
The two classical vacua are \(\pm\Phi_0 \), where \(\Phi_0=m/g \).
Clearly the constant configuration \((\phi,\pi)=(\Phi_0,0) \)
minimizes the value of the Hamiltonian energy functional amongst 
all finite energy configurations which satisfy 
\begin{equation}\label{vacbc}
\lim\limits_{|x|\to\infty}\,\phi(x)\,=\,\Phi_0\,;
\end{equation}
a similar assertion holds for \((-\Phi_0,0)\,. \) Expanding
around these vacua leads to the Hamiltonians
\begin{equation}\label{ham-vac}
  H(\pm\Phi_0+\varphi,\pi)\,=\,\int\,\Bigl[\frac{1}{2}\,\bigl(\pi^2+
  \partial_x\varphi^2+4m^2\varphi^2\bigr)\pm 2mg\varphi^3
\,+\,\frac{1}{2}
g^2\varphi^4\,\Bigr]\,dx\,.
\end{equation}
The quadratic part of this Hamiltonian, namely
\begin{equation}\label{vach}
H^{vac}_0(\varphi,\pi)\,=\,\int\,\frac{1}{2}\,
\bigl(\pi^2+\partial_x\varphi^2+4m^2\varphi^2\bigr)\,dx\,,
\end{equation}
describes the quantum mechanics of non-interacting relativistic scalar 
bosons of mass \(2m\) - these bosons are the fundamental particles 
of the theory.
The cubic and quartic terms describe interactions between
these particles, the strength being determined by the (positive) 
coupling constant \(g\). We will be interested in analysing the dynamics in the limit
\(g\downarrow 0\), and it will in any case always be assumed that \(0<g<1\).

The classical soliton,
\begin{equation}\label{csol}
\Phi_S(x)\,=\,\frac{m}{g}\,\tmx\,,\qquad\Pi_S(x)=0\,,
\end{equation}
is a solution of the classical Hamiltonian equations of
motion
\begin{equation}\label{ceom}
\dot\pi-\partial_x^2\phi+\cU'(\phi)\,=\,0\,,\qquad
\dot\phi-\pi\,=\,0\,.
\end{equation}
The soliton has the property that \(\Phi_S \) interpolates between 
the two vacua as its
asymptotic boundary values, i.e.,
\begin{equation}\label{abv}
\Phi_S(x)\,\to\,\pm\Phi_0\quad\hbox{as }\;x\,\to\,\pm\infty\,.
\end{equation}
These boundary conditions endow the soliton with topological charge
\[
\int_{-\infty}^{+\infty} J^0_{top}\,dx=\int_{-\infty}^{+\infty}\partial_x\Phi_S\,dx
=\Phi_S(+\infty)-\Phi_S(-\infty)=\frac{2m}{g},
\]
and \(\Phi_S\) minimizes the value of the Hamiltonian energy functional amongst 
all finite energy configurations which satisfy these 
boundary conditions. However, the
soliton is not unique due to the action of the translation group: the set of
energy minimizers is \(\{(\Phi_S(\,\cdot\,-\xi),0)\}_{\xi\in\smr}\,. \)
The energy of an energy minimizer equals
the minimum value of \(H \) on the set of finite energy configurations
verifying \eqref{abv}; this minimum value is the classical
rest mass of the soliton, given by 
\begin{equation}\label{mcl}
  \M_{cl}\,=\,\frac{4m^3}{3g^2}\,=\,
  \frac{\scm_{cl}}{g^2}\,,\qquad{\scm}_{cl}\,=\,\frac{4m^3}{3}\,.
\end{equation}
Expanding
around the soliton leads to the Hamiltonian
\begin{align}\label{ham-sol}\begin{split}
  &  H(\Phi_S+\varphi,\pi)\,=\,\frac{\scm_{cl}}{g^2}\,
  +\,
H^{sol}_g({\varphi,\pi})\\
& H^{sol}_g({\varphi,\pi})  =\, \int\,\frac{1}{2}\,
  \bigl(\pi^2+\partial_x\varphi^2+4m^2\varphi^2-6m^2\ssmx\varphi^2\bigr)\,dx
  +\,\int\,\bigl( 2mg\tmx\varphi^3
\,+\,\frac{1}{2}
g^2\varphi^4\,\bigr)\,dx\,.
\end{split}\end{align}
This Hamiltonian describes fluctuations around the basic soliton, centered
at the origin. These
fluctuations are determined infinitesimally by the linearized operator
\(K=-\partial_x^2+4m^2-6m^2\ssmx\). As discussed below, this operator has a one dimesional
kernel which reflects the fact that physically the soliton
is able to move along the orbit of the translation group without any
 ``energetic cost'', i.e. dynamically
the parameter \(\xi=\xi(t) \) becomes time-dependent, and one studies solutions
of the form \begin{equation}\label{cmod}
\Phi_S\bigl(x-\xi(t)\bigr)+\varphi(t,x)\,.
  \end{equation}
Now Lorentz invariance implies the existence of
exact solutions of the classical equations
of motion \eqref{ceom}
of the form \(\Phi_S\bigl(\frac{x-ut-x_0}{\sqrt{1-u^2}}\bigr) \),
in which the soliton moves along a straight line.
More importantly for present purposes, this behaviour is actually stable
generic behaviour in the low energy limit, and the dynamics can be
approximated on appropriate time scales (in the \(H^1\times L^2\) norm)
by the Newtonian equation of motion for a freely moving
particle of
mass \(\scm_{cl} \), i.e.,
\begin{equation}\label{newt}
\dot\eta=\,0\,\qquad\hbox{where}\;\;\eta\,=\,\M_{cl}\,\dot\xi\quad\;\;\hbox{(momentum)}\,.
  \end{equation}
These types of problems, with some representative theorems for higher dimensional gauge theories,
are surveyed in \cite{MR2360179} from a mathematical point of view, and in \cite{MR2068924} from a
physical perspective.
The inclusion of the mass in \eqref{newt} is a matter of convention here,
but in the presence of external potentials is unavoidable. We now discuss
how this picture might be expected to be
modified in the quantum case.

The quantum field theory for the Hamiltonian \eqref{ham-vac} was constructed
by the Hamiltonian method in \cite{MR0247845}. With a spatial cut-off the theory
admits a Schr\"odinger
representation formulation with respect to a Gaussian measure \(\mu_0 \)
on the space of tempered distributions (see Proposition \ref{sch});
see \cite{MR0674511} for a review. Moving to the
soliton sector via \eqref{ham-sol} corresponds essentially to shifting the field by
\(\Phi_S-\Phi_0 \), which is not a Cameron-Martin vector for
\(\mu_0 \), and in measure theoretic
terms leads to a representation supported on a measure which is
singular with respect to the vacuum measure - it will
be called the {\em shifted vacuum} representation. In this representation the field is
\(\Phi_S+\phisch\), where \(\phisch\) is an operator of multiplication on
\(L^2(\mu_0)\), with conjugate field \(\pisch\) as in \eqref{seq}.
Construction
of the quantum field theory using this representation as starting point leads to the 
quantum theory in the {\em soliton sector}, as opposed to the vacuum sector. In fact we will see it is useful to
use two different but equivalent representations in the soliton sector to reveal
the physics. In particular, the unitary equivalence leads both to the precise interpretation \eqref{suml}-\eqref{cba}
of the semiclassical mass shift, and to the introduction of appropriate coordinates to make physical sense of the
dynamics under interaction, see \S\ref{modulth}.
These representations are studied in \S\ref{qsol}. We now consider what the expected physics is in the limit
of small \(g\,. \)

\paragraph{The soliton as a nonrelativistic quantum particle.}
Firstly, it is
to be hoped, that in the limit of small
coupling \(g \) the soliton will behave as a quantum particle of mass
\(\M_{cl}(1+o(1))=\frac{\scm_{cl}}{g^2}(1+o(1))\), 
and thus that the Schr\"odinger equation with Hamiltonian
\(P^2/(2\M_{cl}) \) should appear in the analysis of the limit
\({g\downarrow 0} \), in place of \eqref{newt}.
On account of the \(g \)-dependence of \(\M_{cl} \) displayed in \eqref{mcl}
this indicates that the quantum fluctuations thus described will be suppressed
to be \(O({\M_{cl}}^{-\frac{1}{2}})= O(g\scm_{cl}^{-\frac{1}{2}})) \) in the semiclassical regime, as
would be expected from basic quantum mechanics. For example the analysis of
Gaussian wave packets for the free Schr\"odinger evolution of a particle
with mass \(\M_{cl}\) leads to the conclusion that the width of the wave packet
is
\(\gtrsim\sqrt{\frac{\hbar t}{\M_{cl}}}=g\sqrt{\frac{\hbar t}{\scm_{cl}}}\)), see for example
\cite{MR0129790}*{Chapter VI}. Two conclusions to be kept in mind can be drawn from this:
\begin{itemize}
  \item
in order to ``see'' the Schr\"odinger
equation and the quantum fluctuations in the limit, it is necessary to look
at small scales, which at finite times would be of order \(g \);
\item
  the standard deviation of the spatial fluctuations does however grow linearly in time, and so on longer time scales
  \(\sim g^{-a}\) the fluctuations would be larger, of standard deviation \(\sim g^{1-\frac{a}{2}}\).
  \end{itemize}
  Thus we might hope to be able to analyze solutions to the
quantum field theory in which the field takes the form (at fixed time)
\begin{equation}\label{qmod}
\Phi_S(x-\xi-gQ)+\varphi\,\approx\,\Phi_S(x-\xi)-g\Phi_S'(x-\xi)Q+\varphi
  \end{equation}
where \(\xi \) is a {\em classical} c-number giving the location of the  classical solution
about which we quantize, while \(gQ \) represents an \(O(g) \) quantum fluctuation in its
location. But we should keep in mind that on larger time scales - and it is \(\sim g^{-\frac{1}{2}}\) that will be
particularly relevant - the fluctuation in \(Q\) will be of order \(\sqrt{\langle Q^2\rangle}\sim g^{-\frac{1}{4}}\).  

We will study quantum dynamics around nontrivial classical motions
\(\xi(t)\), so that \(\xi+gQ\) is the soliton position operator.
In favorable circumstances, it is to be hoped that the operator \(Q\) can 
be realized in the Schr\"odinger picture as the operator of multiplication
by \(Q \) on a wave function \(\chi=\chi(t,Q) \) whose evolution can be approximated for
small \(g\) by a modification of the Schr\"odinger equation
\begin{equation}\label{seom}
i\frac{\partial\chi}{\partial t}\,+\,\frac{P^2}{2\scm_{cl}}\chi\,=0\,.
\end{equation}
Here the momentum operator \(P\) conjugate to \(Q\) is \(P=-i\frac{\partial}{\partial Q}\) in the standard case \(L^2(dQ)\),
or \(P=-i\frac{\partial}{\partial Q}+i{\scm_{cl}}\sqrt{\theta}Q\) in the Gaussian case \(L^2(\gamma_\theta(dQ))\) where
\(\gamma_\theta\) is the Gaussian on \(\R\) with variance \(1/(2\scm_{cl}\sqrt{\theta})\), see \S\ref{notn}.

\paragraph{Bosons in the soliton background.}
In addition to this Schr\"odinger particle, there are transverse modes which can be understood by analyzing the quadratic part
of the Hamiltonian, which is obtained by expanding around a kink located at
the origin, namely,
\begin{equation}\label{qh}
H^{sol}_0(\varphi,\pi)\,=\,\int\,\frac{1}{2}\,
\bigl(\pi^2+\partial_x\varphi^2+4m^2\varphi^2-6m^2\ssmx\varphi^2\bigr)\,dx\,.
\end{equation}
This will be quantized in \S\ref{hcr}, firstly in the shifted vacuum representation \(\phisch,\pisch\)
by treating the final term as a perturbation of the free Hamiltonian for
mass \(2m\) bosons, and secondly by developing the quantization based on the operator \(K=-\partial_x^2+4m^2-6m^2\ssmx\)
in place of \(K_0=-\partial_x^2+4m^2\). This latter approach leads to a different {\em solitonic} representation
\(\upphi,\uppi\)
of the Heisenberg relations which
diagonalizes the Hamiltonian, see \ref{simh},  and hence 
reveals the following transverse modes:
\begin{itemize}
\item An assembly of bosons (or mesons) moving in the background
  potential \({u}(x)=-6m^2\ssmx \) created by the soliton itself,
  described in normal form by the Hamiltonian \(\h
  =\int\omega^{}_k a_{k}^\dagger a^{}_{k}
  \,dk\) with \(\omega^{}_k=\sqrt{4m^2+k^2}\), which defines a  self-adjoint operator
  on the Fock space \({\fock}\) defined in \eqref{fock}.
\item An oscillatory mode (pulsation of the soliton)
  of frequency \(\omega^{}_d=\sqrt{3}m\), described by harmonic oscillator
  Hamiltonian \({h}_d=\omega^{}_d a_{d}^\dagger a^{}_{d}\). In the
Schr\"odinger picture this 
  determines a self-adjoint operator acting on
  \(L^2(\R,\gamma_d)\)
  in the usual way, see \eqref{l2g}, with 
the Gaussian measure \(\gamma_d(dq_d)\define\pi^{-\frac{1}{2}}\omega^{\frac{1}{2}}_d\exp{[-\omega^{}_d q_d^2]}\,dq^{}_d \) of covariance
\((2\omega_d)^{-1} \) arising as the square of the ground state \((2\omega_d)^{\frac{1}{4}}\mbchi_0(\sqrt{2\omega_d}q_d)\)
of the frequency \(\omega_d\)  oscillator.
\end{itemize}
This expected picture of the quantum field theory in the soliton sector - a quantum
particle interacting with a quantum field - is broadly similar
to that which appears on quantization of the Abraham model, see \cite{MR2097788}. However
there is a difference that in the case of the Abraham model a particle-field
decomposition is given from the beginnning whereas in the present case these features
have to be derived
using an appropriate choice of solution of the Heisenberg relation. Indeed, use of
the shifted vacuum representation \((\phisch,\pisch) \) of the Heisenberg relation
mentioned previously, (or its Fock space equivalent \((\varphi,\pi) \) in \eqref{cfxi1}-\eqref{cfxi2}), does not 
bring out these features. Instead it is helpful, as just mentioned,
to use another representation \((\upphi,\uppi) \), see \eqref{cfsgext}), which essentially diagonalizes \(H^{sol}_0 \) and will be referred to 
as the {\em solitonic representation}. Generalizing to allow a kink with arbitrary centre \(\xi\in\R\), the 
construction is based on the operator \(K(\xi)=-\partial_x^2+4m^2-6m^2\ssmxi\)
which appears on linearization about \(\Phi_S(\,\cdot\,-\xi)\).
This operator is a non-negative self-adjoint 
operator on \(L^2(\R) \) with domain \(\Dom (K(\xi))=H^2(\R) \).
The spectrum consists of:
\begin{itemize}
\item zero, with a one-dimensional
kernel \(\langle\{ {\tteox}\}\rangle \); 
\item one simple discrete nonzero eigenvalue \(\omega_d^2=3m^2 >0\), 
\[
K(\xi) {\tteonex}\,=\,\omega_d^2 {\tteonex}\,,\qquad \omega^{}_d=\sqrt{3}m
\]
with corresponding spectral subspace 
\(\langle\{ {\tteonex}\}\rangle \); 
\item continuous spectrum \([4m^2,+\infty)\,. \)
\end{itemize}
In addition to the normalizable eigenfunctions \( {\tteox}\in\cs(\R)\) 
and \( {\tteonex}\in\cs(\R)\), there
are generalized (Jost) eigenfunctions \(e_{k\xi}\in L^\infty(\R)\cap C^\infty(\R) \) which satisfy
\[
K(\xi)e_{k\xi}=(k^2+4m^2)e_{k\xi}\,,\qquad \hbox{and}\quad e_{k\xi}(x)\sim e^{ikx}\quad (x\to+\infty)\,.\] 
Explicit formulae for these eigenfunctions,
and the corresponding spectral resolution for \(K(\xi) \), are derived and
displayed in the appendix.
The field \(\upphi\) in this solitonic representation is built from the
the pair \((Q,\upphiperp)\) consisting of the
position operator  \(Q\) for the soliton, and
a transverse field operator \(\upphiperp\) to handle the bosons which arise from quantization of the  transverse degrees of freedom:
\[\upphi(x)\,=\,-{\sqrt{\scm_{cl}}}
   {\tteo}({}x{})\,Q\,
  +\upphiperp(x)\]
see \eqref{cfsgext} for the detailed formulae for this, and also \(\uppi(x)\).
The quadratic part of the Hamiltonian takes the form
\begin{equation}\label{simh}
  \triplecolon H^{sol}_0\triplecolon\,\define\,
  \frac{P^2}{2\scm_{cl}}
  +{h}_d+{{\h}}
\end{equation}
as a self-adjoint operator acting on 
\begin{equation}\label{aos}
  {\mfrH}(\theta)\,=\,
  L^2(\R,dQ)\otimes\mfrF=L^2(\R,dQ;\mfrF)\quad\hbox{where}\quad
  \mfrF=L^2(\gamma_d)\otimes{\fock}\,,
\end{equation}
with \(P=-i\frac{\partial}{\partial Q}+i{\scm_{cl}}\sqrt{\theta}Q\) in the Gaussian case \(L^2(\gamma_\theta(dQ))\), and
the special case \(\theta=0\) is included with the understanding that \(\gamma_\theta(dQ)|_{\theta=0}=dQ\); the dependence on
\(\theta\) will be suppressed when \(\theta=0\).
The triple colons indicate normal ordering with respect
to the solitonic representation, see Remark \ref{triplecolon}.
The Hilbert space \(\mfrF\) is the {\em transverse} Fock space generated by the modes
described in the two
items following \eqref{qh}; this is formulated precisely in Theorem \ref{first},
and explained in detail and explicitly in \S\ref{qsol}. We take as {\em transverse vacuum} the vector
\begin{equation}\Omega'\define\id_{\smr}\otimes\Omega_0\in\mfrF\label{tfv}\end{equation}
where
\(\id_{\smr}\) just means the function identically equal to one in \(L^2(\gamma_d) \), which will be omitted except when required for emphasis.

This representation
is unitarily equivalent to the shifted vacuum representation - there exists a unitary intertwining map \(\bbS^\theta:\mfrH(\theta)\to{L^2(\mu_0)}\)
such that for all Schwartz test functions \(f \)
\[
(\bbS^\theta)^{-1}\circ \Exp[i\phisch(f)] \circ\bbS^\theta\,=\,\Exp[i\upphi(f)]\qquad
(\bbS^\theta)^{-1}\circ \Exp[i\pisch(f)] \circ\bbS^\theta\,=\,\Exp[i\uppi(f)]\,,
\]
and it is proved in Section \ref{regno} that
 \begin{equation}\label{suml}
(\bbS^\theta)^{-1}\circ \doublecolon \pmb{H}^{sol}_{0}\doublecolon \circ\bbS^\theta\,=\,\triplecolon H^{sol}_{0}\triplecolon
\,+\,\Delta\M_{scl}\,,
   \end{equation}
where \(\doublecolon \pmb{H}^{sol}_{0}\doublecolon \) means the normal ordered 
second quantization of \eqref{qh} with respect to the shifted vacuum 
representation, and \(\Delta\M_{scl} \)
  is the Dashen-Hasslacher-Neveu semiclassical
  mass shift
\begin{equation}\label{cba}
  \Delta\M_{scl}
  =\,\frac{m}{2\sqrt{3}}-\frac{3m}{\pi}
  \end{equation}
 which was computed by another method in \cite{Dashen}.
To explain our results in slightly more detail, consider that the definition and construction of
the quantum theory requires three preparatory actions:
\begin{enumerate}
\item Choice of solution of the Heisenberg commutation relation in
  both the vacuum sector
  and the solitonic sector; in fact in the latter case two different
  solutions are useful as discussed above.
\item Ultra-violet regularization of the fields in both sectors, carried out in a {\em consistent} way (see \S\ref{fieldreg}).
  Introduction of a spatial (or infrared) cut-off \({\beefb}\) into the interaction terms of the Hamiltonian, which
  satisfies certain technical conditions given below.
\item Subtraction of the {\em same} counter-terms (see \S\ref{c.t.}) for
  both vacuum sector and solitonic sector Hamiltonians. (The counter-terms used
  correspond to normal ordering for the vacuum sector.)
\end{enumerate}
The regularization is achieved in all representations
via convolution with a smooth function,
the ultra-violet cut-off being determined by
a positive real number \(\kappa\); as \(\kappa\to+\infty\) the cut-off is removed.
Regarding the third point, the counter-terms are chosen by normal
ordering using the vacuum representation \eqref{cf1}-\eqref{cf2}).
Following this through in
the vacuum sector, taking the formal Hamiltonian \eqref{ham-vac} as
starting point, leads to a normal ordered and regularized Hamiltonian
\(\doublecolon \pmb{H}^{vac}_{g,{\beefb},\kappa}\doublecolon \) acting
on \(L^2(\mu_0)\), or equally well the corresponding operator acting on Fock space
(indicated without the bold face), see \eqref{trh}.
In the solitonic sector we take \eqref{ham-sol} as the starting point.
(As we've seen above, there is actually some additional freedom, in that
the expansion in \eqref{ham-sol} can equally well be carried out with the
soliton located at an arbitrary \(\xi\in\R\). It is necessary to take advantage
of this to extend classical modulation theory to describe soliton motion in
dynamically nontrivial situations, see Theorem \ref{main} below.)
As mentioned previously, we use convolution
and subtract the {\em same} counter-terms as in the vacuum sector - this
corresponds to normal ordering in the solitonic sector using the 
shifted vacuum representation
\eqref{cfxi1}-\eqref{cfxi2}, see in particular \eqref{solint} and \eqref{quadsol} in \S\ref{c.t.}.
In the end this leads to the study of
a Hamiltonian
\[\doublecolon \pmb{H}^{sol}_{g,{\beefb},\kappa}\doublecolon
\,=
\,\int  \doublecolon \pmb{\mcH}^{sol}_{g,{\beefb},\kappa}
\doublecolon\,dx\,,
\qquad
\doublecolon\pmb{\mcH}^{sol}_{g,{\beefb},\kappa}\doublecolon\,=\,
\doublecolon \pmb{\mcH}^{sol}_{0,\kappa}\doublecolon
+{\beefb}\doublecolon {\mcH}_{I,g}^{sol}(\phisch_\kappa)\doublecolon\,.
\] 
(The double colon indicates normal
ordering with respect to the shifted vacuum representation
while the triple colon
is used for normal ordering in the solitonic representation.)
We study the Schr\"odinger evolution with initial data
\(\Psi_0\in{\fock}\).
In order to obtain the simple normal form \eqref{simh} for the quadratic part
of the Hamiltonian, and hence uncover the dynamics in the
limit \({g\downarrow 0}\), it is necessary to move to the representation
\eqref{cfsgext} on the Hilbert space \(\mfrH(\theta)\),
via the unitary
transformation \(\bbS^\theta:\mfrH(\theta)\to{L^2(\mu_0)}\) obtained in
Theorem \ref{mac}. Even in the absence of
an external field this has dynamical consequences, yielding a precise interpretation of the
Dashen-Hasslacher-Neveu semiclassical mass correction formula.
\begin{theorem}\label{first}
  In the limit \(\kappa\to+\infty\)
  the operators \(\doublecolon \pmb{H}^{sol}_{g,{\beefb},\kappa}\doublecolon \)
  determine a
  self-adjoint operator \(\doublecolon \pmb{H}^{sol}_{g,{\beefb}}\doublecolon \)
  on \({L^2(\mu_0)}\)
  which is bounded below and determines a strongly continuous one-parameter
  unitary group via the Stone theorem.
  Let \([-t_1(g),t_1(g)] \) be a time interval given for each \(g>0 \) which
  satisfies \(\lim_{{g\downarrow 0}}\sqrt{g}\,t_1(g)=0\), then for all initial
  values \(F=\bbS^\theta\hat F\in L^2(\mu_0)\)
  \begin{equation}\label{f2}
\lim_{{g\downarrow 0}}\,\sup_{|t|\leq t_1(g)}\,\Bigl\|\,e^{it\Delta\M_{scl}}\,
\Exp[-it\doublecolon \pmb{H}^{sol}_{g,\beefb}\doublecolon ]\,F
\,-\,\bbS^\theta\Exp[-it \triplecolon H^{sol}_0\triplecolon]\,\hat F\Bigr\|\,=\,0\,,
  \end{equation}
Here  \(\triplecolon H^{sol}_0\triplecolon\) is defined in \eqref{simh} while
  \(\Delta\M_{scl}\) is as above in \eqref{suml} and \eqref{cba}.
\end{theorem}
\noindent
This is proved in Section \ref{5p1}.
\begin{remark}\label{onehalf}
  To see the significance of the condition \(\sqrt{g}t_1(g)=o(1)\) on the time-scale, one can compare with the corresponding statement
  in the vacuum case in which the condition is \(gt_0(g)=o(1)\), see Theorem \ref{1st}. The reason for the difference is the presence of
  the zero mode corresponding to translation of the kink, which spreads in the usual manner of quantum dispersion and the presence of
  \(Q^2\) terms quadratic in the position fluctuation operator force this restriction on \(t_1\). In order to compute the
  semiclassical motion on time-scales of \(O(1/\sqrt{g})\) it will be necessary to treat nonperturbatively terms of this type, as will be seen
  in the statement of the main theorem below and in its proof in \S\ref{5p1}.
  \end{remark}
\begin{remark}\label{clarity}
  It should be emphasized that while the quadratic
  Hamiltonian \({\h} \)
  describing bosons in the soliton background looks the same as that for free
  bosons, the interactions in physical space are affected by the presence of the
  soliton. This shows up, for example, in the formula \(
  \int{({4\pi\omega^{}_k})^{-\frac{1}{2}}}a_{k}^\dagger \tilde U(k;\xi)dk\) for the
  operator creating a boson in state determined by a physical space Schwartz function \(U \)
  (in the continuous spectral subspace); see \S\ref{qsol} for this and
  related formulae.
  The notation \(\tilde U\) indicates the
  {\em distorted Fourier transform} \eqref{dft}, which appears in place of the
  Fourier transform in the free case. Now \(\tilde U \) is
  constructed from the scattering analysis of the linearized
  operator \(K \), see
  \cite{MR2108588} and the Appendix, and
  depends on the background
  soliton. Insertion of such dependencies into
  the Hamiltonian means that the actual physical space dynamics of
  the bosons does depend on the presence of the soliton, even without including effects from the interaction part of the Hamiltonian.
  \end{remark}
\subsection{Modulation Theory in the presence of an external electric field}\label{modulth}
Now turning on the electric field (nonzero \(\lambda\)), the theory described by the action
\(S_\lambda\) is put into Hamiltonian form in \S\ref{5p2}, leading to
the classical Hamiltonian
\[
  H^{\lambda}(\pi,\phi)\,=\,\int_{\smr}\,\mcH^{\lambda}(\pi,\phi)\,dx\,,\quad
  \mcH^{\lambda}(\pi,\phi)\,=\,\frac{1}{2}\,\bigl(
\pi^2\,+\,\partial_x\phi^2\bigr)\,+\,\cU(\phi)
\,-\,\lambda {\elec}\,\phi\,,
\]
determining the interaction of the soliton with the electric field \({\elec}(t,x)\) \eqref{defelec}, which will be assumed to be
a \(C^2\) function of \(t\) into \(\cs(\R)\) for simplicity.
Under quantization this leads to an additional {\em time-dependent}
term arising from the external field, namely
\(\,-\,\lambda\,\int {\elec}(t,x)\,\bigl(\Phi_S+\varphi(x)\bigr)\,dx\,,
\) so that we work with the Hamiltonian with spatially cut-off interaction, namely
\begin{equation}\label{totalham1}
  \doublecolon \pmb{H}^{sol,{\elec}}_{g,{\beefb}}(t)\doublecolon\,=\,\doublecolon \pmb{H}^{sol}_0\doublecolon-\lambda\phisch({\elec}(t))\,
  +\doublecolon H^{sol}_{I,g,{\beefb}}(\phisch)\doublecolon
\end{equation}
where \(\phisch({\elec}(t))=\langle\phisch(\cdot),\elec(t,\cdot)\rangle\,,\) is the distributional pairing, and
the interaction is obtained by normal ordering
\(H^{sol}_{I,g,{\beefb}}\) from \eqref{defhs}.
The quadratic Hamiltonian \(\doublecolon \pmb{H}^{sol}_0\doublecolon\) is defined precisely in Theorem \ref{sadj2},
together with its Fock space version (which is written without bold face).
For now we
work on the Hilbert space \(L^2(\mu_0) \), using the Schr\"odinger representation of the fields; this is
related to the Fock space formulation by the unitary equivalence \(\I\) from Proposition \ref{sch} to
\eqref{cfxi1}-\eqref{cfxi2} to obtain the representation
\[
\Phi(x{})=\Phi_S(x)+\phisch(x{})\,,\qquad
\Pi(x{})=\pisch(x{})\,,
\]
for the fields, where \((\phisch,\pisch)\) are as in \eqref{seq}.
Substitution of these into \eqref{totalham1} yields the relevant Hamiltonian, which is actually time-dependent due to
the middle term in \eqref{totalham1}.
In order to prove that this Hamiltonian generates an evolution, we use the fact that at each frozen time \(t\)
it
is self-adjoint on \(L^2(\mu_0) \) and apply Kato's method from \cite{MR0407477} to produce a family of evolution operators
\(\{{{\bevot}}(t,s)\}\), see
Theorem \ref{extexist2}. We now consider how to extract from these operators the dynamics of the soliton. In so doing it
is helpful to keep in mind both classical modulation theory for solitary waves, e.g. from \cite{MR1186038}-\cite{MR0783974},
and the following solvable quantum mechanics problem.
\begin{example}
  In the following \(E_j\in C(\R)\) for \(j=0,1\) and \(v_{tr}(t,y)=E_1(t)y\). Consider the initial value problem for
  a wave function \(\psi=\psi(t,x,y)\in \C\) for \((t,x,y)\in\R\times\R^2\)
  \[
  i\partial_t\psi=-\frac{1}{2\scm}\partial_x^2\psi-\frac{1}{2}\partial_y^2\psi+\frac{1}{2}\omega^2 y^2\psi
  +(E_0(t)x+v_{tr}(y))\psi\,,\qquad\psi(0,x,y)=e^{i\eta_0x}\varphi(x-\xi_0)\mbchi_0(\sqrt{2\omega}y)\,.
  \]
(We use notation following \eqref{hgwp} for the eigenfunctions of the quantum oscillator.)
  This Hamiltonian is separable: for present purposes it is helpful to interpret \(\psi\) as
  the wave function describing a quantum particle
  moving along the \(x\)-axis, with additional transverse degrees of freedom along the \(y\)-axis. It moves
  under the influence of a time-dependent
electric field \(E_0\), while undergoing transverse oscillations in the potential \(\omega^2 y^2/2+v_{tr}(t,y)\).
Forming the ansatz
\begin{equation}\label{anz} \psi(t,x,y)=e^{i\eta(t)x+i\theta(t)}\phi(t,x-\xi(t),y)\end{equation}
we find that if \(t\mapsto (\xi(t),\eta(t),\theta(t))\) are chosen so that
\[
\dot\eta=-E_0\,,\,\dot\xi=\eta/\scm\quad\hbox{and}\quad\dot\theta=-\frac{1}{2\scm}\eta^2
\]
then
\[
i\partial_t\phi=-\frac{1}{2\scm}\partial_x^2\phi-\frac{1}{2}\partial_y^2\phi+(\frac{1}{2}\omega^2 y^2
  +v_{tr}(y))\phi\,,
  \]
  (so the effect of the electric field in the \(x\) direction has been incorporated into the ``classical'' quantities \(\xi,\eta\) which
  evolve as a classical particle of unit charge and mass \(\scm\) in an electric field \(E_0\).)
Now specialize to \(v_{tr}(t,y)=E_1(t)y\) with the transverse degrees of freedom initially in the ground state \(\mbchi_0(\sqrt{2\omega}y)\) with respect to the transverse oscillator potential \(\omega^2 y^2/2\), so that the initial condition is
\(\psi(0,x,y)=e^{i\eta_0x}\phi(0,x-\xi_0)\mbchi_0(\sqrt{2\omega}y)\,.\)
This initial value problem has solution
  \begin{equation}\label{mpsol}
\psi(t,x,y)=e^{i\eta(t)x+i\Theta(t)}\,\phi(t,x-\xi(t))D_{c(t)}\mbchi_0(\sqrt{2\omega}y)\,.
\end{equation}
where \(\phi(t,Q)=\Exp[\frac{it}{2\scm}\partial_Q^2]\phi(0,Q)\) solves the {\em free} Schr\"odinger equation
\(i\partial_t\phi+\frac{1}{2\scm}\partial_Q^2\phi=0\).
Given \(c_1+ic_2\in\C\),  the (unitary) displacement operator \(D_c:L^2(dy)\to L^2(dy)\) acting on the transverse
Hilbert space is defined by
\begin{equation}\label{defdisc}
D_c\chi(y)=\exp\Bigl[i\sqrt{2\omega}c_2(y-\sqrt{\frac{2}{\omega}}c_1)\Bigr]\chi\Bigl(y-\sqrt{\frac{2}{\omega}}c_1\Bigr)
\end{equation}
and the parameters evolve according to
\(\dot\eta=-E_0\,,\,\dot\xi=\eta/\scm\,,\,\dot c_1=\omega c_2\,,\,\dot c_2=-\omega c_1-E_1/\sqrt{2\omega}\) and
\(\dot\Theta=2\dot c_1c_2-\sqrt{2/\omega}E_1c_1-\eta^2/(2\scm)-\omega(c_2^2+c_1^2)/2-\omega/2\).
\end{example}
In this example, the multiplication operator \(x=\xi+Q\) describes the position along the \(x\)-axis of the particle which is also undergoing transverse oscillations in the \(y\) direction. We aim to give such a description of the soliton in analogy to this: as a quantum particle in the semiclassical limit,
in which there is an underlying classical motion described by position/momentum parameters \(\xi,\eta\),
quantum fluctuations \(Q\) in the particle position around \(\xi\),
and with a displacement operator like \(D_c\) to handle transvere fluctuations, as in \eqref{mpsol}, which however now
span the infinite dimensional transverse Fock space \(\mfrF\) of
\eqref{aos}.
The new features to be incorporated
in transferring this framework to the soliton include:
{
\renewcommand{\theenumi}{({Qu\arabic{enumi}})}
\begin{enumerate}[label=(Qu\,\arabic*), wide=10pt, leftmargin=*]

\item how do we define, starting from the quantum field, a
  quantum position fluctuation operator for the soliton to take the role of \(Q\)?
\item what operator corresponds to the shift \(x\to x-\xi(t)\) in \eqref{anz} describing the ``classical'' component of the soliton's motion?
\item in order to reveal the physics
  it is necessary to transform into the Hilbert space \(\mfrH(\theta)=L^2(\gamma_\theta(dQ))\otimes\mfrF\) via operators
which take into account the location of the soliton: how are these to be defined?
\item the transverse degrees of freedom now span an infinite dimensional space \(\mfrF\), the
  transverse Fock space defined in \eqref{aos}: how does the displacement operator \(D_c\) in \eqref{defdisc}
  generalize as an operator on \(\mfrF\)?
\end{enumerate}
}

If we think in terms of the modulation theory of solitary waves in classical field theory,
the expectation is indeed that, just as in the preceding example, the electric field will induce an evolution
\(t\mapsto (\xi(t),\eta(t))\) of the soliton parameters which will differ from Newton's equation \eqref{newt} by an additional force determined (to highest order) by the electric field, projected along the
translational zero mode of the kink \({\tteox}\).
Quantum mechanically one would anticipate the introduction of quantum fluctuations \(Q\),
so that \(\xi+gQ\) is the overall position operator for the soliton. (The semiclassical scaling factor \(g\) is
motivated in the discussion leading to \eqref{qmod}). Properly speaking, the operator \(Q\)
is defined by the representations \eqref{cfsgextxi} of the Heisenberg relations, according to which
it is appropriate to identify \(Q\) with \(-\scm_{cl}^{-1/2}\phisch({\tteox})\), and this answers
Qu 1 above. Accepting this, to answer Qu 2 we 
work in analogy to \eqref{anz}: in terms of the standard quantum mechanical Weyl operators
on \(L^2(dx;L^2(dy))\), namely,
\[
V(\epsilon)=\Exp[i\epsilon x]\qquad\hbox{and}\qquad U(\epsilon)=\Exp[i\epsilon(-i\partial_x)]\,,
\]
the formula \eqref{anz} can be written
\[
\psi=e^{i\theta}V(\eta)U(-\xi)\phi\,=e^{i\theta}\Exp[i\eta x]U(-\xi)\phi
\]
thus to transport this to our problem
we are led to the introduction of unitary quantum modulation operators on \(L^2(\mu_0)\)
\begin{equation}\label{defdis}
\Dis(t)=\Dis(t;g)=e^{i\Theta_0}\Exp\bigl[{{-ig\frac{\eta}{\sqrt{\scm_{cl}}}\phisch({\tteox})}}\bigr]\,\mbU(\delta_{\xi}\Phi_S)\biggr|_{(\xi,\eta)=(\xi(t),\eta(t))}\qquad\hbox{with}\quad \delta_\xi\Phi_S=\Phi_S-\Phi_S(\cdot-\xi)\in\cs(\R)\,,
\end{equation}
and where  \(\mbU\) (resp.  \(\U\))
are the Weyl field displacement operators which act on \(L^2(\mu_0)\) (resp. Fock space \(\fock\))
as in \eqref{displace}, and the phase \(t\mapsto\Theta_0(t)=\int_0^t\dot\Theta_0\in\R\) will be chosen in the course of the proof to cancel various phases which arise in the construction. The
operators \(\Dis(t)\) answer Qu 2. To explain more fully: the theorems of Cameron-Martin and Shale
imply that the {\em soliton sector is  not unitarily related to the vacuum sector}, but
the various Hilbert representation spaces one might consider by quantizing around
a soliton \(\Phi_S(\cdot-\xi)\) {\em for various} \(\xi\) {\em are all unitarily related}, or in measure theoretic terms can be described as \(L^2\) spaces formed from equivalent (i.e., mutually absolutely continuous) measures.
This is because
\(\delta_\xi\Phi_S=\Phi_S-\Phi_S(\cdot-\xi)\) is a Schwartz
  function, and as such lies in the Cameron-Martin space. As a consequence, the
  displacement of the field involved in comparing the representations
  \eqref{cfsgextxi} for arbitrary \(\xi\) and for a specific
  value, say \(\xi=0\)
  or indeed any other value, induces an equivalent measure and
  the transformation is unitarily implementable on Fock space
  via the Weyl operator
  \(\U(\delta_\xi\Phi_S)\), which acts on Fock space as
  \begin{equation}\label{displace}
\U(\delta_\xi\Phi_S)\circ\varphi\circ \U(\delta_\xi\Phi_S)^*\,=\,\varphi+\delta_\xi\Phi_S\,,
  \end{equation}
  roughly speaking moving from a representation where the fields are relative to a soliton located at arbitrary \(\xi\)
  to a representation relative to a soliton
  located at \(\xi=0\), as well as shifting the momentum.
  In the Schr\"odinger representation the
  corresponding operator, \(\mbU(\delta_\xi\Phi_S)\), acts in the corresponding way, and is constructed from
the square root of
  the Radon-Nikodym derivative, see \eqref{rndis} and the equation following.

The operators which deal with Qu 3
\begin{equation}\label{wdw}
  \bbS^\theta(\xi):\mfrH(\theta)\to L^2(\mu_0)\,,\qquad\xi\in\R
  \end{equation}
are introduced in \eqref{defmbs}, using work in \S\ref{sss}, and are constituted from
generalized second quantization operators, constructed using the spectral decomposition about the soliton centred at
\(\xi\),
together with a Radon-Nikodym factor. 
The final query is dealt with straightforwardly, but requires some notation 
to fully describe all the transverse degrees of freedom (which were introduced in the discussion surrounding
\eqref{qh}-\eqref{tfv}). In analogy to \(D_c\),
a unitary displacement operator \(\D_{\cee,\eff}:\mfrF\to\mfrF\) is introduced which acts
on the field as \(\upphi\mapsto
 \D_{\cee,\eff}\circ\upphi\circ \D_{\cee,\eff}^*=\upphi-
  \upphi_{scl}\,,
  \) and also on the transverse Fock vacuum as
  \(\Omega'\mapsto \Omega_{\cee,\eff}=\D_{\cee,\eff}\Omega'\), see \eqref{disvac}.
  Here,
\begin{itemize}\item
  \(\cee=\cee_1+i\cee_2\) determines the discrete mode with centre
  \(\sqrt{\frac{2}{\omega_d}}\cee_1\) moving with \(\dot\cee_1=\omega_d \cee_2\),
  and \item \(k\mapsto \eff(k)\in\C\) determines
  in a similar way the dynamics of the oscillatory modes;
\end{itemize}
  see the discussion in \S\ref{limdyn} for detailed formulae.
  Explicitly, in terms of the eigenfunctions introduced prior to \eqref{simh} and in Appendix \ref{eigen},
the  field displacement is given by the following ``{\em semiclassical field},'' which is given at fixed time by
    \begin{align}\label{defiscl}
\upphi_{scl}(x;\xi,\cee,\eff)    =\,
  \frac{2\cee_1}{\sqrt{2\omega^{}_d}}\,
{\tteonex}(x) 
+\frac{1}{\sqrt{2\pi}}\,\int\,\frac{1}{\sqrt{2\omega^{}_k}}\,
\bigl(\eff(k)e_{k\xi}(x{})+\overline{\eff(k)} {e_{-k\xi}(x)}\bigr)\,dk\,,
    \end{align}
    representing an averaged effect of the motion and electric field on the transverse degrees of freedom.
    \begin{remark}\label{reality}
      The semiclassical field depends only on the real parts \(\cee_1\) and \(\eff_1\) of the transverse degree of
      freedom coordinates. Here, in the same way as under the Fourier transform, the complex conjugate of \(\eff\) is
      \(\eff^\flat(k)=\overline{\eff(-k)}\), so the real part is
      \(\Re\eff(k)=\eff_1(k)=(\eff(k)+\overline{\eff(-k)})/2\) and the imaginary part is
      \(\Im\eff(k)=\eff_2(k)=(\eff(k)-\overline{\eff(-k)})/2i\). Observe also that by unitary properties of the
      eigenfunction expansion (see \S\ref{eigen} and the discussion of the distorted Fourier transform \eqref{dft})
      there holds
      \begin{equation}\label{bdscl}
\|\upphi_{scl}\|_{L^2}\leq const.\,\bigl(|\cee_1|+\|\eff_1/\omega_{\bullet}\|_{L^2}\bigr)\leq const.\,\bigl(|\cee_1|+\|\eff_1\|_{L^2}\bigr)\,.
      \end{equation}
      \end{remark}
We will establish that the soliton parameters evolve according to
equations of the form
\begin{equation}\label{newtpert}\dot\xi=g^2{\scm_{cl}}^{-1}\eta+gV_{-1}(\xi,\dot\xi,\cee,\eff)\,,\quad \hbox{and} \quad
\dot\eta=-\frac{1}{g}\sqrt{\scm_{cl}}\lambda({\elec},{\tteox})_{L^2}+
\frac{1}{g}V_{1}(\xi,\dot\xi,\cee,\eff)
\,,\end{equation}
where
\begin{align}\label{defv1xi}
  &{{V_{-1}}}(\xi,\dot\xi,\cee,\eff)\define\dot\xi \frac{1}{\sqrt{\scm_{cl}}}\bigl(\upphi_{scl}, {\tteox}'\bigr)_{L^2}\qquad\hbox{and}\\
  &{V_1}(\xi,\dot\xi,\cee,\eff)\define\dot\xi \scm_{cl}^{1/2}
\Bigl(  \cee_2\sqrt{2\omega_d}\bigl(\tteox,\partial_\xi {\tteonex}\bigr)_{L^2}
    +\bigl(\tteox,\partial_\xi\mcF_{u_\xi}^{-1}\sqrt{2\omega_\bullet}\eff_2\bigr)_{L^2}\,
    \Bigr)\,.\label{defv1eta}
\end{align}
These equations are derived in \S\ref{repcalc}.
The additional perturbative terms
describe the interaction of the soliton with the transverse oscillatory modes (which determine the mean field
displacement \(\upphi_{scl}\) in the transverse Fock space \(\mfrF\)); the
corresponding coordinates
  \( (\cee(t),\eff(t,\cdot))\)
evolve according to
\begin{align}\begin{split}
  i\dot \cee-\omega_d \cee=\beed&\define\frac{\lambda}{\sqrt{2\omega_d}}\,\int {\tteonex}(x){\elec}^{eff}(t,x)\,dx\,,\qquad\qquad\hbox{and}\\
  i\dot \eff-\omega^{}_k \eff={\beek}&\define\frac{\lambda}{\sqrt{2\pi}}\,\int\,\frac{1}{\sqrt{2\omega_k}}\,e_{-k\xi}(x)\,{\elec}^{eff}(t,x)\,dx\,
  \,,
  \end{split}
  \label{defeeze1}\end{align}
where the transverse modes are acted on by an effective electric field given by
\begin{equation}\label{defeff}{\elec}^{eff}={\elec}-\frac{g\dot\xi\eta}{\lambda\sqrt{\scm_{cl}}}{{\tteox}'}
          \,.\end{equation}
          
So the picture is of the soliton behaving as a quantum particle with fluctuations about the classical evolution \eqref{newtpert} determined by the
Schr\"odinger equation \eqref{opsch}, while the transverse modes undergo an oscillatory motion with the displacement coordinates \(\cee(t),\eff(t,k)\)
oscillating around their slowly evolving {\em real} mean values
\begin{equation}\label{meanfieldtrans}
\cee_0(t)=-{\beed}(t)/\omega_d\qquad \eff_0(t,k)=-{\beek}(t)/\omega_k\,.
\end{equation}
This oscillatory character of the transverse dynamics, stated precisely in \eqref{asex}-\eqref{imcloser},  turns out to be important in proving Theorem \ref{main}. 
The reader will find some motivation for this picture in \S\ref{heuristics}, but of course the proper justification for these considerations lies in the analysis of solutions of the
system \eqref{newtpert}-\eqref{defeeze1} in \S\ref{limdyn} and in the statement and proof of Theorem \ref{main} which follows.

          The main theorem compares this limiting dynamics with the full evolution under the following hypotheses:
{
\renewcommand{\theenumi}{({H\arabic{enumi}})}
\begin{enumerate}[label=(H\,\arabic*), wide=10pt, leftmargin=*]
\item The electric field \(t\mapsto{\elec}(t,\cdot)\in\cs(\R)\) is twice
  continuously differentiable into Schwartz space, with
  \[
  \int_{\smr}\,\Bigl(|\elec(t,x)|^2+\frac{1}{{g}}\,|\partial_t\elec(t,x)|^2
  +\frac{1}{g^2}\,|\partial_t^2\elec(t,x)|^2\,\Bigr)dx\leq M_1<\infty
  \]
\item The infrared cut-off function \({\beefb}:\R\to[0,1]\)
  is of the form \({\beefb}(x)={\beefb}_0(x/R_g)\) where \({\beefb}_0\) is an even Schwartz function,
  nonincreasing on \([0,\infty)\)  and equal to one on \([-1,1]\) which verifies \eqref{lowerbc}.
  In addition \(mR_g=\ln (m/g)^{N}\) with \(N\) is sufficiently large ( \(N>7\)),
  and restrict attention to \(g\) sufficiently small that \(R_g<1\), so that \eqref{lowerbc} holds for
  all \(g\).
\end{enumerate}
}
The following theorem, which is the main result of the paper,
refers to families of solutions of the quantum theory in which there is a soliton moving
along the trajectory \(t\mapsto\xi(t)\) from \eqref{newtpert}, with quantum fluctuations described by a wave function \(\chi\) which is initially Gaussian,
and where the transverse degrees of freedom traverse adiabatically a succession of
vacua \(\Omega_{\cee,\eff}=\D_{\cee,\eff}\Omega'\) displaced according to \((\cee,\eff)\) from
\eqref{defeeze1}.
(The solutions depend on the parameter \(g\to 0+\): this
dependence on \(g\) is indicated explicitly in the upcoming statement,
but will be left implicit in the proof to avoid cluttering up the formulae.) A heuristic discussion
and derivation of some of the approximate equations is presented after the proof in \S\ref{heuristics}.
\begin{theorem}\la{main} Under the hypotheses (H1)-(H2) and for sufficiently small \(g>0\), assume there to be 
  given initial values
\begin{itemize}\item
  \((\xi^g(0),\eta^g(0))=(\tilde\xi_0,g^{-\frac{3}{2}}\tilde\eta_0)\in\R^2\), with \((\tilde\xi_0,\tilde\eta_0)\) independent of \(g\), and
\item
  \((\cee^g(0),\eff^g(0,\cdot))\in \C\times \cs(\R;\C)\) with \(|\cee^g(0)-\cav^g(0)|+\|\omega_k(\eff^g(0,k)-\fav^g(0,k))\|_{L^2(dk)}
  =O(\sqrt{g})\).
  \end{itemize}
  Then there exist \(\tau_2>0\), independent of \(g\),
  and continuously differentiable functions 
  \(t\mapsto (\xi^g(t),\eta^g(t))\in\R^2\)
and  \( t\mapsto (\cee^g(t),\eff^g(t,\cdot))\in \C\times \cs(\R;\C)\)     
which satisfy \eqref{newtpert} and \eqref{defeeze1} on a time interval \(0\leq t\leq \tau_2/\sqrt{g}\),
on which interval they also obey the bounds in Corollary \ref{cproof} and
\begin{equation}\label{moav}
 |\cee^g(t)-\cav^g(t)|+\|\omega_k(\eff^g(t,k)-\fav^g(t,k))\|_{L^2(dk)}
  =O(\sqrt{g})\,,
\end{equation}
with \(\cav^g,\fav^g\)  as in \eqref{meanfieldtrans}.

Furthermore, there exists a Kato evolution operator \(\bevot^g(t,s)\) on \(L^2(\mu_0)\) which is
   generated by the time-dependent
   Hamiltonian \eqref{totalham1} and has properties listed in Theorem \ref{extexist2}. 
   We consider families of initial data in \(L^2(\mu_0)\) given by
   \[
\Psi^g(0)=\Delta(0)
\,\bbS^\theta(\xi^g(0))\chi^g(0,\cdot)\otimes\Omega_{\cee^g(0),\eff^g(0)}
   \]
where \(\chi^g(0,Q)\) is Gaussian with variance \(\sigma^g(0)^2=a g^{-\frac{1}{2}}\,,a>0\).
There exist continuously differentiable \(t\mapsto\Theta_0(t;g)\) in \eqref{defdis} such that
\begin{equation}\label{f2e}
\lim_{{g\downarrow 0}}\,\sup_{0\leq t\leq\frac{\tau_2}{\sqrt{g}}}\,\Bigl\|\,
\bevot^g(t,0)\Psi^g(0)
-\Delta(t;g)\bbS^\theta(\xi^g(t))\,\,\chi^g(t,\cdot)\otimes\Omega_{\cee^g(t),\eff^g(t)}\Bigr\|=0\,,
\end{equation}
where 
\begin{equation}\label{opsch}
i\frac{\partial \chi^g}{\partial t}(t,Q)=\bigl(\frac{P^2}{2\scm_{cl}}+V_2(t) Q^2\bigr)\chi^g(t,Q)\,,\qquad\Bigl(
\hbox{with } V_2 \hbox{ as defined in \eqref{opp}}\Bigr)\,.
\end{equation}
More generally the same conclusion holds for families of initial data
\(\chi^g(0,Q)\) for which the solutions satisfy
bounds \[\|Q^rP^l\chi^g(t,Q)\|_{L^2}^2\leq c_{r,l}g^{-(r-l)/2}\]
for \(l\in\{0,1\}\) and \(r\in\{0,1,\dots 6\}\) on the time interval
\(0\leq t\leq\frac{\tau_2}{\sqrt{g}}\).
\end{theorem}
\begin{proof}
As commented already, {\em from now on we drop the family index \({}^g\) to avoid clutter.}
  The theorem asserts a limiting relationship between the solution of the quantum field theory generated by \eqref{totalham} in \(L^2(\mu_0)\), and the limiting dynamics on the space \(\mfrH(\theta)\), consisting of the one-particle
  evolution \eqref{opsch} and transverse quantum fluctuations around a mean determined by \eqref{defeeze1}. This limiting relationship is mediated by the one-parameter unitary transformations \(\Delta(t)\)
    and the change of representation \eqref{wdw}.
  Theorem \ref{extexist2} ensures the existence of the Kato evolution operator
generated by \eqref{totalham},
with the continuity and differentiability properties listed there. As just said, in order to prove
the approximation theorem in the limit \(g\downarrow 0\) we
  will apply unitary transformations to put \eqref{totalham1} into a form
  where it can be successfully compared with the effective Hamiltonian, namely
\begin{align}\label{effh0}\begin{split}
  {\rm H}^{eff}_0 (t)\,\define\,& h_{\mbox{\tiny 1P}}+{h}_{{\cav},{\fav}}\qquad\hbox{where}\\
  &h_{\mbox{\tiny 1P}}
=\frac{P^2}{2\scm_{cl}}+V_2 Q^2
  ,\qquad\hbox{and}\\
  &{h}_{{\cav},{\fav}}=\,{h}_d
+
{\beed}(a^{}_d+a_d^\dagger)+
{{\h}}+\int\Bigl({{\beekm}}^{}a^{}_k+{\beek}a_k^\dagger\Bigr)dk\,;
\end{split}\end{align}
or more accurately we want to compare the exact evolution operator \(\bevot\) with the
evolution operator \(\bevot_{scl}\) generated
by the effective Hamiltonian according to \eqref{defld}.
Here \((\cav,\fav)\) are as in \eqref{meanfieldtrans}, and because they determine the ground state of the
effective transverse Hamiltonian \({h}_{{\cav},{\fav}}\) defined in \eqref{effh0},
they are a convenient label for this Hamiltonian which will be used from now on.
The limiting dynamics are described in \S\ref{limdyn}, and an existence for the coupled
system \eqref{newtpert}-\eqref{defeeze1} which determines these dynamics is proved in
Theorem \ref{lse}. This produces a curve
\(t\mapsto (\xi,\eta,\cee,\eff)|_{t}\in\R\times\R\times\C\times\cs(\R)\) which determines
the transformations used to prove \eqref{f2e},
and it will become apparent that
the equations \eqref{newtpert}-\eqref{defeeze1}
are chosen precisely to ensure the error terms introduced in the process can be
controlled, leading to the proof of the theorem. The details of this are now given, based on
the results of calculations presented in \S\ref{repcalc}.

The first thing is to allow the soliton to move:
define, given a \(C^1\) function \(t\mapsto (\xi(t),\eta(t))\), unitary operators \(\{\Dis(t)\}_{t\in\smr}\)
as in \eqref{defdis},
and thence a modified solution operator
\begin{equation}\label{modsol}
\tilde\bevot(t,s)=\Dis(t)^*\bevot(t,s)\Dis(s)\,.
\end{equation}
The following formula summarizes the calculation of the derivative with respect to \(s\)
of this operator
    \begin{equation}\label{tottdi}
      \frac{d}{ds}\tilde\bevot(t,s)\,F=i\tilde\bevot(t,s)\Bigl[
  \doublecolon \pmb{H}^{sol}_{0\xi}\doublecolon+\doublecolon H^{sol}_{I,g,\xi,{\beefb}}(\phisch)\doublecolon
-\lambda\phisch({\elec}^{eff})+\frac{d}{ds}\Theta_0+\frac{d}{ds}\Theta_1
+\frac{\sqrt{\scm_{cl}}}{g}(\dot\xi-\frac{g^2\eta}{\scm_{cl}})\pisch( {\tteox})
-\frac{{g}\dot\eta}{\sqrt{\scm_{cl}}}\phisch( {\tteox})
+Err^0_{\mbox{\tiny IR}}
  \Bigr]\,F      \,,
  \end{equation}
valid for \(F\) in the generator's domain,
and where, see Theorem \ref{sadj2},
  \begin{equation}\label{arbloc}
  \doublecolon  {\pmb H}^{sol}_{0\xi}\doublecolon \,=\,\doublecolon  {\pmb H}^{vac}_{0}\doublecolon 
-\frac{1}{2}\doublecolon \int\,6m^2\,\ssmxi\,\phisch(x)^2\,dx\doublecolon \;,
\end{equation}
\begin{equation}\label{hintxi}H^{sol}_{I,g,\xi,\beefb}(\phisch)\,=\,
\int\,\Bigl[\,2mgb(x)\tmxi\phisch^3
\,+\,\frac{1}{2}b(x)
g^2\phisch^4\,\Bigr]\,dx\qquad\hbox{and}
\end{equation}
\begin{equation}
\label{pf1}i\frac{d}{ds}\Theta_1=+i\Big\langle\delta_\xi\Phi_S\,,\,\lambda{\elec}^{eff}(s)
  +g\frac{\dot\eta}{\sqrt{\scm_{cl}}} {\tteox}\Big\rangle
  +i\frac{g^2\eta^2}{2{\scm_{cl}}}
  \,.
  \end{equation}
  The final equation defines \(\Theta_1\) together with the initial condition \(\Theta_1(0)=0.\)
Next, as explained above we need to make use of the operator defined in
  \eqref{wdw} in order to make explicit and to control the dynamics on the Hilbert space
  \(\mfrH(\theta)\) in the solitonic representation \eqref{cfsgextxi}.
  This leads to the following two formulae which give the effective generator of the evolution in this representation,
   in which the
   position and momentum operators \(Q,P\) for the soliton are defined by
   \({\sqrt{\scm_{cl}}} Q=-\upphi( {\tteo})\) and \({P}=-{\sqrt{\scm_{cl}}}\uppi( {\tteo})\).
   We now transfer the Hamiltonian opertors to \(\mfrH(\theta)\)
   via the unitary change of representation \(\bbS_s\define \bbS^\theta(\xi(s))\) at time \(s\),
and making use of \eqref{chrepeq} generalized to allow for arbitrary location of the soliton in the obvious way:
\begin{align}\label{chrepeqxi}\begin{split}
    &(\bbS^{\theta }(\xi))^*\circ \Bigl(\doublecolon {\pmb H}^{sol}_{0\xi}\doublecolon -\lambda\phisch({\elec}^{eff})\Bigr)\circ \bbS^{\theta}(\xi)\,
=\,\frac{P^2}{2\scm_{cl}}
+{h}_{{\cav},{\fav}}\,+\,\Delta\M_{scl}\,=\,{\rm H}^{eff}_0-V_2Q^2
\,+\,\Delta\M_{scl}\,,\\
&(\bbS^{\theta}(\xi))^*\circ\doublecolon \pmb{H}^{sol}_{I,g,\xi,{\beefb}}(\phisch)\doublecolon\circ (\bbS^{\theta}(\xi))\,
=\,{\rm H}^{sol}_{I,g,\xi,{\beefb}}(Q,\upphiperp)\,;
   \end{split}\end{align}
  see \eqref{intixi}-\eqref{intixit} for explicit formulae. This leads to
   \begin{align}\label{dds2}\begin{split}
\Bigl(\frac{d}{ds}\tilde\bevot(t,s)\Bigr)\bbS_sF\,&=\,
\tilde\bevot(t,s)\bbS_s
\Bigl[i {\rm H}^{eff}_0\,+i{\hat{\rm H}}^{sol}_{I}(Q,\upphiperp)-iV_2Q^2+
i\frac{d}{ds}\sum_{j=0}^2\Theta_j+
-\frac{iP}{g}\Bigl(\dot\xi-\frac{g^2\eta}{\scm_{cl}}\Bigr)+ig\dot\eta Q+\eirup
\Bigr]\,F\,.\end{split}
\end{align}
  Here an upright font is used to indicate that the relevant operators have been transferred to the Hilbert space
  \(\mfrH(\theta)\), and 
  the interaction Hamiltonian, now written \({\hat{\rm H}}^{sol}_{I}(Q,\upphiperp)\),  has been slightly modified
  by transfer of the \(Q^3\)
  term into the infrared error
  \(\eirup\), see \eqref{ffir} for the exact formulae.
  The linear term \(-\lambda\phisch({\elec}^{eff}(s))\) in the Hamiltonian has been absorbed in the effective
  quadratic Hamiltonian \({\rm H}^{eff}_0\) in \eqref{effh0}, and the resulting dynamics is investigated in \S\ref{limdyn}.
The new phase contribution is \(\Theta_2(s)=s\Delta\M_{scl}\), arising from \eqref{intixi}.
  However, this is not quite the final
  transformation needed, because
  the time dependence of
  \(\bbS_s\) introduces an additional error term \({\rm Err}_{\mbox{\tiny TD}}\), given in \eqref{errtd},
  which acts as an additional
  apparent contribution to the Hamiltonian. We compute this in the case that \(F\) is the time-dependent function
  \begin{equation}\label{snf}
  s\mapsto F(s)=
  \chi(s,Q)\otimes\D_{\cee(s),\eff(s)}\,\Omega'\in\mfrH(\theta)\,,
  \end{equation}
  where \(\chi\) solves \eqref{opsch}.
  By \eqref{evo} we have
\(\frac{d}{ds}\chi(s,\cdot)\D_{\cee(s),\eff(s)}\,\Omega'=
\bigl(i\dot\Theta_3-i{\rm H}_0^{eff}\bigr)\chi(s,\cdot)\D_{\cee(s),\eff(s)}\,\Omega'
\,,\)
  so there is a cancellation of \({\rm H}_0^{eff}\), leading to
  \begin{equation}\label{tottd}
    \frac{d}{ds}\Bigl(\tilde\bevot(t,s)\bbS(s)\,\chi(s,\cdot)\D_{\cee(s),\eff(s)}\,\Omega'\Bigr)
    \,=\,i\tilde\bevot(t,s)\bbS(s)\,\Bigl(\,{\hat{\rm H}}^{sol}_{I}(Q,\upphiperp)-V_2Q^2
    +
\eirup
    +\eirtd+\frac{d}{ds}\sum_{j=0}^4\Theta_j\Bigr)\,\chi(s,\cdot)\D_{\cee(s),\eff(s)}\,\Omega'\,.
  \end{equation}
  The free phase \(\Theta_0\) is chosen to cancel the other phases, so that
  \(\frac{d}{ds}\sum_{j=0}^4\Theta_j=0\) and zero initially. More importantly,
 \(V_2\) has to be chosen to
 cancel the problematic terms alluded to in Remark \ref{onehalf}. Precisely
  \begin{equation}\label{opp}V_2=V_{2,1}+V_{2,2}+V_{2,3}\,,\end{equation}
  and we pair \(V_{2,1}\) with \({\hat{\rm H}}^{sol}_{I}\)
 to cancel out the
 {\em mean} of the \(Q^2\) term in the interaction, since this turns out to be nonperturbative - see \eqref{annav},
   while
   \(V_{2,2}\) and \(V_{2,3}\) are chosen in \eqref{defv22} and \eqref{defv23} to cancel
   similar terms which appear after integration by parts of \(\eirtd\)  in \eqref{revduh2},
as now explained.
  
  The modulation equations for \(\dot\xi,\dot\eta\) are chosen to cancel the
 {\em averages with respect to the transverse variables}
  of the terms which are linear in
  the kink quantum operators \(Q,P\). The crucial point is that this procedure leaves interaction terms
  which can be effectively bounded by averaging.
  Computations in \S\ref{repcalc}-\ref{controlf} give the following for the additional errors
\begin{equation}\label{cleaner}i\eirtd\chi\,\D_{\cee,\eff}\Omega'=\dot\xi\,
\D_{\cee,\eff}\Bigl(\chi\Xi^0 \Omega'+\,(Q\chi)\Xi^1 \Omega'+iP\chi\Xi^2 \Omega'\Bigr)
\end{equation}
  introduced
  by the time dependence of \(\bbS_s=\bbS^\theta(\xi(s))\). With the cancellations taken into account, the operators
  \(\Xi^j\) are generalized Wick polynomials given by
  \begin{align}\label{errtd2}\begin{split}&\Xi^0(\upphi(\cdot;\xi))=-\frac{1}{2}\,\triplecolon \bigl(\upphiperp,\partial_\xi K^\theta(\xi)^{\frac{1}{2}}\upphiperp\bigr)\triplecolon\,+\upphiperp\bigl(
K^\theta(\xi)^{\frac{1}{2}}\partial_{\xi}\upphi_{scl}
-
    ic_2\sqrt{2\omega_d}\tteonex'
+i\partial_\xi\mcF_{u_\xi}^{-1}(\sqrt{2\omega_\bullet}f_2)\bigr)
      \,,
      \\
&\Xi^1(\upphi(\cdot;\xi))=\scm_{cl}^{1/2}\upphiperp\bigl(K^\theta(\xi)^{\frac{1}{2}}{{\tteox}'}\bigr)\;
      \quad\hbox{and}\quad\Xi^2(\upphi(\cdot;\xi))=\scm_{cl}^{-1/2}\upphiperp({{\tteox}'})\,.\end{split}\end{align}
  (The dependence on the representation and \(\xi\) will be suppressed when confusion is unlikely.) 
  In particular the \(\Xi^j\Omega'\in\Omega^{',\perp}\) so that these error terms takes values in the subspace of
  the transverse Fock space which is {orthogonal to the transverse vacuum} \(\Omega'\), which allows an integration by parts
  to prove the final bounds.
  These bounds are obtained by  combining \(\eirtd\) with the interaction Hamiltonian \({\hat{\rm H}}^{sol}_{I}(Q,\upphiperp)\) and the
  infrared error term \(\eirup\) and then using the Duhamel formula
  \begin{align}\notag
  \bigl({\tilde\bevot}(t,0)-\bevot_{scl}(t,0)\bigr)\bbS_0\psi(0,Q)\Omega_{\cee(0),f(0)}
  &=\,i\,\int_0^t\,\tilde{\bevot}(t,s)\bbS_s
\D_{\cee,\eff}
  \Bigl[
    {\hat{\rm H}}^{sol}_{I,g,{\beefb}}(Q,\upphiperp+\upphi_{scl},\xi)-V_{2,1}Q^2\\\label{revduh2}
    &\phantom{=\,i\,\int_0^t\,\tilde{\bevot}(t,s)\bbS^\theta(\xi(s))+ddddddddd}
+\eirup(Q,\upphiperp+\upphi_{scl},\xi)
       \\
       &\phantom{=\,i\,\int_0^t\,\tilde{\bevot}(t,s)\bbS^\theta(\xi(s))}
       +\dot\xi(s)\,
(\Xi^0\,+\,\Xi^1Q\,+\,i\Xi^2P)-V_{2,2}Q^2-V_{2,3}Q^2
\Bigr]
  \chi(s,Q)\Omega'
  \,ds\,,
\notag\end{align}
to control their effect on the evolution. The orthogonality property just mentioned allows the final line to be integrated by parts using
Lemma \ref{5g0}. (To justify \eqref{revduh2} one applies the fundamental theorem of calculus to \(\tilde{\bevot}(t,s)\bevot_{scl}(s,0)\Psi(0)\),
and the crucial point is that under the limiting dynamics \eqref{defld} maps \(\Psi(0)\) to a sufficiently nice vector of the form
\eqref{snf}, as in the
corresponding stage of the proof of Theorem
\ref{first} -  details are given there.)
The final stage of the proof is therefore to bound these error terms, using the bounds already established in
\eqref{moav} and Corollaries \ref{cproof} and \ref{imclosercorl}. This is carried out in Lemmas
\ref{esthi}-\ref{finest},  the results of which are now summarized, in the situation of Theorem \ref{main}:
\begin{enumerate}
\item To estimate
  \(\|\int_0^{{\tau_2}/{\sqrt{g}}}\,\tilde{\bevot}(t,s)\bbS_s
\D_{\cee,\eff}
    \bigl({\hat{\rm H}}^{sol}_{I,g,{\beefb}}(Q,\upphiperp+\upphi_{scl},\xi)-V_{2,1}Q^2\bigr)
    \chi(s,Q)\Omega'
    \,ds\,,\|\) we read off from \eqref{annav} that it can be bounded by the sum of the bounds in
    Lemmas \ref{esthi} and Lemma \ref{finest}, under the assumptions of the theorem. The right side of
    the inequality in \eqref{esthi} is
\(
    =O\Bigl(|\frac{\tau_2}{\sqrt{g}}\|{\beefb}\|_{L^2}|\bigl(g(1+g^{-\frac{1}{4}})+g^2(1+\dots g^{-1}\bigr)\Bigr)=O(g^{\frac{1}{4}} \ln\frac{1}{g}).\)
Similarly, the right side of \eqref{btg} is \(O(\sqrt{g}\ln\frac{1}{g})\).
\item To control the infrared error term, the assumption (H2) is used to ensure that the term exponential in the
  cut-off length is smaller than the various negative powers in the coupling constant, to be precise subject to the
  choice \(N>7\):
  \[\|\int_0^{{\tau_2}/{\sqrt{g}}}\,\tilde{\bevot}(t,s)\bbS_s
\D_{\cee,\eff}
\eirup(Q,\upphiperp+\upphi_{scl},\xi)
    \chi(s,Q)\Omega'
    \,ds\|=O\Bigl(|\frac{\tau_2}{\sqrt{g}}|\bigl(g^{-2}g^{N/2}(1+\dots g^{-1}\bigr)\Bigr)
=o(1)\,.
    \]
  \item By Lemmas \ref{esterrtd} and \ref{a2} and Corollary \ref{imclosercorl},
    the final line of \eqref{revduh2} can be estimated in a similar way to the previous items
    to be \(O\bigl(g^{\frac{1}{4}}(1+\ln\frac{1}{g})\bigr)\); the only new feature arises in the
    third from final line, which is seen to be
    \[
    O\Bigl(|\frac{\tau_2}{\sqrt{g}}\|(\id+{\N})^{-1/2}\delta h\||\sqrt{g}\bigl(1+g^{-\frac{1}{4}}\bigr)\Bigr)\,,
    \]
    which can in turn be further controlled to be \(O(g^{\frac{1}{4}})\) using Corollary \ref{imclosercorl},
    which is an averaging theorem for the transverse dynamics, exploiting the fact that the transverse
    oscillations take place on a faster time-scale than the soliton motion.
\end{enumerate}
This completes the proof.
\end{proof}
\subsubsection{Some heuristics}\label{heuristics}
Theorem \ref{main} can helpfully be thought of as a Born-Oppenheimer type of approximation (see
\cite{MR2905846} for a textbook discussion and \cite{MR0588684} for a mathematical treatment
of the time dependent method):
the soliton (which is heavy for small \(g\)) plays the role of the nucleus, and the transverse bosonic modes (often referred to as mesons
in \cite{cole} and \cite{MR0503137}) play the role of electrons.
A useful heuristic approach to understanding soliton dynamics, and guessing some of the formulae which are included in the conclusion of the theorem, is via the method of averaged, or effective, action. In the present case one can insert
\[
\phi(t,x)=\Phi_S(x-\xi(t))+\varphi\,,\qquad\phi_t(t,x)=-\dot\xi\Phi_S'(x-\xi(t))+\varphi_t
  \]
  into the action, and expand in \(\varphi\), leading to
  \begin{equation}
S_\lambda=S_\lambda^{cl}+S_\lambda^1+\dots
  \end{equation}
  with
  \[S_\lambda^{cl}=\int\Bigl[\frac{1}{2}
\frac{\scm_{cl}}{g^2}\dot\xi^2-\frac{\scm_{cl}}{g^2}+\lambda\int\E(t,x)\Phi_S(x-\xi)dx\Bigr]\,dt
\]
and
\[
  S_\lambda^1=\int\Bigl[
  \frac{1}{2}\varphi_t^2-\frac{1}{2}\varphi K(\xi)\varphi
  +\bigl(\lambda\E(t,x)+\partial_t(\dot\xi\Phi_S'(x-\xi))\bigr)\varphi\,
  \Bigr]\,dxdt\,.
\]
Recalling that \(\Phi_S'(x-\xi)={\sqrt{\scm_{cl}}}\tteo(x-\xi)/g\), we find
the Euler-Lagrange equation of motion for \(S_\lambda^{cl}\) is
\[
\ddot\xi=-\frac{\lambda g}{\sqrt{\scm_{cl}}}\,\int\E(t,x)\tteo(x-\xi)dx\,,
\]
which agrees with \eqref{newtpert} to highest order, with the definition of the soliton momentum as
\(\eta=\scm_{cl}\dot\xi/g^2\). This choice of \(\ddot\xi\) essentially removes the component of \(\elec\) along the zero mode,
in the sense that it implies (to highest order)
\[
\E(t,x)+\partial_t(\dot\xi\Phi_S'(x-\xi))=\E(t,x)+\frac{{\sqrt{\scm_{cl}}}}{\lambda g}\Bigl(\ddot\xi\tteo(x-\xi)-\dot\xi^2\tteo'(x-\xi)\Bigr)
=\elec^{eff}(t,x)
\]
{\em on the subspace \(\langle\tteo(\cdot-\xi)\rangle^\perp\) orthogonal to the zero mode}. The transverse dynamics is then
determined by the Euler-Lagrange equation for \(S^1_\lambda\), which reads
\(
\varphi_{tt}+K(\xi)\varphi-\lambda\elec^{eff}=0\,.
\)
The level of discussion here is completely heuristic, but is perhaps useful in that it does lead to the classical modulation equation \eqref{newtpert}
and indicates a simple origin for the effective electric field \eqref{defeff} which drives the transverse mode dynamics,
to highest order.

In reading the main theorem it is helpful to consider a rescaling of the system \eqref{newtpert}-\eqref{defeeze1}, which explains the
            appropriate conditions under which an interesting semiclassical limiting dynamics can be uncovered. Introduce a slow time variable
            \(\tau=\sqrt{g} t\), and rescale \(\eta\to\tilde\eta=g^{-\frac{3}{2}}\eta\) while leaving \(\xi=\tilde\xi\) unchanged, and we find
            \begin{equation}\label{newtperttilde}\frac{d\tilde\xi}{d\tau}={\scm_{cl}}^{-1}\tilde\eta+{g}
V_{-1}(\tilde\xi,\frac{d\tilde\xi}{d\tau},\cee,\eff)
            \quad \hbox{and} \quad
\frac{d\tilde\eta}{d\tau}=-\sqrt{\scm_{cl}}\lambda\bigl({\elec},{\tteo}(\cdot-\tilde\xi)\bigr)_{L^2}+
\sqrt{g}V_{1}(\tilde\xi,\frac{d\tilde\xi}{d\tau},\cee,\eff)
\,,\end{equation}
while the transverse degrees of freedom oscillate rapidly (on the slow timescale of \(\tau\)). This suggests we can hope to obtain nontrivial soliton dynamics
with this scaling, and an important question is what additional assumptions are needed to understand and control the quantum fluctuations. One aspect is that the
electric field should be varying on the slow timescale, so that the transverse degrees of freedom can be understood via averaging - see the hypothesis
(H1). Regarding the semiclassical dynamics of the soliton itself, it is as usual necessary to consider the width of the wave packet: recalling the discussion around \eqref{qmod}, the most natural
assumption seems to be to consider wave packets of standard deviation \(\sqrt{\langle Q^2\rangle}=O(g^{-\frac{1}{4}})\), and this assumption is built into the assumptions on the initial data in the main result, Theorem \ref{main}. (This corresponds to fluctuations of order
\(\sqrt{\langle (gQ)^2\rangle}=O(g^{\frac{3}{4}})\) in the original unscaled spatial coordinates about the classical motion \(\xi\).)
\subsection{Notation}
\label{notn}
      {\em Inner products:} For vectors in a Hilbert space the inner product will be written \((\cdot,\cdot)\) when no confusion as to the
      inner product is likely, and a subscript to indicate the Hilbert space in question will only be used for emphasis when necessary. Thus on
      functions of \(x\in\R\) the inner product
      \((f_1,f_2)_{L^2}=\int\overline{f_1(x)}f_2(x)dx\) is written \((f_1,f_2)\) when no confusion is likely,  and this inner product should be
      assumed for such functions of \(x\in\R\) unless otherwise indicated. For vectors in Fock space the same notation
      \((\Phi,\Psi)\) means the Fock space inner product, as defined following \eqref{fock}.
      Similarly for norms, the notation \(\|\cdot\|\) will mean either the \(L^2\) norm for functions of the Fock space
      norm, and no confusion should be possible.
      Other inner products on functions \(\R\to\C\) are determined from a
nonnegative self-adjoint operator \(A\) with domain
\(\Dom{A}\subset L^2(\R)\) by the formula
\((f,g)_A=(f,Ag)_{L^2}=(A^{\frac{1}{2}}f,A^{\frac{1}{2}}g)_{L^2}\).
The corresponding
symmetric
bilinear form is defined on \(\Dom{A^{\frac{1}{2}}}\times\Dom{A^{\frac{1}{2}}}\). In the
particular case \(A=K_0=4m^2-\partial_x^2\) this gives the Sobolev \(H^1\)
inner product, and fractional powers give the general \(H^s\) Sobolev inner products. In
particular, the case \(H^{\frac{1}{2}}\) arises from the inner product
\begin{equation}\label{h12ip}\bigl(
\phi,\psi\bigr)_{K_0^{\frac{1}{2}}}
=\bigl(
\phi,\psi\bigr)_{\frac{1}{2}}=\int\overline{\hat\phi}(\xi)
  (4m^2+\xi^2)^{\frac{1}{2}}\hat\psi(\xi)d\xi\,,
\end{equation}
which, together with its dual inner product defined as
\((f,g)_{C_0^{\frac{1}{2}}}=(f,g)_{-\frac{1}{2}}\) where \(C_0=K_0^{-1}\),
appears in the Schr\"odinger representation for the free field of
mass \(2m\).
\\
  {\em Operators:} The Hessian operator determining small oscillations around the kink soliton centered at
  \(\xi\in\R\) is
  \(K(\xi)=K_0+u_\xi(x)\) with \({u}_\xi(x)=-6m^2\ssmxi\); when \(\xi=0\) it is suppressed.
\\
      {\em Distributions:}
Schwartz space \(\cs(\R)\) is topologized by the seminorms \(\oldnorm{f}_{N}=\sum_{m_1+m_2\leq N}\sup_{x}|x^{m_1}\partial_x^{m_2} f(x)|\); in the absence of a contrary statement, Schwartz functions should be taken to be real-valued.
The space of complex-valued Schwartz functions will be indicated \(\cs_\smc(\R)\). The operation of complex conjugation
\(f\to\overline{f}\) converts under (distorted) Fourier
transformation into the operation
\(g\mapsto g^\flat\), where we
write \(g^\flat(k)=\overline{g(-k)}\), so that on the (distorted) Fourier side a real test function is one satisfying
\(g^{\flat}=g\) - the context should make it clear what is meant.

The pairing between a tempered distribution \(\Phi\in\cs'(\R)\) and a test function
\(f\in\cs(\R)\) is written either \(\Phi(f)\) or \(\langle\Phi,f\rangle\); the latter will also
be used for complex-valued functions to indicate the complex bilinear form
\(\langle f_1,f_2\rangle=\int f_1(x)f_2(x)dx\), as opposed to
the Hermitian inner product \((f_1,f_2)\) (although of course they agree for real functions). 
\\
{\em Transforms:} We write the {\em Fourier transform} 
as
\(
  \hat f(k)=(2\pi)^{-1/2}\int\,e^{-ikx}f(x)\,dx\,
  , \) and the {\em distorted Fourier transform} \(U\mapsto\tilde U \) is given in
  \eqref{dft}.
  \\
  {\em Gaussian measures:}
\begin{itemize}
\item If \(\sigma_0\) is a positive number then
  \(\gamma(\sigma_0^2)\) is the measure \(\gamma(\sigma_0^2)=(2\pi \sigma_0^2)^{-1/2}\exp[-x^2/2\sigma_0^2]dx\) on \(\R\),
and  for the special case \(\sigma_0^{2}=1/(2\scm_{cl}\sqrt{\theta})\)
we write \(\gamma_\theta=\gamma(\sigma_0^2)\) and, using wave packet notation below,
\(\gamma_\theta(dQ)=\mbChi_\theta^2(Q)dQ\), so that the function identically equal to one
\(\id_{\smr}(Q)\) represents a particle whose wave function is \(\mbChi_\theta\in L^2(dQ)\);
\item If \(\omega_d>0\) is the frequency of the discrete mode \({\tteone}\), then we use the
  Schrodinger representation on \(L^2(\R;\gamma_d(dq_d))\) where
  \(\gamma_d\define\pi^{-\frac{1}{2}}\omega^{\frac{1}{2}}_d\exp{[-\omega^{}_d q_d^2]}\,dq^{}_d \) in which the
  ground state is \(\id_{\smr}(q_d)\) and the excited eigenstates
  are scaled Hermite polynomials in \(q_d\in\R\);
\item if \(C\) is a continuous and nondegenerate bilinear form on
\(\cs(\R)\), then \(\bfga(C)\) is the measure on \(\cs'(\R)\) with
Fourier transform \(\cs'(\R)\ni f\mapsto \exp[-(f,Cf)/2]\).
\end{itemize}
{\em Quantum fields in various representations:}
We write the quantum fields describing 
fluctuations around the soliton as 
\((\varphi,\pi) \) in the shifted vacuum representation \eqref{cfxi1}-
\eqref{cfxi2}, 
but
\((\upphi,\uppi) \) in the solitonic representation \eqref{cfsgext}, according to which the
field is constituted from a pair \((Q,\upphiperp)\), the position operator \(Q\) of the soliton
and a transverse field operator \(\upphiperp\).
The two representations are unitarily related via \(\bbS  \) introduced in \S\ref{qsol}.
The two Fock spaces, \({\fock} \) and \(\mfrF \)
are defined in \eqref{fock} and
\eqref{newfock} and \eqref{l2g}, with number operators \(\N_0\) and \(\hat{\N}\) in
\eqref{numb0} and \eqref{numb}.
Schr\"odinger representation versions of both representations, indicated
by using corresponding bold face fonts 
\(\phisch,\pisch,\bfupphi,\bfuppi,\mfrbF_0,\mfrbF\dots\).
for the fields and Fock spaces. The double colon 
\(\doublecolon O\doublecolon \) (resp. triple colon 
\(\triplecolon O\triplecolon \)) 
is used to indicate an operator normal ordered with respect to the shifted
vacuum (resp. solitonic) representation. 
\({\Poly},{\Poly(\varphi)},{\Poly(\phisch)},{\Poly}(\upphi),{\Poly}(\bfupphi)\dots
\)
  are dense subspaces defined 
just after \eqref{fock} and in \S\ref{qsol}, and a variant
\(\widehat{\Poly}\) is defined in \S\ref{qtt}. We refer to {\em transverse polynomials} as those
built from sums of products of \(\upphi(f)\) with \(f\) orthogonal to the zero mode \(\tteo\), or
more generally, \(\tteox\), so they can equivalently be thought of as polynomials in the transverse field
\(\upphiperp\) of \S\ref{subsubq}.
Regularized fields \(\varphi_\kappa,
\upphi_\kappa\dots \) are all defined by convolution with an approximate identity
\(\delta^{[\kappa]} \) as in \S\ref{fieldreg}, and this induces regularized
operators \(K_{0,\kappa},C_{0,\kappa},K_{\kappa}\dots \) as in \eqref{regcov}
and \eqref{regfc}.

\noindent
{\em Hamiltonians:}
\(\h
=\int\omega^{}_k a_{k}^\dagger a^{}_{k}\,dk\) with \(\omega^{}_k=\sqrt{4m^2+k^2}\)
is the second quantized Hamiltonian
determined by a dispersion relation \(k\mapsto\omega_k \), as in the
discussion following  \eqref{qh}; a regularized version is \(\h_\kappa\) defined in 
Remark \ref{regen}.
Next \({h}^{}_d={h}(\omega^{}_d)\) is the Hamiltonian for a one dimensional
oscillator with frequency \(\omega^{}_d\).
Generally a Hamiltonian density is written in calligraphic, as
for example in the expression
\(\doublecolon\pmb{H}^{sol}_{0\xi}\doublecolon=\doublecolon\int{\pmb{\mcH}}^{sol}_{0\xi}(x)\,dx\doublecolon \)
with \(2{\pmb{\mcH}}^{sol}_{0\xi}=\pisch^2+\phisch K(\xi)\phisch\),
for the free (quadratic) Hamiltonian
arising from expanding about the soliton centred at \(\xi\in\R\). (In \S\ref{regno} a regularization parameter \(\kappa\) also
appears as a suffix on Hamiltonians, for example in \eqref{therel}, but its meaning can be distinguished from the soliton centre \(\xi\)
by the presence of a comma, so no confusion should be possible.) Finally, the electric field is included in the Hamiltonian
\({\pmb H}^{sol,{\elec}}_{0\xi}\), see \eqref{includelec} for example.

\noindent
{\em Wave packets:}
The Gauss-Hermite wave packet solutions to the free
Schr\"odinger equation \eqref{seom},
derived in \cite{andrews}, are given in terms of the Hermite polynomials
\(\He_n(x)=(-1)^ne^{\frac{x^2}{2}}\partial_x^n e^{-\frac{x^2}{2}}\) by
\begin{equation}\label{hgwp}
\mbChi_{n\sigma_0}(t,Q)=\frac{1}{\sqrt{n!\sqrt{2\pi}}}\,
  \sqrt{\frac{2\scm_{cl}\sigma_0}{t-2i\scm_{cl}\sigma_0^2}}\,
  \biggl(\frac{{t+2i\scm_{cl}\sigma_0^2}}{{t-2i\scm_{cl}\sigma_0^2}}\biggr)^{\frac{n}{2}}
  \exp\biggl[\frac{itQ^2}{8\scm_{cl}\sigma_0^2\sigma(t)^2}-\frac{Q^2}{4\sigma(t)^2}
    -\frac{i(2n+1)\pi}{4}\biggr]\,
  \He_n\Bigl(\frac{Q}{\sigma(t)}\Bigr)
  \end{equation}
  where \(\sigma_0\) is a real positive constant and
\(\sigma(t)^2=\sigma_0^2+{t^2}/({4\scm_{cl}^2}{\sigma_0}^2)\)
is the variance which increases with \(t\).
The combinatorial factor
  ensures normalization \(\int|\mbChi_{n\sigma_0}(t,Q)|^2dQ=1\) at all times \(t \), and
  the \(\{\mbChi_{n\sigma_0}(t,Q)\}_{n=0}^\infty\) form an
  orthonormal basis for \(L^2(dQ)\) at each fixed \(t \). The phase factor is chosen
  so that \(\mbChi_{n\sigma_0}(0,Q)\) is real. In particular
  \(
\mbChi_{n\sigma_0}(0,Q;\sigma_0)={\sigma_0}^{-\frac{1}{2}}\mbchi_n(Q/{\sigma_0})\) where
\(\mbchi_n(y)\define(2\pi)^{-\frac{1}{4}}(n!)^{-\frac{1}{2}}\exp[-y^2/4]\He_n(y)\,.
\)
The functions \((2\omega)^{\frac{1}{4}}\mbchi_n(\sqrt{2\omega}x)\) are orthonormal set
of eigenfunctions for the oscillator Hamiltonian \((-\partial_x^2+\omega^2x^2)/2\) with
eigenvalues \((n+\frac{1}{2})\omega\).

When \(\sigma_0^{2}=1/(2\scm_{cl}\sqrt{\theta})\)
it will be convenient to write
\begin{equation}\label{defbchi}
\mbChi_\theta(Q)=\mbChi_{0\sigma_0}(0,Q)
  \end{equation}
and will also
work with the Hilbert space
\begin{equation}\label{aost}
  {\mfrH}(\theta)\,=\,
  L^2(\R,\gamma_\theta(dQ))  \otimes\mfrF\,.
\end{equation}
where \(\gamma_\theta(dQ)=\scm_{cl}^{\frac{1}{2}}\theta^{\frac{1}{4}}\pi^{-\frac{1}{2}}e^{-\scm_{cl}\sqrt{\theta}Q^2}dQ=\mbChi_\theta(Q)^2dQ\), with the
understanding that if \(\theta=0\) then \(\gamma_0(dQ)=dQ\).

There are also Gaussian wave packet solutions to the equation
\begin{equation}\label{seomq}
  i\frac{\partial\Psi}{\partial t}\,+\,\frac{1}{2\scm_{cl}}\frac{\partial^2\Psi}{\partial Q^2}\,
  -w(t)Q^2\Psi\,=\,0\,,\qquad w:\R \to\R\;\;\hbox{continuously differentiable}\,,
\end{equation}
of the form \(\Psi(t,Q)=A(t)\exp[-Q^2/(4\sigma(t)^2)]\) where \(i\dot A=A/(4\scm_{cl}\sigma^2)\) and
\begin{equation}\label{siggeq}
\frac{d}{dt}\sigma^2=\frac{i}{2\scm_{cl}}-4iw\sigma^4\,.
\end{equation}
We will be interested in the case \(w=O(g)\),  where \(g\) is a small positive parameter, in which case
it is natural to introduce \(\tau=\sqrt{g}t\) and \(\tilde y=\sqrt{g}\sigma^2\), which will solve
\[
\frac{d\tilde y}{d\tau}=\frac{i}{2\scm_{cl}}-4i\tilde w\tilde y^2\,,
\]
with \(g\tilde w(\tau)=w(g^{-1/2}\tau)\). Given real initial data \(\tilde y(0)>0\) there will be
a unique complex solution \(\tilde y(\tau)\) with positive real part
on some time interval \(|\tau|\leq\tau_{loc}\) depending on
\(\tilde y(0)\) and (the maximum on the interval of) \(\tilde w\).
Note also that \(\tilde y^{-1}\) obeys a differential equation of the same form, and so we may assume that
\(\max_{|t|\leq\tau_{loc}}\bigl(|\tilde y(\tau)|+|\tilde y(\tau)^{-1}|\bigr)=M<\infty\,.\)
From this the definition
\(\sigma(t)=g^{-1/4}\sqrt{\tilde y(\sqrt{g}t)}\) gives for small positive \(g\) a continuously differentiable solution of
\eqref{siggeq} on the (long) interval \(|t|\leq\tau_{loc}/\sqrt{g}\), with bounds above and below
by a multiple of \(g^{-1/4}\). Substituting to solve for \(A\), and thence defining \(\Psi\) as above yields a solution of
\eqref{seomq} which is initially Gaussian with real variance parameter \(\sigma(0)=g^{-1/4}\sqrt{\tilde y(0)}>0\).

Now taking the real part of the
equation for \(\tilde y=\tilde y_1+i\tilde y_2\) we deduce that
\[
\tilde y_1(\tau)=\tilde y_1(0)\exp[g\int_0^\tau\tilde w(\tau')\tilde y_2(\tau')d\tau']
  \]
  so we may assume without loss of generality, by adjusting \(\tau_{loc}\), that for any positive number \(\iota>1\),
  \begin{equation}
\iota^{-1}\sigma(0)^2\leq\Re\sigma(t)^2\leq \iota\sigma(0)^2\,.
\end{equation}
This is useful for controlling the expectation values of powers of \(Q\), under the assumption
that \(\sigma(0)=g^{-1/4}\sqrt{\tilde y(0)}>0\) for some \(\tilde y(0)\) independent of \(g\). On account of the formula
\[
|\Psi(t,Q)|^2dQ=|A(t)|^2\exp\Bigl[-\frac{\Re\sigma(t)^2Q^2}{2|\sigma(t)|^4}\Bigr]\,dQ\,,
\]
conservation of
probability gives, for normalized wave packets with \(\|\Psi\|_{L^2(dQ)}=1\),
\[
\sqrt{2\pi}|A(t)|^2\frac{|\sigma(t)|^2}{\sqrt{\Re\sigma(t)^2}}=1\,.
\]
From this, and the bounds above, it follows directly by scaling that we have (for some numbers \(c_r(\iota,M),b(M)\) independent of \(g\))
\begin{equation}
\|Q^r\Psi(t)\|_{L^2(dQ)}^2\leq c_rg^{-r/2} \qquad\|P^r\Psi(t)\|_{L^2(dQ)}^2\leq b_rg^{r/2}
\qquad\hbox{ and }\|Q^rP\Psi(t)\|_{L^2(dQ)}^2\leq d_rg^{-(r-1)/2}
\label{varbounds}\,
\end{equation}
for \(r=0,1,2\dots\) and  \(|t|\leq\tau_{loc}/\sqrt{g}\). Note also that \(L^2\) bounds like this will remain valid for wave packets written using the
Schr\"odinger representation determined by the Gaussian measure \(\gamma_\theta(dQ)\), i.e., under the unitary equivalence
\begin{align}
  L^2(\R,d{Q})&\equiv 
  L^2(\R,\mbChi_\theta(Q)^2d{Q})=L^2(\R,\gamma_\theta(dQ))\\
  f({Q})&\mapsto\mbChi_{\theta}(Q)^{-1}f({Q})
\end{align}
under which the momentum operator \(P=-i\partial_Q\) is mapped into the operator
\(Pf= -i\frac{df}{d{Q}}({Q})+{i\scm_{cl}\sqrt{\theta}}f({Q})\)\,.

\section{The Heisenberg Commutation Relations (CCR)}
\setcounter{equation}{0}
\la{hcr}
To solve the quantum field theory associated to the Hamiltonian
\eqref{ham} it is necessary to find a Hilbert space \(\mfrH \), such that
the classical fields, \(\phi\) and \(\pi \), are replaced by 
operator-valued distributions acting on \(\mfrH\). These 
operator-valued distributions - called (Heisenberg) quantum fields, and 
denoted \(\Phi^H\) and \(\Pi^H\) - are required to verify the
Heisenberg equal time commutation relation, namely,
\(
[\Phi^H(t,x)\,,\,\Pi^H(t,y)]\,=\,i\delta(x-y)\,,
\)
as well as 
the equations of motion \eqref{ceom}, appropriately interpreted. This
is the quantum theory in the Heisenberg picture. In the Schr\"odinger 
picture, one instead works with time-independent (time-zero) quantum fields
\(\Phi,\Pi\), which are operator-valued distributions, verifying the canonical commutation relation (CCR)
\begin{equation}\label{ccr}
[\Phi(x)\,,\,\Pi(y)]=i\delta(x-y)\,,
\end{equation}
interpreted distributionally. These fields generate families of unitary operators
\(\{\Exp[i\Phi(f)]\}_{f\in\cs}\) and \(\{\Exp[i\Pi(f)]\}_{f\in\cs}\) which verify the Weyl relations
\begin{equation}\label{weyl}
\Exp[i\Pi(f_2)]\,\Exp[i\Phi(f_1)]=e^{i(f_1,f_2)_{L^2}}\Exp[i\Phi(f_1)]\,\Exp[i\Pi(f_2)]\,,
\end{equation}
(for real-valued test functions \(f_1,f_2\)). 
These fields are then used to build, starting from the
formal expression \eqref{ham}, a {\em self-adjoint} Hamiltonian
operator acting on \(\mfrH\,. \)
Once this is achieved, the theorem of Stone provides a 
strongly continuous one-parameter group
of unitary transformations, i.e., a collection \(\{\Exp[-itH]\}_{t\in\R} \) of
linear mappings
constituting a  one-parameter unitary group which defines the
quantum dynamics and also connects
the Heisenberg and Schr\"odinger pictures, through the formal relations 
\(\Phi^H(t,x)=\Exp[+itH]\Phi(x) \Exp[-itH]\,,\) and 
\(\Pi^H(t,x)=\Exp[+itH]\Pi(x) \Exp[-itH]\,,\) etc. 
We will work in the Schr\"odinger
picture, so that a proof of an existence theorem for the quantum dynamics
consists of fixing a representation of \eqref{ccr} for time-independent
fields, and then proving self-adjointness of the Hamiltonian obtained
by substituting these fields into \eqref{ham} - this latter process
requires regularization and taking limits. 
\subsection{Quantization in the vacuum sector.}\label{vacquant}
We first recall from \cite{MR810217} the quantization procedure
in the case of the topologically trivial boundary conditions \eqref{vacbc}.
Write the classical field as \(\Phi_0+\varphi \), 
where the field \(\varphi \) is subject to
the boundary condition \(\lim_{|x|\to\infty}\varphi(x)=0\,. \) 
The classical Hamiltonian is now
\begin{align}\label{hamvac}
  H^{vac}({\varphi,\pi})\,=\,
  H(\Phi_0+\varphi,\pi)\, & =\,\int\,\Bigl[\frac{1}{2}\,
\bigl(\pi^2+\partial_x\varphi^2+4m^2\varphi^2\bigr)\,+\,2mg\varphi^3
\,+\,\frac{1}{2}
g^2\varphi^4\,\Bigr]\,dx\,, \\
                          & =\,H^{vac}_{0}\,+\,H^{vac}_{I,g}
\,.
\end{align}
Here
\begin{equation}\label{defh0}
H^{vac}_{0}({\varphi,\pi})\,
=\,\frac{1}{2}\,\int\,\Bigl[\,
  \pi^2\,+\,\varphi\,K_0\varphi\,\Bigr]\,dx\qquad\hbox{and}\quad
H^{vac}_{I,g}(\varphi)\,=\,\int\,2mg\varphi^3
\,+\,\frac{1}{2}
g^2\varphi^4\,dx\,
\end{equation}
and \(K_0=(-\partial_x^2+4m^2)\,. \)
Later we will also make use of the associated covariance operator
\[
C_0\,=\,K_0^{-1}\,=\,(-\partial_x^2+4m^2)^{-1}\,,
\]
and its square root. We now recall the standard solution of \eqref{ccr}
for the vacuum sector fields \((\Phi,\Pi)=(\Phi_0+\varphi,\pi)\) and the 
operators obtained by substituting these into the classical expressions
for the Hamiltonian, giving sufficient detail for what we will need below.
\paragraph{Fock Space.}
Now to 
define a corresponding pair of quantum fields, still denoted \(\varphi,\pi \),
we introduce Fock space, \(\mfrH_0\), defined as the (complete) Hilbert direct sum of the
symmetric n-fold tensor powers of \(L^2(\R) \), defined with
Lebesgue measure \(dk\), i.e.,
\begin{equation}\label{fock}
  \mfrH_0\,\define\,\Symhatl\,=\,\bigoplus\limits_{n=0}^\infty\Symn (L^2(\R,dk))\,.
\end{equation}
(The use of \(k\) indicates that we will be using this on the Fourier side, i.e. for momentum space wave functions.)
For \(n=0 \) it is to be understood that \({\rm Sym}^0(L^2(\R))=\C\,.\)
A typical element, \(\Psi\in{\fock} \), is a sequence of
functions \(\{\Psi_n\}_{n=0}^\infty \), where \(\Psi_n\in L^2(\R^n) \)
is symmetric with respect to interchange of any pair of coordinates:
\[
\Psi(k_1,\dots ,k_i,\dots ,k_j,\dots ,k_n)= 
\Psi(k_1,\dots ,k_j,\dots ,k_i,\dots ,k_n)\,.
\] 
The Fock space norm is \(\|\Psi\|^2=\sum\|\Psi_n\|_{L^2(\smr^n)}^2\,. \)
The element with \(\Psi_0=1 \) and \(\Psi_n=0 \) for \(n\geq 1 \)
is called the vacuum, and will be denoted \(\Omega^{}_0 \), or
\(|\,0\,\rangle\,. \)

A useful dense subspace,  \({\Poly}\), or written \({\Poly(\varphi)}\) if the field representation needs to be indicated,
is the algebraic span of the {\em symmetric}
tensor products \(\hat f_1\odot \hat f_2\odot \dots \odot \hat f_n\define\Symn\prod_{j=1}^n \hat f_j(k_j) \) of 
Schwartz functions \(\hat f_1,\dots \hat f_n\) (which products themselves lie in \(\cs(\R^n)\cap\Symn (L^2) \)).
\({\Poly}\) is a subspace of the finite particle subspace, \({\Fin_0(\cs)}\subset\fock\) of vectors with
\(\Psi_n=0\) for \(n>N\) for some
integer \(N\) and \(\Psi_n\in\cs(\R^n)\forall n\); there is a generalization of this in the transverse Fock space
defined in Remark \ref{lpart}.
For each \(k\in\R \) and \(\Psi\in{\Poly(\varphi)}\), the annihilation and creation operators are
given, respectively, by
\begin{align}
(a^{}_k\Psi)_{n-1}(k_1,\dots,k_{n-1})\,      & =\,\sqrt{n}\Psi_n(k,k_1,\dots,k_{n-1})\,,\;\hbox{and} \\
(a_k^\dagger\Psi)_{n+1}(k_1,\dots,k_{n+1})\, & =\,
\sum_{j=1}^{n+1}\,\frac{1}{\sqrt{n+1}}\,\delta(k-k_j)\Psi_n
(k_1,\dots,\widehat{k_j},\dots,k_{n+1})\,.
\end{align}
(The hat indicates an omitted argument.) Recall that the domain
of \(a_k^\dagger \) consists only of the zero vector, and properly
speaking the expression above gives rise to a densely defined 
bilinear form
on \({\fock} \), rather than a densely defined operator. 
Alternatively, these formal expressions can be regarded
as defining operator-valued distributions, and it can be checked that
they satisfy \([a^{}_k,a_l^\dagger]=\delta(k-l) \), interpreted
appropriately,  see (\cite{MR0493420}). 
We recall the basic estimates for Wick Operators from \cite{MR0674511}*{\S 4}.
  Given a function or distribution \(w\in\cs'(\R^{m+n})\), a corresponding Wick
  operator on Fock space is given  formally by
  \begin{equation}\label{wickop}
    \Op(w)\,=\,\int_{\R^{n+m}}\,
    a^\dagger_{k_m}\dots a^\dagger_{k_1}\,w(k_1,\dots k_m,k'_1,\dots k_n')
    \, a^{}_{k_1'}\dots a^{}_{k_n'}\prod_{j=1}^mdk_j\prod_{j=1}^ndk_j'\,.
    \end{equation}
  Writing 
\begin{equation}\label{numb0}{\N_0}=\int a^\dagger_k a^{}_k\,dk \end{equation} 
for the number operator, the following bounds hold in the case that the kernel is
  square integrable:
  \begin{equation}\label{best}
\|(\id+{\N_0})^{-m/2}\Op(w)(\id+{\N_0})^{-n/2}\|\,\leq\,\|w\|
\end{equation}
where on the left hand side \(\|\,\cdot\,\| \) means Fock space
operator norm, while on the right hand side \(\|w\| \) means the
operator norm of the mapping \({\rm Sym}^n(L^2(\R))\to
{\rm Sym}^m(L^2(\R))\) determined by the kernel \(w\). More generally, the identity (on finite particle vectors)
\begin{equation*}
\Op(w)(\id+{\N_0})^{\alpha}=(\id+{\N_0}+n-m)^{\alpha}\Op(w)\qquad\qquad(\alpha\in\R)
\end{equation*}
implies that for \(a+b\geq m+n \),
\begin{equation*}
  \|(\id+{\N_0})^{-a/2}\Op(w)(\id+{\N_0})^{-b/2}\|
  \,\leq\,(1+|m-n|)^{|m-a|/2}\|w\|\,,
\end{equation*}
so in particular if \(l=m+n\) we can choose \(a=l,b=0\) or \(a=0,b=l\) to obtain
\begin{equation}\label{best2}
  \max\Bigl\{\|(\id+{\N_0})^{-l/2}\Op(w)\|\,,\,\|\Op(w)(\id+{\N_0})^{-l/2}\|\Bigr\}
  \,\leq\,(1+l)^{l/2}\|w\|\,.
\end{equation}
\noindent
As a final point, for square integrable \(w\),
the operator
\(\Op(w)\) is closable and the
the dense subspace \({\Poly(\varphi)}\) is a core, see \cite{MR0493420}*{Theorem X.44}.

Introducing the dispersion relation
\(\omega^{}_k=\sqrt{k^2+4m^2} \), we define the fields
\begin{align}\la{cf1}
\varphi(x)\, & =\,\frac{1}{\sqrt{2\pi}}\,\int\,\frac{1}{\sqrt{2\omega^{}_k}}\,
\bigl(a^{}_ke^{ikx}+a_k^\dagger e^{-ikx}\bigr)\,dk\,,\hbox{ and } \\
\pi(x)\,     & =\,\frac{1}{\sqrt{2\pi}}\,\int\,-i\,\sqrt{\frac{\omega^{}_k}{2}}\,
\bigl(a^{}_ke^{ikx}-a_k^\dagger e^{-ikx}\bigr)\,dk\,.\la{cf2}
\end{align}
Again, these expressions really define operator-valued distributions, e.g., if 
\(f\in\cs(\R) \) then \(\varphi(f) \) is the unbounded, densely defined
operator given by
\begin{equation}\label{cf1f}
\varphi(f)\,=\,\int\,\frac{1}{\sqrt{2\omega^{}_k}}\,
\bigl(a^{}_k\,\hat f(-k)+a_k^\dagger\,\hat f(k)\bigr)\,dk\,,
\end{equation}
where \(\hat f(k)=\mcF (f)(k)=(2\pi)^{-1/2}\int\,e^{-ikx}f(x)\,dx \) is the
Fourier transform. 
Another useful way of expressing the above is to introduce
operator-valued distributions
\begin{equation}\label{alpha0}
g\mapsto a(g)\,=\,\int\,\overline{g(k)}a^{}_k\,dk\,,\quad
g\mapsto a^\dagger(g)\,=\,\int\, g(k)a^\dagger_k\,dk\,,\qquad\qquad g\in\cs(\R) \,,
\end{equation}
and also corresponding Fourier transforms
\begin{equation}\label{alpha0f}
\alpha(f)\,=\,a(\hat f)\qquad\alpha(f)\,=\,
a^\dagger(\hat f) \qquad f\in\cs(\R) \,.
\end{equation}
These are formally adjoint to one another; notice that \(f\mapsto\alpha(f)\) is complex anti-linear whereas
\(f\mapsto \alpha^\dagger(f)\) is complex linear. Now complex conjugation \(f\to\overline{f}\) maps under Fourier
transformation \(\mcF\) to the operation
\(\hat f\to\hat f^\flat\) where, for any function \(g:\R\to\C\)
we write \(g^\flat(k)=\overline{g(-k)}\).
With these definitions it is possible to write the above defined fields as
\begin{equation}\label{segal}
\varphi(f)\,=\,\frac{1}{\sqrt{2}}\,\Bigl(\alpha(\overline{K_0^{-1/4}{f}})+\alpha^\dagger(K_0^{-1/4}{f})
\Bigr)
\,,
\qquad
\pi(f)\,=\,-\frac{i}{\sqrt{2}}\,\Bigl(\alpha (\overline{K_0^{1/4}f})-\alpha^\dagger(K_0^{1/4}{f})
\Bigr)\,,
\end{equation}
and the Heisenberg relation is a consequence of the only non-zero
commutator \([\alpha(f),\alpha^\dagger(g)]=\int \overline{ {\hat f}(k)}{\hat g}(k)dk=\int \overline{f(x)}g(x)dx, \)
valid for Schwartz test functions.
Note that (i) \(\alpha^\dagger(f)\,\hat f_1\odot\dots \odot \hat f_n=\sqrt{n+1}\hat f\odot \hat f_1\odot\dots \odot \hat f_n\);
(ii) \(\alpha^\dagger(f)\) and \(\alpha(f)\) are formally adjoint to one another;  and (iii) both field operators
in \eqref{segal} are \(\C\)-linear in \(f\), but it
is only for real \(f\) that these expressions lead to self-adjoint operators; in the case of real
\(f\) the complex conjugate inside \(\alpha ( )\) is redundant.

One can now check that the pair \((\Phi,\Pi)=(\Phi_0+\varphi,\pi) \) solves
\eqref{ccr}, again interpreted appropriately. After discarding
an (infinite) constant, the free Hamiltonian
is (\cite{MR810217}*{\S III.I.4}):
\begin{equation}\label{hamfree}
\doublecolon H^{vac}_{0}\doublecolon\,
=\,\frac{1}{2}\,\int\,\doublecolon \,
\pi^2\,+\,\varphi\,K_0\,\varphi\,\doublecolon \,dx\,
=\,\int\,\omega^{}_k \, a_k^\dagger\, a^{}_k
\,dk\,.
\end{equation}
(In fact, to obtain the semiclassical correction to the soliton
mass, we will keep track of a regularized version of the discarded
constant and compare it with the corresponding quantity in the
solitonic quantization.)
As usual, colons indicate the {\em normal ordered product}
of the field operators, obtained by moving all the annihilation operators
to the right. The final expression
\begin{equation}\label{assemb}
\h\,=\,\int\omega^{\phantom{\dagger}}_k\, a_k^\dagger\, a^{\phantom{\dagger}}_k\,dk
  \end{equation}
is the Hamiltonian for an
assembly of noninteracting bosons with dispersion relation
\(\omega^{}_k=\sqrt{4m^2+k^2}\); on the \(n\)-particle wave function it acts as multiplication by
the positive function \(\sum_{i=1}^n
  {\omega_{k_i}}\), and so defines a self-adjoint operator with domain
\begin{equation}\label{domhvac}
\Dom(\h)=\Dom(\doublecolon H^{vac}_0\doublecolon)
\define\Bigl\{\Psi\in\bigoplus\limits_{n=0}^\infty\Symn (L^2(\R)):\sum_{n}\|(\sum_{i=1}^n
  {\omega_{k_i}})\Psi_n(k_1,\dots k_n)\|^2_{L^2}<\infty\Bigr\}\,.
\end{equation}
Finally, with regard to the free field,
Fock space has provided a representation of \eqref{weyl}: the essentially self-adjoint operators \eqref{segal}
yield unitary operators written \(\V(f)=\Exp[i\varphi(f)]\) and \(\U(g)=\Exp[i\pi(g)]\) which verify
the Weyl relation \(\U(g)\V(f)=e^{i(f,g)}\V(f)\U(g)\) for real Schwartz functions \(f,g\).
These two families of unitary operators can also be usefully combined to define the
{\em complex} displacement operator determined by complex-valued \(f=f_1+if_2\in\cs_\smc(\R)\):
\begin{equation}\label{cdis0}
\D_0(\hat f_1+i\hat f_2)=e^{-i(f_1,f_2)}\U(-\sqrt{2}K_0^{-\frac{1}{4}}f_1)\,\V(\sqrt{2}K_0^{\frac{1}{4}}f_2)\,,
\end{equation}
which has a particularly simple form \eqref{crel00} when written in terms of creation/annihlation operators on
\(\fock\), and paramterized by the Fourier transforms \(\hat f_1,\hat f_2\).
We will make use of the following commutation formulae.
\begin{prop} The relations
\begin{align}\label{araki}\begin{split}
  &[\doublecolon H^{vac}_0\doublecolon\,,\,\U(f)]\,u\,=\,\U(f)\Bigl(
  -\varphi(K_0f)+\frac{1}{2}\langle f,K_0f\rangle\Bigr)\,u\\
    &[\doublecolon H^{vac}_0\doublecolon\,,\,\V(f)]\,u\,=\,\V(f)\,\Bigl(
\pi(f)+\frac{1}{2}\langle f,f\rangle\Bigr)\,u\,.\end{split}
\end{align}
hold for \(u\in \Dom(\doublecolon H^{vac}_0\doublecolon)\).
\end{prop}
\proof
These formulae appear in various forms and frameworks in the articles \cite{MR0127821,MR0180855,MR0180856}; for completeness, and to make precise the domain, we will derive them by differentiation of the Heisenberg field, concentrating on the second formula, (a similar argument works for the first).
To start with we know the time-dependent free Heisenberg field is given by
\begin{equation}\label{cf3}
\Exp[+it\doublecolon H^{vac}_0\doublecolon]\varphi(f) \Exp[-it\doublecolon H^{vac}_0\doublecolon]\,=\,\varphi^H(t,f)\,=\,
\int\,\frac{1}{\sqrt{2\omega^{}_k}}\,
\bigl(a^{}_k\,\hat f(-k)e^{-i\omega^{}_kt}+a_k^\dagger\,\hat f(k)e^{+i\omega^{}_kt}\bigr)\,dk\,,
\end{equation}
which holds applied to arbitrary vectors \(u\in{\Poly(\varphi)}\). By the analytic vector theorem, both \(\varphi(f)\) and \(\varphi^H(t,f)\)
are essentially self-adjoint on \({\Poly(\varphi)}\) (see \cite{MR0493420}) and so generate unitary groups which verify the exponentiated form of \eqref{cf3}, i.e.,
\begin{equation}\label{diffme}
\Exp[+it\doublecolon H^{vac}_0\doublecolon]\,\Exp[i\varphi(f)]\, \Exp[-it\doublecolon H^{vac}_0\doublecolon]\,=\,\Exp[i\varphi^H(t,f)]\,.
\end{equation}
Now \(\varphi^H(t,f) u\to\varphi(f) u\) as \(t\to 0\) for all \(u\in{\Poly(\varphi)}\), and so by \cite{MR751959}*{Theorem VIII.21-VIII.25}
  \(\Exp[i\varphi^H(t,f)]\to \Exp[i\varphi(f)]\) in the strong operator sense. In fact more is true:
  \begin{equation}\label{diffh}
\lim_{t\to 0}\frac{1}{t}\left(\Exp[i\varphi^H(t,f)] - \Exp[i\varphi(f)]\right)\, u\,=\,i\Exp[i\varphi(f)]\left(\pi(f)+\frac{1}{2}(f,f)\right)\, u\,,
    \end{equation}
for \(u\in{\Poly(\varphi)}\). Accepting this temporarily, we use it to differentiate \eqref{diffme} and hence derive \eqref{diffh}. Recall the Duhamel formula
\[
\frac{1}{t}\left(\Exp[i\varphi^H(t,f)] - \Exp[i\varphi(f)]\right)u=i\int_0^1
\,\Exp[i(1-{t'})\varphi^H(t,f)]\frac{1}{t}\left(\varphi^H(t,f)- \varphi(f)\right)\Exp[i{t'}\varphi(f)]\,u\,d{t'}\,.
\]
This identity holds for \(u\in {\Poly(\varphi)}\) by the fundamental theorem of calculus, because
\[
{t'}\mapsto \Exp[i(1-{t'})\varphi^H(t,f)]\frac{1}{t}\left(\varphi^H(t,f)- \varphi(f)\right)\Exp[i{t'}\varphi(f)]\,u\in\fock
\]
is differentiable because \(u\in {\Poly(\varphi)}\subset \Dom(\varphi(f))\) and (see below)  \(\Exp[i{t'}\varphi(f)]\,u\in\cap_{s}\Dom(\N_0^s)\subset
\Dom(\varphi^H(t,f))\).
Now we want to compare the difference quotient \(\frac{1}{t}\left(\varphi^H(t,f)- \varphi(f)\right)\) with
\(\pi(f)\). To do this we also differentiate the Weyl relation: so replace \(g\) by \(\epsilon g\) in the Weyl relation and consider difference quotients
\[
\frac{1}{\epsilon}\bigl(\U(\epsilon g)-\id\bigr)\V(f)u=\frac{1}{\epsilon}\bigl(e^{i\epsilon(f,g)}-1\bigr)\V(f)\U(\epsilon g)u
+\V(f)\frac{1}{\epsilon}\bigl(\U(\epsilon g)-\id\bigr)u
\]
to deduce that \(\V(f)\) maps the domain of \(\pi(g)\) into itself, and
\[
\pi(g)\V(f)u=\V(f)\pi(g)u+(f,g)\V(f)u\,,
\]
for \(u\in\Dom(\pi(g))\). Putting \(g={t'} f\) for \(0\leq{t'}\leq 1\), and reverting to the path-ordered exponential
notation for \(\U({t'} f)\) for clarity, we deduce that (for \(u\in\Dom(\pi(f))\))
\begin{align}
  \Exp[i\varphi(f)]\,\pi(f)\,u
  &=\int_0^1
\,\Exp[i(1-{t'})\varphi(f)]\Exp[i{t'}\varphi(f)]\pi(f)u\,d{t'}\,\\
  &=\int_0^1
\,\Exp[i(1-{t'})\varphi(f)]\left(\pi(f)-(f,f){t'} \right)\Exp[i{t'}\varphi(f)]\,u\,d{t'}\,\\
  &=\lim_{t\to 0}\int_0^1
\,\Exp[i(1-{t'})\varphi^H(t,f)]\pi(f)\Exp[i{t'}\varphi(f)]\,u\,d{t'}\,-\frac{1}{2}(f,f),\Exp[i\varphi(f)]\,u
\end{align}
the last equality holding on account of \(\int_0^1{t'} d{t'}=\frac{1}{2}\) and
the strong operator topology convergence \(\Exp[i\varphi^H(t,f)]\to \Exp[i\varphi(f)]\) (as \(t\to 0\)).
Combining with the Duhamel formula above gives
\begin{align*}
\frac{1}{t}\left(\Exp[i\varphi^H(t,f)] - \Exp[i\varphi(f)]-it\Exp[i\varphi(f)]\pi(f)\right)u&=i\int_0^1\Bigl[
  \,\Exp[i(1-{t'})\varphi^H(t,f)]\frac{1}{t}\left(\varphi^H(t,f)- \varphi(f)-t\pi(f)\right)\\
&\qquad\qquad\qquad\qquad\qquad  \times\Exp[i{t'}\varphi(f)]\,u\,\Bigr]d{t'}
+\frac{i}{2}(f,f),\Exp[i\varphi(f)]\,u\,.
\end{align*}
By \eqref{cf1},\eqref{cf2} and \eqref{cf3}, and with reference to \eqref{best2}, we have the following bound in Fock space norm
\[
\|\frac{1}{t}\left(\varphi^H(t,f)- \varphi(f)\right)U - \pi(f)U\|\leq \const |t|\|\N_0^{\frac{1}{2}}U\|
\qquad\hbox{where}\quad U=\Exp[i{t'}\varphi(f)]u\,,
\]
as \(t\to 0\), since the function \(k\mapsto (e^{it\omega_k}-1-it\omega_k)\hat f(k)\) is \(O(t^2)\) in each Schwartz seminorm.
But for \(u\in{\Poly(\varphi)}\), \(U\in\cap_{s}\Dom(\N_0^s)\); to see this, write
\[
\Exp[i{t'}\varphi(f)]=e^{-\frac{1}{4}\|g\|^2}\,\Exp[\frac{1}{\sqrt{2}}a^\dagger(g)]\,
\Exp[\frac{1}{\sqrt{2}}a(g^\flat)],
\]
where \(g=\mcF(K_0^{-1/4}{f})\in\cs(\R)\),
and observe that, since the final factor takes \({\Poly(\varphi)}\) to itself, it suffices to bound the middle factor on an
arbitrary \(J\)-particle vector
\(u_J\). But this can be done using \eqref{best}:
\[
\|\N_0^s a^\dagger(g)^Nu_J\|\leq(J+N)^s\bigl((J+N)(J+N-1)\dots(J+1)\bigr)^{\frac{1}{2}}\|g\|^N\|u_J\|\,,
\]
and so (on account of the square root in this formula and the \(1/N!\) in the exponential series)
the series defining \(\Exp[\frac{1}{\sqrt{2}} a^\dagger(g)]\) converges,
as does that for \(\N_0^s\Exp[\frac{1}{\sqrt{2}}a^\dagger(g)]\) for any \(s\in 0,1,2\dots\), allowing us to conclude
that \(\Exp[i{t'}\varphi(f)]u\in\cap_{s}\Dom(\N_0^s)\) for \(u\in{\Poly(\varphi)}\). (This fact was also used in the justification of the Duhamel formula above.)
This completes the derivation of \eqref{diffh}, and hence of the second equation of \eqref{araki}, on \({\Poly(\varphi)}\).
This formula then
automatically extends to the domain \(\Dom(\doublecolon H^{vac}_0\doublecolon)\) since the operators in
question are closed and
\[
\Dom(\pi(f))\supset\Dom(\N_0^{\frac{1}{2}})\supset\Dom(\doublecolon H^{vac}_0\doublecolon)\,,
\]
with bounds \(\|\pi(f)\Psi\|\leq const. (\|\Psi\|+\|\N_0^{\frac{1}{2}}\Psi\|)\) and
\(\|\N_0\Psi\|\leq\|(\doublecolon H^{vac}_0\doublecolon)\Psi\|\). (This last step shows that the second formula of \eqref{araki}
is to be interpreted as saying, in addition to the given commutation relation, that for \(f\in\cs\) the Weyl operator \(\V(f)\) leaves
invariant \(\Dom(\doublecolon H^{vac}_0\doublecolon)\).)
\qed
\paragraph{The Schr\"odinger representation.}
There is an alternative representation
of the Heisenberg relations \eqref{ccr} in which the Hilbert space is a Gaussian
space. To be precise, let 
\begin{equation}\label{defmu0}
  \mu_0 \,=\,\bfga({\frac{1}{2\sqrt{K_0}}})
\end{equation}
be the Gaussian measure on \(\cs'(\R) \)
with covariance \(\frac{1}{2\sqrt{K_0}}=\frac{1}{2}{C_0}^{\frac{1}{2}}\), where
\(C_0^{\frac{1}{2}}\) is the operator with integral kernel
\begin{equation}\label{defC0}
C_0^{\frac{1}{2}}(x,y)\,=\,(-\Delta+4m^2)^{-\frac{1}{2}}(x,y)\,=\,
\frac{1}{2\pi}\,\int_{\smr}\,
\frac{e^{ik(x-y)}}{(k^2+4m^2)^{\frac{1}{2}}}
\,dk
\,=\,
\frac{1}{2\pi}\,\int_{\smr}\,
\frac{e^{ik(x-y)}}{\omega^{}_k}
\,dk\,
,
\end{equation}
and form
the Gaussian Hilbert space \(L^2(\cs'(\R),\mu_0) =L^2(\mu_0)\). Write (in boldface) \(\phisch\)
for a typical point of \(\cs'(\R)\), so
that the coordinate functions are the functions 
\(\phisch\mapsto\phisch(f)\) for \(f\in\cs(\R)\);
we use the same notation to indicate the corresponding multiplication
operators on \(L^2(\mu_0) \). Addition
and multiplication of such coordinate functions generates the polynomials,
which correspond to \({\Poly(\varphi)}\) under the unitary equivalence
\(\I\) which is about to be introduced.
Recall from \cite{MR1474726}*{Chapter 2} the Wiener chaos
orthogonal decomposition, which yields a collection \(\{{\rm Pr}_n\}_{n=0}^\infty\)
of mutually orthogonal projection operators with \(\oplus{\rm Pr}_n=\id\), the
range of \({\rm Pr}_n\) being the orthogonal complement of the
closed linear span of polynomials of degree \(n-1\) within the closed linear
span of polynomials of degree \(n\) (see also \cite{MR887102}*{\S 6.3}).
\begin{prop}
\label{sch}
There exists a unitary map \(\I\) taking \({\fock}\) onto \(L^2(\mu_0) \),
such that \(\I\Omega^{}_0=1 \) (i.e., the function on \(\cs'\)
identically equal to one), and if \(f_j\in\cs(\R)\forall j\)
\[
\I \doublecolon \varphi(f_1)\varphi(f_2)\dots\varphi(f_N)\doublecolon \Omega^{}_0
\,=\,{\rm Pr}_N\,\phisch(f_1)\phisch(f_2)\dots\phisch(f_N)\,\,\,.
\]
In addition \(\I\) induces the following action on the operators:
\begin{align}\label{seq}\begin{split}
\I\,\circ\,\varphi(f)\,\circ\,\I^{-1}\, & =\,\phisch(f) \hbox{  (multiplication operator)}\,, \\
\I\,\circ\,\pi(f)\,\circ\,\I^{-1}\,     & =\,\pisch(f)=-iD_{f} +{i}\phisch({C_0}^{-\frac{1}{2}} f)
\,,
\end{split}
\end{align}
where \(D_f \) is the directional derivative operator (along
\(f\in\cs(\R)\)) given by \(D_f A(\phisch)
=\lim_{\epsilon\to 0}\frac{A(\phisch+\epsilon f)-A(\phisch)}{\epsilon}\)
on an appropriate domain (which includes the polynomials, i.e. the algebra
generated by the coordinate functions).
\end{prop}
The Cameron-Martin space for the measure \(\mu_0\) is \(H^{\frac{1}{2}}\), and
so the operation of displacement of the field \(\bfdel_g:\phisch\to\phisch+g\)
(i.e. translation in the space \(\cs'\)) produces by push-forward
an equivalent measure (i.e., \((\bfdel_g)_*\mu_0\) is mutually
absolutely continuous with \(\mu_0\)) if and only if \(g\in H^{\frac{1}{2}}\),
see \cite{MR1642391}*{Theorem 2.4.5}.
In the case \(g\in H^{\frac{1}{2}}\) the Radon-Nikodym derivative is
given by
\begin{equation}\label{rndis}
  \frac{d(\bfdel_g)_*\mu_0}{d\mu_0}
  \,=\,
  \Exp\Bigl[-2\phisch(K_0^\frac{1}{2}g)
    -\bigl(g,K_0^\frac{1}{2}g\bigr )_{L^2}\Bigr]\,,
\end{equation}
and there is a corresponding unitary operator \(\mbU(g)\) of field displacement
which acts on polynomials as
\[
\mbU(g)\Psi(\phisch)=\Psi(\phisch+g)\,\sqrt{  \frac{d(\bfdel_g)_*\mu_0}{d\mu_0}}\,.
\]
(One can check that if \(f\) is a real test function, then \(-i\) times the infinitesimal generator of the unitary group
\(\{\mbU(\epsilon f)\}_{\epsilon\in\smr}\) is \(\pisch(f)\) on the subspace of polynomials, which subspace is a domain of essential self-adjointness for
\(\pisch(f)\). Therefore \(\mbU(f)=\Exp[i\pisch(f)]\).)
The \(\mbU(f)\)  combine with the Weyl operator \(\mbV(f)\Psi=\Exp[{i\phisch(f)}]\Psi\)
to give a representation of the Weyl relations \eqref{weyl}
\(\mbU(g)\mbV(f)=e^{i(f,g)}\mbV(f)\mbU(g)\,.\)
(In \eqref{rndis}, for \(\tilde g=K_0^\frac{1}{2}g\in H^{-\frac{1}{2}}\), the function
 \(\phisch\mapsto\phisch(\tilde g)\) is defined
in \(L^2(\mu_0(d\phisch))\)
as the measurable extension of the function
\(\{\phisch\mapsto\phisch(f)\}\), which is well defined for \({f\in\cs}\), see the discussion in the proof of
Theorem \ref{mac}).


\paragraph{Self-adjointness.}
We recall a result on self-adjointness
from \cite{MR0282243} and \cite{MR0247845}; see also
\cite{MR810217}*{Theorem II.3.1.3}
and \cite{MR0674511}. In the following \(\doublecolon H^{vac}_{0}\doublecolon\)
is the self-adjoint operator discussed above with domain \eqref{domhvac}.
\begin{theorem}\label{sadj}
Given \({\beefb}\in L^1(\R)\cap L^2(\R)\)
the operator obtained by substitution of \eqref{cf1}
into 
  \begin{equation}\label{defhSco}
H^{vac}_{I,g,{\beefb}}(\varphi)\,=\,\int\,\Bigl[\,2mg{\beefb}(x)\varphi^3
\,+\,\frac{1}{2}
g^2{\beefb}(x)\varphi^4\,\Bigr]\,dx\,,
\end{equation}
normal ordering and forming 
\( \doublecolon H^{vac}_{g,{\beefb}}\doublecolon\,=\,\doublecolon H^{vac}_0\doublecolon\,+\,\doublecolon H^{vac}_{I,g,{\beefb}}\doublecolon \) 
  defines an operator which is  bounded below and
  self-adjoint on
  \[
  \Dom(\doublecolon H^{vac}_{g,{\beefb}}\doublecolon)
  =\Dom(\doublecolon H_0^{vac}\doublecolon)\bigcap\Dom(\doublecolon H^{vac}_{I,g,{\beefb}}\doublecolon)\,.
  \]
  \end{theorem}

\subsection{Quantization in the solitonic sector.}\label{qsol}In order to
describe 
the quantum field theory in the solitonic sector,
we take as starting point the expression \eqref{ham-sol} for the classical Hamiltonian expanded
around the soliton, and introduce a spatial cut-off \({\beefb}:\R\to\R\), leading to:
\begin{align}\label{defhs}\begin{split}
    & \frac{\scm_{cl}}{g^2}\,+\,
    H^{sol}_{g,{\beefb}}({\varphi,\pi})\,  =\,  \frac{\scm_{cl}}{g^2}\,+\,H^{sol}_{0}({\varphi,\pi})
  \,+\,H^{sol}_{I,g,{\beefb}}(\varphi)\,,\quad\hbox{where}                  \\
                                 &\qquad H^{sol}_{0}({\varphi,\pi})\;\,=\,\int\,\mcH^{sol}_{0}(\varphi,\pi)\,dx\,=\,
  \frac{1}{2}\,\int\,\Bigl[\,
    \pi^2\,+\,\varphi\,K\,\varphi\,\Bigr]\,dx\,\quad\hbox{and} \\
                                 & \qquad H^{sol}_{I,g,{\beefb}}(\varphi)\,=\,\int \mcH^{sol}_{I,g,{\beefb}}(\varphi)\,dx\,=\,
\int\,\Bigl[\,2mg{\beefb}(x)\tmx\varphi^3
\,+\,\frac{1}{2}{\beefb}(x)
g^2\varphi^4\,\Bigr]\,dx\,,
  \end{split}
\end{align}
which is to be taken as the definition of \(H^{sol}_{g,{\beefb}},\mcH^{sol}_{0}\dots\) etc. This expression
replaces \eqref{defh0} and \eqref{defhSco} in the vacuum case, where
\begin{equation}\label{def-k}
K\,=\,-\partial_x^2+4m^2-6m^2\ssmx=K_0-6m^2\ssmx\,.
\end{equation}
In quantizing this Hamiltonian
two natural possibilities present themselves:
\begin{itemize}
  \item[(i)] treat the \(\ssmx\) term which appears in
    \(H^{sol}_{0}\) perturbatively, and base the quantization on the same
    vacuum sector solution \eqref{cf1}-\eqref{cf2} of the Heisenberg CCR
    \eqref{ccr};
  \item[(ii)] form another soliton sector  solution of the Heisenberg CCR based on
    the operator \(K\) in place of \(K_0\).
 \end{itemize}

The first option is based on the quantum field Hamiltonian
\(\doublecolon H^{vac}_0\doublecolon +\doublecolon \tilde H^{sol}_{I,g,{\beefb}}\doublecolon\), where the latter operator is obtained by substitution of \eqref{cf1}
and then normal ordering the formal expression
\begin{equation}\label{delH}
               \tilde H^{sol}_{I,g,{\beefb}}(\varphi)=H^{sol}_{I,g,{\beefb}}(\varphi)
  -3m^2\int\,\ssmx\,\varphi^2\,dx\,;
  \end{equation}
this is convenient for existence theory, and leads to self-adjointness results etc
as a direct consequence of the classic results reviewed in \cites{MR0674511,MR810217}
just like Theorem \ref{sadj}. The second option
allows an explicit analysis of the semiclassical limit, so we will make
use of both. It is important that
these two solutions of \eqref{ccr} are unitarily equivalent, so that both
quantizations refer to the same theory - this issue is addressed below in
Theorems \ref{mac} and \ref{uniteq}. Next we describe the two approaches
in detail.

\subsubsection{Soliton quantization using vacuum sector solution of CCR.}

In this approach we continue to use the same solution \eqref{cf1}-\eqref{cf2}
of the CCR, but shifted by the classical soliton profile \(\Phi_S\).
Explicitly these are, in full,
\begin{align}\la{cfxi1}
\Phi(x{})   & =\Phi_S(x)+\varphi(x{})\,=\,\Phi_S(x)+\frac{1}{\sqrt{2\pi}}\,\int\,\frac{1}{\sqrt{2\omega^{}_k}}\,
\bigl(a^{}_ke^{ik{x}}+a_k^\dagger e^{-ik{x}}\bigr)\,dk\,,\hbox{ and } \\
\Pi(x{})    & =\pi(x{})\,=\,\frac{1}{\sqrt{2\pi}}\,\int\,-i\,\sqrt{\frac{\omega^{}_k}{2}}\,
\bigl(a^{}_ke^{ik{x}}-a_k^\dagger e^{-ik{x}}\bigr)\,dk\,,\la{cfxi2}
\end{align}
acting on the Hilbert space \({\fock}\) defined in
\eqref{fock}.
An alternative way of giving the representation, which is useful for comparison
with other representions, is to use the definitions \eqref{alpha0} and pair them with a real Schwartz function
\begin{align}\notag
\Phi(f{})\, & =\,\int\Phi_S(x)f(x)dx\,+\,\int\,\frac{1}{\sqrt{2\omega^{}_k}}\,
\Bigl(a^{}_k\,\hat f(-k{})+a_k^\dagger\,\hat f(k{})\Bigr)\,dk\\& =\int\Phi_S(x)f(x)dx\,+
\,\frac{1}{\sqrt{2}}\,\Bigl(\alpha({K_0^{-1/4}f})+\alpha^\dagger(K_0^{-1/4}f)
\Bigr)
\,,\label{cfxialt1}
\\
\Pi(f{})\,  & =-i\,\int\,\sqrt{\frac{\omega^{}_k}{2}}\,
\Bigl(a^{}_k\,\hat f(-k{})-a_k^\dagger\,\hat f(k{})\Bigr)\,dk\,=
\,-\frac{i}{\sqrt{2}}\,\Bigl(\alpha ({K_0^{1/4}f})-\alpha^\dagger(K_0^{1/4}f)\Bigr)
\,.\label{cfxialt2}
\end{align}

\begin{theorem}[Self-adjointness]\label{sadj2}
  (i) The quadratic Hamiltonian in the solitonic sector obtained by
normal ordering the classical expression
\(      
\frac{1}{2}\,
\int(
\pi^2\,+\,\varphi K\varphi)\,dx
\)
with respect to the representation \eqref{cfxi1}-\eqref{cfxi2},
namely,
  \[\doublecolon H^{sol}_{0}\doublecolon \,=\,\doublecolon H^{vac}_{0}\doublecolon +v(\varphi)\,,\qquad
v(\varphi)\,=\,-\,\doublecolon \int\,3m^2\,\ssmx\,\varphi(x)^2\,dx\doublecolon 
  \] is well-defined on the domain
  \(\Dom(\doublecolon H^{vac}_{0}\doublecolon)\subset{\fock}\), and is 
essentially self-adjoint on this domain with self-adjoint
  extension (also written \(\doublecolon H^{sol}_0\doublecolon\)). This operator 
  verifies the lower bound
\(\doublecolon H^{sol}_{0}\doublecolon\geq\Delta\M_{scl}=\frac{m}{\sqrt{3}}-\frac{6m}{\pi}\,.\)
The domains of the self-adjoint extensions satisfy
\begin{equation}\label{domains}
\Dom(\doublecolon H^{sol}_{0}\doublecolon)\cap\Dom(\N_0)\subset\Dom(\doublecolon H^{vac}_{0}\doublecolon)\,.
  \end{equation}

(ii) Let \({\beefb}\in L^1(\R)\cap L^2(\R)\) and assume there exists a positive constant \(\delta\) such
  that \({\beefb}(x)\geq \delta\,\ssmx\) for all \(x\).
  Then the formal Hamiltonian \eqref{defhs}
defines, after substituting \eqref{cfxi1}-\eqref{cfxi2}, normal ordering with  and taking the operator sum,
an unbounded operator which equals
\(\doublecolon H^{sol}_{0}\doublecolon+\doublecolon H^{sol}_{I,g,{\beefb}}(\varphi)\doublecolon\)
on
\(\Dom(\doublecolon H^{sol}_0\doublecolon )\cap\Dom( \doublecolon H^{sol}_{I,g,{\beefb}}(\varphi)\doublecolon)
\)
and in particular on the polynomial subspace \(\Poly(\varphi)\).
It is bounded below and has a self-adjoint extension \(\doublecolon {H}^{sol}_{g,\beefb}\doublecolon\)
with domain \(\Dom(\doublecolon H^{vac}_0\doublecolon )\cap\Dom(
\doublecolon \tilde H^{sol}_{I,g,{\beefb}}(\varphi)\doublecolon)\).

(iii) The isomorphism \(\I\) in Proposition \ref{sch} maps these to self-adjoint operators on
\(L^2(\mu_0)\), which will be denoted \(\doublecolon {\pmb H}^{vac}_0\doublecolon\),
\(\doublecolon {\pmb H}^{sol}_{0}\doublecolon\), \(\doublecolon \pmb{H}^{sol}_{g,\beefb}\doublecolon\)
and \( \doublecolon \tilde H^{sol}_{I,g,{\beefb}}(\phisch)\doublecolon
\) etc.

(iv) Identical results hold after expanding about a soliton centered at arbitrary \(\xi\in\R\) and lead to
operators written \(\doublecolon\pmb{H}^{sol}_{0\xi}\doublecolon\) etc.
\end{theorem}
\proof
(i) The essential self-adjointness assertion is a consequence of
\cite{MR751959}*{Theorem X.14} given that it follows from \eqref{best} that the operator
\(\int\,6m^2\,\doublecolon\,\ssmx\,\varphi(x)^2\,\doublecolon\,dx\)
is bounded on \(\Dom({\N_0})\supset\Dom(\doublecolon H^{vac}_{0}\doublecolon)\), and so is a relatively
bounded perturbation of \(\doublecolon H^{vac}_0\doublecolon\) and is well defined on
\(\Dom(\doublecolon H^{vac}_{0}\doublecolon)\); it is also a consequence of
the result in this reference that any core for \(\doublecolon H^{vac}_{0}\doublecolon\) is a core
for \(\doublecolon H^{sol}_{0}\doublecolon\).
The precise lower bound is proved in \S\ref{mass-shift}, together with a determination of the
domain of the self-adjoint extension in Theorem \ref{scconv}. The inclusion
\eqref{domains} can be deduced from the
Duhamel formula
\begin{equation}\label{duh0}
  \Exp[-t\doublecolon H^{vac}_{0}\doublecolon]u_0\,=\,
  \Exp[-t\doublecolon H^{sol}_{0}\doublecolon]u_0\,-\,
  \int_0^t
\Exp[-(t-s)\doublecolon H^{sol}_{0}\doublecolon]
\bigl(\doublecolon H^{sol}_{0}\doublecolon-\doublecolon H^{vac}_{0}\doublecolon\bigr)
\Exp[-s\doublecolon H^{vac}_{0}\doublecolon]u_0\,ds\,,
  \end{equation}
as follows. Now the operator
\((\doublecolon H^{sol}_{0}\doublecolon-\doublecolon H^{vac}_{0}\doublecolon)\N_0^{-1}\)
is bounded by \eqref{best}, and the number representation simultaneously diagonalizes the operators
\(\N_0\) and \(\doublecolon H^{vac}_{0}\doublecolon\) so that
\(\Exp[-s\doublecolon H^{vac}_{0}\doublecolon]:\Dom(\N_0)\to\Dom(\N_0)\) and
\([\N_0,\Exp[-s\doublecolon H^{vac}_{0}\doublecolon]=0\) in the strict sense,
so by strong continuity of the semigroups we can take the limit of  \(t^{-1}\times\)
the final term in \eqref{duh0}
\[
\lim_{t\to 0+}\,t^{-1}\int_0^t
\Exp[-(t-s)\doublecolon H^{sol}_{0}\doublecolon]
\bigl(\doublecolon H^{sol}_{0}\doublecolon-\doublecolon H^{vac}_{0}\doublecolon\bigr)
\Exp[-s\doublecolon H^{vac}_{0}\doublecolon]u_0\,ds\,=\,
\bigl(\doublecolon H^{sol}_{0}\doublecolon-\doublecolon H^{vac}_{0}\doublecolon\bigr)u_0
\]
for any \(u_0\in\Dom(\N_0)\). It then follows from \eqref{duh0} that for such \(u_0\)
\(\lim_{t\to 0+}(\Exp[-t\doublecolon H^{sol}_{0}\doublecolon] -\id)u_0\) exists if and only if
\(\lim_{t\to 0+}(\Exp[-t\doublecolon H^{vac}_{0}\doublecolon ]-\id)u_0\) exists, which implies \eqref{domains}.
(It should be said that \eqref{duh0} holds under the assumption only that \(u_0\in\Dom(\N_0)\) by approximation
from the case \(u_0\in\Dom(\doublecolon H^{vac}_{0}\doublecolon)\subset \Dom(\N_0)\) - the possibility of which
approximation follows from the aforementioned simultaneous diagonalizability of these two operators).

(ii) As for Theorem \ref{sadj}, the assertion in (ii) follows from
classic self-adjointness results,
applied to the Hamiltonian \(H^{vac}_0+\tilde H^{sol}_{I,g,{\beefb}}(\varphi)\) after normal ordering,
see \eqref{delH}.
The semi-boundedness condition on the perturbing polynomial
now takes the requirement that
\(-6m^2\ssmx\varphi^2+2mg{\beefb}(x)\tmx\varphi^3
+\frac{1}{2}g^2{\beefb}(x)\varphi^4\)
be bounded below. This will be met if
\begin{equation}\label{lowerbc}
  {\beefb}(x)\geq\delta\ssmx\quad\hbox{holds everywhere, for some}\; \delta>0\,.
\end{equation}
Under this condition
there is a lower bound \eqref{klb} for the regularized interaction which is
sufficient to ensure that \(\Exp[-t\doublecolon \tilde H^{sol}_{I,g,{\beefb}}\doublecolon]\) is integrable
(in the Schr\"odinger representation) and hence that the closure of the
operator sum defines a self-adjoint operator, see
\cite{MR0493420}*{Theorem X.59}, and
it is bounded below by Theorem X.58 in the same reference, while the domain
of self-adjointness can be determined from \cite{MR0247845}, as in Theorem \ref{sadj}.
\qed
\begin{remark}
  \label{trans1}
  Notice that the representation \eqref{cfxi1}-\eqref{cfxi2}
  differs from \eqref{cf1}-\eqref{cf2} by a displacement by the
  classical soliton profile \(\Phi_S\).
  This operation is not unitarily implementable
  because \(\Phi_S\) is not in the Cameron-Martin
  space \(H^{\frac{1}{2}}\), see the  discussion following Proposition \ref{sch}).
  \end{remark}

\subsubsection{Soliton quantization using soliton sector solution of CCR.}\label{subsubq}
The formulae \eqref{segal} 
defining the free field have to be modified to take account of the 
different spectral properties of \(K(\xi) \) as compared to \(K_0\,. \)
For present purposes it is clearest to work with the special case \(\xi=0\), and necessary
modifications for general \(\xi\) can be made as needed. (Overall translation invariance means that
if the soliton is translated by \(\xi\in\R \), as explained
prior to \eqref{mcl}, then the corresponding eigenfunctions are translated by \(\xi \) also.
The Jost eigenfunctions take on an additional phase factor to maintain the
normalization \(e_{k\xi}(x)\sim e^{ikx}\). Overall,
this means that, for each \(\xi\in\R \), the spectral resolution of
the operator 
\(K(\xi)=(-\partial_x^2+4m^2-6m^2\ssmxi) \)
can be deduced immediately from that of the operator \(K=K(0)\), on
which we concentrate. (Also note that \(K_0\), with zero as subscript, refers to the vacuum operator
\(-\partial_x^2+4m^2\), not \(K(0)\)!)
The operator \(K \) is a non-negative self-adjoint 
operator on \(L^2(\R) \) with domain \(\Dom (K)=H^2(\R) \).
Recall, from the discussion preceding \eqref{simh}, that the spectrum of \(K \) consists of
(i) a one-dimensional
kernel \(\langle\{ {\tteo}\}\rangle \); 
(ii) one simple discrete nonzero eigenvalue \(3m^2 >0\), 
\[
K {\tteone}\,=\,\omega_d^2 {\tteone}\,,\qquad \omega^{}_d=\sqrt{3}m
\]
with corresponding spectral subspace 
\(\langle\{ {\tteone}\}\rangle \) and (iii) continuous spectrum \([4m^2,+\infty)\,. \)
In addition to the \(L^2(\R) \) eigenfunctions \( {\tteo}\in\cs(\R)\) 
and \( {\tteone}\in\cs(\R)\), there
are generalized eigenfunctions \(e_k\in L^\infty(\R)\cap C^\infty(\R) \) which satisfy
\[
Ke_k=(k^2+4m^2)e_k\,,\qquad \hbox{and}\quad e_k(x)\sim e^{ikx}\quad (x\to+\infty)\,.\] 
See the appendix for explicit formulae.
Spectral decomposition
provides an integral representation for \(U\in L^2(\R) \), which can be
given explicitly as
\begin{align}\label{gfe}
U(x)\,=\,\bigl( {\tteo}{},U\bigr)_{L^2} {\tteo}(x{})
 & +\bigl( {\tteone}{},U\bigr)_{L^2} {\tteone}(x{}) \\
 & \quad+\frac{1}{2\pi}\,\iint_{\smr\times\smr}
\,{e_{-k}}(y{})U(y)e_k(x{})\,dy\,dk\,.
\notag\end{align}
It is useful to define, associated to the potential \(u(x)=-6\ssmx\),
the {\em distorted Fourier transform}
\(\mcF_u:U\mapsto\tilde U \) by
\begin{equation}\label{dft}
  \tilde U(k)\,=\,
\mcF_u(U)(k)
\,=\,  (2\pi)^{-1/2}\int\,{e_{-k}}(x{})U(x)\,dx \,,
  \end{equation}
  where \(e_{-k}\) is the Jost eigenfunction introduced in the appendix;
the same works for the translated potential \(u_\xi(x)=-6\ssmxi\),
to define \(\mcF_{u_\xi}\) using \(e_{-k\xi}\), see \S\ref{diffrep}.
The distorted Fourier transform maps the Schwartz space into itself.
Restricting for simplicity to \(\xi=0\),
\(\mcF_u\) admits as right inverse
the map \(\mcF_u^{-1}:f\mapsto \breve{f}\), where
\begin{equation}\label{dfti}
  {\breve f}(x)\,=\,
\mcF_V^{-1}(f)(x)\,=\,
  (2\pi)^{-1/2}\int\,e_{k}(x{})f(k)\,dk\,,
\end{equation}
which also maps the Schwartz space into itself
and which extends, by \eqref{comprel}, to define a partial isometry 
whose initial space is \(L^2(dk) \) 
and whose final
space is the subspace
\(\P_cL^2(\R)=\langle\{ {\tteo}, {\tteone}\}\rangle^\perp \subset L^2(\R)\), i.e.,
the \(L^2\)-orthogonal complement
of the discrete spectral subspace. As with the Fourier transform, the reality
of \(U\) shows up as the condition \(\overline{\tilde U}(-k)=\tilde U(k)\). Of course
all this extends to general \(\xi\in\R\).

These facts form the basis for the construction of
a set of solutions of the Heisenberg
relations \eqref{ccr} of the form
\begin{equation}\label{nsh}
\Bigl(\Phi(x),\Pi(x)\Bigr)=\Bigl(\Phi_S(x{})+\upphi(x),\uppi(x)\Bigr) 
  \end{equation}
different to \eqref{cfxi1}-\eqref{cfxi2}.
Before giving the full expression \eqref{cfseg1}-\eqref{cfseg2},
it is useful to explain how
this solution is built up. It is supposed to describe a quantum mechanical
particle (the kink) with momentum
\(P \), interacting with the oscillatory mode of frequency \(\sqrt{3}m \)
and the radiation modes associated to the continuous spectrum
\([4m^2,+\infty)\,. \) 
To describe an isolated quantum particle, we could make use of
a pair of operators \((Q,P) \) which satisfy \([Q,P]=i \), and act
on the space \(L^2(\R,dQ) \) with \(Q \) as the (unbounded) operator of
coordinate multiplication, i.e. \(Q:g(Q)\mapsto Qg(Q) \),
and \(P=-i\partial_Q \); slightly more generally \([Q,\eta-i\partial_Q]=i \) for
any constant \(\eta\).
For the case at hand, we will use such a pair of 
operators to describe the kink; its centre being described by the operator  
of multiplication by \(Q\),
and represents quantum mechanical fluctuations around the
classical location of the kink at the origin; these are small in an appropriate sense
when \(g\) is small.
The remaining modes are described by the ``fluctuation'' fields around the
soliton \((\upphi,\uppi) \), given by formulae analogous to 
\eqref{segal}.
Define a new Fock space as the complete
direct sum
\begin{equation}\label{newfock}
\bigoplus\;\Symn\P_0^\perp(L^2(\R,dx))\,,
\end{equation}
constructed this time out of the Hilbert space of square integrable 
fluctuations of the kink which are
orthogonal to the infinitesimal translations \( {\tteo} \), i.e., \(\P_0\)
is the orthogonal projector onto this subspace so that
\[
\P_0^\perp(L^2(\R,dx))\,=\,\langle\{ {\tteo}\}\rangle^\perp\,\subset\, L^2(\R,dx)\,,
\]
and the vacuum in \eqref{newfock} is \((1,0,\dots)\).
The creation/annihilation operators
\(\upalpha{\phxi}^\dagger,\upalpha{\phxi} \) act on \(\mfrF\), and the corresponding
{\em transverse} fluctuation fields are given by
\begin{equation*}
\upphiperp(f)\,=\,
\frac{1}{\sqrt{2}}\,\Bigl(\upalpha{\phxi}(K{\phxi}^{-1/4}f)
+\upalpha{\phxi}^\dagger(K{\phxi}^{-1/4}f)\Bigr)\quad\hbox{  and  }
\quad\uppiperp(f)\,=\,
-\frac{i}{\sqrt{2}}
\Bigl((\upalpha{\phxi}(K{\phxi}^{1/4}f)-\upalpha{\phxi}^\dagger(K{\phxi}^{1/4}f)\Bigr)\,,
\end{equation*}
for \(f\in \cs(\R)\cap \P_0^\perp(L^2(\R))\), in analogy to \eqref{segal}. Crucially,
\(\P_0^\perp(L^2(\R))\) is an invariant spectral subspace for \(K\) on which it is
strictly positive with bounded inverse \(C:\P_0^\perp(L^2(\R))\to \P_0^\perp(L^2(\R))\), so working on 
this subspace \(K^{-1}=C\). The symbols \(\upalpha\,\upalpha^\dagger\), which are written in an upright font,
are the generalization of the symbols in \eqref{alpha0} to the solitonic representation, see
\eqref{compme}.

Now we form a solution of the Heisenberg relations \eqref{ccr}.
This is achieved by the following definition
of quantum fields given, as operator-valued distributions, by
\begin{align}\la{cfseg1}
\upphi(f)\,            & =\,-\bigl( {\tteo},f\bigr)_{L^2}{\sqrt{\scm_{cl}}}\,Q\,+\,
\frac{1}{\sqrt{2}}\,\Bigl(\upalpha{\phxi}(C{\phxi}^{1/4}f_{\perp})
+\upalpha{\phxi}^\dagger(C{\phxi}^{1/4}f_{\perp})\Bigr)\,,
                                                                                         \\
\uppi(f)\,             & =\,-\frac{P}{\sqrt{\scm_{cl}}}\bigl( {\tteo},f\bigr)_{L^2}
-\frac{i}{\sqrt{2}}
\Bigl((\upalpha{\phxi}(K{\phxi}^{1/4}f_{\perp})-\upalpha{\phxi}^\dagger(K{\phxi}^{1/4}f_{\perp})\Bigr)\,,
\la{cfseg2}
\end{align}
where \(f\in\cs(\R) \) and 
\(f_{\perp}=f-\bigl(f, {\tteo}\bigr)_{L^2} {\tteo} \)
is the component of \(f\) in \(\P_0^\perp(L^2(\R))\,. \)
The commutation relation reads
\begin{align*}
[\upphi(f),\uppi(g)]\, & =\,
\Bigl([Q,P]\bigl(f, {\tteo}\bigr)_{L^2}
\bigl(g, {\tteo}\bigr)_{L^2}
+\frac{i}{2}\bigl[\upalpha{\phxi}(C{\phxi}^{1/4}f_{\perp}),\upalpha{\phxi}^\dagger(K{\phxi}^{1/4}g_{\perp})\bigr] \\
                       & \phantom{=\,
\Bigl([Q,P]\bigl(f, {\tteo}\bigr)_{L^2}
\bigl(g, {\tteo}\bigr)_{L^2}}
\qquad-\frac{i}{2}
\bigl[\upalpha{\phxi}^\dagger(C{\phxi}^{1/4}f_{\perp}),\upalpha{\phxi}(K{\phxi}^{1/4}g_{\perp})\bigr]
\Bigr)
                                                                                         \\
\,                     & =\,i\,\bigl(f,g\bigr)_{L^2}\,.
\end{align*}
Notice here that a quantum fluctuation operator for the position of the kink has been introduced as
  \(
  Q=-{\scm_{cl}}^{-\frac{1}{2}}\upphi( {\tteo})\);
  basically up to scaling \(Q\) is identified with
  the field paired against the zero mode
  \( {\tteo}\).
A more explicit form
is obtained from \eqref{cfseg1}-\eqref{cfseg2} by extracting the
test functions, leading to:
\begin{align}\la{cfsgext}\begin{split}
  \upphi(x)\,          & =\,-{\sqrt{\scm_{cl}}}
   {\tteo}({}x{})\,Q\,
  +\frac{1}{\sqrt{2\omega^{}_d}}
 {\tteone}({}x{})\,(a^{}_{d}+a_{d}^\dagger)                                          \\
                       & \phantom{=\,\Phi_Sx{}-X {\tteo}x{}}
+\frac{1}{\sqrt{2\pi}}\,\int\,\frac{1}{\sqrt{2\omega^{}_k}}\,
\bigl(a^{}_{k}e_k({}x{})
+a_{k}^\dagger {{e_{-k}}}({}x{})\bigr)\,dk\,,                      \\
\uppi(x)\,             & =\,-\frac{1}{\sqrt{\scm_{cl}}}
 {\tteo}({}x{})\,P\,-
i\sqrt{\frac{\omega^{}_d}{2}}
(a^{}_{d}-a_{d}^\dagger) {\tteone}({}x{})\\
                       & \phantom{===\,\Phi_Sx{}-pppp}+
\frac{1}{\sqrt{2\pi}}\,\int\,-i\sqrt{\frac{\omega^{}_k}{2}}\,
\bigl(a^{}_{k}e_k({}x{})
-a_{k}^\dagger {{e_{-k}}}
({}x{}\bigr))\,dk
\,.
\end{split}\end{align}
The operators \(a^{}_{d},a_{d}^\dagger \) are annihilation and
creation operators for the discrete mode with frequency \(\omega^{}_d\).
The \(a^{}_{k},a_{l}^\dagger \) satisfy \([a^{}_{k},a_{l}^\dagger]=\delta(k-l) \)
(which holds in the same sense as the corresponding relation in the vacuum sector).
The operators 
\(a^{}_{d},a_{d}^\dagger,a^{}_{l},a_{l}^\dagger \) 
can be related to the \(\upalpha{\phxi}^\dagger,\upalpha{\phxi} \)
via the formulae (which define operator-valued distributions):  
\begin{align}\label{compme}
\upalpha(f)\,=\,( {\tteone},f)\,a^{}_d\,+\,\int\,\tilde f(-k)a^{}_k\,dk\,,\qquad\qquad
\upalpha^\dagger(f)\,=\,( {\tteone},f)\,a^\dagger_d\,+\,\int\,\tilde f(k)a^\dagger_k\,dk\,,\qquad\qquad f\in\cs(\R): ( {\tteo},f)=0 \,,
\end{align}
in which the {\em distorted Fourier transform} replaces the ordinary Fourier transform appearing in the
equation preceding \eqref{segal}.
The operators \(\upalpha(f)\,\upalpha^\dagger(f)\)
retain analogous versions of the three properties noted after \eqref{segal}.
The Heisenberg relation is a consequence of the completeness relation \eqref{comprel}:
\begin{align}
[\Phi(x),\Pi(y)]\,                                                                            & =\,i\, {\tteo}(x) {\tteo}(y)\,
+\,i\, {\tteone}(x) {\tteone}(y)\,+\,\frac{i}{2\pi}\,\int_{\smr}e_k(x)
{{e_{-k}}}(y)\,dk\,\notag \\ & =\,i\delta(x-y)\,.\notag
\end{align}
For comparison with \eqref{cfxialt1}-\eqref{cfxialt2}, the representation can be
written, after pairing with a Schwartz test function \(U\),
\begin{align}\la{cfsgextalt}
  \upphi(U)\, & =\,-{\sqrt{\scm_{cl}}}
  \bigl( {\tteo},U\bigr)\,Q
  +\frac{1}{\sqrt{2\omega^{}_d}}
\bigl( {\tteone},U\bigr)\,(a^{}_{d}+a_{d}^\dagger) \\
              & \phantom{=\,\Phi_Sx-X {\tteo}x mmmmmmmmmmm}
+\,\int\,\frac{1}{\sqrt{2\omega^{}_k}}\,
\Bigl(a^{}_{k}{\tilde U(-k)}
+a_{k}^\dagger \tilde U(k)\Bigr)\,dk\,,\notag      \\
\uppi(U)\,    & =\,-\frac{1}{\sqrt{\scm_{cl}}}
\bigl( {\tteo},U\bigr)\,P\,
-i\sqrt{\frac{\omega^{}_d}{2}}
\bigl( {\tteone},U\bigr)\,(a^{}_{d}-a_{d}^\dagger)
\la{cfsg22alt}                                     \\
              & \phantom{===\,\Phi_Sx-\xi-mmmmmmmmmmmmmm}+
\,\int\,-i\sqrt{\frac{\omega^{}_k}{2}}\,
\Bigl(a^{}_{k}{\tilde U(-k)}
-a_{k}^\dagger\tilde U(k)
\Bigr)\,dk\notag
\,.
\end{align}
Observe that if \(U \) is orthogonal to \( {\tteo} \) and \( {\tteone}\) then the field
operator creates  particle in this state through the operator
\(
\int{({2\omega^{}_k})^{-\frac{1}{2}}}
a_{k}^\dagger \tilde U(k)dk\)
(in place of
  \(\int\,{(2\omega^{}_k)}^{-\frac{1}{2}}a_k^\dagger\,\hat U(k)dk\)
in the vacuum representation).
The replacement in the integral of the Fourier transform by
the distorted Fourier transform \eqref{dft} provided
by scattering theory indicates that this representation is describing the same quantum particles
(bosons) as in the free case \eqref{cfxialt1}-\eqref{cfxialt2}
by using the scattering theory to map them on to free bosons. (The remaining 
terms (in addition to the integral) in these formulae give the
standard  quantum mechanical quantization of
the discrete spectrum.)

The linear space of fluctuations about the soliton
\(
\P_0^\perp(L^2(\R,dx))\,=\,\langle\{ {\tteo}\}\rangle^\perp\,\subset\, L^2(\R)
\)
admits, by \eqref{gfe}-\eqref{dft}, an isometric isomorphism
\begin{align}\label{nxi}
  \P_0^\perp(L^2(\R))\, & \to\,\R\oplus L^2(\R,dk) \\
  U                     & \mapsto \Bigl(\bigl( {\tteone},U\bigr),\tilde U(k)\Bigr)\,.\notag
  \end{align}
Introducing the coordinate operator \(q_d\propto\upphi( {\tteone})\) and applying
second quantization shows that \(\mfrF \)
can be realized as
a tensor product space: there is a unitary equivalence
\begin{equation}\label{l2g}
\bigoplus\limits_{n=0}^\infty\Symn \P_0^\perp(L^2(\R,dx))
  \,\to\,L^2(\R,\pi^{-\frac{1}{2}}\omega^{\frac{1}{2}}_d\exp{[-\omega^{}_d q_d^2]}\,dq^{}_d)
\,\otimes\,\bigoplus\limits_{n=0}^\infty\Symn L^2(\R,dk)
\,=\,L^2(\R,\gamma_d)\otimes{\fock}\define\mfrF\,,
\end{equation}
under which \(a^{}_{d}\,\) maps to \(\frac{1}{\sqrt{2\omega^{}_d}}\partial_{q_d}\,,\)
\(a_{d}^\dagger\,\) maps to \(\frac{1}{\sqrt{2\omega^{}_d}}
(2\omega^{}_d \,q^{}_d
-\partial_{q_d})\,\) and the vacuum in \eqref{newfock}
maps to \(\Omega'=\id_{\smr}\otimes\Omega_0\) where \(\id_{\smr}=\id_{\smr}(q_d)\) is the function identically equal to \(1 \)
in \(L^2(\gamma_d)\); this is the {\em transverse } vacuum. We introduce
a number operator
\begin{equation}\label{numb}
{\N}\,=\,a_{d}^\dagger a^{}_{d}\,+\,\int a^\dagger_k a^{}_k\,dk \,.
\end{equation}

\begin{remark}\label{lpart} The generalization of the finite particle subspace \({\Fin_0(\cs)}\subset\fock\) is
  \(\Fin(\cs)=\C[q_d]\otimes{\Fin_0(\cs)}\subset\mfrF\),
  the algebraic span of \(q_d^{n_d}\Psi_n\) with symmetric \(\Psi_n\in\cs(\R^n)\) and \(l=n+n_d\) finite, and
  this generalizes to \(\Fin(L^2)\) in an obvious way. We shall say
  \(I\ni s\mapsto F(s)\in\Fin(L^2)\) is \(C^1\) on an interval \(I\) if the preceding finite particle condition
  holds for all \(s\in I\) with the same
  \(l\) and with \(s\mapsto \Psi_n(s)\in\L^2(\R^n)\) and all coefficient functions \(C^1\).
A slightly smaller subspace than \(\Fin(\cs)\) is 
\(\Poly(\upphiperp)\subset\mfrF\) which is also the algebraic
span of \(\He_{n_d}(\sqrt{2\omega_d}q_d)\Symn\prod_{j=1}^n f_j(k_j)\, \) where
the \(\{f_j\} \) are Schwartz functions; the spanning elements lie in the \(l\)-particle subspace of \(\mfrF\) for
\(l=n_d+n\) (the kernel of \(\N-l\)). Tensoring with appropriate wave packets
in the soliton coordinate \(Q\) gives a subspace closely related to the
subspace spanned by polynomials in the Schr\"odinger representation, see Corollary \ref{uniteqcor}.
The Fock space polynomial bounds \eqref{best}-\eqref{best2} can be applied to physical space
transverse field polynomials given the isometric property of the
distorted Fourier transform. Thus consider, for symmetric \(v\in L^2\)
\[P(\upphiperp)=\triplecolon\int v(x_1,\dots x_l)\prod_{j=1}^l\upphiperp(x_j)dx_j\triplecolon\,,\]
which leads us to consider a sum of terms of the form \((a_d^{\dagger})^{m_d}a_d^{n_d}\Op(w)\) with
notation from \eqref{wickop}, and \(m_d+n_d+m+n=l\), with for example
\[
w(k_1,\dots k_m)=(2\omega_d)^{-r}\prod_{j=1}^m(4\pi\omega_{k_j})^{-1/2}\int v(x_1,\dots x_l)
e_{-k_{1}}(x_1)\dots e_{-k_{m}}(x_m)
\tteone(x_{m+1})\dots\tteone(x_l)\prod_{j=1}^l dx_{j}
\]
when \(m_d=n_d=r,n=0,m=l-2r\), and analogous formulae in the other cases. In all cases the  isometric properties
of \eqref{dft} imply \(\|w\|_{L^2}\leq const. \|v\|_{L^2}\). This in turn implies the following non-optimal bounds for
the operator norm on \(\mfrF\), generalizing
\eqref{best2}:
\begin{equation}\label{best25}
  \max\Bigl\{\|(\id+{\N})^{-l/2}P(\upphiperp)\|\,,\,\|P(\upphiperp)(\id+{\N})^{-l/2}\|\Bigr\}
  \,\leq\,const.\,(1+l)^{l/2}\|v\|\,.
\end{equation}
where, on the right hand side \(\|v\| \) means the \(L^2\) norm of the symmetric
function \(v\) which determines the polynomial as above.
\end{remark}
Now to describe the full quantization of the soliton, using
these two ingredients,
we form the total Hilbert space as in \eqref{aos}.
Substituting \eqref{cfsgext} into the Hamiltonian and normal ordering gives (formally)
\(\frac{\scm_{cl}}{g^2}+\triplecolon H^{sol}_{0}\triplecolon +O(g) \)
with
\begin{equation}\triplecolon H^{sol}_{0}\triplecolon 
\,=\,\frac{1}{2}\,\int\,\triplecolon\Bigl[\,
  \uppi^2\,+\,\upphi\,K\,\upphi\,\Bigr]\triplecolon\,dx\,.\label{fsh}\end{equation}
\begin{remark}\label{triplecolon}
  Normal ordering in the solitonic representation is only applied in the transverse degrees of freedom, i.e.,
  to \(\upphiperp,\uppiperp\), leaving alone the quantum variables describing the soliton \(Q,P\). This will
  be maintained in the solitonic Schr\"odinger representation.
  \end{remark}
\begin{lemma}\label{wbf}
  Substitute regularized versions of \eqref{cfsgext} into \eqref{fsh}
  and interpret the resulting expression as a bilinear form valued integral on \({\Poly(\varphi)}\times{\Poly(\varphi)}\),
  and take the limit in the weak topology. Then
  \begin{align}\label{bilinear}
    \begin{split}
\triplecolon H^{sol}_{0}\triplecolon\, & 
=\,
\frac{P^2}{2\scm_{cl}}\,+\,\omega^{}_d a_{d}^\dagger a^{}_{d}
\,+\,
\int\omega^{}_k a_{k}^\dagger a^{}_{k}
\,dk                                                          \\
                                       & =\,\frac{P^2}{2\scm_{cl}}\,+\,{h}^{}_d
\,+\,
    {{\h}}\,,
    \end{split}
\end{align}
where \({h}^{}_d={h}(\omega^{}_d)\) is the Hamiltonian for a one dimensional
oscillator with frequency \(\omega^{}_d\) and \(\h\) is as in \eqref{assemb}.
\end{lemma}
\proof
This expression could be obtained in the same way as the corresponding result for the
vacuum representation, \cite{MR0674511}*{Theorem 4.4}, but making
use of the properties of the eigenfunction expansion given in \S\ref{eigen} in place
of the Fourier transform. In the context of this paper it is most natural to carry out the derivation
as stated, but defining the Hamiltonian by taking the limit of regularized expression defined via
regularized fields, see Step two in the proof of Theorem \ref{scconv} in \S\ref{qtt} for the details. \qed

The operator appearing the  previous Lemma is quadratic and generates a unitary evolution on the
space \(L^2(\R,dQ)\otimes\mfrF\) which, under the description above,
can be given as
\begin{equation}\label{sclev}
\Exp[-it\triplecolon H^{sol}_{0}\triplecolon]\,=\,
\Exp[-it\frac{P^2}{2\scm_{cl}}]\otimes
\exp[-it\omega^{}_d]\otimes
\Gamma\Bigl(\exp[-it\omega^{}_\bullet]\Bigr)\,,
\end{equation}
where the \(\Gamma\) notation in the final line stands for the
transformation on \(\bigoplus\Symn L^2(\R,dk)\) induced by the
map \(\exp[-it\omega^{}_\bullet]:f(k)\mapsto \exp[-it\omega^{}_k]f(k)\),
which is unitary on \( L^2(\R,dk)\), see \cite{MR0489552}*{Chapter 1}.
\begin{remark}\label{msa}
  The (closure of the) operator \(\triplecolon H^{sol}_{0}\triplecolon\) is self-adjoint. To specify its
  domain decompose, as in Remark \ref{lpart},  simultaneously with respect to the operators \(h^{}_d\) and
  the number operator \({\N_0}\), so that \(\Psi\) corresponds to the sequence
  \(
  \{\sum_l\Psi_{n,l}\He_l(\sqrt{2\omega_d}q_d)\}
  \)   
  and each \(\Psi_{n,l}=\Psi_{n,l}(Q,k_1,\dots k_n)\) is symmetric in \(k_1,\dots k_n\).
  Then \((h^{}_d+{{\h}})\Psi\) corresponds to
  the sequence 
  \(
  \{\sum_l(l\omega_d+\sum_{i=1}^n\omega_{k_i}^{})\Psi_{n,l}\He_l(\sqrt{2\omega_d}q_d)\}
  \)
  and
  \begin{align}\label{domhtrip}
  \Dom\triplecolon H^{sol}_{0}\triplecolon\,=\,
  \biggr\{
  \Psi:\sum_{n,l}\Bigl(\|(l\omega_d+\sum_{i=1}^n\omega_{k_i}^{})\Psi_{n,l}\|_{L^2(dQdk)}^2
  +\|(l\omega_d+\sum_{i=1}^n\omega_{k_i}^{})^{\frac{1}{2}} & \frac{d\Psi_{n,l}}{dQ}\|_{L^2(dQdk)}^2 \\
                                                           & +\|\frac{d^2\Psi_{n,l}}{dQ^2}\|_{L^2(dQdk)}^2
  \Bigr)<\infty
  \biggl\}\,.
  \end{align}
  The Fock space \(\mfrF\) is spanned by \(\He_l(\sqrt{2\omega_d}q_d)F_n(k_1,\dots k_n)\)
  for square integrable \(F_n\) symmetric in \(k_1,\dots k_n\), and
  \((h^{}_d+{{\h}})\) is strictly positive on its domain and diagonal on this spanning set, so that (for example) its resolvent, \((\lambda+ h^{}_d+{{\h}})^{-1}\) for \(\lambda>0\)
  can be written explicitly:
  \[
  (\lambda+h^{}_d+{{\h}}))^{-1}\He_l(\sqrt{2\omega_d}q_d)F_n(k_1,\dots k_n)=
  \frac{\He_l(\sqrt{2\omega_d}q_d)F_n(k_1,\dots k_n)}{\lambda+l\omega_d+\omega^{}_{k_1}+\dots\omega^{}_{k_n}}\,.
  \]
  \end{remark}
\noindent
  Translation invariance means that the operator \(\triplecolon H^{sol}_{0}\triplecolon\) does not have a vacuum
  (ground state) vector. However, it is useful to consider a minor modification of the solution of the Heisenberg relations in which
  vectors which are formed as a tensor product of a Gaussian wave packet in \(Q\) and \(\Omega'\) can be used
  in place of this nonexistent vacuum: the next remark makes this explicit.
\begin{remark}\label{gaussch}
It will be useful to consider some alternative solutions of the Heisenberg relations based on Gaussian probability measures on \(\R\ni Q\). Making use of the formula
for the Gaussian wave packet \eqref{hgwp}, in particular
\begin{equation}
\mbChi_{0\sigma}(0,Q)=\frac{1}{(2\pi\sigma^2)^{\frac{1}{4}}}\exp[-\frac{Q^2}{4\sigma^2}]\,,
\label{gh0}\end{equation}
there is (for each \(\sigma>0\)) a probability measure
  \(\mbChi_{0\sigma}(0,Q)^2dQ\) on the real line, and on the corresponding 
  \(L^2\) space
  there is a solution of the Heisenberg relation
  \([Q,P]=i\) in which \(Q\) is represented by multiplication by \({Q}\) and \(P\)
  by the operator
  \(f({Q})\mapsto -i\frac{df}{d{Q}}({Q})+\frac{i}{2{\sigma}^2}{Q}f({Q})\);
  these are all unitarily equivalent
  to the standard Schr\"odinger representation via the unitary equivalence
  \(L^2(\R,d{Q})\ni f\mapsto \mbChi_{0\sigma}(0,Q)^{-1}f\in
  L^2(\R,\mbChi_{0\sigma}(0,Q)^2d{Q})\), which
  yields \(-i\frac{d}{dQ}\) as the operator representing \(P\).
  We can include this unitary equivalence into the field theoretic situation
using the following representation of the Heisenberg relations
  \begin{align}\la{cfseg1t}
\upphi(f)\,                                                                                         & =\,-\bigl( {\tteo},f\bigr)_{L^2}{\sqrt{\scm_{cl}}}\,Q\,+\,
\frac{1}{\sqrt{2}}\,\Bigl(\upalpha{\phxi}(C^{\frac{1}{4}}f_\perp)
+\upalpha{\phxi}^\dagger(C^{\frac{1}{4}}f_\perp)\Bigr)\,,
                                                                                                                                \\
                                                                                         \uppi(f)\, & 
                                                                                         =\,-\frac{1}{\sqrt{\scm_{cl}}}
                                                                                         \bigl(
-i\frac{d}{dQ}+\frac{i}{2{\sigma}^2}Q
\bigr)\bigl( {\tteo},f\bigr)_{L^2}
-\frac{i}{\sqrt{2}}
\Bigl((\upalpha{\phxi}(K{\phxi}^{1/4}f_\perp)-\upalpha{\phxi}^\dagger(K{\phxi}^{1/4}f_\perp)\Bigr)\,,
\la{cfseg2t}
\end{align}
  where for \(f\in\cs(\R) \) we write
  \(f_\perp=\P_0^\perp f=f-\P_0 f =f-( {\tteo},f)_{L^2} {\tteo}\).
  We extend the definition of the \(\upalpha \) operators from \(\P_0^\perp(L^2\cap\cs)\)
  to all of
  \(L^2\cap\cs\) with the formulae
    \[
    \upalpha^{}( {\tteo})=-{\sigma}\frac{d}{dQ}\;\hbox{ and }\;
    \upalpha^\dagger( {\tteo})=+{\sigma}\frac{d}{dQ}
    -\frac{1}{{\sigma}}Q\,.
      \]
      These formulae can of course be used for any value of \(\sigma>0\), in particular for the value \(\sigma_0\)
      connected to \(\theta>0\) by \(\sqrt{\theta}=1/(2{\sigma_0}^{2}\scm_{cl})\) so that
       \(\mbChi_\theta(Q)=\mbChi_{0\sigma_0}(0,Q)\) and \(\gamma_\theta(dQ)=\mbChi_\theta(Q)^2dQ\) for
      the corresponding measure, as in \S\ref{notn}. Introduce
      \({{K^\theta}}=\theta\P_0+K\) and \({C^\theta}=\theta^{-1}\P_0+C^\perp\), then 
\begin{align}
            \upphi(f)\,             & \,=\,
         \frac{1}{\sqrt{2}\theta^{\frac{1}{4}}}\,( {\tteo},f)_{L^2}\Bigl(\upalpha{\phxi}({\tteo})
         +\upalpha{\phxi}^\dagger({\tteo})\Bigr)\,
         \,+\,
\frac{1}{\sqrt{2}}\,\Bigl(\upalpha{\phxi}((C^\perp)^{\frac{1}{4}}f_\perp)
+\upalpha{\phxi}^\dagger((C^\perp)^{\frac{1}{4}}f_\perp)\Bigr)\,,
\notag                                                                           \\ \label{cfseg1tt}
\,                                  & =\,
\frac{1}{\sqrt{2}}\,\Bigl(\upalpha{\phxi}((C^\theta)^{\frac{1}{4}}f)
+\upalpha{\phxi}^\dagger((C^\theta)^{\frac{1}{4}}f)\Bigr)\,,
                                                                                 \\\notag
                         \uppi(f)\, & \,=\,
                         -\frac{i}{\sqrt{2}}\theta^{\frac{1}{4}}\,( {\tteo},f)_{L^2}\Bigl(\upalpha{\phxi}({\tteo})
         -\upalpha{\phxi}^\dagger({\tteo})
\Bigr)
-\frac{i}{\sqrt{2}}
\Bigl((\upalpha{\phxi}(K{\phxi}^{1/4}f_\perp)-\upalpha{\phxi}^\dagger(K{\phxi}^{1/4}f_\perp)\Bigr) \\
                                    & \,=\,
                  \,-\,\frac{i}{\sqrt{2}}
                  \Bigl((\upalpha{\phxi}((K^\theta)^{\frac{1}{4}}f)-\upalpha{\phxi}^\dagger((K^\theta)^{\frac{1}{4}}f)
                  \Bigr)\,.
\label{cfseg2tt}
\end{align}
The vector \(\id_{\smr}(Q)\Omega'\in L^2(\gamma_\theta(dQ))\otimes\mfrF\)
can be used in place of the nonexistent vacuum of
\(\triplecolon H^{sol}_{0}\triplecolon\). Under the above unitary equivalence with \eqref{cfseg1}-\eqref{cfseg2}
it corresponds to \(\mbChi_\theta(Q)\Omega'\in L^2(dQ)\otimes\mfrF\). The domain of the operator
\(\triplecolon H^{sol}_{0}\triplecolon\) acting as a self-adjoint operator on
\(\mfrH(\theta)\) is in obvious analogy to \eqref{domhtrip}.
  \end{remark}
\subsubsection{Schr\"odinger representation in the solitonic sector.}\la{sss}
There are two approaches to this: in the first,
the representation \eqref{cfxi1}-\eqref{cfxi2}
is equivalent to a Schr\"odinger representation in which \(\varphi \) is
the multiplication operator \(\phisch\), again written in boldface, acting on the space \(L^2(\mu_0) \), exactly as in the vacuum case (Proposition \ref{sch}). The second
approach is to construct a completely new Schr\"odinger representation based on
\eqref{cfsgext} and attuned to the dynamics in the presence of the soliton.
In naive analogy to the vacuum case, such a new Schr\"odinger representation might
be expected to be
based on a Gaussian measure with covariance \(\frac{1}{2} K^{-\frac{1}{2}}\), with \(K\)
as in \eqref{def-k}. However, recalling the discussion around \eqref{nxi}, \(K\) has a one dimensional kernel
\(\Ker K=\P_0(L^2(\R))=\langle\{ {\tteo}\}\rangle \), so this is not immediately
applicable
and modifications are needed:
we will work with
the operators introduced in Remark \ref{gaussch}, namely,
\({{K^\theta}}=\theta\P_0+K\), which is  strictly positive for \(\theta>0 \),
and its inverse \({C^\theta}=\theta^{-1}\P_0+C^\perp\) where \(C^\perp=\P_0^\perp K^{-1}\P_0^\perp \) is the
covariance operator obtained by restricting to the spectral subspace \(\P_0^\perp(L^2)\) (on which
\(K\) is strictly positive and invertible, with inverse \(C^\perp\)); the \(\perp\) in the latter operator
will be suppressed unless it seems helpful for emphasis.
It is actually useful to carry out the construction of the measure also for \(\theta>0\), but we first describe the
\(\theta=0\) case. 

The Bochner-Milnos theorem allows construction of a measure
whose covariance is \(\frac{1}{2}(C^\perp)^{\frac{1}{2}}\) (which will be written as \(\frac{1}{2}C^{\perp,\frac{1}{2}}\)),
and thence the measure \(\mu\) defining
a Schr\"odinger representation based on
\eqref{cfsgext}.
References for what is needed here are \cite{MR0162118}*{Chapter V},
\cite{MR1741419}*{Chapters III.7,IV}) and \cite{MR2098271}*{Section  IX.10}.
Introducing the
  subspace of tempered distributions which annihilate the zero mode, i.e.,
  \[
\cs'_0(\R)\,\define\,\{\bfupphi\in\cs'(\R):\bfupphi( {\tteo})=0\}\,,
\]
we will use an identification
\(
\cs'(\R)\,\cong \,\R\,{\tteo}\,\oplus\,\cs'_0(\R)
\), see below;
here and in what follows it is to be understood that \({\tteo}\) defines a distribution via integration
against test functions
in the standard way.
Now \(\cs'_0\)  is the dual of the quotient space
\(\cs_0(\R)\define \cs(\R)/\langle\{ {\tteo}\}\rangle \) ; as a quotient by a closed subspace of a nuclear Frechet space, \(\cs_0\) is itself
a nuclear Frechet space, and so reflexive.
Also, \(\cs_0\) can be identified via a linear homeomorphism with
\(\{f\in\cs(\R):(f, {\tteo})_{L^2}=0\} \), the \(L^2\)-orthogonal complement of
\(\langle\{ {\tteo}\}\rangle\) in the space of Schwartz functions.
On \(\cs_0\)  the operator \(K \) descends to define a strictly positive and invertible operator
 \(K^\perp=\P_0^\perp K\P_0^\perp \), with inverse
 \(C^\perp=\P_0^\perp K^{-1}\P_0^\perp\).
To apply
the Bochner-Minlos theorem it suffices then to observe the continuity of the Fourier transform
\[
\cs_0\ni f\mapsto \exp\bigl[-\frac{1}{4}(f,(C^\perp)^{\frac{1}{2}} f)_{L^2}\bigr]\,.
\]
To conclude, there exists
a Gaussian measure on \(\cs'_0(\R) \) with covariance
\(\frac{1}{2}C^{\perp,{\frac{1}{2}}}\). This gives the space
\begin{equation}\label{mbfock}
\mfrbF\define L^2\bigl(\cs'_0(\R),\bfga(\frac{1}{2}C^{\perp,\frac{1}{2}})\bigr)
\end{equation}
which is the Schr\"odinger representation version of the transverse Fock space \(\mfrF \),
see Corollary \ref{uniteqcor}. The transverse vacuum \(\Omega'\in\mfrF\) maps to
\(\id_{\cs'_0}\), the function identically equal to one in \(\mfrbF\).
Following from \eqref{cfsgext} we introduce an operator
\(\mbQ\) by
\(
  {\sqrt{\scm_{cl}}}\mbQ=-\bfupphi( {\tteo})\) as the coordinate multiplication operator corresponding to the
  zero mode.
Since
we can write
\begin{equation}\label{defperp}\bfupphi(f)=\bfupphi(( {\tteo},f){\tteo}+\P_0^\perp f)
=( {\tteo},f)\bfupphi( {\tteo})+\bfupphi(\P_0^\perp f)
=\bfupphi(\P_0 f)+\bfupphiperp(f)
=-({\tteo}, f){\sqrt{\scm_{cl}}}\mbQ
+\bfupphiperp(f)
\end{equation}
there is an isomorphism
\begin{align}\label{isoid}\begin{split}\cs'(\R)                                               & \,=\,\R\oplus\cs'_0(\R) \\
  \bfupphi                                                           & \mapsto\bigl(\mathbf{Q},\bfupphi(\P_0^\perp(\,\cdot\,))\bigr)
  \end{split}\end{align}
(with \(\mathbf{Q}\) then to be identified with the coordinate \(Q\) on \(\R\)), which allows us to generate, for \(\theta\geq 0\),
a product measure according to
\begin{equation}\label{muth0}
\gamma_\theta(dQ)\otimes\bfga(\frac{1}{2}C^{\perp,\frac{1}{2}})\,,
\end{equation}
where \(\gamma_\theta(dQ)=
{\pi^{-\frac{1}{2}}}{\theta^{\frac{1}{4}}}\scm_{cl}^{\frac{1}{2}}\exp[-\sqrt\theta\scm_{cl}Q^2]dQ
=\mbChi_{\theta}(Q)^2dQ\) and for \(\theta=0\) it is to be understood that \(\gamma_0(dQ)=dQ\). (See also \S\ref{notn} for notation).
We will see that for {\em positive} \(\theta\) the isomorphism above identifies this  measure with the Gaussian measure \(\mu(\theta) \) on \(\cs'(\R)\) with
covariance \({\frac{1}{2}{(C^\theta)}^{\frac{1}{2}}}\):
\begin{align}\label{muth}\begin{split}
&\mu(\theta)=\bfga({\frac{1}{2}{(C^\theta)}^{\frac{1}{2}}})\cong\gamma_\theta(dQ)\otimes
\bfga(\frac{1}{2}C^{\perp,\frac{1}{2}})
\\
&L^2(\mu(\theta))\cong L^2(\gamma_\theta(dQ))\otimes\mfrbF
\,.
\end{split}\end{align}

\begin{notation}\label{qdef}
The bold face \(\mbQ\) will be used
when it is useful to emphasize we are referring to the operator of multiplication by
\(
  -{{\scm_{cl}}^{-1/2}}\bfupphi( {\tteo})\) (resp. \(
  -{{\scm_{cl}}^{-1/2}}\phisch( {\tteo})\) )
on
\(L^2(\cs';\mu(\theta))\) (resp. \(L^2(\mu_0)\)), as opposed to the coordinate on the real line in the identification \eqref{isoid}, but
usually  the meaning is clear by context and just \(Q\) may be used for both.
For \(\theta>0\) we write the space of polynomials as
\({\Poly}(\bfupphi)\subset L^2(\mu(\theta)) \) and \({\Poly}(\upphi)\) for the corresponding subspace of
\(L^2(\gamma_\theta(dQ))\otimes\mfrF\) determined according to the equivalence \eqref{muth}.
For \(\theta=0\) we abuse notation slightly and use the same notation, \({\Poly}(\upphi)\subset
L^2(dQ)\otimes\mfrF\) for the subspace which corresponds to the polynomials in \(L^2(\gamma_\theta(dQ))\otimes\mfrF\)
under the standard unitary equivalence \(L^2(\gamma_\theta(dQ))\to L^2(dQ)\), given by
multiplication by \(\mbChi_\theta(Q)\); thus for \(\theta=0\) the subspace \({\Poly}(\upphi)\) is the algebraic span of expressions of the type appearing
on the left of \eqref{lhs}.
As in Remark \ref{triplecolon}, normal ordering indicated with \(\triplecolon\) is applied only in the transverse
space \eqref{mbfock}, leaving alone \(Q,P\).
\end{notation}
The Schr\"odinger representation determined by \(\mu(\theta)\) in \eqref{muth0}
and the
(generalized) Fock representation \eqref{cfsgextalt}-\eqref{cfsg22alt} as in Remark \ref{gaussch} are
related by a unitary equivalence (for \(\theta> 0\))
\begin{equation}\label{uniteqsol}
  \J^{\theta}:L^2(\gamma_\theta(dQ))\otimes\mfrF
\to L^2(\gamma_\theta(dQ))\otimes\mfrbF\cong L^2(\cs'(\R),\mu(\theta))\,,
\end{equation}
under which the vector
\(\id_{\smr}(Q)\Omega'\in L^2(\gamma_\theta(dQ))\otimes\mfrF\) 
from Remark \ref{gaussch}
corresponds to the function \(\id_{\cs'}\in L^2(\mu(\theta))\) identically
equal to one.
The equivalence extends to \(\theta=0\) by taking the product measure \eqref{muth0} with \(\theta=0\)
to be the definition of the measure
\(\mu=\mu(0)\) on \(\cs'\) via \eqref{isoid}.
With this definition it is evident that the measures \(\mu(\theta)\) and \(\mu\) are equivalent
since \(dQ\) and \(\gamma_\theta(dQ)\) are
equivalent measures on \(\R\);
the requirement that 
\(\mbChi_\theta(Q)\Omega'\in L^2(dQ)\otimes\mfrF\) corresponds to \(\id_{\cs'}\in L^2(\mu(\theta))\),
fixes the unitary equivalence of the corresponding \(L^2\) spaces, and this vector may be used in place of the nonexistent vacuum of \(\triplecolon H^{sol}_{0}\triplecolon\).
%
More substantially, the measures \(\mu(\theta)\) are all equivalent to the vacuum measure \(\mu_0\) defined by
the free covariance, as will now be proved in the next theorem. 
The crucial thing is to prove equivalence of \(\mu(\theta)\) and \(\mu_0\) for positive \(\theta\),
and then obtain the factorization in \eqref{muth} in (iv) of the theorem.

\begin{theorem}\label{mac}
 For each \(\theta>0\), the Gaussian measure \(\mu(\theta) \,=\,\bfga({\frac{1}{2}{(C^\theta)}^{\frac{1}{2}}})\)
  on \(\cs'(\R)\) with covariance \(\frac{1}{2}{(C^\theta)}^{\frac{1}{2}}\)
  is equivalent (in the sense of mutual absolute continuity) to
  the vacuum Gaussian measure \(\mu_0 \,=\,\bfga({\frac{1}{2}C_0^{\frac{1}{2}}})\)
  of covariance \({\frac{1}{2}C_0^{\frac{1}{2}}}\). The Radon-Nikodym derivative
  is formally
  \begin{equation}\label{rnd}
  \frac{d\mu(\theta)}{d\mu_0}\,=\,\det(\id+\mbO(\theta))^{\frac{1}{2}}\exp\bigl[
    -\bigl(\phisch,(C_0^{\frac{1}{2}}{(K^\theta)}^{\frac{1}{2}}-\id)\phisch\bigr)_{\frac{1}{2}}\,\,
    \bigr]\,,
  \end{equation}
  where \(\mbO(\theta):L^2(\R)\to L^2(\R)\) is given by
  \[\mbO(\theta)\,=\,C_0^{\frac{1}{4}}({(K^\theta)}^{\frac{1}{2}}-K_0^{\frac{1}{2}})C_0^{\frac{1}{4}}\,.\] (The precise meaning of \eqref{rnd} is given in the course of the proof.)
\begin{itemize}\item[(i)]The operator \(\mbO(\theta)\) is trace-class on \(L^2(\R)\), or equivalently, the operator
  \((C_0^{\frac{1}{2}}{(K^\theta)}^{\frac{1}{2}}-\id)\) is trace-class on \(H^{\frac{1}{2}}\), the Sobolev
  space determined by the inner product
  \((\phi,\psi)_{\frac{1}{2}}\) defined in \eqref{h12ip}. 
\item[(ii)]  The expression \eqref{rnd}
  defines an element of \(L^{p_*/2}(d\mu_0)\) for some \(p_*>2 \).
\item[(iii)] The square root of \eqref{rnd} defines \(\bfOm^\theta\in L^{p_*}(d\mu_0)\), and under the
  mapping \(\I\) in Proposition \ref{sch} corresponds to
a smooth vector for \(\N_0\) in Fock space, i.e.
  \(\bbvac^\theta\define\I^{-1}\sqrt{\frac{d\mu(\theta)}{d\mu_0}}\in\cap_{s=0}^\infty\Dom(\N_0^s)\subset\fock\), given explicitly by
  \[
\Omega^\theta\,=\,\exp\Bigl[-\sum\,
\frac{\lambda_n/2}{1+\lambda_n/2}\,\frac{\alpha_n^\dagger \alpha_n^\dagger}{2}
  \Bigr]\Omega_0\,,
\]
where, using notation as in \eqref{alpha0}, \(\alpha_n^\dagger=\alpha^\dagger(f_n)\), where \(\{f_n\}\) is an orthonormal basis of
\(L^2\) consisting of the eigenfunctions of \(\mbO(\theta)\) with eigenvalue \(\lambda_n=\lambda_n(\theta)\).
(The vector \(\bfOm^\theta\in L^{p_*}(d\mu_0)\) corresponds to \(\id_{\smr}(Q)\Omega'\in L^2(\gamma_\theta(dQ))\otimes\mfrF\equiv\mfrH(\theta)\), see Corollary \ref{uniteqcor}.)
\item[(iv)] \(\mu(\theta)\) as defined above verifies \eqref{muth}.
  \end{itemize}
\end{theorem}
\noindent
Statement (i) of this theorem will be deduced from
\begin{theorem}\label{tc}
  If \(\theta>0\) the operator
  \(\kq(\cv-\cvo)\kq\) is trace-class on \(L^2(\R)\).
\end{theorem}
\noindent
This is proved below in \S\ref{tcprop} by an analysis of integrals arising from explicit expressions
for the kernels of the covariance operators.
    \paragraph{Proof of Theorem \ref{mac}} (assuming Theorem \ref{tc}.) Statement (i) follows from an application of the theorem of Shale,
    in the form given in \cite{MR1178936}*{Theorem 45} (see also
    \cite{MR0489552}*{Theorem I.23},  or
    \cite{MR1642391}*{Chapter 6}), and hinges on the trace-class property just mentioned. Given (i),
    statements (ii) and (iii) follow from results in the literature (referenced below) on
unitary implementability and Bogoliubov transformations.
We start by expanding on the statement of the theorem
in terms of measures on the space of tempered distributions so as to be able to work with
    the limiting expression for the Radon-Nikodym derivative, which does not 
have an a priori meaning, but is defined by a limiting process; this eventually leads to \eqref{muth}.

    The tempered distribution \(f\mapsto\phisch(f)\) is defined as a continuous map
    on the space of Schwartz test functions \(f\in\cs(\R)\), and in its turn the map
    \(\phisch\mapsto\phisch(f)\) is continuous on \(\cs'(\R)\) (endowed with the weak-* topology)
    for all such \(f\). But the formula
    \begin{equation}\label{si}
    \|\phisch(f)-\phisch(g)\|_{L^2(\mu_0)}^2
    \,=\,\frac{1}{2}\bigl(C_0^{\frac{1}{4}}(f-g),C_0^{\frac{1}{4}}(f-g)\bigr)_{L^2}
    \,=\,\frac{1}{2}(f-g,f-g)_{{\frac{1}{2}}}
    \end{equation}
  determines a unique extension of \(\phisch(K_0^{\frac{1}{4}}f)\) as a 
  measurable function of \(\phisch\)
  in the space \(L^2(\mu_0)\) for any \(f\in L^2(\R)\), i.e.
  \(\phisch(\chi)\) is so defined for \(\chi\in H^{-\frac{1}{2}}(\R)\,.\) Now if \(\{f_n\}\)
  is an orthonormal basis of \(L^2(\R)\) then \(\{K_0^{\frac{1}{4}}f_n\}\) is an orthonormal
  basis for \(H^{-\frac{1}{2}}(\R)\), and we can expand
  \(\chi=\sum\chi_nK_0^{\frac{1}{4}}f_n\in H^{-\frac{1}{2}}(\R)\), with 
\(\chi_n=(f_n,C_0^{\frac{1}{4}}\chi)_{L^2}\). This
  induces a dual expansion
  \begin{equation}\label{dualexp}
  \phisch(\chi)\,=\,\sum\phisch_n\chi_n\,
  =\,\bigl\langle\,\sum\phisch_n\,C_0^{\frac{1}{4}}f_n,\chi\bigr\rangle
\end{equation}
where \(\phisch_n=\phisch(K_0^{\frac{1}{4}}f_n)\in L^2(\mu_0)\) are well-defined for all \(n \) by
the preceding discussion, and satisfy
\((\phisch_n,\phisch_m)_{L^2(d\mu_0)}=\frac{1}{2}\delta_{nm}\,; \) the expansion \eqref{dualexp} converges in
\(L^2(\mu_0)\) by \eqref{si}. With the \(\phisch_n\)
as coordinates we identify 
\(L^2(\mu_0) \) with the space \(\R^\infty \) 
with the infinite product probability measure
\(\prod_n \Bigl(\pi^{-\frac{1}{2}}\exp[-\phisch_n^2]d\phisch_n\Bigr)\).
This allows an interpretation of the formal exponential factor in \eqref{rnd}
as \(\exp[-\sum_{m,n}\phisch_m\phisch_n(f_m,\mbO(\theta) f_n)] \), which in turn
suggests choosing \(\{f_n\}\) to be an orthonormal basis of eigenfunctions of
\(\mbO(\theta)\), with eigenvalues \(\lambda_n \), which satisfy
\(\sum|\lambda_n|<\infty \) (under the condition that \(\mbO(\theta)\) is trace-class). Considering
the directional derivative along \(C_0^{\frac{1}{4}}f_n\) of a functional which is a polynomial in the
\(\{\phisch_n\}\), i.e., \(F(\phisch)=P(\phisch_1,\phisch_2,\dots\phisch_N)\), we find (see \eqref{seq})
\begin{align}
  \,\pisch_nF=\pisch(C_0^{\frac{1}{4}}f_n)F=-i\frac{d}{d\epsilon}\Bigl|_{\epsilon=0}F(\phisch+
  \epsilon C_0^{\frac{1}{4}}f_n)+{i}\phisch({C_0}^{-\frac{1}{2}} C_0^{\frac{1}{4}}f_n)F
=-i\frac{\partial P}{\partial\phisch_n}+i\phisch_nP\,,
\end{align}
the first equality being the definition of \(\pisch_n\).
Using notation as in \eqref{alpha0}, but with boldface to indicate
Schr\"odinger representation,
\(\bfalpha_n=\bfalpha(f_n),\bfalpha_n^\dagger=\bfalpha^\dagger(f_n)\), we have
\begin{equation}\label{alphan}
  \phisch_n=\frac{1}{\sqrt{2}}(\bfalpha_n+\bfalpha^\dagger_n)\,,\qquad
  \pisch_n=\frac{-i}{\sqrt{2}}(\bfalpha_n-\bfalpha^\dagger_n)\,\qquad
  \frac{\partial }{\partial\phisch_n}=\phisch_n+i\pisch_n=\sqrt{2}\bfalpha_n\,.
  \end{equation}

Now consider the following expression
\[
\exp\bigl[
    -\bigl(\phisch,(C_0^{\frac{1}{2}}{(K^\theta)}^{\frac{1}{2}}-\id)\phisch\bigr)_{\frac{1}{2}}\,\,
    \bigr]\,=\,\lim_{N\to\infty}\prod_{n=1}^N\exp\bigl[
-(\phisch_n)^2(f_n,\mbO(\theta) f_n)
  \bigr]\,,
\]
to be the putative {\em definition} of the exponential factor in \eqref{rnd}.
In fact this limit does exist by \cite{MR0489552}*{Lemma I.24}.
To establish the trace-class property, note that by
Theorem \ref{tc} the operator
\(\mbB=\kq(\cv-\cvo)\kq\) is trace-class, while \(\kq\cv\kq=\id+\mbB\) and
\((\id+\mbB)^{-1}\) are bounded by Proposition \ref{ksb}.
It follows that  \((\id+\mbB)^{-1}-\id=-\mbB(\id+\mbB)^{-1}\)
is also trace-class, i.e. \(C_0^{\frac{1}{4}}{(K^\theta)}^{\frac{1}{2}}C_0^{\frac{1}{4}}=\id+\mbO(\theta)\)
with \(\mbO(\theta)\) trace-class, as required to establish (i).
It follows that the square root determinant in the formula for the
Radon-Nikodym derivative is well-defined and equals \(\prod_n(1+\lambda_n)^{1/2} \),
so that the expression \eqref{rnd} is to be interpreted as
\[
\lim_{N\to\infty}\prod_{n=0}^N\Bigl((1+\lambda_n)^{\frac{1}{2}}\exp\bigl[
-\lambda_n(\phisch_n)^2
  \bigr]\Bigr)\,.
\]
By a result of Segal, a proof of which appears in \cite{MR0489552}*{\S\,I.6}, 
this expression is known to converge in \(L^p\bigl(\prod_n (\pi^{-\frac{1}{2}}\exp[-\phisch_n^2]d\phisch_n)\bigr)\) for some
\(p>1\) to give the stated formula for the Radon-Nikodym derivative and assertion (ii).
Next, to prove (iii): the vacuum is characterized as being the vector annihilated by
\[
\frac{1}{\sqrt{2}}\Bigl(\,\frac{\partial}{\partial\phisch_n}+\lambda_n\phisch_n\,\Bigr)\,
=\,\I\,\Bigl((1+\lambda_n/2)\,\bfalpha_n\,+\,{\lambda_n/2}\,\bfalpha_n^\dagger\,\Bigr)\I^{-1}\,;
\]
this vector can be seen to be (in the Fock representation) a scalar multiple of
\[
\exp\Bigl[-\sum_{n=0}^\infty\,
\frac{\lambda_n/2}{1+\lambda_n/2}\,\frac{\alpha_n^\dagger \alpha_n^\dagger}{2}
\Bigr]\Omega_0\,,
\]
where \(\bfalpha_n^{\dagger}\) is the creation operator for the state \(f_n\), see \eqref{alphan}.
The argument given  in \cite{MR516713}*{\S 4}
then implies smoothness with respect to the number operator.

Finally we need to prove (iv). The construction above has depended upon a choice of a positive real number
\(\theta \), related to the variance \({\sigma_0}^2 \)
by
\(\sqrt{\theta}=1/(2{\sigma_0}^{2}\scm_{cl}) \); however, as far as the 
Radon-Nikodym factor and the unitary equivalences are concerned this dependence
all but drops out, as can be seen by making a particular choice of basis in the proof above.
Referring to \eqref{lhs},
we need to check the \(\theta \) dependence
which seeps in
via the operator \(\mbO(\theta) \); explicitly:
\[
\mbO(\theta)=\mbO(0)+\,\sqrt{\theta}\,C_0^{\frac{1}{4}}\P_0C_0^{\frac{1}{4}}\,.
\] 
This suggests that in making sense of \eqref{rnd}
we work with the expression  \(\exp[-\sum_{m,n}\phisch_m\phisch_n(f^0_m,\mbO(\theta) f^0_n)] \)
where \(\{f_n^0\} \) is an \(L^2\)-orthonormal set of
eigenfunctions of \(\mbO(0) \), rather than of \(\mbO(\theta) \) as in the proof.
In particular it is easy to check that
 \(K_0^{\frac{1}{4}} {\tteo} \) is an eigenfunction with eigenvalue \(\lambda_0^0=-1 \), and we define
 \(f^0_0\) to be \(K_0^{\frac{1}{4}} {\tteo}/\|K_0^{\frac{1}{4}} {\tteo}\|_{L^2} \); let the
remaining eigenvalues, none of which equal \(-1 \), be written
\(\lambda_1^0,\lambda_2^0,\dots \) then with this choice the Radon-Nikodym 
factor becomes
\begin{align}\label{defrn}
{\frac{d\mu(\theta)}{d\mu_0}}
\,&=\,\det(\id+\mbO(\theta))^{\frac{1}{2}}\exp\bigl[
-{\sqrt{\theta}}\phisch( {\tteo})^2    
+\phisch_0^2
-\sum_{n=1}^\infty\lambda_n^0\,\phisch_n^2
    \bigr]\,,
\\\notag
\,&=\,\det(\id+\mbO(\theta))^{\frac{1}{2}}\exp\bigl[
-{\sqrt{\theta}}\phisch( {\tteo})^2    
+\phisch_0^2\bigr]\,
\prod_{n=1}^\infty\exp\bigl[
-\lambda_n^0\phisch_n^2
    \bigr]\,.
\end{align}
Here \(\phisch_n=\phisch(K_0^{\frac{1}{4}}f^0_n)\); observing that the \(L^2\) inner products \(({\tteo},K_0^{\frac{1}{4}}f^0_n)=0\), we
identify the infinite product here with the transverse measure, at least after normalization.
The determinant can be found by normalization:
 \[
  \det(\id+\mbO(\theta))^{\frac{1}{2}}
  \,=\,\frac{\theta^{1/4}}{\|K_0^{\frac{1}{4}} {\tteo}\|}\,\prod_{1}^\infty(1+\lambda_n^0)^{1/2}\,,
  \]
  and observe that all the \(\theta\) dependence here is in the \(\theta^{\frac{1}{4}}\).
[This formula can be proved by approximating
\[
\phisch({\tteo})=\lim_{N\to\infty} \sum_{n=0}^N c_n\phisch_n\,,\qquad
c_n=(\phisch_n,\phisch({\tteo}))_{L^2(\mu_0)}=\|K_0^{1/4}{\tteo}\|(f_0^0,C_0^{1/2}f_n^0)_{L^2},
\]
where the limit converges in \(L^2(\mu_0)\) and so subsequentially \(\mu_0\)-a.e.;  then use the
bounded convergence theorem and evaluate the resulting finite dimensional Gaussian integrals. Note that
the formula for \(c_n\) specializes to \(c_0=\|K_0^{1/4}{\tteo}\|^{-1}\), and the change of variables
\((\phisch_0,\phisch_1,\dots\phisch_N)\to(\phisch'_0,\phisch'_1,\dots\phisch'_N)=
(\sum_{n=0}^N c_n\phisch_n,\phisch_1,\dots\phisch_N)\) has Jacobian \(c_0\).]

The
transverse measure is defined by the formula (for arbitrary Borel sets \(A\subset \R^k\) and \(k\in\N\))
\begin{align}
\bfga(\frac{1}{2}C^{\perp,\frac{1}{2}})\bigl(\{(\phisch_1,\dots\phisch_k)\in A\}\bigr)
&=\int_{(\phisch_1,\dots\phisch_k)\in A}\,\prod_{1}^\infty(1+\lambda_n^0)^{1/2}
\prod_{n=1}^\infty\exp\bigl[
-\lambda_n^0(\phisch_n)^2
\bigr]\,d\mu_0(\phisch)\\
&=\int_{(\phisch_1,\dots\phisch_k)\in A}\,\prod_{1}^k(1+\lambda_n^0)^{1/2}
\pi^{-{1/2}}\exp\bigl[
-(1+\lambda_n^0)(\phisch_n)^2
\bigr]\,d\phisch_n
\end{align}
(with the understanding that the products in the first line do in fact converge,
according to the preceding proof.) This leads to
\begin{align}\label{illusion}\begin{split}
\mu(\theta)&=\pi^{-\frac{1}{2}}\,\frac{\theta^{1/4}}{\|K_0^{\frac{1}{4}} {\tteo}\|}
\exp\bigl[
  -{\sqrt{\theta}}\phisch( {\tteo})^2\bigr]\,d\phisch_0\otimes\bfga(\frac{1}{2}C^{\perp,\frac{1}{2}})\,,\\
&=\pi^{-\frac{1}{2}}{\sqrt{\scm_{cl}}}{\theta^{1/4}}
\exp\bigl[
  -{\scm_{cl}}{\sqrt{\theta}}\mbQ^2\bigr]\,d\mbQ\otimes\bfga(\frac{1}{2}C^{\perp,\frac{1}{2}})\,,
\end{split}\end{align}
with \(\mbQ\) obtained from the field by \({\sqrt{\scm_{cl}}}
\mbQ=-\phisch( {\tteo})\).
Thus we have obtained the factorization in \eqref{muth0}-\eqref{muth}, with \(Q\) as a coordinate on the factor
\(\R\) of the decomposition \eqref{isoid}
and \(\gamma_\theta(dQ)=\mbChi_{\theta}(Q)^2dQ\).
\qed


\begin{remark}
The presence of the eigenvalue \(\lambda_0^0=-1\)
in the spectrum of \(\mbO(0) \) is the reason it is necessary to deform
the covariance operator by  the parameter \(\theta \), and arises directly from
the presence of the zero mode \( {\tteo} \) in the spectral analysis 
of \(K\), which 
is itself a consequence of translation invariance, the relation being
made plain by the fact that the corresponding eigenfunction of \(\mbO(0) \) is 
\(K_0^{\frac{1}{4}} {\tteo} \).
\end{remark}
\begin{remark}
To connect the preceding discussion up with the
formula given in \cite{MR1178936}*{Theorem 4.5} note that
the term in the exponential can be rewritten
\[-\bigl(\phisch,(C_0^{\frac{1}{2}}{(K^\theta)}^{\frac{1}{2}}-\id)\phisch\bigr)_{\frac{1}{2}}
  \,=\,
  -\bigl(\phisch,
  (T'T-\id)\phisch\bigr)_{\frac{1}{2}}\,,\]
  where \(T=C_0^{\frac{1}{4}}{(K^\theta)}^{\frac{1}{4}}\) and \('\) means adjoint in the
  \((\;\;)_{\frac{1}{2}}\) inner product, (so that
  \(T'=C_0^{\frac{1}{2}}T^*{K_0^{\frac{1}{2}}}\) where \(T^*\) is the ordinary \(L^2\)
  adjoint, so that \(T'T=C_0^{\frac{1}{2}}{(K^\theta)}^{\frac{1}{2}}\).)
      \end{remark}
\begin{theorem}\label{uniteq}
(i)  For \(\theta>0\) there is a unitary isomorphism
  \begin{align}\label{sth}\begin{split}
    \mbS^{\theta} \,:\,L^2(\cs'(\R),\mu(\theta))\,
    & \to\, L^2(\cs'(\R),\mu_0) \\
    \Psi\,                                                                                          & \mapsto\,\sqrt{\frac{d\mu(\theta)}{d\mu_0}}\Psi
  \end{split}\end{align}
  with \(\bfOm^\theta=\sqrt{\frac{d\mu(\theta)}{d\mu_0}}\in L^{p_*}(\mu_0) \) for some \(p_*>2 \),
   by the assertion about \eqref{rnd}
in Theorem \ref{mac}. 
  which induces an equivalence between the vacuum Schr\"odinger
  representation \eqref{seq} of the Heisenberg relations 
  and the corresponding solution
  in which the field operator, being the operator of
  multiplication by
  \(\phivs(f)= (\mbS^{\theta})^{-1}\phisch(f)\mbS^{\theta}=\phisch(f)\) is unchanged,
  but the conjugate momentum  is now represented by
  \[\pivs(f) =(\mbS^{\theta})^{-1}\pisch(f)\mbS^{\theta}=-iD_{f} +{i}\phivs({{(K^\theta)}}^{\frac{1}{2}} f)
  \,.\]
  For \(f\) a real Schwartz function, these operators are essentially self-adjoint on \({\Poly}(\bfupphi) \)
  in the field. 
  The creation/annihilation operators are given by
  \begin{equation}\label{crann}
  \bfupalpha^\theta(f)=\frac{1}{\sqrt{2}}\Bigl(
  \phivs\bigl((K^\theta)^{\frac{1}{4}}f\bigr)+i
  \pivs\bigl((C^\theta)^{\frac{1}{4}}f\bigr)
  \Bigr)\,\hbox{  and }\;
  (\bfupalpha^\theta(f))^\dagger=\frac{1}{\sqrt{2}}\Bigl(
  \phivs\bigl((K^\theta)^{\frac{1}{4}}f\bigr)-i
  \pivs\bigl((C^\theta)^{\frac{1}{4}}f\bigr)
  \Bigr)\,.
  \end{equation}

  (ii) For \(\theta=0\), recalling the definition of \(\mfrbF\) above and the remark following
  \eqref{uniteqsol}, there is a unitary isomorphism
  \[
\mbS=\mbS^0\,:\,L^2(\R,dQ)\otimes\mfrbF\to L^2(\mu_0)\,,
\]
whose action on \({\Poly}(\bfupphi) \) is
\begin{align}
    Q^{n_0}\mbChi_{\theta}(Q)\prod_{j=1}^N\bfupphi( g_j)^{n_j} & \mapsto
    ({-\sqrt{\scm_{cl}}})^{-n_0}
    \phisch({ {\tteo}})^{n_0}\prod_{j=1}^N\phisch( g_j)^{n_j}
    \sqrt{\frac{d\mu(\theta)}{d\mu_0}}\,.
    \label{lhs}\end{align}
where \(g_1,
g_2\dots\) are a countable set of Schwartz test functions orthogonal to
\({\tteo}\), the zero mode eigenfunction \(K\).
     In particular, the state in which there are no bosons present and the
  soliton is described by a Gaussian wave packet \(\mbChi_{\theta}(Q) \) corresponds
  to the state \(\sqrt{\frac{d\mu(\theta)}{d\mu_0}}\) in the vacuum Schr\"odinger
  representation. The right hand side of \eqref{lhs} actually lies
in \(L^p(\mu_0) \) for all \(p<{p_*} \).
  \end{theorem}
\begin{remark}\label{budd}
  It is worth noting that the apparent dependence on \(\theta\) in \eqref{lhs} is illusory, as can be seen
by comparing the formula \eqref{illusion} with the expression for \(\mbChi_\theta\).
\end{remark}
We can also form Fock space versions of these maps,
firstly going to the Fock space \(\mfrF\) on the domain,
we have
\[
\bbS^{\theta}\define\mbS^{\theta}\circ\J^{\theta}:\mfrH(\theta)\to L^2(\mu_0)\,,
\]
and we drop the \(\theta\) superscript in the case \(\theta=0\). Composing further on the left with
\(\I^{-1}\) we obtain the Fock space valued
version of \(\bbS\):
\[
\mfrS^{\theta}\define\I^{-1}\circ\bbS^{\theta}:\mfrH(\theta)\to\fock\,,\]
and these provide unitary equivalences between the corresponding solutions of the Heisenberg relations:
\begin{corollary}\label{uniteqcor}
The solutions
\eqref{cfxi1}-\eqref{cfxi2} and \eqref{cfsgext}-\eqref{cfsg22alt}, and more generally
\eqref{cfseg1t}-\eqref{cfseg2t},
of the Heisenberg relations \eqref{ccr} are unitarily equivalent via the unitary
isomorphism just defined from the Hilbert space \(\mfrH(\theta)\define L^2(\gamma_\theta(dQ))\otimes\mfrF\)
to \(\fock\).
It intertwines the corresponding fields, i.e.,
  \[
  \upphi(f)=({\mfrS}^\theta)^{-1}\varphi(f){\mfrS^\theta} \qquad\hbox{and}\qquad
  \uppi(f)=({\mfrS}^\theta)^{-1}\pi(f){\mfrS^\theta} \,,
\]
for \(\theta\geq 0\).
For \(\theta>0\) (resp. \(\theta=0\)) the isomorphism \(\mfrS^\theta\) maps
\(\id_{\smr}(Q)\Omega'\in L^2(\gamma_\theta(dQ))\otimes\mfrF\) 
from Remark \ref{gaussch}
(resp. \(\mbChi_{\theta}(Q)\Omega'\in L^2(dQ)\otimes\mfrF\)) to \(\Omega^\theta\) defined in item (iii) of Theorem \ref{mac}.
Under the isomorphism of \( L^2(\R,dQ)\otimes\mfrbF \) onto the Hilbert space
\(\mfrH(\theta)\), the
subspace \({\Poly}(\bfupphi)\)
maps to a subspace \({\Poly}(\upphi)\) which is the algebraic span
of  \(Q^{n_0} \mbChi_{\theta}(0,Q)q_d^{n_1}\Symn\prod_{j=1}^n g_j(k_j)\, \)
with \(n_0,n_1\) ranging over the
nonnegative integers,
and the \(g_j\in\cs(\R)\) and \(L^2\)-orthogonal to the discrete eigenfunctions \({\tteo},{\tteone}\).
  \end{corollary}
  \proof
  From Remark \ref{gaussch} we obtain
  unitary equivalence with the Schr\"odinger representation
  on
  \(L^2(\cs'(\R),d\mu(\theta))\) by mapping
  vacuum to vacuum and \(\upalpha^{},\upalpha^\dagger\) (defined in \eqref{compme})
  to the bold-faced annihilation/creation operators
  (defined in Theorem \ref{uniteq}), as in the standard proof of uniqueness of Fock
  representations in \cite{MR887102}*{Section 6.3}. Since the covariance
  operator for the representations in \eqref{crann} and \eqref{cfseg1t}-\eqref{cfseg2t} are the same this unitary equivalence extends to one between the
  fields \((\upphi,\uppi)\) and \((\phivs,\pivs)\).
  But this latter representation is in turn equivalent to that
  determined by the vacuum measure \(\mu_0 \), by Theorem \ref{uniteq},
  and the result is proved.\qed
\begin{remark}\label{unchanged}
  Although in the Schr\"odinger representation the field is unchanged,
as \((\mbS^\theta)^{-1}\phisch(f)\mbS^{\theta}=\phisch(f)\),
  it is convenient to use the
  upright \(\phivs(f)\) font to distinguish the representation; 
also, in the corresponding Fock space representations the fields
of course act on quite different Hilbert spaces and need to be distinguished.
The momentum is also indicated with the upright font \(\pivs\), and does change under \(\mbS^\theta\).
As indicated, we write
\({\Poly}(\bfupphi) \) for the dense subset of \(L^2(\R,dQ)\otimes\mfrbF \)
obtained by taking finite linear combinations of expressions
as in the left hand side of \eqref{lhs}, essentially polynomials in
the field \(\bfupphi\), really the same subspace as \({\Poly}(\upphi)\).
The boldface just serves to
indicate when the Schr\"odinger representation is being used when it is
necessary to emphasize this. 
\end{remark}

\subsubsection{Trace-class properties - proof of Theorem \ref{tc}}\label{tcprop}
Since \(\P_0\) is the spectral projection onto the kernel of \(K\),
the positive square
root of \({{K^\theta}}\) is given by \({(K^\theta)}^{\frac{1}{2}}=K^{\frac{1}{2}}+\theta^{\frac{1}{2}}\P_0\).
We introduce for comparison the operator \(K_0^{\frac{1}{2}}+\theta^{\frac{1}{2}}\P_0\),
which is also a strictly positive self-adjoint operator with domain \(H^1(\R)\),
with inverse \(\cvot\) which is bounded on \(L^2\).
Recalling that the trace-class
operators form an ideal (within the algebra of bounded operators) characterized
by having finite trace norm, we see that the theorem is a consequence of the
following two lemmas and the triangle inequality for the trace norm.\qed
\begin{lemma}\label{tcl2}
If \(\theta>0\) the operator
  \(\kq\bigl(\cvot-\cvo\bigr)\kq\) is trace-class on \(L^2(\R)\).
\end{lemma}
\begin{lemma}\label{tcl1}
If \(\theta>0\) the operator
  \(\kq\bigl((\cv-\cvot\bigr)\kq\) is trace-class on \(L^2(\R)\).
\end{lemma}
To prove these we will make use of the following trace-class 
criterion.
\begin{theorem}[\cite{MR0246142}*{Section III.10}]\label{gk}
    An integral operator \(Af(x)=\int\,A(x,y)f(y)dy\) is a trace-class
    operator on \(L^2(\R)\) if
\begin{enumerate}
    \item[(i)] the function \((x,y)\to A(x,y)\) is continuous,
    \item[(ii)] \((f,Af)_{L^2}\geq 0\) for all continuous and compactly supported \(f\), and
    \item[(iii)] \(\int_{\smr} A(x,x)dx<\infty\).
    \end{enumerate}
  \end{theorem}
\noindent
    {\em Proof of Lemma \ref{tcl2}.}
    This simple result follows from the following explicit formulae, which are displayed as they will be of use below:
    \[
    \bigl(\cvot-\cvo\bigr)f\,=\,-\,\theta^{\frac{1}{2}}\,
    \frac{(C_0^{\frac{1}{2}} {\tteo},f)_{L^2}C_0^{\frac{1}{2}} {\tteo}}{1+\theta^{\frac{1}{2}}
    (C_0^{\frac{1}{2}} {\tteo}, {\tteo})_{L^2}}\,,
    \]
    so that
    \begin{equation}\label{r2}
    K_0^{\frac{1}{4}}\bigl(\cvot-\cvo\bigr)K_0^{\frac{1}{4}}f\,=\,-\,\theta^{\frac{1}{2}}\,
    \frac{(C_0^{\frac{1}{4}} {\tteo},f)_{L^2}C_0^{\frac{1}{4}} {\tteo}}{1+\theta^{\frac{1}{2}}
    (C_0^{\frac{1}{2}} {\tteo}, {\tteo})_{L^2}}\,.
    \end{equation}
    Now choose an orthonormal basis \(\{e_j\}_{j=1}^\infty\) with
    \({\tteone}\) proportional to \(C_0^{\frac{1}{4}} {\tteo}\), and
    recall that a self-adjoint bounded non-negative operator \(A\)
    on \(L^2\) is trace-class
    if and only if \(\sum_{j}|(Ae_j,e_j)_{L^2}|<\infty\) for some o.n. basis.
    \qed
    
\noindent
{\em Proof of Lemma \ref{tcl1}.}
In what follows recall the
operator monotonicity of inversion and taking square roots.
The operator inequality \(0\leq K<K_0\), which is evident by inspection, implies
that
\(0\leq K^{\frac{1}{2}}<(K_{0})^{\frac{1}{2}}\), and hence for any \(\theta>0\)
\[\cvot<((K^{\theta})^{\frac{1}{2}})^{-1}={(C^\theta)}^{\frac{1}{2}}\,.\]
It follows that
\begin{equation}\label{opmon}
\kq\bigl(\cv-\cvot\bigr)\kq\,>\,0\,.
\end{equation}
Writing \[\kq\bigl(\cv-\cvot\bigr)\kq=\kq\bigl(\cv-\cvo\bigr)\kq
-\kq\bigl(\cvot-\cvo\bigr)\kq\,,\]
it is sufficient to show that the two operators on the right hand side
satisfy the conditions (i) and (iii) in Theorem \ref{gk}. This is clearly true of
the second operator on the right, since by formula \eqref{r2} it has a kernel
proportional to
\(C_0^{\frac{1}{4}} {\tteo}\otimes C_0^{\frac{1}{4}} {\tteo} \),
a tensor product of a Schwartz function with itself. It therefore remains to 
prove the same for the operator \(A_2=\kq\bigl(\cv-\cvo\bigr)\kq\). We write
this as \(A_2=\kq A_3\kq\), with \(A_3=\bigl(\cv-\cvo\bigr)\), and use the following
two items to analyze these operators.
\begin{enumerate}
\item[(a)] The action of the operator \(\kq\) can be realized as
a pseudodifferential operator acting on the kernel \(A_3(x,y)\),
so that the Fourier transforms of the integral kernels of \(A_2\) and \(A_3\), i.e.,
\[\hat A_j(k,l)=(2\pi)^{-1}\int\exp[-ikx-ily]A_j(x,y)dxdy\qquad  (j=2,3,)\]
are
related by
\[
\hat A_2(k,l)\,=\,(4m^2+k^2)^{\frac{1}{4}}\hat A_3(k,l)
(4m^2+l^2)^{\frac{1}{4}}\,,
\]
or, in a convenient notation,
\[
A_2(x,y)=(4m^2-\partial_x^2)^{\frac{1}{4}}(4m^2-\partial_y^2)^{\frac{1}{4}}
A_3(x,y)\,.
\]
\item[(b)]
The work in the Appendix yields an explicit formula for the integral kernel
of the operator \(A_3\):
\begin{align}\label{fc1}
A_3(x,y)=\bigl(\cv-\cvo\bigr)(x,y)\, & =\,\frac{1}{\sqrt 3m} {\tteone}(x) {\tteone}(y)\,
+\,\frac{1}{\sqrt \theta} {\tteo}(x) {\tteo}(y) \\
                                     & \quad\,+\,\frac{1}{2\pi}\,\int_{\smr}\,\notag
\frac{\Bigl[\overline{{\ttF}(k,y)}{\ttF}(k,x)-
  (k^2+m^2)(k^2+4m^2)\Bigr]e^{ik(x-y)}}
{(k^2+m^2)(k^2+4m^2)^{\frac{3}{2}}}
\;dk\,,\notag
\end{align}
where \({\ttF}(k,x)=(k^2+3imk\tmx-2m^2+3m^2\ssmx)\).
\end{enumerate}

The first two terms in \eqref{fc1} give no difficulty: as tensor products of Schwartz functions with themselves, even after the action of the pseudodifferential operators as in item (a) they produce smooth kernels which decrease rapidly along the diagonal, and so
satisfy the requirements of (i) and (iii) in Theorem \ref{gk}.
So we concentrate on the contributions from the integral in \eqref{fc1}.

Firstly, notice that away from the diagonal \(x=y\) the integral
defines a smooth function since it is a well-behaved oscillatory integral.
Thus it is sufficient to restrict to
the positive quadrant \(\{x>0,y>0\}\) and the negative quadrant
\(\{x<0,y<0\}\), in checking that this contribution to the kernel verifies
(i) and (iii) in Theorem \ref{gk}.
Write \(\tmx=\mp 1+\tmx\pm 1\), for \(x\lessgtr 0\)
and note that \(1\pm\tmx\) and its derivatives approach zero as \(x\to\mp\infty\) etc.,
 and similarly for \(y\).
Expanding out, there is a cancellation
of the \(k^4\) in the numerator, leading to an expression of the form 
\[\int_{\smr}\,\notag
\frac{\Bigl[\overline{{\ttF}(k,y)}{\ttF}(k,x)-
  (k^2+m^2)(k^2+4m^2)\Bigr]e^{ik(x-y)}}
{(k^2+m^2)(k^2+4m^2)^{\frac{3}{2}}}
\;dk\,=\,\sum_{j=0}^3\sum_{\alpha_j=1}^{N_j}\,
f^{\alpha_j}_j(x)g^{\alpha_j}_j(y)I_j(x-y)\]
where for each \(j\in{\{1,2,3\dots\}}\), the functions \(\{f_j^{\alpha_j}, g_j^{\alpha_j}\}\) are all
either constants or smooth functions
which together with their derivatives, decay exponentially at infinity, and
\begin{equation}\label{ij}
I_j(x-y)=\int_{\smr}\,
\frac{k^j\,e^{ik(x-y)}}
{(k^2+m^2)(k^2+4m^2)^{\frac{3}{2}}}
\;dk\,=\,(-\partial^2+4m^2)^{-1}(-i)^jN^{(j)}(x-y)\,,
\end{equation}
with \(N\in C^1\), see Appendix \ref{bessel}. For example, the \(j=3\) term, analysis of which is critical to the argument, is given
by
\[
\bigl(3im(\tmx\pm 1)-3im(\tmy\pm 1)\bigr)I_3(x-y)\,.
\]
(As indicated above, the \(\pm\) signs should be chosen according to whether we work
in the positive of negative quadrant of the plane. The \(\pm 1\)s actually cancel, and
are irrelevant in bounded regions - they are only put in to ensure exponential decay
when \((x,y)\to (\pm\infty,\pm\infty)\).)

\qed

\begin{lemma}\label{est}
  Let \(f,g\) be Schwartz functions and \(I_j\) as in \eqref{ij}.
\begin{itemize}
\item For \(j\in\{0,1,2\}\) the integral
  \[\Gamma_{1,j}(x,y)\define(4m^2-\partial_x^2)^{\frac{1}{4}}(4m^2-\partial_y^2)^{\frac{1}{4}}
  \bigl(f(x)I_j(x-y)g(y)\bigr)\]
  defines a continuous function which decays rapidly
  along the diagonal \(y=x\) so that \(\int_{\smr}|\Gamma_{1,j}(x,x)|dx<\infty\,.\)
\item For \(j=3\) the integral
  \[\Gamma_2(x,y)\define(4m^2-\partial_x^2)^{\frac{1}{4}}(4m^2-\partial_y^2)^{\frac{1}{4}}
  \Bigl[\bigl(f(x)-f(y)\bigr)I_j(x-y)\Bigr]\]
  defines a continuous function and \(\int_{\smr}|\Gamma_{2}(x,x)|dx<\infty\,.\)
\end{itemize}
\end{lemma}
\proof
First some heuristics: observe that the generalized integral
\begin{equation}\label{ijgen}
I_{a,b}(z)=\int_{\smr}\,
\frac{k^a\,e^{ikz}}
{(k^2+m^2)(k^2+4m^2)^{b}}
\;dk
\end{equation}
is absolutely convergent and defines a continuous function of \(z\) for
\(a<1+2b\), so that all the integrals appearing in both assertions themselves
define continuous functions of \(z\). However, the pseudodifferential operators
\((4m^2-\partial_{x/y}^2)^{\frac{1}{4}}\) acting on these integrals
are of order \(\frac{1}{2}\), and so
their combined effect is heuristically another power of \(k\), which takes the
second integral, i.e. \(\Gamma_2\), out of the regime of absolute convergence.
We discuss this case first. The point is that
continuity holds, the just-mentioned lack of smoothness notwithstanding, 
due to the presence
of the factor \(f(x)-f(y)\), which
serves to restore continuity near the diagonal \(x=y \), which (as previously noted) is
the only region of difficulty. To actually prove this we use a Fourier representation
\[
\Gamma_2(x,y)\,=\,(2\pi)^{-\frac{1}{2}}\iint\,e^{i(kx+ly)}(4m^2+k^2)^{\frac{1}{4}}
(4m^2+l^2)^{\frac{1}{4}}\biggl(
\frac{\hat f(k+l)}{\nu(-l)}-\frac{\hat f(k+l)}{\nu(k)}
\biggr)\,dkdl\,,
\]
where
\[
\frac{1}{\nu(k)}\,=\,\frac{k^3}{(k^2+m^2)(k^2+4m^2)^{\frac{3}{2}}}\,.
\]
Over a common denominator the integrand is \((2\pi)^{-\frac{1}{2}}e^{i(kx+ly)}\) times
\begin{equation}\label{fact}
\hat f(k+l)\Biggl[\frac{k^3(l^2+m^2)(l^2+4m^2)^{\frac{3}{2}}
    +l^3(k^2+m^2)(k^2+4m^2)^{\frac{3}{2}}}
  {(k^2+m^2)(l^2+m^2)(k^2+4m^2)^{\frac{5}{4}}(l^2+4m^2)^{\frac{5}{4}}}\Biggr]\,.
\end{equation}
To analyze this integral we change variables \((k,l)\to (u=k+l,l)\), so \eqref{fact}
becomes
\begin{equation}\label{fact2}
\hat f(u)\Biggl[\frac{(-l+u)^3(l^2+m^2)(l^2+4m^2)^{\frac{3}{2}}
    +l^3((-l+u)^2+m^2)((-l+u)^2+4m^2)^{\frac{3}{2}}}
  {((-l+u)^2+m^2)(l^2+m^2)((-l+u)^2+4m^2)^{\frac{5}{4}}(l^2+4m^2)^{\frac{5}{4}}}\Biggr]\,.
\end{equation}
Now
divide the domain \(\R^2=\mcR_1\cup\mcR_2\) with, given a small number \(\epsilon\),
\[
\mcR_1=\{(u,l):|u|\geq\epsilon|l|\}\qquad\hbox{and}\qquad
\mcR_2=\{(u,l):|u|\leq\epsilon|l|\}\,\subset \{(k=u-l,l):|k|\geq (1-\epsilon)|l|\}.
\]
In \(\mcR_1\) the inequality \(|u|\geq\epsilon|l|\) and
the rapid decay of \(\hat f\) (which is a consequence of the assumption
that \(f\in\cs(\R)\)) implies that for arbitrarily large positive integers \(N_1,N_2\)
there exists a constant such that
\[
|\hat f(u)|\leq C_1(1+|u|)^{-N_1}(1+\epsilon |l|)^{-N_2}\,.
\]
This implies absolute integrability over \(\mcR_1\) with respect to \(dudl\).
For absolute integrability
over \(\mcR_2\) it is necessary to make use of a cancellation arising from
the ``\(f(x)-f(y)\)'' structure. The factor \(\hat f(u)\) in \eqref{fact2}
ensures rapid decay in \(u\), so we need only consider growth in \(l\).
The inequality
\(|k|\geq (1-\epsilon)|l|\) implies the denominator in \eqref{fact}, and hence
\eqref{fact2}, is
\(\geq C_2(1+|l|^9)\), and the highest, and only dangerous,
power of \(l \) {\em in the numerator} arises
(after expanding out the polynomial parts)
solely from the term \(\hat f(u)\) times
\[
l^5\Bigl(\bigl((-l+u)^2+4m^2\bigr)^{\frac{3}{2}}-
\bigl(l^2+4m^2\bigr)^{\frac{3}{2}}\Bigr)
=\frac{l^5\Bigl(\bigl((-l+u)^2+4m^2\bigr)^{3}-
\bigl(l^2+4m^2\bigr)^{3}\Bigr)}{\Bigl(\bigl((-l+u)^2+4m^2\bigr)^{\frac{3}{2}}+
\bigl(l^2+4m^2\bigr)^{\frac{3}{2}}\Bigr)}\,.
\]
Although this is formally \(\sim \hat f(u)\times O(l^8)\), a cancellation arising from the
``\(f(x)-f(y)\)'' structure is now manifest, and
the inequality \(|u|\leq\epsilon|l|\) implies that
the numerator in \eqref{fact2} is in fact \(\leq C_3|\hat f(u)|(1+|u||l|^7)\), ensuring
absolute integrability with respect to \(dudl\) over \(\mcR_2\). Continuity of \(\Gamma_2\) is now a consequence of
the dominated convergence theorem, since the integrand in the formula for
\(\Gamma_2 \) is bounded by an absolutely integrable function which is
independent of \(x,y\), thus ensuring continuity of the function
\((x,y)\mapsto \Gamma_2(x,y)\).

It is now possible to take the limit \(y\to x\), and conclude
that \((2\pi)^{\frac{1}{2}}\Gamma_2(x,x)\) is equal to
\begin{equation}\label{simple}
\iint\,e^{iux}
\hat f(u)\biggl(\frac{(-l+u)^3(l^2+m^2)(l^2+4m^2)^{\frac{3}{2}}
    +l^3((-l+u)^2+m^2)((-l+u)^2+4m^2)^{\frac{3}{2}}}
  {((-l+u)^2+m^2)(l^2+m^2)((-l+u)^2+4m^2)^{\frac{5}{4}}(l^2+4m^2)^{\frac{5}{4}}}\biggr)
\,dudl\,,
  \end{equation}
By the same argument, this is a continuous function of \(x\).
To establish integrability, integrate by parts twice
to deduce that
\((2\pi)^{\frac{1}{2}}\,(ix)^2\,\Gamma_2(x,x)\) is equal to
\[
\iint\,e^{iux}\biggl(\frac{d}{du}\biggr)^2
\Biggl[\hat f(u)\biggl(\frac{(-l+u)^3(l^2+m^2)(l^2+4m^2)^{\frac{3}{2}}
    +l^3((-l+u)^2+m^2)((-l+u)^2+4m^2)^{\frac{3}{2}}}
  {((-l+u)^2+m^2)(l^2+m^2)((-l+u)^2+4m^2)^{\frac{5}{4}}(l^2+4m^2)^{\frac{5}{4}}}\biggr)\Biggr]
\,dudl
\]
which can be shown to be finite exactly as previously, so that
\(\int|\Gamma_2(x,x)|dx<\infty\), as claimed.

To establish the conclusions for \(\Gamma_{1,j} \) is easier. If either
\(f \) or \(g \) is a constant, the preceding argument works but
with \(\nu \) replaced by
\[
\frac{1}{\nu_j(k)}\,=\,\frac{k^j}{(k^2+m^2)(k^2+4m^2)^{\frac{3}{2}}}\,.
\]
in the formulae. This obviates
the need to search for any cancellation since the integrand is immediately
of sufficiently rapid decay to apply dominated convergence. For the
case that both \(f \) and \(g \) are Schwartz we use the Fourier representation
\begin{align*}
\Gamma_{1,j}(x,y)\,&=\,
(2\pi)^{-1}
\iint\,e^{i(kx+ly)}(4m^2+k^2)^{\frac{1}{4}}
(4m^2+l^2)^{\frac{1}{4}}\biggl(
\frac{\hat f(k-u)\hat g(u+l)}{\nu_j(u)}
\biggr)\,dkdldu\\
&=\,
(2\pi)^{-1}
\iint\,e^{i(k'x+l'y)}(4m^2+(k'+u)^2)^{\frac{1}{4}}
(4m^2+(l'-u)^2)^{\frac{1}{4}}\biggl(
\frac{\hat f(k')\hat g(l')}{\nu_j(u)}
\biggr)\,dk'dl'du
\end{align*}

with the change variables \((u,k,l)\to(u,k',l')=(u,k-u,l+u)\). Now
notice that since \(j=0,1,2\) absolute integrability follows easily from
the bound \(|\nu_j(u)^{-1}|\leq const.(1+|u|)^{j-5} \) and the fact that
\(\hat f,\hat g \) are Schwartz. The remainder of the argument is the same.
\qed

\section{Regularization and normal ordering of the Hamiltonian}\label{regno}
In this section we compare Hamiltonian operators
formed from the classical expressions in \eqref{defhs} by insertion of either the vacuum representation
\eqref{cfxi1}-\eqref{cfxi2} and normal ordering, or the solitonic representation \eqref{cfsgext}
and normal ordering. As emphasized by Coleman in \cite{cole}, it is important that both of these normal
orderings are induced by subtraction of the same counter-terms as used in the vacuum sector of the theory, and
regularized in the same way, before taking limits.
Insisting upon these points leads to the precise interpretation of the DHN mass shift in the
first line of \eqref{chrepeq}.
The other important conclusion of this chapter is the explicit formula for the interaction Hamiltonian
in the solitonic representation, see \eqref{iii} and \eqref{inti}. We explain a little more fully before
stating the main results in Theorem \ref{chrep} and working through the details in \S\ref{fieldreg}-\S\ref{intt}.

The existence of dynamics in the solitonic sector theory (with spatial cutoff) as a unitary evolution on
\(L^2(\mu_0)\) follows from Theorem \ref{sadj2}, asserting the self-adjointness of the operator
\(\doublecolon {\pmb H}^{sol}_{0}\doublecolon+\doublecolon H^{sol}_{I,g,{\beefb}}(\phisch)\doublecolon\) which arises from first of these representations. On the other hand, to gain an explicit understanding of the soliton dynamics it is necessary to use the solitonic representation, and the main conclusions of this section allow a precise comparison of the operators obtained from inserting these two representations in \eqref{defhs}. Indeed at the quadratic level use of \eqref{cfsgext} leads
to the operator \(\triplecolon H^{sol}_{0}\triplecolon\) of \eqref{simh} acting on \(\mfrH(\theta)\) in place of
the operator \(\doublecolon {\pmb H}^{sol}_{0}\doublecolon\) acting on \(L^2(\mu_0)\) of Theorem \ref{sadj2}.
Recall from Theorem \ref{uniteq} the unitary equivalence of these two representations via the operator
\(\bbS^\theta:\mfrH(\theta)\to L^2(\mu_0)\). In order to make the comparison, we specify precisely the regularization procedure used to prove self-adjointness, and this procedure
must be the same for both representations - convolution with an approximate identity,
see \eqref{regf}.
For the interaction, an infrared regularization, or spatial cut-off, is introduced by inserting
the factor \({\beefb}={\beefb}(x)\) into the formula for \(H^{sol}_{I,g,{\beefb}}\) in \eqref{defhs}.
It is crucial to specify a counter-term density \(\tilde{\mcH}^{sol}_{c.t.}\)
to be subtracted from the
Hamiltonian density before taking the  limit which removes the regularization - these should be essentially the same
counter-terms as for the Hamiltonian construction of the vacuum sector theory, 
see \S\ref{c.t.}.
Inserting the same two representations leads, after regularization and taking limits,
to the relation \eqref{chrepeq} between the quadratic Hamiltonian operators, as well as
to the interaction Hamiltonians: \(\doublecolon H^{sol}_{I,g,{\beefb}}(\phisch_\kappa)\doublecolon\)
(defined by normal ordering \eqref{defhs} in \(\fock \) or \(L^2(\mu_0)\))
and \({\rm H}^{sol,\kappa}_{I,g,{\beefb}}\) (defined in \eqref{intteq}). The following summarizes the conclusions, using the
definitions for regularization in \S\ref{fieldreg}.


\begin{theorem}\label{chrep}\begin{enumerate}
  \item[(i)] On \(L^2(\mu_0)\): in the limit \(\kappa\to+\infty\) both \(\int\mcH^{sol}_0(\phisch_\kappa,\pisch_\kappa)+
    \mcH_{I,g,{\beefb}}^{sol}(\phisch_\kappa)+
    \tilde{\mcH}^{sol}_{c.t.}(\phisch_\kappa)\,dx\) and
    \[\int\left( \frac{1}{2}\left(
    \pisch_\kappa^2+\phisch_\kappa\,(K\phisch)_\kappa\right)+
\mcH_{I,g,{\beefb}}^{sol}(\phisch_\kappa)+
    \tilde{\mcH}^{sol}_{c.t.}(\phisch_\kappa)\right)dx\] converge as
    bilinear forms on  \({\Poly}(\phisch)\times {\Poly}(\phisch)\) to the form related to the self-adjoint operator
    \(\doublecolon {\pmb H}^{sol}_{0}+ H^{sol}_{I,g,{\beefb}}(\phisch)\doublecolon\) of Theorem \ref{sadj2};
  \item[(ii)] On \(\mfrH(\theta)\): similarly, in the limit \(\kappa\to+\infty\)
    \(\int\left( \frac{1}{2}\left(
\uppi_\kappa^2+\upphi_\kappa\,(K\upphi)_\kappa\right)+
\mcH_{I,g,{\beefb}}^{sol}(\upphi_\kappa)+\tilde{\mcH}^{sol}_{c.t.}(\upphi_\kappa)\right)dx\) converges
  as a 
    bilinear form on  \({\Poly}(\upphi)\times {\Poly}(\upphi)\) to the form related to
    \(\Delta\M_{scl}+\triplecolon H^{sol}_0\triplecolon+{\rm H}^{sol}_{I,g,{\beefb}}\)  where the operators are defined in
    \eqref{simh} and \eqref{iii}-\eqref{inti};
  \item[(iii)] Under the unitary equivalence of Remark \ref{budd}, there holds
    \begin{align}\label{chrepeq}\begin{split}
&\bbS^{\theta *}\circ \doublecolon {\pmb H}^{sol}_{0}\doublecolon \circ \bbS^{\theta}\,=\,\triplecolon H^{sol}_{0}\triplecolon
\,+\,\Delta\M_{scl}\,,\qquad \Delta\M_{scl}=\frac{m}{\sqrt{3}}-\frac{6m}{\pi}\,,\\
&\bbS^{\theta *}\circ\doublecolon H^{sol}_{I,g,{\beefb}}(\phisch)\doublecolon\circ \bbS^{\theta}\,=\,{\rm H}^{sol}_{I,g,{\beefb}}\,.
   \end{split}\end{align}
\end{enumerate}
  \end{theorem}
\proof
After regularization of the fields, \eqref{regf}, the resulting operators are well defined on
\(\Poly(\phisch)\) and can be substituted into the Hamiltonian density, \eqref{defhs} in the present case,
leading to \(\mcH^{sol}_0(\phisch_\kappa,\pisch_\kappa)\), see \eqref{quadsol};
this can then be normal ordered,
leading to a regularized quadratic Hamiltonian density.
Carrying out the same procedure with the solitonic representation fields
\eqref{cfsr1}-\eqref{cfsr2} produces first \(\mcH^{sol}_0(\upphi_\kappa,\uppi_\kappa)\) given in \eqref{www},
related to the preceding density by
\begin{equation}\label{precden}
\bbS^{\theta *}\circ \mcH^{sol}_0(\phisch_\kappa,\pisch_\kappa)\circ\bbS^{\theta}\,=\,
\mcH^{sol}_0(\upphi_\kappa,\uppi_\kappa)\,,
\end{equation}
and similarly (using a slightly different definition of regularized quadratic Hamiltonian density which
turns out to be often more convenient to work with):
\begin{equation}\label{precden2}
\bbS^{\theta *}\circ\frac{1}{2}\left(
    \pisch_\kappa^2+\phisch_\kappa\,(K\phisch)_\kappa\right)\circ\bbS^{\theta}
    \,=\,    \frac{1}{2}\left(
\uppi_\kappa^2+\upphi_\kappa\,(K\upphi)_\kappa\right)\,.
\end{equation}
Normal ordering of the regularized Hamiltonian in the solitonic sector is achieved by
the introduction of certain counter-terms \eqref{solctmod} into the Hamiltonian, which are induced from
the vacuum sector normal ordering. Introducing \eqref{solctmod} onto both sides of \eqref{precden2}
gives the relation
\begin{equation}\label{therel}
\bbS^{\theta *}\circ \doublecolon {\pmb\mcH}^{sol}_{0,\kappa}\doublecolon \circ\bbS^{\theta}\,=\,\mathrm{O}^\kappa+
\Delta\tilde{\cal M}_{scl,\kappa}\,,\qquad \hbox{where}\quad{\pmb\mcH}^{sol}_{0,\kappa}\define\frac{1}{2}\left(
    \pisch_\kappa^2+\phisch_\kappa\,(K\phisch)_\kappa\right)\,.
\end{equation}
Here \begin{itemize}
\item \(\mathrm{O}^\kappa\) is the normal ordering with respect to the representation
  \eqref{cfsgext} of the right side of \eqref{precden2}; \(\int \mathrm{O}^\kappa(x)dx\)
  converges to the operator \eqref{bilinear} as described in Lemma \ref{wbf}.
  \item \(\Delta\tilde{\cal M}_{scl,\kappa}\) is the sum of the final three terms appearing on the right side of \eqref{summ},
and its integral has
limit \(\Delta\M_{scl}\) as \(\kappa\to+\infty\), see \S\ref{qtt}.
\end{itemize}
This implies the first line of \eqref{chrepeq} - to be precise, Theorem \ref{scconv} allows us to take
the limit of the (spatially) integrated form of the preceding relation in the sense of
bilinear forms. For the left side,
use \(\I\) to transfer to Fock space
and consider
\[
  \doublecolon  {H}^{sol}_{0,\kappa}\doublecolon=
\I^{*}\doublecolon \pmb{H}^{sol}_{0,\kappa}\doublecolon\I=
\int\doublecolon \I^{*}\pmb{\mcH}^{sol}_{0,\kappa}\doublecolon\I\,dx
=\doublecolon H^{vac}_{0,\kappa}\doublecolon +v(\varphi_\kappa)\,.
\]
Referring to Remark \ref{regen} for the first term and
\cite{MR0674511}*{\S4-5} for the second, we see that
\begin{equation}\label{reglim1}
  \lim_{\kappa\to\infty}(\Psi_1,\doublecolon  {H}^{sol}_{0,\kappa}\doublecolon\Psi_2)=
  (\Psi_1,\doublecolon  {H}^{sol}_{0}\doublecolon\Psi_2)
\end{equation}
for \(\Psi_1,\Psi_2\) polynomials or even in 
\(\Dom(\sqrt{\doublecolon H^{vac}_{0}\doublecolon})\).
The right side of \eqref{therel} requires more care: it is possible to take the limit as \(\kappa\to+\infty\)
of its spatial integral
as a bilinear
form on \({\Poly(\upphi)}\times{\Poly(\upphi)}\,,\)
giving the right side of the first line of \eqref{chrepeq}.
This corresponds to the consideration of the
left side as a bilinear form on \(\bbS^\theta{\Poly(\upphi)}\times\bbS^\theta{\Poly(\upphi)}\,.\) It is
established in the proof of Theorem \ref{scconv} that \(\bbS^\theta{\Poly(\upphi)}\) is a core for both
\(\doublecolon \pmb{H}^{sol}_0\doublecolon\) and \(\doublecolon \pmb{H}^{vac}_0\doublecolon\), where
\(\pmb{H}^{sol}_0\) is as in Theorem \ref{sadj2}, and that taking the limit
on \(\bbS^\theta{\Poly(\upphi)}\times\bbS^\theta{\Poly(\upphi)}\) gives precisely the
form associated to this operator. This proves the first line of \eqref{chrepeq} and the second is established in
\S\ref{intt}.
\qed

\subsection{Regularization of the fields}\label{fieldreg}

The full Hamiltonian is constructed as a perturbation of
the free Hamiltonian, and the crucial step in establishing self-adjointness
is to prove a uniform bound below for a family of regularized
Hamiltonians. We regularize {\em consistently} in both
the vacuum and the solitonic sector, and use this to obtain a comparison
of two representations, see Theorem \ref{scconv} in particular,
which leads to the semiclassical mass shift. 
To introduce an appropriate regularization we will
make use of an approximate identity, defined as follows. Let
\({\delta^{[1]}}\in C_0^\infty(\R) \) be a non-negative, even function with 
\({\delta^{[1]}}(x)=0 \) for \(|x|\geq 1 \), and satisfying
\(\int{\delta^{[1]}}(x)\,dx=1 \). 
For \(\kappa>0 \) define \({\delta^{[\kappa]}}(x)=\kappa{\delta^{[1]}}(\kappa x). \)
Then, as \(\kappa\to+\infty \), 
the convolution operators 
\[f\mapsto  f_\kappa:= {\delta^{[\kappa]}}\,*\,f\] 
tend to the identity, both as operators
on \(L^p \), for \(p<\infty \), and also pointwise (resp. locally uniformly)
in regions of continuity (resp. uniform continuity) of the function \(f \). 
Now define the regularized fields at a point \(x \) by
\begin{equation}\label{regf}
\varphi_\kappa(x{})\,=\,\varphi({\delta^{[\kappa]}}(\cdot-x))\,,\qquad
\pi_\kappa(x{})\,=\,\pi({\delta^{[\kappa]}}(\cdot-x))\,.
\end{equation}
\noindent
The same definition holds equally well for Fock or Schr\"odinger representations; to be concrete
we work with the former for now.
Analogous to \eqref{cf1}-\eqref{cf2}, or \eqref{cfxi1}-\eqref{cfxi2},
are the following equivalent formulae for the regularized (vacuum quantization)
fields
\begin{align}\la{cfr1}
  \varphi_\kappa(x{})\,                                          & =\,
  \int\,\frac{{\widehat{\delta^{[1]}}(k/\kappa)}}{\sqrt{2\omega^{}_k}}\,
\bigl(a^{}_ke^{ikx{}}+a_k^\dagger e^{-ikx{}}\bigr)\,dk\,,\hbox{ and }        \\
\pi_\kappa(x{})\,                                                & =\,
\int\,-i{\widehat{\delta^{[1]}}}(k/\kappa)\,\sqrt{\frac{\omega^{}_k}{2}}\,
\bigl(a^{}_ke^{ikx{}}-a_k^\dagger e^{-ikx{}}\bigr)\,dk\,,\la{cfr2}
\end{align}
\noindent
where
\(\widehat{\delta^{[1]}}(k)=(2\pi)^{-1/2}\int e^{-ikx}{\delta^{[1]}}(x)dx\).
These latter formulae indicate that the regularization amounts to a smooth
momentum cut-off at scales large compared to \(\kappa \).
The regularization \eqref{regf} determines ultraviolet
regularized
Hamiltonian operators in the vacuum sector,
and also a regularized covariance operator:
\begin{align}\notag
\langle\,0\,|\varphi_\kappa(x{})\varphi_\kappa(y{})|\,0\rangle\, & =\,
\frac{1}{2}C_{0,\kappa}^{\frac{1}{2}}(x,y)\,=
\,\frac{1}{4\pi}\,\iiint_{\smr\times\smr\times\smr}\,
\frac{
{\delta^{[\kappa]}}(y-y')e^{ik(x'-y')}
{\delta^{[\kappa]}}(x-x')
}{(k^2+4m^2)^{\frac{1}{2}}}
\;dx'\,dy'\,dk\,.                                                            \\
                                                                 & =\,
\frac{1}{2}\,\int_{\smr}\,
\frac{|{\widehat{\delta^{[1]}}}(k/\kappa)|^2e^{ik(x-y)}}{(k^2+4m^2)^{\frac{1}{2}}}
\,dk\,,\qquad\hbox{and more generally the expression}\label{regcov}          \\\notag
  F(K_0)_\kappa(x,y)\,                                           & =
\,\frac{1}{2\pi}\,\iiint_{\smr\times\smr\times\smr}\,
F(k^2+4m^2){\delta^{[\kappa]}}(x-x')e^{ik(x'-y')}
{\delta^{[\kappa]}}(y-y')
\;dk\,dx'\,dy'                                                               \\\,                                               & =\,\int_{\smr}\,
|{\widehat{\delta^{[1]}}}(k/\kappa)|^2e^{ik(x-y)}F(k^2+4m^2)
\,dk\,,
\notag\end{align}
defines the integral kernel of an operator
\(F(K_0)_\kappa \), which is a regularization of the operator \(F(K_0) \)
(for appropriate functions \(F\)).
%
\begin{remark}\label{regen} The regularized fields are not actually bounded,
but by \eqref{best} both \(\varphi_\kappa({\N_0}+1)^{-\frac{1}{2}}\) and
\(\pi_\kappa({\N_0}+1)^{-\frac{1}{2}}\) are bounded (as operators on \(\fock\)).
Insertion of the regularized fields above into the free Hamiltonian \eqref{vach}
and normal ordering leads to the regularized free Hamiltonian 
\(  \doublecolon  H^{vac}_{0,\kappa}\doublecolon\, \) which can be computed 
directly to be
\begin{equation}\label{cdtb}
{{\h}_\kappa} =\int 2\pi|{\widehat{\delta^{[1]}}}(k/\kappa)|^2\omega^{}_k a_{k}^\dagger a^{}_{k}
\,dk
\end{equation}
with \(\omega^{}_k=\sqrt{4m^2+k^2}\) is the second quantized Hamiltonian with regularized dispersion relation
\(k\mapsto 2\pi|{\widehat{\delta^{[1]}}}(k/\kappa)|^2\omega^{}_k\). [This
can be checked by applying the expression directly to 
\(\Psi_n\in\Symn L^2(\R,dk) \), and using Fourier inversion in the form
\(\int e^{ix(z-z_0)}dx=2\pi\delta(z-z_0) \)
(as an \(\cs'\)-valued integral).] Now 
\(\sqrt{2\pi}|{\widehat{\delta^{[1]}}}(k/\kappa)|\) is a continuous function equal to one at \(k=0\),
and less than or equal to one everywhere, so the formula above implies by the bounded convergence theorem that
if \(\Psi\in\Dom(\sqrt{\doublecolon H^{vac}_{0}\doublecolon})\) then \[
\lim_{\kappa\to\infty}(\Psi,\doublecolon  H^{vac}_{0,\kappa}\doublecolon\Psi)=(\Psi,\doublecolon  H^{vac}_{0}\doublecolon\Psi)\,,\]
which implies \eqref{reglim1} by polarization.
\end{remark}
In order to compute the energy of the soliton it is essential to
regularize consistently the fields
in the solitonic sector quantization. Precisely, we use the regularized fields
as \(\upphi({\delta^{[\kappa]}}(\cdot-x))\) and
\(\uppi({\delta^{[\kappa]}}(\cdot-x))\,,\)
leading to the following definition of regularized versions of the fields
(compare with \eqref{cfsgext}):
  \begin{align}\la{cfsr1}
    \upphi_\kappa(x)\,                                           & =\,-\sqrt{\scm_{cl}}{Q}
     {\tteokap}(x{})+\frac{1}{\sqrt{2\omega^{}_d}}
(a^{}_{d}+a_{d}^\dagger) {\tteonekap}(x{})                      \\
\notag                                                           & \qquad\qquad+  \frac{1}{\sqrt{2\pi}}\,\iint\,\frac{{\delta^{[\kappa]}}(x-x')}{\sqrt{2\omega^{}_k}}\,
\bigl(a^{}_{k}e_k(x'{})+a_{k}^\dagger {{e_{-k}}}(x'{})\bigr)\,dx'dk\,, \\
\uppi_\kappa(x)\,                                                & =\,-\frac{P}{\sqrt{\scm_{cl}}} {\tteokap}(x{})-i\sqrt{\frac{\omega^{}_d}{2}}
(a^{}_{d}-a_{d}^\dagger) {\tteonekap}(x{})\la{cfsr2}            \\
                                                                 & \qquad\qquad+\frac{1}{\sqrt{2\pi}}\,\iint\,-i\sqrt{\frac{\omega^{}_k}{2}}\,{\delta^{[\kappa]}}(x-x')\,
\bigl(a^{}_{k}e_k(x'{})-a_{k}^\dagger {{e_{-k}}}(x'{})\bigr)\,dx'dk\,.
\notag\end{align}
We will also use
\begin{equation}\la{cfsr3}
    (K\upphi)_\kappa(x)\,  =\,
\sqrt{\frac{\omega^{{3}}_d}{2}}
(a^{}_{d}+a_{d}^\dagger) {\tteonekap}(x{})                             
+  \frac{1}{\sqrt{2\pi}}\,\iint\,
\sqrt{\frac{\omega^{3}_k}{2}}\,\delta^{[\kappa]}(x-x')
\bigl(a^{}_{k}e_k(x'{})+a_{k}^\dagger {{e_{-k}}}(x'{})\bigr)\,dx'dk\,.
\end{equation}
\begin{remark}\label{simp}
Since \( {\tteo}, {\tteone}\) are Schwartz functions there is nothing to be gained
from applying the regularization procedure to them, except for consistency, and
the corresponding formulae with \( \tteokap={\tteo}*{\delta^{[\kappa]}}\) replaced by \( {\tteo}\) etc will
give the same results in the limit \(\kappa\to+\infty\).
  \end{remark}

Notice that, as a consequence of the fact that linearization about the soliton breaks translation
invariance, the formulae analogous to \eqref{cfr1}-\eqref{cfr2} actually define different
regularizations which we write as
  \begin{align}\la{cfsr1alt}
    &\upphi^{alt}_\kappa(x)=\,-\sqrt{\scm_{cl}}{Q}
     {\tteo}(x{})+\upphi^{\perp,alt}_\kappa(x)\,,\quad\hbox{where}\\
&\upphi^{\perp,alt}_\kappa(x)=\,    \frac{1}{\sqrt{2\omega^{}_d}}
(a^{}_{d}+a_{d}^\dagger) {\tteone}(x{})+  \,\int\,\frac{{\widehat{\delta^{[1]}}}(k/\kappa)}{\sqrt{2\omega^{}_k}}\,
\bigl(a^{}_{k}e_k(x{})+a_{k}^\dagger {{e_{-k}}}(x{})\bigr)\,dk\,,\quad\hbox{and}                      \\
&\uppi^{alt}_\kappa(x)=\,-\frac{P}{\sqrt{\scm_{cl}}} {\tteo}(x{})-i\sqrt{\frac{\omega^{}_d}{2}}
(a^{}_{d}-a_{d}^\dagger) {\tteone}(x{})\la{cfsr2alt}                                             \\
                             & \qquad\qquad+\,\int\,-i\sqrt{\frac{\omega^{}_k}{2}}\,{\widehat{\delta^{[1]}}}(k/\kappa)\,
\bigl(a^{}_{k}e_k(x{})-a_{k}^\dagger {{e_{-k}}}(x{})\bigr)\, dk\,.
\notag\end{align}
Use of these in place of \eqref{cfsr1}-\eqref{cfsr2} is generally not permissible: for example,
only \eqref{cfsr1}-\eqref{cfsr2} give rise to the correct Dashen-Hasslacher-Neveu semiclassical
mass shift (which was originally computed by taking the limit of the problem in a sequence of increasing intervals
in \cite{Dashen}). On the other hand \eqref{cfsr1alt}-\eqref{cfsr2alt} can be useful as
an intermediate approximation in the analysis of Wick polynomials in the field, see \S\ref{intt}.
  \subsection{Counter-terms and semi-boundedness}\label{c.t.}
  We will specify explicitly the counter-terms in item (i) in Theorem \ref{chrep}, which lead
  to semi-boundedness \eqref{klb}.
  The appropriate counter-terms are determined by normal ordering of the Hamiltonian
with respect to the (regularized) covariance. As noted already, the same subtractions
should be made for both the vacuum and solitonic sectors - otherwise it would not
be possible to make any meaningful statements about the mass of the
kink. (Here the ``same subtractions'' means same in terms of 
the original field \(\phi \) in \eqref{ham}; to express them
in terms of \(\varphi \) it is necessary to
take account of the different shifts of the field used in defining the theory
via \eqref{ham-vac} and \eqref{ham-sol} in the
vacuum and soliton sectors.) So we start with
the Hamiltonian on \(\fock\) in the form \eqref{hamvac}, and consider the effect of 
normal ordering with respect to the covariance
\(
\frac{1}{2}C_{0}^{\frac{1}{2}}\).
Defining the number (independent of \(x \))
\[
\gamma_\kappa\,=\,
\frac{1}{2}C_{0,\kappa}^{\frac{1}{2}}(x,x)\,=\,
\frac{1}{4\pi}\,\iiint_{\smr\times\smr\times\smr}\,
\frac{
{\delta^{[\kappa]}}(x-y')e^{ik(x'-y')}
{\delta^{[\kappa]}}(x-x')
}{(k^2+4m^2)^{\frac{1}{2}}}
\;dx'\,dy'\,dk\,=\,
\frac{1}{2}\,\int_{\smr}\,
\frac{|{\widehat{\delta^{[1]}}}(k/\kappa)|^2}{(k^2+4m^2)^{\frac{1}{2}}}
\,dk\,.
\]
We recall the formulae
\begin{align}
                             & \doublecolon \varphi_\kappa^2\doublecolon =\,\varphi_\kappa^2
  -\gamma_\kappa\,,\quad\doublecolon \varphi_\kappa^3\doublecolon \,=\,\varphi_\kappa^3-3\gamma_\kappa\varphi_\kappa
  \quad\hbox{and}                                                                            \\
                             & \doublecolon \varphi_\kappa^4\doublecolon \,=\,\varphi_\kappa^4-6\gamma_\kappa\varphi_\kappa^2+3\gamma_\kappa^2\,.
\end{align}
In the {\em vacuum sector} 
the total {\em regularized Hamiltonian} with infrared cutoff \({\beefb}\) is
\begin{equation}\label{trh}
\doublecolon H^{vac}_{g,\beefb,\kappa}\doublecolon\,\define\,\doublecolon H^{vac}_{0,\kappa}\doublecolon +\doublecolon H^{vac}_{I,g,\beefb}(\varphi_\kappa)\doublecolon =\int \doublecolon \mcH^{vac}_{0,\kappa}\doublecolon +\beefb(x)\doublecolon \mcH^{vac}_{I,g}(\varphi_\kappa)\doublecolon \,dx\,,\end{equation}
where
\begin{align}\begin{split}\label{vacint}
\doublecolon \mcH^{vac}_{0,\kappa}\doublecolon \,
&=\,\frac{1}{2}\Bigl[\,
\pi_\kappa^2\,+\,\varphi_\kappa\,K_0\varphi_\kappa\,
-\,
K_{0,\kappa}^{\frac{1}{2}}(x,x)\,\Bigr]\,=\frac{1}{2}\doublecolon \bigl(\pi_\kappa^2\,+\,\varphi_\kappa\,K_0\varphi_\kappa\bigr)\doublecolon \,,
\qquad\hbox{and}\\
  \doublecolon \mcH^{vac}_{I,g}(\varphi_\kappa)\doublecolon \,
                             & =\,\Bigl[\,
  2mg\bigl(\varphi_\kappa^3-3\gamma_\kappa\varphi_\kappa\bigr)
  \,+\,\frac{1}{2}g^2\bigl(\varphi_\kappa^4
-6\gamma_\kappa\varphi_\kappa^2
+3\gamma_\kappa^2\bigr)
\,\Bigr]\,                                                                                   \\
                             & =\,
  2mg\,\doublecolon \varphi_\kappa^3\doublecolon 
  \,+\,\frac{1}{2}g^2\,\doublecolon \varphi_\kappa^4\doublecolon 
\,.\end{split}
\end{align}
Thus we have introduced the following counter-terms (or subtractions)
in the definition of the regularized Hamiltonian {\em density}:
\[
{\mcH}^{vac}_{c.t.}(\varphi_\kappa)\,=\,
-3g^2{\beefb}\gamma_\kappa\varphi_\kappa^2-6mg{\beefb}\gamma_\kappa\varphi_\kappa-
\frac{1}{2}K_{0,\kappa}^{\frac{1}{2}}(x,x)\,+\,\frac{3g^2{\beefb}}{2}\gamma_\kappa^2 \,\,.
\]
In order to derive a comparable Hamiltonian in the solitonic sector we
have to take into account the shift \(\varphi_\kappa\to -\Phi_0+\Phi_S+\varphi_\kappa\)
to obtain the corresponding counter-terms (using the representation \eqref{cfxi1}-\eqref{cfxi2}):
\begin{equation}\label{solct}
{\mcH}^{sol}_{c.t.}(\varphi_\kappa)\,=\,
-3g^2{\beefb}\gamma_\kappa(-\Phi_0+\Phi_S+\varphi_\kappa)^2
-6mg{\beefb}\gamma_\kappa(-\Phi_0+\Phi_S+\varphi_\kappa)-
\frac{1}{2}K_{0,\kappa}^{\frac{1}{2}}(x,x)\,+\,\frac{3g^2{\beefb}}{2}\gamma_\kappa^2 \,\,.
\end{equation}
(Notice that, on account of this shift, the quadratic ``mass renormalization''
and linear  counter-terms  induce \(O(g^0) \) modifications of the
soliton sector Hamiltonian due to the \(g\) dependence of
\(\Phi_0\) and \(\Phi_S\). We do not bother to regularize \(\Phi_S\) since it is smooth, although it would be strictly consistent to do so. ) All together this
leads to the following regularized Hamiltonian density:
\begin{align}\notag
\,                           & {\;}\,\frac{1}{2}\,\Bigl[\,
\pi_\kappa^2\,+\,\varphi_\kappa\,K\varphi_\kappa\,
-6m^2\gamma_\kappa {\beefb}\,\bigl(\ttmx\,-\,1\bigr)\,-\,
K_{0,\kappa}^{\frac{1}{2}}(x,x)\,\Bigr]\,+\,{\beefb}\doublecolon \mcH_{I,g}^{sol}(\varphi_\kappa)
\doublecolon \\
\label{m2}
                             & \quad =\,\frac{1}{2}\,\Bigl[\,
\pi_\kappa^2\,+\,\varphi_\kappa\,(K\varphi)_\kappa\,
+6m^2\gamma_\kappa {\beefb}\,\ssmx\,-\,
K_{0,\kappa}^{\frac{1}{2}}(x,x)\,+
\,\varphi_\kappa\,\bigl(K\varphi_\kappa-(K\varphi)_\kappa\bigr)
\Bigr]
\,+\,\doublecolon {\beefb}\mcH_{I,g}^{sol}(\varphi_\kappa)\doublecolon,                                       \\
\intertext{where}
\label{solint}
                             & \doublecolon \mcH_{I,g}^{sol}(\varphi_\kappa)\doublecolon \,
                                                   =\,\,
  2mg\tmx\bigl(\varphi_\kappa^3-3\gamma_\kappa\varphi_\kappa\bigr)
  \,+\,\frac{1}{2}g^2\bigl(\varphi_\kappa^4
-6\gamma_\kappa\varphi_\kappa^2
+3\gamma_\kappa^2\bigr)
\,                                                                                           \\
                             & 
                                                   \phantom{
\doublecolon \mcH_{I,g}^{sol}(\varphi_\kappa)\doublecolon \,
                                                     }
                                                   =\,\,
  2mg\tmx\,\doublecolon \,\varphi_\kappa^3\,\doublecolon 
  \,+\,\frac{1}{2}g^2\,\doublecolon \,\varphi_\kappa^4\,\doublecolon 
\,\,.\notag
\end{align}
\begin{remark}\label{ircob}
  The infrared cutoff is included in \eqref{m2}, but to make contact with the DHN mass shift formula it will be set identically equal to one, since that
  formula gives the mass shift in the infinite volume limit. Observe that if one ignores the infrared cutoff for purposes of defining the counter-terms, and
  then re-inserts \({\beefb}\) in front of the interaction terms \(\mcH_{I,g}^{sol}(\varphi_\kappa)\) at the end, we end up instead with the following density
  \begin{equation}\label{m3}
\mcH^{sol,alt}_{g,\beefb,\kappa}\,=\,\frac{1}{2}\,\Bigl[\,
\pi_\kappa^2\,+\,\varphi_\kappa\,(K\varphi)_\kappa\,
+6m^2\gamma_\kappa \,\ssmx\,-\,
K_{0,\kappa}^{\frac{1}{2}}(x,x)\,+
\,\varphi_\kappa\,\bigl(K\varphi_\kappa-(K\varphi)_\kappa\bigr)
\Bigr]
\,+\,\doublecolon \beefb\mcH_{I,g}^{sol}(\varphi_\kappa)\doublecolon
  \end{equation}
  the only effect on the actual Hamiltonian is to shift it by
  an constant equal to the integral of
  \[\delta_{{\beefb},\kappa}=3m^2\gamma_\kappa \int\,({\beefb}(x)-1)\,\ssmx\,dx
  \]
  which vanishes if \({\beefb}(x)\to 1\) boundedly; the most convenient thing is to adjust the counter term density:
  \begin{equation}\label{solctmod}
\tilde{\mcH}^{sol}_{c.t.}={\mcH}^{sol}_{c.t.}-\delta_{\beefb,\kappa}\,,
\end{equation}
to take account of this, as in the statement of Theorem \ref{chrep}.
We therefore find it convenient to use as regularized Hamiltonian
\begin{equation}\label{tsrh}\doublecolon H^{sol}_{g,\beefb,\kappa}\doublecolon\,\define\,
\doublecolon H^{sol}_{0,\kappa}\doublecolon +\doublecolon H^{sol}_{I,g,\beefb}(\varphi_\kappa)\doublecolon
\end{equation}
where \(\doublecolon H^{sol}_{0,\kappa}\doublecolon=
\int\doublecolon \mcH^{sol}_{0,\kappa}\doublecolon\,dx
\)
with
\(
\mcH^{sol}_{0,\kappa}\define
\frac{1}{2}\bigl(
\pi_\kappa^2\,+\,\varphi_\kappa\,(K\varphi)_\kappa\bigr)
\)
and
\(\doublecolon H^{sol}_{I,g,{\beefb}}(\varphi_\kappa)\doublecolon=\int {\beefb}\doublecolon \mcH_{I,g}^{sol}(\varphi_\kappa)\doublecolon \,dx\,.\)
We'll see in Proposition \ref{agreesil} that in the limit \(\kappa\to +\infty\) this agrees with the Hamiltonian defined by \eqref{m3}.
Notice that the subtractions here, being induced from those made
in the vacuum sector involve the same covariance operator, and so the
normal ordering symbol \(\doublecolon \) has the same
meaning. This form for the Hamiltonian will
lead to the DHN formula for the mass shift in \S\ref{mass-shift}, once we compare with the solitonic representation.
\end{remark}

\begin{remark}\label{exc.t.}
  The situation with the counter-terms is clarified by displaying them in relation to the \(\phi^4\)
  Hamiltonian \(H(\phi,\pi)\) in \eqref{ham}, i.e., before shifting to vacuum or soliton sector. Also
  put \({\beefb}\equiv 1\) for simplicity in this remark, then
  the Hamiltonian density is (up to  constant)
  \[
  \frac{1}{2}\bigl(\pi^2+(\partial_x\phi)^2+4m^2\phi^2\bigr)+\frac{1}{2}g^2\phi^4-3m^2\phi^2\,,
  \]
  and
  quantizing in the space \(L^2(\mu_0)\) and regularizing, the standard counter-terms to be included are
  \[
  -\frac{1}{2}K_{0,\kappa}^{\frac{1}{2}}(x,x)
-3g^2\gamma_\kappa\phi_\kappa^2+\frac{3}{2}\gamma_\kappa^2g^2+3m^2\gamma_\kappa
\]
which correspond (for \({\beefb}\equiv 1\)) to the counter-terms \({\mcH}^{vac}_{c.t.}\) (resp. \({\mcH}^{sol}_{c.t.}\)) above
after making the shift \(\varphi+\Phi_0=\phi\) (resp. \(\varphi+\Phi_S=\phi\)) as in
\eqref{ham-vac} (resp. \eqref{ham-sol}). This is discussed in \cite{cole}.
  \end{remark}

For the determination of the relation between the 
free Hamiltonians in the vacuum
and solitonic sectors, it is necessary to 
make a precise definition of the regularized free Hamiltonian, 
and take a limit. In this connection,
recall from \cite{MR0674511}*{Theorem 4.4}
that the Fock space form for the free Hamiltonian can be fixed by looking at
the expression for the Hamiltonian in terms of fields as a bilinear
form on \({\Poly(\varphi)}\times{\Poly(\varphi)}\,,\) so we follow the same procedure here.
As indicated in \eqref{m2} and \eqref{notonly}, it does make a difference
for finite \(\kappa \) whether 
we just regularize the field \(\varphi \) and then apply the
operator \(K\), or regularize \(K\varphi \) (as in the definition of \(
\doublecolon \mcH^{sol}_{0,\kappa}\doublecolon
\) above):
\begin{equation}\label{notonly}
K\varphi_\kappa(x)-(K\varphi)_\kappa(x)\,=\,
-6m^2\ssmx\,\varphi_\kappa(x)
+6m^2\int
{\delta^{[\kappa]}}(x-x')\ssmx'\varphi(x')
\,dx'
\end{equation}
To check that other choices
lead to the same answer in the limit \(\kappa\to+\infty\), we first compute
\begin{align}  \label{quadsol}
\doublecolon \mcH^{sol}_{0,\kappa}\doublecolon
\,   & =\,\frac{1}{2}\,\doublecolon\,\Bigl[\,
\pi_\kappa^2\,+\,\varphi_\kappa\,(K\varphi)_\kappa\,\bigr]\doublecolon \\\notag
\,   & =\,\frac{1}{2}\,\doublecolon\,\Bigl[\,
\pi_\kappa^2\,+\,\varphi_\kappa\,(K_0\varphi)_\kappa\,
-6m^2\int
\,\varphi_\kappa(x)
{\delta^{[\kappa]}}(x-x')\ssmx'\varphi(x')
\, dx'
\bigr]\doublecolon
                                                                       \\\notag
\,   & =\,
\frac{1}{2}\,\Bigl[\,
\pi_\kappa^2\,+\,\varphi_\kappa\,(K\varphi)_\kappa\,
-\,
K_{0,\kappa}^{\frac{1}{2}}(x,x)\,
+\,6m^2\,\langle\,\Omega_0\,,\int
\varphi_\kappa(x)(\ssm x'\varphi(x'))\delta^{[\kappa]}(x-x')dx'
\,\Omega_0\,\rangle\,\Bigr]
                                                                       \\
\begin{split}
\,   & =\,
\frac{1}{2}\,\Bigl[\,
\pi_\kappa^2\,+\,\varphi_\kappa\,(K\varphi)_\kappa\,
-\,
K_{0,\kappa}^{\frac{1}{2}}(x,x)\,
+\,  6m^2\gamma_\kappa\,\ssmx\,\Bigr]
                                                                       \\
  \, & \qquad-\,
\Bigl[  3m^2\gamma_\kappa\,\ssmx\,
-\,\frac{3m^2}{4\pi}\iiint
  \frac{e^{ik(x'-y)}}{(k^2+4m^2)^{\frac{1}{2}}}\delta^{[\kappa]}(x-y)\delta^{[\kappa]}(x'-x)
  \ssm x'dx'dydk\,
    \Bigr]\,.\end{split}\label{quadsol2}
\end{align}
This indicates that not only \eqref{notonly}, but also the consequent 
normal ordering adjustments arising from the choice in \eqref{quadsol} 
potentially
affect the regularized Hamiltonian for finite \(\kappa \).
Nevertheless, the following lemmas indicate
that these effects vanish in the limit \(\kappa\to+\infty \) in the sense of convergence of quadratic forms:
\begin{prop}\label{agreesil} Referring to \eqref{m3}, define  \(H^{sol,alt}_{g,{\beefb},\kappa}=\int\,\mcH^{sol,alt}_{g,{\beefb},\kappa}\,dx\), then
  \[\doublecolon H^{sol}_{g,{\beefb},\kappa}\doublecolon\,-H^{sol,alt}_{g,{\beefb},\kappa}
  +\delta_{{\beefb},\kappa}\,\to\,0\]
  in the limit \(\kappa\to+\infty \)
as a bilinear
form on \({\Poly(\varphi)}\times{\Poly(\varphi)}\,,\) with the understanding that the integrals are interpreted as
bilinear form valued integrals.
  \end{prop}
This follows from the following lemmas. The first shows that the penultimate term in \eqref{m2} has vanishing contribution
to the Hamiltonian in the sense just mentioned.
\begin{lemma}\label{altregham}
  \[
\lim_{\kappa\to+\infty}\,\int\,\varphi_\kappa\,\bigl(K\varphi_\kappa-(K\varphi)_\kappa\bigr)\,dx\,=\,0
  \]
in the sense of convergence as a bilinear form on \({\Poly(\varphi)}\times{\Poly(\varphi)}\).
  \end{lemma}
\noindent
This is proved in Appendix \ref{arh}.  A closely related calculation, 
also given in Appendix \ref{arh}, gives the following result.
\begin{lemma}\label{crf}

(a) In Fock space operator norm 
\[\lim_{\kappa\to+\infty}\Bigl\|({\N_0}+1)^{-\frac{1}{2}}
\Bigl(\int
\doublecolon \ssmx\,\varphi_\kappa(x)^2\doublecolon \,dx
-\iint
\doublecolon \varphi_\kappa(x)
{\delta^{[\kappa]}}(x-x')\ssmx'\varphi(x')
 \doublecolon\, dxdx'
\Bigr)({\N_0}+1)^{-\frac{1}{2}}\Bigr\|\,=\,0\,.\]
 
(b) In the Schr\"odinger representation, both
 \(\int\ssmx \doublecolon \phisch_\kappa(x)^2\doublecolon \,dx\)
 and
 \[\iint
\doublecolon \phisch_\kappa(x)
{\delta^{[\kappa]}}(x-x')\ssmx'\phisch(x')
 \doublecolon \,dx'dx
 \]
 converge to  \(\int\ssmx \doublecolon \phisch(x)^2\doublecolon \,dx\) in
\(L^p(\mu_0)\) for every \(p<\infty\).
  \end{lemma}
\noindent
The next result, proved in the same appendix,
deals with the error in the zero point energy correction
in \eqref{quadsol}.
\begin{lemma}\label{zpd}
In the limit \(\kappa\to+\infty\)
  \[
\int\,\Bigl[
  6m^2\gamma_\kappa\,\ssmx\,-\,\frac{6m^2}{4\pi}\iiint
  \frac{e^{ik(x'-y)}}{(k^2+4m^2)^{\frac{1}{2}}}\delta^{[\kappa]}(x-y)\delta^{[\kappa]}(x'-x)\ssm x'\,\,dx'dydk
    \Bigr]\,dx=O\Bigl(\frac{\ln\kappa}{\kappa}\Bigr)\,\,.\]
  \end{lemma}

\paragraph{Semiboundedness.} To obtain the existence theory we use the Hamiltonian in the form \eqref{delH}, which leads us to consider
the corresponding regularized spatially cut-off Hamiltonian density
\begin{equation}\label{delHreg}
  \doublecolon \tilde\mcH_{I,g,\beefb}^{sol}(\varphi_\kappa)\doublecolon \,
\define\,\bigl[\,-3m^2\ssmx\doublecolon \,\varphi_\kappa^2\,\doublecolon\,+
  2mg{\beefb}(x)\tmx\,\doublecolon \,\varphi_\kappa^3\,\doublecolon 
  \,+\,\frac{1}{2}g^2{\beefb}(x)\,\doublecolon \,\varphi_\kappa^4\,\doublecolon 
\,\bigr]\,,
\end{equation}
which admits, pointwise, the lower bound
\begin{align*}
\doublecolon \tilde\mcH_{I,g,\beefb}^{sol}(\varphi_\kappa)\doublecolon \,                                 & \geq\,
\Bigl(\frac{1}{2}g^2{\beefb}-\frac{3}{4}\epsilon_1^{\frac{4}{3}}-\frac{1}{2}\epsilon_2^2
-\frac{1}{2}\epsilon_3^2-\frac{1}{4}\epsilon_4^4\Bigr)\varphi_\kappa^4 \\
                                                                                          & \quad+\,
\biggl[
3m^2\gamma_\kappa\ssmx+\frac{3}{2}\gamma_\kappa^2 g^2 {\beefb}-\frac{(2mg{\beefb}\tmx)^4}{4\epsilon_1^4}
-\frac{(3g^2\gamma_\kappa {\beefb})^2}{2\epsilon_2^2}                         \\
                                                                                          & \phantom{3m^2\gamma_\kappa\ssmx+\frac{3}{2}\gamma_\kappa^2 g^2 {\beefb}-}
  -\frac{(3m^2\ssmx)^2}{2\epsilon_3^2}
-\frac{3(6mg{\beefb}\gamma_\kappa\tmx)^{4/3}}{4\epsilon_4^{4/3}}
\biggr]\,,
\end{align*}
for any choice of \(\epsilon_1,\epsilon_2,\epsilon_3,\epsilon_4\) (possibly depending on \(x\) - proved by use
of \(ab\leq \frac{\epsilon^pa^p}{p}+\frac{b^q}{q\epsilon^q}, p^{-1}+q^{-1}=1,\epsilon>0\)
for \(p=4/3,2\).) Now for small \(\epsilon_0\) choose
\(\epsilon_1^{\frac{4}{3}}=\epsilon_2^2=\epsilon_3^2=\epsilon_4^4=\epsilon_0g^2{\beefb}\) to deduce that, with the condition
\[
{\beefb}(x)\geq const.\ssmx
\]
there exist \(C_1,C_2>0\), independent of \(x,\kappa,{\beefb}\) but depending upon \(g\), such that
\begin{equation}\label{klb}
\doublecolon \tilde\mcH^{sol}_{I,g,\beefb}(\varphi_\kappa)\doublecolon\,\geq\,
-C_1{\beefb}-C_2{\beefb}\gamma_\kappa^2\,.
  \end{equation}
(The dependence on \(g\) is no worse than \(O(g^{-2})\), in the sense that \(|g^2C_1|+|g^2C_2|\) is bounded for
\(|g|\leq 1\).) The lower bound \eqref{klb} 
implies that the operator \( \doublecolon H^{vac}_0\doublecolon + 
\int\,\doublecolon \tilde\mcH_{I,g,\beefb}^{sol}(\varphi_\kappa)\doublecolon\,dx  \) is bounded
below uniformly in \(\kappa \) and determines a self-adjoint operator 
\(\doublecolon H^{sol}_{g,\beefb}\doublecolon \)  as remarked in the discussion of Theorem \ref{sadj2}.
\subsection{Change of representation - quadratic terms}\label{qtt}
in this section we work out some details of the relation between the vacuum representation
 \eqref{cfxi1}-\eqref{cfxi2} and the representation
\eqref{cfsgext}, and exploit the fact that the latter diagonalizes the quadratic part of the
Hamiltonian in the solitonic sector to obtain precise information about
\(\doublecolon {\pmb H}^{sol}_{0}\doublecolon \) which is not otherwise evident.
So consider the effect on the quadratic part of the Hamiltonian of the change of representation, as described in
Corollary \ref{uniteqcor}.
There is a unitary
isomorphism
\(\mbS^{\theta}:L^2(\gamma_\theta(dQ))\otimes\mfrbF\cong L^2(\mu(\theta))\to L^2(\mu_0)\)
or equivalently
\[
\bbS^\theta:\mfrH(\theta)\to L^2(\mu_0)
\]
which is defined on the solitonic Hilbert space \eqref{aost}. It intertwines the regularized fields
\[
\bbS^{\theta *}\circ \phisch_\kappa\circ\bbS^{\theta}=\upphi_\kappa\,,
\qquad
\bbS^{\theta *}\circ\pisch_\kappa\circ\bbS^{\theta}=\uppi_\kappa\,,
\]
from which follows \eqref{precden}-\eqref{precden2}. We now normal order
using \(K,K_0,C_0=K_0^{-1} \) as in \eqref{regcov} and \eqref{regfc}).
Consider the first term on the right side of \eqref{precden2}, i.e., 
\begin{equation}\label{www}
\frac{1}{2}\,\Bigl[\,
\uppi_\kappa^2\,+\,\upphi_\kappa\,(K\upphi)_\kappa\,\Bigr]\,,
\end{equation}
and normal order with respect to the representation \eqref{cfsgext} (indicating with a colon). This
produces
\[\frac{1}{2}\omega_d {\tteonekap}(x)^2+
  \frac{1}{4\pi}\int\omega_k\delta^{[\kappa]}(x-x_2')\delta^{[\kappa]}(x-x_1')e_{-k}(x_1')e_{k}(x_2')dx_1'dx_2'dk
=\frac{1}{2}K^{\frac{1}{2}}_{\kappa}(x,x)\,,
  \]
with  \(K_\kappa \) as in \eqref{regfc}. Similarly write \(K=K_0-6m^2\ssmx\) and normal order  
the left of \eqref{precden2} with respect to \eqref{cfxi1}-\eqref{cfxi2}, indicating this normal ordering in
\(\mfrH(\theta)\) with a triple colon. All together this
leads to
\begin{align}\label{summ}\begin{split}
&  \bbS^{\theta *}\,
\doublecolon \pmb{\mcH}^{sol}_{0,\kappa}\doublecolon
  \bbS^{\theta}\,=\,
  \bbS^{\theta *}\doublecolon
  \left(\frac{1}{2}\left(
    \pisch_\kappa^2+\phisch_\kappa\,(K\phisch)_\kappa\right)\right)
  \doublecolon \bbS^{\theta}\,                     =\,\triplecolon  \left(\frac{1}{2}\left(
\uppi_\kappa^2+\upphi_\kappa\,(K\upphi)_\kappa\right)\right)\triplecolon
\,+\,\Delta\tilde{\cal M}_{scl,\kappa}\qquad\hbox{where}\\
&\Delta\tilde{\cal M}_{scl,\kappa}(x)\,=\,
\frac{1}{2}K^{\frac{1}{2}}_{\kappa}(x,x)
\,-\,\frac{1}{2}K^{\frac{1}{2}}_{0,\kappa}(x,x)
                                                                                          \,+3m^2\,\left(\,\Omega_0\,,\int
\varphi_\kappa(x)(\ssm x'\varphi(x'))\delta^{[\kappa]}(x-x')dx'
\,\Omega_0\,\right)_{\mfrH_0}\,
.
\end{split}
\end{align}
Now let \(\Omega'\) be the vacuum in the transverse Fock space \(\mfrF\), and
  let \(F\) be smooth with \(\int |F(Q)|^2\gamma_\theta(dQ)=1\), then 
\begin{align*}
  \left(\,\bbS^\theta F\otimes\Omega'\,,\,\doublecolon{\pmb\mcH}^{sol}_{0,\kappa}(x)\doublecolon\,
  \bbS^\theta F\otimes
  \Omega'\right)_{L^2(\mu_0)}\, & =\,
  \frac{\tteokap(x)^2}{2{\scm_{cl}}}\|(F'(Q)-\scm_{cl}\sqrt{\theta}QF(Q))\|_{L^2(\gamma_\theta(dQ))}^2\,+\,
\Delta\tilde{\cal M}_{scl,\kappa}(x)
\,.
\notag
\end{align*}
This indicates that the integral of \(\Delta\tilde{\cal M}_{scl,\kappa}(x)\)
gives the infimum of
the quadratic part of the energy. In the limit \(\kappa\to +\infty\) 
we can replace the final term in \(\Delta\tilde{\cal M}_{scl,\kappa}(x)\) by the expression in Lemma \ref{zpd}, and 
thence compute
that the sum of the three
terms integrated over \(x\in\R\) has the following nonzero limit \(\Delta\M_{scl}\):
\begin{lemma}\label{mass}
 \begin{equation}\label{dahn}
  \Delta\M_{scl}\,\define\,
\lim_{\kappa\to+\infty}\,\int\,\frac{1}{2}\Bigl(\,K^{\frac{1}{2}}_\kappa(x,x)
-K^{\frac{1}{2}}_{0,\kappa}(x,x)
+3m^2\ssmx C_{0,\kappa}^{\frac{1}{2}}(x,x)\Bigr)\,dx\,=\,-m\,(\frac{3}{\pi}-
\frac{1}{2\sqrt{3}})\,,
\end{equation}
and furthermore \(\int\Delta\tilde{\cal M}_{scl,\kappa}\,dx\) has the same limiting value as \(\kappa\to\infty\).
\end{lemma}
\proof It is to be understood here that the Lemma
is asserting the existence of the limit in \eqref{dahn};
the proof is in \S\ref{mass-shift}. The last assertion is a consequence of Lemma \ref{crf}. \qed

We now turn to the limit of \eqref{summ} (integrated)
as \(\kappa\to+\infty\), in particular relating the limit of
\(\doublecolon {\pmb H}^{sol}_{0,\kappa}\doublecolon=\int\doublecolon \pmb{\mcH}^{sol}_{0,\kappa}\doublecolon\) to the limit of
the first term on the right hand side, which 
is the operator \(\triplecolon H^{sol}_0\triplecolon\) in \eqref{bilinear}.
This latter operator is
itself self-adjoint with domain \(\Dom(\triplecolon H^{sol}_{0}\triplecolon)\)
given in Remarks \ref{msa} and \ref{gaussch}.
  This leads to the precise definition (referenced in Theorem \ref{sadj2}) of the self-adjoint operator
  \(\doublecolon {\pmb H}^{sol}_0\doublecolon\) on \(L^2(\mu_0)\).
  \begin{theorem}\label{scconv}
  \(\bbS^{\theta *}\circ\doublecolon {\pmb H}^{sol}_{0,\kappa}\doublecolon \circ \bbS^{\theta}\) converges in the limit \(\kappa\to\infty\), as a bilinear form on
  \({\Poly}(\upphi)\times {\Poly}(\upphi)\), to 
  \(\Delta\M_{scl}+\triplecolon H^{sol}_0\triplecolon\). This limit defines a closable quadratic form whose
  closure is associated to the self-adjoint operator \(\Delta\M_{scl}+\triplecolon H^{sol}_{0}\triplecolon\), and
  whose form domain is \(\Dom\bigl((\triplecolon H^{sol}_{0}\triplecolon)^{\frac{1}{2}}\bigr)\). Applying the unitary transformation \(\bbS^{\theta}\), the
  quadratic form defined by the limit of \(\doublecolon {\pmb H}^{sol}_{0,\kappa}\doublecolon \)
  is closable and defines a self-adjoint operator \(\doublecolon {\pmb H}^{sol}_{0}\doublecolon\)  with
  domain \(\bbS^{\theta}\Dom(\triplecolon H^{sol}_{0}\triplecolon)\), which equals
  \( \doublecolon {\pmb H}^{vac}_{0}\doublecolon-\int 3m^2\ssmx\doublecolon \phisch(x)^2\doublecolon\,dx\) on the dense subspace \({\bbS^{\theta}\,}{\Poly}(\upphi)\,, \) which is a domain of essential
  self-adjointness.
The operator \(\doublecolon {\pmb H}^{sol}_{0}\doublecolon\) so defined is
  the self-adjoint operator referred to in Theorem \ref{sadj2}, and
\(\bbS^{\theta *}\circ \doublecolon {\pmb H}^{sol}_{0}\doublecolon \circ\bbS^{\theta}\,=\,\triplecolon H^{sol}_{0}\triplecolon
\,+\,\Delta\M_{scl}\,.\)
\end{theorem}
  \proof{\em Step One.} Recall from Lemma \ref{wbf} that if the expression obtained by substitution of
  \eqref{cfsgext} into \eqref{fsh} is formally interpreted as a weak bilinear form valued integral,
  then it equals the bilinear form defined by the expression \eqref{simh}, on the domain
  \({\Poly}(\upphi)\times{\Poly}(\upphi)\). In the next step we show the weak limit of
  the corresponding regularization, \(\int\,\mathrm{O}^\kappa\,dx\), defined by substitution of \eqref{cfsr1}-\eqref{cfsr2} into \eqref{www}
  and normal ordering,
  gives the same expression for \(\triplecolon H^{sol}_{0}\triplecolon\)\,,
completing the proof of Lemma \ref{wbf}.\\
  \noindent
  {\em Step Two.}
Consider the convergence, as \(\kappa\to\infty\),
of \(\int\,\mathrm{O}^\kappa\,dx\),
in the sense of a weak convergence (pointwise bilinear form convergence) on the domain
\({\Poly}(\upphi)\times{\Poly}(\upphi)\). For positive \(\kappa\)
this operator has a meaning as an integral of terms involving normal ordered pairs of
creation/annihlation operators, which always define a bilinear form.
Terms involving the discrete modes do not present difficulties, so we
concentrate on those involving the continuous modes,
of which we consider as representative that involving two
annihilation operators, namely
\[
\frac{1}{4\pi}\int\Bigl[\int\,a_ka_l\Bigl(\frac{\omega_k(\omega_k-\omega_l)}{\sqrt{\omega_k\omega_l}}\Bigr)
{\delta^{[\kappa]}}(x-x'){\delta^{[\kappa]}}(x-x'')e_k(x')e_l(x'')dx'dx''dkdl\Bigr]dx\,.
\]
to be understood as a weak bilinear form valued integral.
We want to show that this converges in the weak sense, as \(\kappa\to+\infty\), to
\[
\frac{1}{4\pi}\int\,a_ka_l\Bigl(\frac{\omega_k(\omega_k-\omega_l)}{\sqrt{\omega_k\omega_l}}\Bigr)
e_k(x)e_l(x)dkdldx\,,
\]
and then show that this is zero.
The weak interpretation above means taking the matrix element of the integrand between
two elements of \({\Poly}(\upphi)\), which will reduce the above integral to one of
the form
\[
\frac{1}{4\pi}\int\,f(k,l)\Bigl(\frac{\omega_k(\omega_k-\omega_l)}{\sqrt{\omega_k\omega_l}}\Bigr)
{\delta^{[\kappa]}}(x-x'){\delta^{[\kappa]}}(x-x'')e_k(x')e_l(x'')dx'dx''dkdldx\,.
\]
with \(f\) a Schwartz function.
We recall the fact that \({\delta^{[\kappa]}}*U(x)\to U(x)\) at points of continuity of \(U\), (and in fact 
uniformly on intervals of uniform continuity of \(U\)), and apply the dominated convergence theorem.
Restricting the integral to bounded intervals of \(x\) this gives convergence immediately, so that
\begin{align}\notag
\lim_{\kappa\to+\infty}                                                                              & \frac{1}{4\pi}\int\,\id_{\{|x|\leq 10\}}(x)\,f(k,l)\Bigl(\frac{\omega_k(\omega_k-\omega_l)}{\sqrt{\omega_k\omega_l}}\Bigr)
{\delta^{[\kappa]}}(x-x'){\delta^{[\kappa]}}(x-x'')e_k(x')e_l(x'')dx'dx''dkdldx                  \\\,                  & =\,
\frac{1}{4\pi}\int\,\id_{\{|x|\leq 10\}}(x)\,f(k,l)\Bigl(\frac{\omega_k(\omega_k-\omega_l)}{\sqrt{\omega_k\omega_l}}\Bigr)
e_k(x)e_l(x)dkdldx\,.
\notag\end{align}
For infinite intervals additional integration by parts arguments suffice to establish convergence.
Referring to \eqref{ekdef}, we see that it is sufficient to consider the case that
\(e_k(x')\) is replaced by \(g(x')h(k)e^{ikx'}\) and
\(e_l(x'')\) is replaced by \(\tilde g(x'')\tilde h(l)e^{ilx''}\) with
\(h(k)\) a polynomial in \(k \) divided by \({\sqrt{(k^2+m^2)(k^2+4m^2)}} \), and similarly
for \(\tilde h(l) \), and \(g,\tilde g\) either identically equal
to 1, or otherwise one of the functions \(\sech^2 m(\cdot)\) or \(\tanh m(\cdot)\).
It follows that 
\(G(k,l)=h(k)\tilde h(l)f(k,l)\) is a Schwartz function, and that
it is sufficient to establish that for such \(g,\tilde g,G \)
\begin{align}\begin{split}\label{mpm}
\lim_{\kappa\to +\infty}\,\int_{\R^5}\,                                                              & \id_{\{|x|\geq 10\}}(x)G(k,l)
{\delta^{[\kappa]}}(x-x'){\delta^{[\kappa]}}(x-x'')g(x')\tilde g(x'')e^{ikx'+ilx''}dx'dx''dkdldx \\\, & =\,
\int_{\R^3}\,\id_{\{|x|\geq 10\}}(x)G(k,l)
g(x)\tilde g(x)e^{i(k+l)x}dkdldx\,.
\end{split}\end{align}
After two integration by parts (in \(k\) and \(l\)), the right hand side can be written as
\begin{align}
\int_{\R^3}                                                                                          & \,\id_{\{|x|\geq 10\}}(x)G(k,l)
g(x)\tilde g(x)e^{i(k+l)x}dkdldx\notag                                                           \\\,                                                           & =\,
\int_{\R^3}\,\id_{\{|x|\geq 10\}}(x)\frac{\partial^2_{k,l}G(k,l)}{(ix)^2}
g(x)\tilde g(x)e^{i(k+l)x}dkdldx\,
\label{rhsom}\end{align}
Carrying out the same integration by parts on the left hand side of \eqref{mpm} leads to
\begin{equation}\label{lhsom}
\int_{\R^3}\id_{\{|x|\geq 10\}}(x)\Bigl[\int_{\R^2}\frac{\partial^2_{k,l}G(k,l)}{(ix')(ix'')}
{\delta^{[\kappa]}}(x-x'){\delta^{[\kappa]}}(x-x'')g(x')\tilde g(x'')e^{ikx'+ilx''}dx'dx''\Bigr]dkdldx
\end{equation}
Noting that for \(\kappa\) large the function \({\delta^{[\kappa]}}(x-x')\)
vanishes unless \(|x-x'|\leq\kappa^{-1}<1\), the integrand over the outer \(\R^3\)
integral can be bounded by
\[const.\id_{\{|x|\geq 10\}}(x)\int_{\R^2}\frac{|\partial^2_{k,l}G(k,l){\delta^{[\kappa]}}(x-x'){\delta^{[\kappa]}}(x-x'')|}{1+x^2}dx'dx''
\leq const.' \id_{\{|x|\geq 10\}}(x)\frac{|\partial^2_{k,l}G(k,l)|}{1+x^2}
  \]
which is integrable and independent of \(\kappa\). Hence by the dominated convergence theorem
the limit of \eqref{lhsom} exists and equals \eqref{rhsom}, establishing \eqref{mpm}. Combining this with
the argument for \(|x|\leq 10\), we have proved that
\begin{align}\notag
\lim_{\kappa\to+\infty}                                                                              & \frac{1}{4\pi}\int\,f(k,l)\Bigl(\frac{\omega_k(\omega_k-\omega_l)}{\sqrt{\omega_k\omega_l}}\Bigr)
{\delta^{[\kappa]}}(x-x'){\delta^{[\kappa]}}(x-x'')e_k(x')e_l(x'')dx'dx''dkdldx                  \\\,                  & =\,
\frac{1}{4\pi}\int\,f(k,l)\Bigl(\frac{\omega_k(\omega_k-\omega_l)}{\sqrt{\omega_k\omega_l}}\Bigr)
e_k(x)e_l(x)dkdldx\,,
\notag\end{align}
(which is actually zero by the orthogonality relations for the \(e_k\) in the appendix).
The proof that the other terms (involving both creation/annihlation and only creation operators) give rise to
the corresponding terms in \eqref{bilinear} is similar.                                                        \\
\noindent
{\em Step Three.} Since the limiting
expression defines the easy to understand self-adjoint operator \(\triplecolon H^{sol}_{0}\triplecolon\) in \eqref{fsh},
we can use the limit of \eqref{summ}
    to {\em define}
    a self-adjoint operator \(O\) on \({\fock}\),
    whose domain is \(\bbS^{\theta}
    \Dom(\triplecolon H^{sol}_{0}\triplecolon)\) and such that
    \(\bbS^{\theta *}\circ O\circ\bbS^{\theta}\) equals the right hand side of
    \eqref{suml}.
    It remains to relate
    \(O\) to the operator \(\doublecolon H^{sol}_0\doublecolon \), or
\(\doublecolon {\pmb H}^{sol}_0\doublecolon \),
    (as defined in Theorem \ref{sadj2}).
    For this purpose it is useful to work
at the level of quadratic forms, interchangeably using the
Schr\"odinger and Fock solitonic representations, indicating the latter with boldface,
and writing \({\Poly}(\bfupphi) \) (resp. \({\Poly}(\upphi)\)) for the dense sets generated by the polynomials
in the field in either case as described in Corollary \ref{uniteqcor}.

The self-adjoint
operator \(\doublecolon {\pmb H}^{sol}_{0,\kappa}\doublecolon \) is related to the spatial integral of the
quadratic form
\((\Psi,\doublecolon {\pmb\mcH}^{sol}_{0,\kappa}\doublecolon\Psi)\),
which converts under the unitary transformation \(\bbS^{\theta}\)
as in \eqref{summ}.
Transferring to the corresponding Schr\"odinger representations
the relation between \(\Psi\) and \(\hat\Psi\) is as described in
the proof of Corollary \ref{uniteq}, and the Radon-Nikodym
  derivative \eqref{rnd} which appears there itself lies in 
\(L^{p_*/2}(\mu_0)\) for some \(p_*>2\).
Now on Fock space we have the formula
\[
  \doublecolon  H^{sol}_{0,\kappa}\doublecolon\,=\,\doublecolon H^{vac}_{0,\kappa}\doublecolon
  -3\int\ssmx \doublecolon \varphi_\kappa(x)^2\doublecolon \,dx
  \,=\,{{\h}}_{\kappa}-3\int\ssmx \doublecolon \varphi_\kappa(x)^2\doublecolon \,dx\,,
  \]
in which the regularized dispersion relation is as in Remark \ref{regen}.
Recall (from \eqref{assemb}) that
  \(\doublecolon H^{vac}_0\doublecolon={{\h}}\) is self-adjoint with domain
  defined in \eqref{domhvac}, while for finite \(\kappa\) the corresponding regularized
  operator  \(\doublecolon H^{vac}_{0,\kappa}\doublecolon={{\h}}_{\kappa}\) is bounded on \(\Dom({\N_0})\).
  Writing \(v(\phisch)=-3\int\ssmx \doublecolon \phisch(x)^2\doublecolon \,dx\)
we get
  \[
  (\Psi,{{\h}}_{\kappa}\Psi)
  =\left({\hat\Psi},\int\left(\triplecolon  \frac{1}{2}\left(
\uppi_\kappa^2+\upphi_\kappa\,(K\upphi)_\kappa\right)\triplecolon
\,+\,\Delta\tilde{\cal M}_{scl,\kappa}\right)\,dx\,{\hat\Psi}\right)
  -\left(\Psi,v(\phisch_\kappa)\Psi\right)\,.
  \]
  Now
  consider the limits of the three terms in the above equation.
\begin{itemize}
\item
  Referring to Remark \ref{regen} and noting that
  \(\omega_{k,\kappa}\nearrow \omega_k\) monotonically as \(\kappa\nearrow +\infty \), we deduce
  by the monotone convergence theorem that
  the left hand side converges as \(\kappa\) goes to infinity:
  \[
  \lim_{\kappa\to+\infty}\,(\Psi,\doublecolon H^{vac}_{0,\kappa}\doublecolon\Psi)_{{\fock}}=\lim_{\kappa\to+\infty}\,
  (\Psi,{{\h}}_{\kappa}\Psi)=(\Psi,{{\h}}\Psi)
  =(\Psi,\doublecolon H^{vac}_{0}\doublecolon \Psi)\,,
  \]
  actually for any \(\Psi\in{\fock}\), with a finite limit occurring precisely when
  \(\Psi\in\Dom(\sqrt{\doublecolon H^{vac}_{0}\doublecolon})\).
\item  We have already noted in Step two that the first term on the right side
  converges for \({\hat\Psi}\in{\Poly}\) to the quadratic form
  \(\bigl({\hat\Psi},(\triplecolon H^{sol}_{0}\triplecolon+\Delta\M_{scl}){\hat\Psi}\bigr)\).
\item
  For the second term on the right, move to the Schr\"odinger representation via
  Proposition \ref{sch}. Lemma \ref{crf} then implies that \(\iint
\doublecolon \phisch_\kappa(x)
{\delta^{[\kappa]}}(x-x')\ssmx'\phisch(x')
 \doublecolon \,dx'dx \) converges
to  \(\int\ssmx \doublecolon \phisch(x)^2\doublecolon \,dx\) 
 as \(\kappa\to+\infty \)
in every \(L^p(\mu_0),p<\infty\), see
  \cite{MR0674511}*{Section 5}.
  Therefore the second term on the right also converges, with limit \(-(\Psi,v(\phisch)\Psi)\) 
  as long as \(\Psi\in L^p\) for some
  \(p\in(2,\infty)\). But \({\bbS^{\theta}\,}{\Poly}\subset \cup_{p>2}L^p \) for some
\(p>2\) by Theorem \ref{uniteq} and the subsequent remark, and so convergence holds for 
\({\Psi}\in{\bbS^{\theta}\,}{\Poly}\).
\end{itemize}
{\em Step Four.}  To conclude we have established that the self-adjoint operator \(O\) satisfies
  \begin{equation}\label{polar}
  (\Psi,\doublecolon H^{vac}_{0}\doublecolon \Psi)+(\Psi,v(\phisch)\Psi)\,=\,
  (\Psi,O\Psi)
    \end{equation}
    for \(\Psi\in{\bbS^{\theta}\,}{\Poly}\); we now claim that this latter subspace is a core for
    both \(\doublecolon H^{vac}_{0}\doublecolon\)
    and \(\doublecolon H^{vac}_{0}\doublecolon+v(\phisch)\).
    Substituting for \(O\) from \eqref{suml} and polarizing \eqref{polar} with \(\bbS^{\theta}\hat\chi=\chi \) then
    taking a supremum over \({\chi\in{\bbS^{\theta}\,}{\Poly}:\|\chi\|=1}\) yields
\[
\|\doublecolon H^{vac}_{0}\doublecolon \Psi\|
=\sup(\chi,\doublecolon H^{vac}_{0}\doublecolon \Psi)=
\sup\Bigl(
  (\hat\chi,\triplecolon H^{sol}_{0}\triplecolon{\hat\Psi})+\Delta\M_{scl}
  -(\chi,v(\phisch)\Psi)\Bigr)
  \]
  which is finite (by the aforementioned \(L^p \) properties of
  \(v(\phisch)\) and \(\Psi\in{\bbS^{\theta}\,}{\Poly}
  \)). Now transfer to Fock space via \(\I\) in Proposition \ref{sch}, and
\(  \doublecolon H^{vac}_{0}\doublecolon\) is a symmetric operator expressible as
a direct sum of  operators of multiplication by \(\sum_{j=1}^N\omega_{k_j}\) in the
\(N^{th}\) slot in
standard Fock space form. So if \(\Psi_\nu\to\Psi\in  \Dom(\doublecolon H^{vac}_{0}\doublecolon)\) and
\(\doublecolon H^{vac}_{0}\doublecolon\Psi_\nu\to F\) it is immediate that
\(\Psi\in  \Dom(\doublecolon H^{vac}_{0}\doublecolon)\) and \(\doublecolon H^{vac}_{0}\doublecolon\Psi=F\)
since there is a.e. convergence in each slot.
All together this implies that
 \({\bbS^{\theta}\,}{\Poly}
 \)
is a core for \(\doublecolon H^{vac}_{0}\doublecolon\).
Next,
since \(v(\phisch)\) is relatively bounded with respect to \(\doublecolon H^{vac}_0\doublecolon\) by \eqref{best},
W\"ust's Theorem \cite{MR751959}*{Theorem X.14} implies that
  \(\doublecolon \pmb{H}^{sol}_0\doublecolon=\doublecolon \pmb{H}^{vac}_0\doublecolon+v(\phisch)\)
  is essentially self-adjoint on any core for \(\doublecolon \pmb{H}^{vac}_0\doublecolon\),
  and so in particular on \({\bbS^{\theta}\,}{\Poly}\). Furthermore,
  \eqref{suml} holds on the whole domain,
  thus identifying the operator \(O\) defined above with the operator
  \(\doublecolon H^{sol}_{0}\doublecolon\) defined in Theorem \ref{sadj2}.  \qed

A slight variation of this result in the case \(\theta=0\) which will be useful can be read off
as a corollary of the proof. Let \(\widehat{\Poly}(\upphi)\) be defined as the 
space of finite complex linear combinations of 
functions \(g(Q)h(q_d)\Symn\prod_{j=1}^n f_j(k_j)\in L^2(dQ)\otimes\mfrF \) where
all the \(h,\{f_j\} \) are Hermite and Schwartz functions exactly 
as before, but \(g\) are now allowed to run through functions of the form
\begin{equation}\label{defwidehat}
g_n(Q;\sigma)=
  \exp\biggl[\frac{i\alpha Q^2}{4\Sigma^2}-\frac{Q^2}{4\Sigma^2}
    \biggr]\,
  \He_n\Bigl(\frac{Q}{\Sigma}\Bigr)
  \end{equation} 
for all \(\Sigma>0 \) and real \(\alpha \).
This is useful because it is invariant under the action of the unitary group
\(\Exp[-it\triplecolon H^{sol}_0\triplecolon] \). (See formula
\eqref{seeme}, and also observe that by shifting time by arbitrary \(t_0\)
it is possible to realize any real values of \(\alpha\) and \(\Sigma\) in \eqref{defwidehat} with
the initial values of \eqref{hgwp} by choosing
\(\sigma_0^2=\Sigma^2/(1+\alpha^2)\) and \(t_0=2\scm_{cl}\alpha\Sigma^2/(1+\alpha^2)\).)
Now the expression \eqref{muth} indicates that
\(\bbS F\in L^p(\mu_0) \) for any such \(F\in\widehat{\Poly}(\upphi) \)
and by the argument in the preceding proof 
\({\bbS\,}\widehat{\Poly}\subset  \Dom(\doublecolon {\pmb H}^{vac}_{0}\doublecolon)
\subset  \Dom(\doublecolon {\pmb H}^{sol}_{0}\doublecolon)\). It then follows from
\cite{MR751959}*{Theorem VIII.11} that \({\bbS\,}\widehat{\Poly}\) is a core
for \(\doublecolon \pmb{H}^{sol}_{0}\doublecolon\). To summarize:
\begin{corollary}\label{core}
The space \(\widehat{\Poly}(\upphi) \) is invariant under the unitary evolution
generated by \(\triplecolon H^{sol}_0\triplecolon\) and
\({\bbS\,}\widehat{\Poly}\) is a core for \(\doublecolon {\pmb H}^{sol}_0\doublecolon\).
\end{corollary}

  \subsection{Computation of the mass shift - proof of Lemma \ref{mass}}\label{mass-shift}

For the main calculation we ignore the factor \(\frac{1}{2}\)
and will reinsert it at the end. From \eqref{regfc} we have the formulae:
\begin{align}\label{kk12}
 & K^{\frac{1}{2}}_{\kappa}(x,y)\,=\,\sqrt{3}m \tteone(x)
 \tteone(y)\,
                                             \\
   & \quad\qquad\qquad\,+\,\frac{1}{2\pi}
\iiint_{\smr\times\smr\times\smr}\,
\Bigl[(-k^2+3imk\tmy'+2m^2-3m^2\ssmy')
{\delta^{[\kappa]}}(y-y')e^{ik(x'-y')}\notag \\
   & \qquad\qquad\qquad\qquad\quad\qquad\qquad\qquad\times
{\delta^{[\kappa]}}(x-x')\frac{(-k^2-3imk\tmx'+2m^2-3m^2\ssmx')}
{(k^2+m^2)(k^2+4m^2)^{\frac{1}{2}}}\Bigr]
\;dk\,dx'\,dy'\,,\notag                      \\
\label{k012}
   & K^{\frac{1}{2}}_{0,\kappa}(x,y)\,=
\,\frac{1}{2\pi}\,\iiint_{\smr\times\smr\times\smr}\,(k^2+4m^2)^{\frac{1}{2}}
{\delta^{[\kappa]}}(y-y')e^{ik(x'-y')}{\delta^{[\kappa]}}(x-x')
\;dx'\,dy'\,dk\,
\qquad\hbox{and}                             \\
\label{c012}
   & C_{0,\kappa}^{\frac{1}{2}}(x,y)\,=
\,\frac{1}{2\pi}\,\iiint_{\smr\times\smr\times\smr}\,
\frac{
{\delta^{[\kappa]}}(y-y')e^{ik(x'-y')}
{\delta^{[\kappa]}}(x-x')
}{(k^2+4m^2)^{\frac{1}{2}}}
\;dx'\,dy'\,dk\,.
\end{align}
When the regularization is removed, i.e., when \(\kappa=+\infty \), the
first two integrals are quadratically divergent, while the third
is logarithmically divergent. The fact that the final answer, \eqref{dahn},
is finite is due to cancellations. It is necessary to handle these
carefully, because the actual limit is {\em not} the naive \(\kappa=+\infty\) limit which is 
defined by combining the three integrals \eqref{kk12},\eqref{k012} and
\eqref{c012} into one and then 
replacing \({\delta^{[\kappa]}} \) by the
delta function \(\delta \)
and performing cancellations. Doing this
leads to 
\begin{equation}\label{dhnaive}
\Delta\M_{scl}^{naive}\,=\,\sqrt{3}m\,+
\,\frac{1}{2\pi}\,
\iint_{\smr\times\smr}\,
\frac{9m^4\ssmx(\ssmx-1)}
{(k^2+m^2)(k^2+4m^2)^{\frac{1}{2}}}
\;\,dk\,dx\,=\,\frac{m}{\sqrt{3}}\,.
\end{equation}

The difference of the first two terms in the integrand
\eqref{dahn} can be written
\[
\Bigl(K^{\frac{1}{2}}_\kappa
-K^{\frac{1}{2}}_{0,\kappa}\Bigr)\Bigl|_{(x,x)} 
\,=\,\sqrt{3}m( \tteone(x))^2\,+\,
\frac{1}{2\pi}\,
\iiint_{\smr\times\smr\times\smr}\,
\frac{{g}(k;x',y')\,{\delta^{[\kappa]}}(x-y')e^{ik(x'-y')}
  {\delta^{[\kappa]}}(x-x')}
     {(k^2+m^2)(k^2+4m^2)^{\frac{1}{2}}}
\;dx'\,dy'\,dk\,.
\]
where
\begin{align*}
{g}(k;x,y)\, & =\,\bigl[(-k^2+3imk\tmy+2m^2-3m^2\ssmy)
(-k^2-3imk\tmx+2m^2-3m^2\ssmx) \\
\,           & \qquad\qquad\qquad -\,(k^2+m^2)(k^2+4m^2)\bigr]\,.
                               \\
\,           & =\,\sum_{j=0}^3\,k^j g_{j}(x,y)\,.
\end{align*}
(Notice the cancellation of the \(k^4 \) term for all \(x,y\) and also the
\(k^3\) term when \(y=x\).)
The limit \(\kappa\to+\infty \)
can be taken through the integral rather directly for \(j=0,1 \), but
for \(j=2,3 \) it is necessary to look more carefully.

\underline{For \(j=0 \)}: define new integration variables
\(\xi=\kappa(x'-x) \) and \(\eta=\kappa(y'-x) \) in place
of \(x',y' \). This leads to the integrand
\begin{align*}
\frac{1}{2\pi}\Bigl(
9m^4\ssm(x+\xi/\kappa)\ssm(x+\eta/\kappa)-6m^4\bigl(
\ssm(x+\xi/\kappa) & +\ssm(x+\eta/\kappa)\bigr)
\Bigr) \\
                   & \qquad\times
\frac{
{\delta^{[1]}}(\xi)e^{ik(\xi-\eta)/\kappa}
{\delta^{[1]}}(\eta)
}{(k^2+m^2)(k^2+4m^2)^{\frac{1}{2}}}\,.
\end{align*}
Since \({\delta^{[1]}} \) is a non-negative, smooth function which is supported
inside \([-1,1] \),
it is easy to see, by considering the cases \(|x|\geq 2/\kappa \) 
and \(|x|\leq 2/\kappa \),
that this integrand is dominated by 
\[
const. e^{-m|x|/2}{\delta^{[1]}}(\xi){\delta^{[1]}}(\eta)(m^2+k^2)^{-3/2}\in 
L^1(dxd\xi d\eta dk) \] 
with \(const. \) a fixed number which is independent of \(\kappa>1 \).
It follows that the limit \(\kappa\to+\infty \)
through the integral can be taken directly
by the dominated convergence theorem, leading to 
\[
\frac{1}{2\pi}\,\iint_{\smr\times\smr}\frac{\bigl(9\ssssm x-12\ssmx\bigr)}
{(k^2+m^2)(k^2+4m^2)^{\frac{1}{2}}}\,dx\,dk\,=\,
-\frac{6m^{-3}}{\pi}\,.
\]
To this should be added the contribution \(\sqrt{3}m \)
from the discrete mode,
and also from the term in \(C_{0,\kappa}^{\frac{1}{2}}(x,x)\) corresponding to \(j=0 \), leading to 
the answer
\[
\sqrt{3}m\,+\,\frac{1}{2\pi}\,\iint_{\smr\times\smr}\frac{\bigl(9m^4\ssssm x-12m^4\ssmx\bigr)}
{(k^2+m^2)(k^2+4m^2)^{\frac{1}{2}}}\,dx\,dk\,
+\,\frac{1}{2\pi}\,\iint_{\smr\times\smr}\frac{3m^4\ssmx}
{(k^2+m^2)(k^2+4m^2)^{\frac{1}{2}}}\,dx\,dk\,=\,
\frac{m}{\sqrt{3}}\,.
\]
This is precisely the naive answer \eqref{dhnaive}. The correct answer
\eqref{dahn} comes from a careful evaluation of the limiting values of the
remaining integrals, whose naive limits are all zero.

\underline{For \(j=1 \)}: the same change of variables
leads to the integrand
\begin{align*}
\frac{1}{2\pi}
\Bigl(\tanh m(x+\eta/\kappa)\bigl(2m^2-3m^2\ssm(x+\xi/\kappa)\bigr) & -
\tanh m(x+\xi/\kappa)\bigl(2m^2-3m^2\ssm(x+\eta/\kappa)\bigr)
\Bigr) \\
                                                                    & \qquad\qquad\times
\frac{3ikm
{\delta^{[1]}}(\xi)e^{ik(\xi-\eta)/\kappa}
{\delta^{[1]}}(\eta)
}{(k^2+m^2)(k^2+4m^2)^{\frac{1}{2}}}\,.
\end{align*}
The only difference with the \(j=0 \) case is that it is 
necessary to write 
\[
\tanh m(x+\xi/\kappa)-\tanh m(x+\eta/\kappa)\,=\,
\int_0^1m\ssm 
\bigl(x+\theta\xi/\kappa+(1-\theta)\eta/\kappa\bigr)\,d\theta\,,
\]
to conclude similarly that the integrand is dominated for \(\kappa>1 \) 
by 
\[const. e^{-m|x|/2}{\delta^{[1]}}(\xi){\delta^{[1]}}(\eta)(m^2+k^2)^{-1}
\in L^1(dxd\xi d\eta dk) 
\] 
so that the limit through the integral can be taken directly,
and this limiting value is zero.

\underline{For \(j=3 \)}:
the integrand is equal to \(\frac{1}{2\pi}\) times
\[
\bigl(3imk^3\tmx'-3imk^3\tmy'\bigr){\delta^{[\kappa]}}(x-y'){\delta^{[\kappa]}}(x-x')
e^{ik(x'-y')}/{(k^2+m^2)(k^2+4m^2)^{\frac{1}{2}}}
\]
so that the integral \(dk \) is naively linearly divergent. However,
writing \(e^{ik(x'-y')}=\frac{1}{i(x'-y')}\frac{d}{dk}e^{ik(x'-y')} \)
and using the change of variables above, we can write
\begin{align}\label{wcw} & 
\frac{3mk^3}{{(k^2+m^2)(k^2+4m^2)^{\frac{1}{2}}}}\frac{(\tmx'-\tmy')}{(x'-y')}
{\delta^{[\kappa]}}(x-y'){\delta^{[\kappa]}}(x-x')\frac{d}{dk}e^{ik(x'-y')}\,dx'\,dy'
 \\
                         & \qquad=
\frac{3mk^3}{{(k^2+m^2)(k^2+4m^2)^{\frac{1}{2}}}}\,
\int_0^1m\,\ssm 
\bigl(x+\theta\xi/\kappa+(1-\theta)\eta/\kappa\bigr)\,d\theta\;
{\delta^{[1]}}(\xi){\delta^{[1]}}(\eta)\,\frac{d}{dk}\,e^{ik(\xi-\eta)/\kappa}
\,d\xi\,d\eta\,.\notag
\end{align}
The integral \(d\xi\,d\eta \) is essentially 
a two dimensional Fourier transform of a smooth compactly supported
function of \(\xi,\eta \), and as such decays rapidly as \(k\to \infty \)
for any fixed \(\kappa>0 \). Therefore, it is permissible to integrate
by parts in \(k \), leading to the integrand
\[
\Bigl(-\frac{d}{dk}\,\frac{3mk^3}{{(k^2+m^2)(k^2+4m^2)^{\frac{1}{2}}}}\Bigr)\,
{\delta^{[1]}}(\xi){\delta^{[1]}}(\eta)\,e^{ik(\xi-\eta)/\kappa}\,
\int_0^1m\,\ssm 
\bigl(x+\theta\xi/\kappa+(1-\theta)\eta/\kappa\bigr)\,d\theta\;
\,.
\]
The limit, as \(\kappa\to\infty \), of this integrated over
\(x,k,\xi,\eta \in\R^4\) is what is needed. It is easy to
check that the integrand is dominated by a function of the same form
as in the cases above, so the limit can be taken through
the integral. The value of the limit is therefore
\[
-\int_{\smr}\,m\,\ssmx\,\times\,\Bigl[\frac{3mk^3}{{(k^2+m^2)(k^2+4m^2)^{\frac{1}{2}}}}
\Bigr]_{-\infty}^{+\infty}\,dx\,=\,-6m^2\,\int\ssmx\,dx\,.
\]
Reintroducing the \(1/(2\pi) \) factor gives the overall contribution 
\(-\frac{6m^2}{2\pi}\ssmx \), in place of 
the naive value of zero from the \(j=3 \) term. Performing the integral
over \(x \) leads to the value \(-6m/\pi \) which is the required correction
to the naive value to give the correct mass shift \eqref{dahn}. It remains 
to show that the remaining terms with \(j=2 \) do not contribute further
corrections.

\underline{For \(j=2 \)}: it is necessary to combine 
the integral involving \({g}_{2} \)
with the corresponding 
naively logarithmically divergent term \(C_{0,\kappa}^{\frac{1}{2}}(x,x) \).
All together this leads to
\begin{align}\label{ibpt}
\iint_{\smr\times\smr}\,
\Bigl[3m^2(         & \ssmx'+\ssmy')-9m^2\smx'\smy'\cosh m(x'-y')+3m^2\ssmx\Bigr] \\
                    & \qquad\qquad\qquad\qquad\qquad\qquad\qquad\qquad\qquad\qquad\qquad\times{\delta^{[\kappa]}}(x-y')e^{ik(x'-y')}
{\delta^{[\kappa]}}(x-x')
\;dx'\,dy'\notag
\end{align}
all multiplied by \(\frac{k^2}{(k^2+m^2)(k^2+4m^2)^{1/2}}\,, \)
and integrated over \((k,x)\in\R\times\R \). Notice that the naive value
of this integral, obtained by everywhere
replacing \({\delta^{[\kappa]}} \) by the delta function
\(\delta \), is zero; we must prove that the limit as \(\kappa\to\infty \)
of the integral really is zero.
Write the quantity in the square brackets in \eqref{ibpt} as
\begin{align}\label{sbr}
\Bigl[\quad\Bigr]\, & =\,9m^2\smx'\smy'\,\Bigl(1-\cosh m(x'-y')\Bigr)\,           \\
                    & \qquad\qquad\qquad +\,3m^2\,\Bigl[
\frac{3}{2}(\smx'-\smy')^2+\frac{\ssmx-\ssmx'}{2}
+\frac{\ssmx-\ssmy'}{2}\,\Bigr]\,.
\notag\end{align}
\noindent
The first two terms in \eqref{sbr} are handled as in \eqref{wcw}, with 
the conclusion that their contribution is zero (due to the quadratic
order of vanishing of \(1-\cosh m(x'-y')\) and \((\smx'-\smy')^2\)
at \(x'=y'\, \)).
To handle the final two terms, write \(e^{ik(x'-y')}=e^{ik(x'-x)}e^{ik(x-y')} \), and
then
\begin{align*}
                    & 3m^2\bigl(\ssmx-\ssmx'\bigr){\delta^{[\kappa]}}(x-y'){\delta^{[\kappa]}}(x-x')
e^{ik(x'-x)}                                                                      \\
                    & \qquad\qquad\qquad=\frac{3m^2}{i}
\frac{(\ssmx'-\ssmx)}{(x'-x)}
{\delta^{[\kappa]}}(x-y'){\delta^{[\kappa]}}(x-x')\frac{d}{dk}e^{ik(x'-x)}\,,
\end{align*}
and similarly with \(x' \) replaced by \(y' \).
Now define, as above, \(\xi=\kappa(x'-x) \) and 
\(\eta=\kappa(y'-x) \), so that
\[
\ssm(x')-\ssmx\,=\,\frac{2m\xi}{\kappa}\rho_{x,\kappa}(\xi)\,,
\qquad\rho_{x,\kappa}(\xi)\,\define\,\int_0^1\,\ssm(x+\theta\xi/\kappa)
\tanh m(x+\theta\xi/\kappa)\,d\theta\,,
\]
and similarly with \(x' \) replaced by \(y' \).
The final two terms in \eqref{sbr} then contribute
\begin{align*}
                    & \frac{3m^2}{i}\,\iint_{\smr\times\smr}\,
\left(
\frac{k^2}{(k^2+m^2)(k^2+4m^2)^{1/2}}
\right)\,                                                                         \\
                    & \quad\qquad\times\,\iint_{\smr\times\smr}\,
{\delta^{[1]}}(\xi){\delta^{[1]}}(\eta)
\Bigl[
e^{-ik\eta/\kappa}\rho_{x,\kappa}(\xi)\frac{d}{dk}e^{ik\xi/\kappa}
-e^{ik\xi/\kappa}\rho_{x,\kappa}(\eta)\frac{d}{dk}e^{-ik\eta/\kappa}
\Bigr]\,
d\xi\,d\eta\,dk\,dx                                                               \\
                    & {}                                                          \\
                    & \quad=\frac{3m^2}{i}\,\iint_{\smr\times\smr}\,
\frac{d}{dk}\left(
\frac{\sqrt{2\pi}\,\widehat{\delta^{[1]}}(k/\kappa)k^2}{(k^2+m^2)(k^2+4m^2)^{1/2}}
\right)\,
\times\Bigl(
\int_{\smr}{\delta^{[1]}}(\xi)\bigl(\rho_{x,\kappa}(\xi)-
\rho_{x,\kappa}(-\xi)\bigr)e^{i\xi k/\kappa}\,d\xi
\Bigr)\,dk\,dx\,.
\end{align*}
where we have used the assumption that \({\delta^{[1]}} \) is even, and
have relabelled the dummy variable \(\eta \) as \(\xi \) in the
second term, to show that the integrand has pointwise 
limit zero as \(\kappa\to+\infty \). To see that this integral has limit
zero we apply the product rule to get
\[
\frac{d}{dk}\left(
\frac{\widehat{\delta^{[1]}}(k/\kappa)k^2}{(k^2+m^2)(k^2+4m^2)^{1/2}}
\right)
\,=\,
\frac{k^2\frac{d}{dk}
\widehat{\delta^{[1]}}(k/\kappa)}{(k^2+m^2)(k^2+4m^2)^{1/2}}\,+\,
\widehat{\delta^{[1]}}(k/\kappa)
\frac{d}{dk}\biggl(\frac{k^2}{(k^2+m^2)(k^2+4m^2)^{1/2}}\biggr)
\]
and consider the resulting two integrals separately.
For the first integral, estimate
\[
\Biggl|\frac{k^2\frac{d}{dk}
\widehat{\delta^{[1]}}(k/\kappa)}{(k^2+m^2)(k^2+4m^2)^{1/2}}
\Biggr|_{L^1(dk)}=O(\kappa^{-1})\,,
\] 
as \(\kappa\to+\infty\,, \) to start with. Next,
observe that
\[
|{\delta^{[1]}}(\xi)\bigl(\rho_{x,\kappa}(\xi)-
\rho_{x,\kappa}(-\xi)\bigr)e^{i\xi k/\kappa}|\,\leq\,
const.|{\delta^{[1]}}(\xi)|e^{-m|x|/2}\,\in\,L^1(dx\,d\xi)\,,
\]
uniformly in \(k,\kappa>1 \). It follows that the  
 the first integral is \(O(\kappa^{-1}) \). 
For the second integral, observe that
\[
\biggl|\widehat{\delta^{[1]}}(k/\kappa)
\frac{d}{dk}\Bigl(\frac{k^2}{(k^2+m^2)(k^2+4m^2)^{1/2}}\Bigr)\biggr|
\,\leq\,\frac{const.}{(k^2+m^2)}\in L^1(dk)\]
with \(const. \) independent of \(\kappa. \) 
But then, since by inspection
\[\lim\nolimits_{\kappa\to\infty}\,
|{\delta^{[1]}}(\xi)\bigl(\rho_{x,\kappa}(\xi)-
\rho_{x,\kappa}(-\xi)\bigr)e^{i\xi k/\kappa}|
\,=0\,,
\]
it follows from
the dominated convergence theorem that the limit
as \(\kappa\to+\infty \)
of the second integral is also zero.

The conclusion of all the above is that the naive limit of the 
logarithmically divergent \(j=2 \) term is equal to the
true limit, but this is not so for the linearly
divergent \(j=3 \) term, whose true limit is equal to
\(-\frac{6m^2}{2\pi}\ssmx \). Reinserting the factor of \(\frac{1}{2}\) leads to
the final answer
\[
\Delta\M_{scl}\,=\,\Delta\M_{scl}^{naive}\,+\,\int_{\smr}
-\frac{3m^2}{2\pi}\ssmx dx\,=\,\frac{m}{2\sqrt{3}}-\frac{3m}{\pi}\,,
\]
as claimed in \eqref{dahn}.
\qed

\subsection{Change of representation - interaction terms}\label{intt}
We now compute the effect of the change of representation \(\bbS^\theta\)
from \eqref{cfxi1}-\eqref{cfxi2} to \eqref{cfsgext} on
the interaction Hamiltonian \eqref{solint}; this amounts to Wick ordering under change of covariance,
and leads to the second formula in Theorem \ref{chrep}. We refer to Remark \ref{triplecolon} for the convention
on normal ordering in the solitonic representation.
As shorthand write, in this section, 
\(Y(x)=-{\sqrt{\scm_{cl}}}{Q}
   {\tteo}(x)\) and \(Y_\kappa(x)=-{\sqrt{\scm_{cl}}}{Q}
   {\tteokap}(x)\) and recalling Remark \ref{simp},
\begin{align}\label{uperpk}
\upphiperp_{\kappa}(x)
\,     & =\,
  +\frac{1}{\sqrt{2\omega^{}_d}}
(a^{}_{d}+a_{d}^\dagger) {\tteonekap}\bigl({}x{}\bigr)      
 \\
\notag & \qquad\qquad+  \frac{1}{\sqrt{2\pi}}\,\iint\,\frac{{\delta^{[\kappa]}}(x-x')}{\sqrt{2\omega^{}_k}}\,
\bigl(a^{}_{k}e_k(x'{})+a_{k}^\dagger {{e_{-k}}}(x'{})\bigr)\,dx'dk\,,
\end{align}
  so that
  \(  \upphi_\kappa(x{})\,  =\,Y_\kappa+\upphiperp_{\kappa}(x)\). We write
\(\upphiperp(x)\) for the corresponding unregularized expression, which
is to be interpreted as an operator valued distribution.
For comparison with \eqref{regcov} we note the formula
\begin{equation}
(\Omega'\,,\upphiperp_{\kappa}(x)\upphiperp_{\kappa}(y)\,\Omega')\,=\,
\frac{ {\tteonekap}(x) {\tteonekap}(y)}{\sqrt{2\omega_d}}
  \,+\,\frac{1}{4\pi}\,\iiint_{\smr\times\smr\times\smr}\,
\frac{
{\delta^{[\kappa]}}(x-x')e_k(x'{}){e_{-k}}(y'{})
{\delta^{[\kappa]}}(y-y')
}{(k^2+4m^2)^{\frac{1}{2}}}
\;dx'\,dy'\,dk\,.
\label{regcovS}
\end{equation}
\noindent
Define \(\tilde\gamma_\kappa(x{})=
\langle\,0\,|\upphiperp_{\kappa}(x)\upphiperp_{\kappa}(x)|\,0\rangle\)
and note 
the fact that \(\delta\gamma_\kappa(x)=\tilde\gamma_\kappa(x{})
-\gamma_\kappa
\)
is uniformly bounded as \(\kappa\to+\infty\), and in this limit converges to
\begin{align}\delta\gamma(x)\,&=\,
  \frac{1}{\sqrt{3} m}\tteone(x)^2
-3m^2\ssmx\int\frac{dk}{(k^2+m^2)(k^2+4m^2)^{\frac{1}{2}}}
+9m^4(\smx)^4\int\frac{dk}{(k^2+m^2)(k^2+4m^2)^{\frac{3}{2}}}\\
\,&=\,
\frac{\sqrt{3}}{2}\ttmx\ssmx+\frac{4\pi\sqrt{3}((\smx)^4-\ssmx)-9(\smx)^4}{6}\,,
\end{align}
by an easier version of the calculations
in \S\ref{mass-shift}. Also there is exponential decay uniformly in \(\kappa\), i.e.
for sufficiently large \(\kappa\) a bound of the form
\(|\delta\gamma_\kappa(x)|\leq const. e^{-m|x|}\), with the constant independent of \(\kappa\), holds.
  Notice that not only is \(\lim_{\kappa\to+\infty}\delta\gamma_\kappa(x)\) finite for each \(x\), but is
  actually a Schwartz function of \(x\).

Now to compute the regularized interaction density \eqref{solint} in the representation
\eqref{cfsgext}, i.e., 
\(\bbS^{\theta *}\circ {\beefb}\mcH^{sol}_{I,g}(\phisch_\kappa)\circ\bbS^\theta\), we need
\begin{equation}\label{whycubed}
  \bbS^{\theta *}\circ \doublecolon \,\phisch_\kappa^3\,\doublecolon \circ\bbS^{\theta}\,=\,
  \mcn_{I,\delta\gamma_\kappa}^3(Y_\kappa,\upphiperp_{\kappa})\,,\qquad\hbox{and}\qquad
  \bbS^{\theta *}\circ \doublecolon \,\phisch_\kappa^4\,\doublecolon \circ\bbS^\theta\,=\,
\mcn_{I,\delta\gamma_\kappa}^4(Y_\kappa,\upphiperp_{\kappa})
  \end{equation}
where
\begin{align}\label{iii} \begin{split}
&\mcn_{I,{\alpha}}^3(Y,\upphi)\,\define\,Y^3+3Y^2\upphi+3Y
\triplecolon\upphi^2\triplecolon+
\triplecolon\upphi^3\triplecolon+
3{\alpha}\upphi+3Y{\alpha}\,,\qquad\hbox{
and}\\
&\mcn_{I,{\alpha}}^4(Y,\upphi)\,\define\,
Y^4+4Y^3\upphi+6Y^2
\triplecolon\upphi^2\triplecolon+
4Y\triplecolon\upphi^3\triplecolon+
\triplecolon\upphi^4\triplecolon+
6Y^2{\alpha}+12Y{\alpha}\upphi+
6{\alpha}\triplecolon\upphi^2\triplecolon
+3{\alpha}^2\,.
\end{split}\end{align}
[These are the formulae for Wick product under change of covariance, but look more complicated due to the
  zero mode part of the field being separated off in \(Y\). To derive them just write down, recalling Remark \ref{triplecolon},  the relevant Wick orderings
  of powers of the fields \(\phisch_\kappa,\upphi_\kappa\) and they follow immediately.]
This leads to the following expression for the spatially cut-off regularized interaction Hamiltonian
\begin{equation}{\rm H}^{sol,\kappa}_{I,g,\beefb}\,\define\,\int\,
2mgb(x)\tmx\,\mcn_{I,\delta\gamma_\kappa}^3(Y_\kappa,\upphiperp_{\kappa})
+\frac{1}{2}g^2b(x)\mcn_{I,\delta\gamma_\kappa}^4(Y_\kappa,\upphiperp_{\kappa})
\,dx\,.
\label{intteq}\end{equation}
These integrals of the densities \(\mcn_{I,\cdot}^j\) involve generalizations of the Wick monomials
in \S\ref{vacquant}, which can be estimated by generalizations of \eqref{best2}, in particular
\eqref{best25} and the following lemma. (Recall the definition of
the number operator \(\N\) in \eqref{numb}.)
\begin{lemma}\label{appint}
  Let \({\beefb}\in L^2(\R)\), then both 
\(\bigl(\int {\beefb}(x)\triplecolon{(\upphiperp)}_{}^n(x)\triplecolon\,dx\bigr)
  (\id+{\N})^{-n/2} \) and the corresponding Wick monomial formed from
\(\upphiperp_{\kappa} \) define bounded operators on \(\mfrF \)
and, in operator norm,
  \[\lim_{\kappa\to+\infty}\Bigl(\int {\beefb}(x)\triplecolon(\upphiperp_{\kappa})^n(x)\triplecolon\,dx-
  \int {\beefb}(x)\triplecolon(\upphiperp_{})^n(x)\triplecolon\,dx\Bigr)
  (\id+{\N})^{-n/2}\,=0\,.
  \]
\end{lemma}
\proof
Consider the monomials formed by inserting the expression for \(\upphiperp \)
into \(\bigl(\int {\beefb}(x)\triplecolon(\upphiperp_{})^n(x)\triplecolon\,dx\bigr)
  (\id+{\N})^{-n/2} \), and similarly for \(\upphi_{\kappa}^{\perp,alt}\) from \eqref{cfsr1alt}.
   Writing (with reference to \eqref{ekdef})
   \begin{align}\label{ekdef2}\begin{split}
&e_k(x)=e^{ikx}y(x;k)\,,\quad y(x;k)=y_0(k)+y_1(k)\tmx+y_2(k)\ssmx\qquad\hbox{where}\\
   & M_*\define \sup_k\bigl(|y_0(k)|+\omega_k|y_1(k)|+\omega_k^2|y_2(k)|\bigr)
   <\infty\,,\end{split}
   \end{align}
and observing that \(\tmx {\beefb}(x) \) and \(\ssmx {\beefb}(x) \) are both in \(L^2(dx) \),
the proof of the boundedness assertion reduces to the standard case \eqref{best}-\eqref{best2} treated in \cite{MR0674511}*{Section 5}.

Next we prove the approximation result.
Using the alternative regularization \(\upphi_{\kappa}^{\perp,alt}\) 
as defined in \eqref{cfsr1alt}, the corresponding assertion
  \[\lim_{\kappa\to+\infty}\Bigl(\int {\beefb}(x)\triplecolon(\upphi_{\kappa}^{\perp,alt})^n(x)\triplecolon\,dx-
  \int {\beefb}(x)\triplecolon(\upphiperp_{})^n(x)\triplecolon\,dx\Bigr)
  (\id+{\N})^{-n/2}\,=0\,.
  \]
  is an essentially immediate consequence of \eqref{best2} via a minor modification of
  the calculation in \cite{MR0674511}*{Proposition 5.8}. In order to establish this result for
  \(\upphiperp_{\kappa}\) we consider the effect of this change of regularization on a typical
  kernel for one of the Wick operators \eqref{wickop} which appear on substitution of the field into
   \(\int {\beefb}(x)\triplecolon(\upphiperp_{\kappa})^n(x)\triplecolon\,dx\). A typical kernel in the
   resultant sum of Wick operators is proportional to 
   \[
\int {\beefb}(x)\prod\frac{{\delta^{[\kappa]}}*e_{k_j}(x)}{\sqrt{2\pi\omega_{k_j}}}\,,
   \]
   while for \(\int {\beefb}(x)\triplecolon(\upphi_{\kappa}^{\perp,alt})^n(x)\triplecolon\,dx\)
   the corresponding kernel is
   \[
\int {\beefb}(x)\prod\frac{{\widehat{\delta^{[1]}}(k_j/\kappa)}e_{k_j}(x)}{\sqrt{\omega_{k_j}}}\,.
   \]
   The difference between an individual pair of factors is proportional to
   \(1/\sqrt{\omega_{k_j}}\) times
   \[g(x,k_j;\kappa)\,\define\,
   {\delta^{[\kappa]}}*e_{k_j}(x)-\sqrt{2\pi}{\widehat{\delta^{[1]}}}(k_j/\kappa)e_{k_j}(x)
   \,=\,\int{\delta^{[1]}}(u)e^{ik_j(x-u/\kappa)}\Bigl[y(x-u/\kappa;k_j)-y(x;k_j)\Bigr]\,du\,.
   \]
 Referring again to \eqref{ekdef2}, the term in square brackets is equal to
   \[
y_1(k)(\tanh mx-\tanh(m(x-u/\kappa)))+y_2(k)(\sech^2 mx-\sech^2(m(x-u/\kappa)))
\]
   and since both functions \(\tmx\) and \(\ssmx\) have derivatives bounded by
   \(const. e^{-m|x|}\),
  there holds
  \[|g(x,k;\kappa)|\leq const.M_*e^{-m|x|/2}/(\kappa \omega_k)\quad\implies\quad
  \|g(x,k;\kappa)\|_{L^2(dk)}\leq const.\sqrt{\pi}M_*e^{-m|x|/2}/(\sqrt{2m}\kappa)\,,
  \]
  by \(\int\omega_k^{-2}dk=\pi/2m\).
  Noting that
  \(\|{\widehat{\delta^{[1]}}}(k_j/\kappa)\omega_{k_j}^{-1/2}\|_{L^2(dk_j)}=O(\ln\kappa)\) we deduce,
\begin{align}
\int {\beefb}(x)\prod\frac{{\delta^{[\kappa]}}*e_{k_j}(x)}{\sqrt{2\pi\omega_{k_j}}}dx
 & \,-\,\int {\beefb}(x)\prod\frac{{\widehat{\delta^{[1]}}}(k_j/\kappa)e_{k_j}(x)}{\sqrt{\omega_{k_j}}}dx\notag \\
 & =\int {\beefb}(x)\prod\frac{{\widehat{\delta^{[1]}}}(k_j/\kappa)e_{k_j}(x)+g(k_j;\kappa)}{\sqrt{\omega_{k_j}}}
\,-\,\int {\beefb}(x)\prod\frac{{\widehat{\delta^{[1]}}}(k_j/\kappa)e_{k_j}(x)}{\sqrt{\omega_{k_j}}}dx\notag
   \end{align}
   can be bounded in \(L^2(\R^n;\prod dk_j)\) by \(const. \|{\beefb}\|_{L^\infty}(1+\ln\kappa)^{n-1}/\kappa\) as \(\kappa\to+\infty\),
   which completes the proof.
   \qed
   
   Taking the \(\kappa\to+\infty\)
   limit of \({\rm H}^{sol,\kappa}_{I,g,{\beefb}}\) (by means of this lemma) gives the interaction Hamiltonian
   \begin{equation}\label{inti}{\rm H}^{sol}_{I,g,\beefb}(Q,\upphiperp)\,\define\,\int\,
2mgb(x)\tmx\,\mcn_{I,\delta\gamma}^3(Y,\upphiperp)
+\frac{1}{2}g^2b(x)\mcn_{I,\delta\gamma}^4(Y,\upphiperp)
\,dx\,,\qquad\hbox{with}\;Y=-{\sqrt{\scm_{cl}}}{Q}
   {\tteo}\bigl({}x{}\bigr)\end{equation}
   which can itself be estimated by the same lemma. As a matter of notation, such operators can be written
   equivalently as \({\rm H}^{sol}_{I,g,{\beefb}}(Q,\upphiperp)\) or as \({\rm H}^{sol}_{I,g,{\beefb}}(\upphi)\), there being no
   essential difference since \(\upphi\) is built from the pair \((Q,\upphiperp)\).

\section{Dynamics}\label{5p1}
In this section an analysis of the dynamics generated by the quantization
of the Hamiltonian \eqref{ham-sol} in the limit \({g\downarrow 0}\) is given.

\subsection{Free Motion.} We first consider the vacuum case \eqref{ham-vac} which is
very simple but worth stating for purposes of comparison with the
solitonic case. The
framework used is that of the standard
representation of the Heisenberg relations \eqref{cf1}-\eqref{cf2}, acting
on the Hilbert space \(L^2(\mu_0)\cong\fock\), and leads to the expected limiting
dynamics, namely a free relativistic field describing an assembly of bosons of mass \(2m\) governed by the quadratic Hamiltonian
\(\doublecolon H^{vac}_{0}\doublecolon \).
\begin{theorem}\label{1st}
  In the limit \(\kappa\to+\infty\)
  the operator \(\doublecolon H^{vac}_{g,{\beefb},\kappa}\doublecolon \)
  determines a
  self-adjoint operator \(\doublecolon H^{vac}_{g,{\beefb}}\doublecolon \) on
  \({\fock}\)
  which is bounded below and determines a strongly continuous one-parameter
  unitary group via the Stone theorem.
As the coupling constant
  \(g\) tends to zero, this one-parameter group satisfies
  \begin{equation}\label{converg}
  \Exp\bigl[-it\doublecolon H^{vac}_{g,\beefb}\doublecolon \bigr]\,\to\,\Exp\bigl[-it\doublecolon H^{vac}_0\doublecolon \bigr]\qquad\qquad \hbox{as}\;{g\downarrow 0}
 \end{equation}
 in the sense of strong pointwise convergence, uniformly for
 time \(|t|\leq t_0(g)\) with
 \(\lim_{{g\downarrow 0}}gt_0(g)=0\). The theorem holds equally well in the Schr\"odinger representation by means of the
 unitary equivalence \(\I\) from Proposition \ref{sch}.
\end{theorem}
\proof
The Duhamel formula
\[
\Exp[-it\doublecolon H^{vac}_{g,{\beefb}}\doublecolon]-\Exp[-it\doublecolon H^{vac}_{0}\doublecolon ]\,=\,
 -i\,\int_0^t\,\Exp[-i(t-s) \doublecolon H^{vac}_{g,{\beefb}}\doublecolon ]\,
 \doublecolon H^{vac}_{I,g,{\beefb}}\doublecolon\,
 \Exp[-is\doublecolon H^{vac}_{0}\doublecolon ]\,ds
  \]
  together with unitarity of the semigroups involved, implies that for any finite particle
  vector \(F\in{\fock}\) there holds
  \[
\bigl\|\Exp[-it\doublecolon H^{vac}_{g,{\beefb}}\doublecolon]F-\Exp[-it\doublecolon H^{vac}_{0}\doublecolon ]F\bigr\|\,\leq\,
\int_0^t\bigl\| \doublecolon H^{vac}_{I,g,{\beefb}}\doublecolon\,\Exp[-is\doublecolon H^{vac}_{0}\doublecolon ]F\bigr\|\,ds\,.
\]
Now if \(F\) is of the form \(\prod_{j=1}^M a^\dagger(\chi_j)\Omega_0\), then
since the Fock vacuum \(\Omega_0 \) is invariant,
\begin{equation}\label{useme}
\Exp[-is\doublecolon H^{vac}_{0}\doublecolon ]F=\prod_{j=1}^M a^\dagger(e^{-is\omega_\bullet}\chi_j)\Omega_0\,,
\quad\hbox{where}\;\;
e^{-is\omega_\bullet}\chi_j(k)=e^{-is\omega^{}_k}\chi_j(k)\;\hbox{and}\;\omega^{}_k=\sqrt{4m^2+k^2}\,,
\end{equation}
so that,
using \eqref{best2}, we can bound
\[
\bigl\|  \doublecolon H^{vac}_{I,g,{\beefb}}\doublecolon\,\Exp[-is \doublecolon
  H^{vac}_{0}\doublecolon ]F\bigr\|\,=\,
\bigl\| \doublecolon H^{vac}_{I,g,{\beefb}}\doublecolon(\id+{\N_0})^{-2}\,(\id+{\N_0})^2
\Exp[-is\doublecolon H^{vac}_{0}\doublecolon ]F\bigr\|\,\leq\,
g\,const.(1+M)^2\sqrt{M!}\prod_{j=1}^M\bigl\|\chi_j\bigr\|\,,
\]
for all \(s\). This implies immediately that
\[
\bigl\|\Exp[-it\doublecolon H^{vac}_{g,{\beefb}}\doublecolon]F-\Exp[-it\doublecolon H^{vac}_{0}\doublecolon ]F\bigr\|
\,\leq\,const.(M) |t|g \prod_{j=1}^M\bigl\|\chi_j\bigr\|\,\]
for \(F\) as above, and hence to \eqref{converg} by the density of the
finite particle vectors and the fact that (by unitarity)
\[
\bigl\|\Exp[-it\doublecolon H^{vac}_{g,{\beefb}}\doublecolon]F_1-\Exp[-it\doublecolon H^{vac}_{0}\doublecolon ]F_1\bigr\|\,\leq\,
\bigl\|\Exp[-it\doublecolon H^{vac}_{g,{\beefb}}\doublecolon]F-\Exp[-it\doublecolon H^{vac}_{0}\doublecolon ]F\bigr\|
+2\bigl\|F_1-F\bigr\|\,.
\qedhere\]

In order to prove an analogous result in the solitonic case,
consider first applying the previous argument using the representation
\eqref{cfxi1}-\eqref{cfxi2}. The difficulty arises in the use of the analogy
to \eqref{useme}, which
introduces factors which are growing in time  into the estimate, due to the
presence of the zero mode in the spectral decomposition
of the operator \(H^{sol}_0\); such an explicitly growing solution
to the linear equation is given in \S\ref{timevo}, see
Remark \ref{zeromod} in particular.
On the time intervals of interest, these factors
become arbitrarily large as \({g\downarrow 0}\) (since each creation operator will
potentially produce a factor)  and so it is essential to find an alternative
approach.
A method to carry out the generalization
successfully is to
employ the representation \eqref{cfsgext}.
This leads to a description of the limiting \(g=0\) dynamics
in terms of the nonrelativistic Schr\"odinger equation for the soliton, in addition to
the assembly of relativistic bosons and a pulsation mode for the soliton,
as in Theorem \ref{first}. However the timescale on which the approximation holds is now shorter for reasons
having to do with quantum dispersion noted earlier in Remark \ref{onehalf}.

\noindent
\proof[Proof of Theorem \ref{first}]
Consider the solution to the Schr\"odinger equation on \(L^2(\mu_0)\), namely,
\[
\Psi_g(t)=\Exp[-it\doublecolon \pmb{H}^{sol}_{g,{\beefb}}\doublecolon ]F,
\]
and use  \(\bbS^\theta:\mfrH(\theta)\to{L^2(\mu_0)}\) to
transform into the representation determined by
\eqref{cfsgext}. We then show that
it is possible to obtain comparison estimates with the evolution
generated by the operator \(\triplecolon H^{sol}_0\triplecolon\) defined in \eqref{simh}.
So define
\(
\hat\Psi_g(t)\,\define\,(\bbS^{\theta})^*
\Exp[-it\doublecolon \pmb{H}^{sol}_{g,{\beefb}}\doublecolon ]F\,,
\)
which is a solution of the equation
  \begin{equation}\label{isasol}
i\frac{\partial}{\partial t}\hat\Psi_g\,=\,(\bbS^\theta)^*\circ
\doublecolon \pmb{H}^{sol}_{g,\beefb}\doublecolon\circ \bbS^\theta\,\hat\Psi_g\,,
\end{equation}
with initial data \(\hat\Psi_g(0)=\hat F\) determined by
\(
\bbS^\theta\hat\Psi_g(0)=\bbS^\theta\hat F=F\,.
\)
Referring to \S\ref{qtt}, and in particular \eqref{chrepeq},
we have
\begin{equation}\label{screp}
(\bbS^{\theta})^*\circ
\doublecolon \pmb{H}^{sol}_{g,\beefb}\doublecolon\circ \bbS^\theta\,
\,=\,\Bigl[\Delta\M_{scl}
  +\,\triplecolon H^{sol}_0\triplecolon+ {\rm H}^{sol}_{I,g,\beefb}\Bigr]\,,
\end{equation}
where \({\rm H}^{sol}_{I,g,{\beefb}}\) is displayed in \S\ref{intt}.
  The operator \((\bbS^{\theta})^*\circ
  \doublecolon \pmb{H}^{sol}_{g,{\beefb}}\doublecolon\circ \bbS^\theta\, \)
  is self-adjoint by Theorem \ref{sadj2} and so generates a unitary group
  \(\{{\evot}_0(t)\}_{t\in\R} \) on the space 
  \({\mathfrak H}(\theta) \) defined in \eqref{aos}, such that
  \(\hat\Psi_g(t)={\evot}_0(t)\hat\Psi_g(0)\).
  For the remainder of the proof we will take
  \(\theta=0\), so that \(\bbS:\mfrH\to L^2(\mu_0)\) acts as in
  the Fock space version of \eqref{lhs}. We also write \(\Theta_2(t)=t\Delta\M_{scl}\) for shorthand. Recall from the proof of
Theorem \ref{scconv} that \({\bbS\,}{\Poly}
\subset \Dom(\doublecolon \pmb{H}^{sol}_{0}\doublecolon)\) and
\({\bbS\,}{\Poly}\subset L^p(\mu_0)\) for some \(p>2\). It follows that
\[{\bbS\,}{\Poly}
  \subset\Dom(\doublecolon \pmb{H}^{sol}_{0}\doublecolon)\cap\Dom(\doublecolon H^{sol}_{I,g,{\beefb}}(\phisch)\doublecolon)\subset\Dom(\doublecolon \pmb{H}^{sol}_{g,{\beefb}}\doublecolon)\]
and hence that
\({\Poly}\subset \Dom(\bbS^{*}\circ
\doublecolon \pmb{H}^{sol}_{g,{\beefb}}\doublecolon\circ \bbS)\). Now recall Corollary \ref{core} and the subspace
\({\bbS\,}\widehat{\Poly} \) defined just prior to it: this is invariant under 
\(\Exp[-is\doublecolon \pmb{H}^{sol}_{0}\doublecolon]\) and is actually a core
for \(\doublecolon \pmb{H}^{sol}_{0}\doublecolon\). By the same argument
\(\widehat{\Poly}(\upphi)\subset\Dom(\bbS^{*}\circ
\doublecolon \pmb{H}^{sol}_{g,{\beefb}}\doublecolon\circ \bbS) \), which we'll now use to 
validate the following Duhamel formula:
\begin{equation}\label{revduh}
  e^{i\Theta_2(t)}\hat\Psi_g(t)-\Exp[-it \triplecolon H^{sol}_0\triplecolon]
  \bbS^{*}\Psi_g(0)\,=\,
   -i\,\int_0^t\,{\evot}_0(t-s)\,
{\rm H}^{sol}_{I,g,{\beefb}}
   \Exp[-is\triplecolon H^{sol}_{0}\triplecolon ]\hat\Psi_g(0)\,ds\,.
\end{equation}
applied to an initial state in \({\Poly}(\upphi) \). To justify this, consider
a tensor product of a wave packet \eqref{hgwp} describing the
location of the kink with a finite particle state describing the bosons: 
\begin{equation}\label{tpg}
  \hat\Psi_g(0)=\mbChi_{n\sigma_0}(0,Q)(a_d^\dagger)^m\prod_{j=1}^M a^\dagger(f_j)\Omega'\,,
  \end{equation}
(where \(\Omega'\) is the transverse Fock vacuum in \(\mfrF \))
  since states in \({\Poly}(\upphi) \) are finite linear combinations of such vectors. Referring to \eqref{hgwp} and
  Appendix \ref{timevo}, we see that
\begin{equation}\label{seeme}
\Exp[-is\triplecolon H^{sol}_{0}\triplecolon]\,\mbChi_{n\sigma_0}(0,Q)
(a_d^\dagger)^m\prod_{j=1}^M {a}^\dagger(\tilde f_j)
\Omega'\,=\,
\mbChi_{n\sigma_0}(s,Q)(e^{i\omega_d s}a_d^\dagger)^m
\prod_{j=1}^M {a}^\dagger(e^{i\omega_\bullet s}\tilde f_j)\Omega'\,.
\end{equation}
This shows that both \(
   \Exp[-is\triplecolon H^{sol}_{0}\triplecolon ]\hat\Psi_g(0) \)
and \({\rm H}^{sol}_{I,g,{\beefb}}\,
   \Exp[-is\triplecolon H^{sol}_{0}\triplecolon ]\hat\Psi_g(0) \)
lie in \(\widehat{\Poly}(\upphi)\subset\Dom(\bbS^{*}\circ
\doublecolon \pmb{H}^{sol}_{g,{\beefb}}\doublecolon\circ \bbS) \). Now
\eqref{revduh} can be proved in the usual way by application of the
fundamental theorem of calculus to
\[
  {\evot}_0(t-s)
   \Exp[-is\triplecolon H^{sol}_{0}\triplecolon ]
   \exp[-i\Theta_2(s)]\bbS^{*}\Psi_g(0) \,.
 \]

By unitarity of all the operators \(\{{\evot}_0(t-s)\}\) we have
\begin{equation}\label{byuni}
\|
e^{i\Theta_2(t)}\hat\Psi_g(t)-\Exp[-it \triplecolon H^{sol}_0\triplecolon]\hat\Psi_g(0)
\|
\,\leq\,
\int_0^t\,
\|\,{\rm H}^{sol}_{I,g,\beefb}
\,\Exp[-is\triplecolon H^{sol}_{0}\triplecolon]\,\hat\Psi_g(0)\,\|\,ds\,.
\end{equation}

In the following \(a^\iota\), with \(\iota\in\{+1,-1\}\),
means either \(a\) if \(\iota=-1\), or \(a^\dagger\) if \(\iota=1\).
Then referring to
\eqref{best}-\eqref{best2} we have the identity 
\[
({{\N}}+1)^2\prod_{j=1}^Ma^{\iota_j}(f_j)=\prod_{j=1}^M\bigl(({{\N}}+1)^{-\frac{1}{2}}a^{\iota_j}(f_j)\bigr)
\Bigl({{\N}}+1+\sum_{1\leq k\leq M}\iota_k\Bigr)^2\prod_{j=1}^M\Bigl({{\N}}+1+
\sum_{j\leq k\leq M}\iota_k\Bigr)^{\frac{1}{2}}\,.
\]

In what follows we use the operator norm bound \(\|({{\N}}+1)^{-\frac{1}{2}}a^\iota(f)\|\leq\|f\|\),
and the fact that, which follows from \eqref{hgwp} by observation,
that
\[|Q^r\mbChi_{n\sigma_0}(t,Q)|^2dQ
={\sigma(t)^{2r}}
|\hat Q^r\mbchi_n(\hat Q)|^2d\hat Q
\]
with \(\hat Q=Q/\sigma(t)\), so that, referring to the discussion following
\eqref{hgwp},
\[
\int|\mbChi_{n\sigma_0}(t,Q)|^2|{Q}|^{2r}dQ\,=\,c_{n,r}2^r
{\sigma_0}^{2r}\bigl(1+\frac{t^2}{4\sigma_0^4\scm_{cl}^2}\bigr)^r
\]
where \(c_{n,r}\) is a number arising from the integration, the precise value of which
is not needed for present purposes.
In the choice of wave packets \eqref{hgwp}, we can introduce \(\tau>0\) arbitrary,
and take \(2{\sigma_0}^2=\tau/\scm_{cl}\).
Referring to the form of the interaction term
\({\rm H}^{sol}_{I,g,{\beefb}}\) in \S\ref{intt} we see that the right hand
side of \eqref{byuni}
can be bounded, for \(|t|\leq t_1(g)\), by
\[
const.\biggl(gt_1(g)\Bigl(1+\bigl(\frac{\tau}{\scm_{cl}}+\frac{t_1(g)^2}{(\tau\scm_{cl})}\bigr)\Bigr)+g^2t_1(g)
\Bigl(1+\bigl(\frac{\tau}{\scm_{cl}}+\frac{t_1(g)^2}{(\tau\scm_{cl})}\bigr)\Bigr)^2
\biggr)\,,
\]
with the constant depending upon \(n,r,M,m\), but independent of \(t,g\).
(Here we are using the fact that the
contribution of the \(Y^3\) term in \eqref{whycubed} vanishes by parity.)
Now choose \(\tau=t_1(g) \) to deduce the result for initial data as in 
\eqref{tpg}.



It follows from the
density of the subspace spanned by initial conditions
of the form \eqref{tpg} that,
given any \(\hat F\) such that \(\bbS^\theta\hat F=F\), there exists a sequence of
\({\hat F}^\nu\) in \({\mfrH}\) of the form \eqref{tpg} converging to \(\hat F\) in the norm of \(\mfrH\),
and furthermore the corresponding solutions \(\hat\Psi_g^\nu\) of \eqref{isasol} verify,
by unitarity,
\[
  \|\hat\Psi_g^\nu(t)-\hat\Psi_g(t)\|
  =\|\hat\Psi_g^\nu(0)-\hat\Psi_g(0)\|=\|\hat F^\nu-\hat F\|
\]
for all times \(t\). Now
for each \(\nu\),
  \[
  \lim_{{g\downarrow 0}}\,\sup_{|t|\leq t_1(g)}\,\Bigl\|\,\hat\Psi^\nu_g(t)
  \,-\,\hat\Psi^\nu_0(t)e^{-i\Theta_2(t)}\Bigr\|\,=\,0\,,
  \]
  where \(\hat\Psi^\nu_0(t)=\Exp[-it \triplecolon H^{sol}_0\triplecolon]
  \hat\Psi_g^\nu(0)=\Exp[-it \triplecolon H^{sol}_0\triplecolon]
  \hat F^\nu\). Again by unitarity \(\|\hat\Psi_0^\nu(t)-\hat\Psi_0(t)\|
  =\|\hat\Psi_0^\nu(0)-\hat\Psi_0(0)\|=\|\hat\Psi_g^\nu(0)-\hat\Psi_g(0)\|=\|\hat F^\nu-\hat F\|\),
  and so
together with the triangle inequality in \(\mfrH\) we have, for arbitrary \(\epsilon>0\),
  \begin{align}\notag
\Bigl\|\,\hat\Psi_g(t)
\,-\,\hat\Psi_0(t)e^{-i\Theta_2(t)}\Bigr\|\, & \leq\,
\Bigl\|\,\hat\Psi^\nu_g(t)
  \,-\,\hat\Psi^\nu_0(t)e^{-i\Theta_2(t)}\Bigr\|\,+\,
  \Bigl\|e^{-i\Theta_2(t)}\bigl(\hat\Psi_0^\nu(t)-\hat\Psi_0(t)\bigr)\Bigr\|\,+\,
  \Bigl\|\hat\Psi_g^\nu(t)-\hat\Psi_g(t)\Bigr\| \\
  \,                                       & \leq\,
  \Bigl\|\,\hat\Psi^\nu_g(t)
  \,-\,\hat\Psi^\nu_0(t)e^{-i\Theta_2(t)}\Bigr\|\,+\,
  2\,\Bigl\|\hat F^\nu-\hat F\Bigr\|<\epsilon
 \,,\end{align}
by first choosing \(\nu\) sufficiently large and then \(g\) sufficiently small.\qed
\noindent
\subsection{Semiclassical limit in the presence of an external field}
\label{5p2}
A Lorentz invariant interaction with a fixed (external)
electromagnetic field \(\A_\mu dx^\mu\) is provided by the action
functional
\[
S_\lambda\,=\,\int\,\Bigl(\frac{1}{2}\partial_\mu\phi\partial^\mu\phi-\cU(\phi)
+\lambda\epsilon_{\mu\nu}\partial_\mu \A_\nu \phi\,\Bigr)\,dxdt;
\]
the interaction term is invariant under gauge transformations
\(\A_\mu\to \A_\mu+\partial_\mu\chi\), so this action is invariant under both 
Lorentz and gauge transformations. Working with the gauge condition \(\A_1=0 \)
leads to the Hamiltonian formulation in which the Hamiltonian is
\begin{equation}\label{hamlam}
  H^{\lambda}(\pi,\phi)\,=\,\int_{\smr}\,\mcH^{\lambda}(\pi,\phi)\,dx\,,\quad
  \mcH^{\lambda}(\pi,\phi)\,=\,\frac{1}{2}\,\bigl(
\pi^2\,+\,\partial_x\phi^2\bigr)\,+\,\cU(\phi)
\,-\,\lambda {\elec}\,\phi\,,
\end{equation}\label{defelec}
in terms of the electric field
\begin{equation}{\elec}(t,x)=-\A_0'(t,x)\,.
  \end{equation}
Under quantization this leads to an additional {\em time-dependent}
term arising from the external field, namely
\[\,-\,\lambda\,\int {\elec}(t,x)\,\bigl(\Phi_S+\varphi(x)\bigr)\,dx\,,
\]
so that,
at the quadratic level, the free evolution is now generated by the
time-dependent Hamiltonian 
\begin{align}
  H^{sol,{\elec}}_0(t)\, & =\,H^{sol}_0-\lambda\int {\elec}(t,x)\varphi(x)\,dx\notag \\
                    & =\,H^{sol}_0-\lambda\varphi({\elec}(t))\,.
\end{align}
The corresponding Heisenberg equation of motion for the quantum field
\(\varphi_{tt}+K\varphi-\lambda{\elec}=0\) is solved explicitly in \S\ref{timevo}.

\subsubsection{Existence of Evolution Operator}
The total Hamiltonian with spatially cut-off interaction is
\begin{equation}\label{totalham}
  \doublecolon \pmb{H}^{sol,{\elec}}_{g,{\beefb}}(t)\doublecolon\,=\,
  \doublecolon \pmb{H}^{sol,{\elec}}_0(t)+H^{sol}_{I,g,{\beefb}}(\phisch)\doublecolon
  \,=\,
  \doublecolon \pmb{H}^{vac}_0\doublecolon -\lambda\phisch({\elec}(t))+\doublecolon \tilde H^{sol}_{I,g,{\beefb}}(\phisch)\doublecolon
\end{equation}
with the interaction terms
\(H^{sol}_{I,g,{\beefb}}\) and \(\tilde H^{sol}_{I,g,{\beefb}}\) as in \eqref{defhs} and \eqref{delH} respectively; we
use the Schr\"odinger representation on \(L^2(\mu_0)\), and the
arguments \((\phisch)\) in the interaction Hamiltonian will be omitted for now. If
\(\lambda=0\) this is a standard problem, the Hamiltonian is self-adjoint and Stone's theorem provides
a unitary one-parameter group. For nonzero \(\lambda\) this picture needs to be extended to
prove existence of an evolution operator, since the
Hamiltonian is now time-dependent, although still self-adjoint at each fixed time.

\begin{theorem}\label{fixt}
\begin{enumerate}
\item[(i)] At each fixed \(t\) the operator \(\doublecolon \pmb{H}^{sol,{\elec}}_0(t)\doublecolon \)
  is essentially self-adjoint on \(\Dom(\doublecolon \pmb{H}^{vac}_0\doublecolon )\).

\item[(ii)] At each fixed \(t\) the operator
  \(\doublecolon \pmb{H}^{sol,{\elec}}_{g,{\beefb}}(t)\doublecolon=
  \doublecolon     \pmb{H}^{sol}_{g,{\beefb}}\doublecolon \,-\,
    \lambda\phisch({\elec}(t))
    \)
is self-adjoint on the domain
\[
\Dom(\doublecolon \pmb{H}^{sol}_{g,{\beefb}}\doublecolon )
\,=\,\Dom(\doublecolon \pmb{H}^{vac}_0\doublecolon )\bigcap\Dom(\doublecolon \tilde H^{sol}_{I,g,{\beefb}}\doublecolon )\,.
\]
\end{enumerate}
\end{theorem}
\proof
Statement (i) is proved as in Theorem \ref{sadj2}.
Statement (ii) follows from the second statement in Theorem
\ref{sadj2}, the Kato-Rellich theorem and the fact that
\eqref{best} implies that \(\forall\epsilon>0\) there exists 
a constants \(C_1,C_2(\epsilon)\) such that
\begin{equation}\label{bestcon}
  \|\varphi(f)\Psi\|\leq C_1\|f\|_{H^{-\frac{1}{2}}}\|(\id+\N_0)^{1/2}\Psi\|
  \leq C_1\|f\|_{H^{-\frac{1}{2}}}(\epsilon\|(\id+\doublecolon \pmb{H}^{vac}_0\doublecolon)\Psi\|+C_2(\epsilon)\|\Psi\|)\,,
\end{equation}
since, together with the inequality
\(\doublecolon \pmb{H}^{vac}_{0}\doublecolon\leq const.\doublecolon \pmb{H}^{sol}_{g,{\beefb}}\doublecolon\)
proved in \cite{MR810217}*{Theorem II.3.1.3},
\eqref{bestcon} implies that \(\phisch({\elec})\) is an infinitesimally bounded perturbation.\qed

The solution operator generated by the quadratic part of the Hamiltonian can be displayed explicitly, see \S\ref{limdyn}.
Given (ii) of Theorem \ref{fixt} it is possible to prove the existence of a solution operator generated by \(\doublecolon \pmb{H}^{sol}_{g,{\beefb}}\doublecolon\) 
in the sense of Kato's theory (as explained for example in \cite{MR0407477}), which gives an analytic framework in which a family of time dependent unbounded operators \(A(t)\) generates a solution operator, which can be written as a path ordered exponential thus:
\[
{\evot}(t,s)\,=\,\Pexp[-\int_s^t A(\sigma)\,d\sigma]\,.
\]
In terms of the Schr\"odinger picture on \(L^2(\mu_0)\) the evolution, now written in bold as
\({\bevot}(t,s)\), is given by the following theorem.
\begin{theorem}\label{extexist2}
  Assume that the mapping \(t\mapsto {\elec}(t,\cdot)\in H^{-\frac{1}{2}}(\R)\) is continuous.
  There exists a solution operator \({\bevot}\), that is to say,
  a family of unitary operators
  \(\{{{\bevot}}(t,s)\}_{(s,t)\in\smr^2}\), which are strongly continuous
  as functions of \(s,t\) (into the space of bounded linear operators
  on the Hilbert space \(\mfrH_0\) given the strong pointwise  topology), such that
  \begin{itemize}
  \item \({{\bevot}}(s,s)=\id\) for all \(s\in\R\);
      \item \({{\bevot}}(t,s){{\bevot}}(s,r)={{\bevot}}(t,r)\) for all \(t,s,r\in\R\);
  \item \(\frac{d}{dt}{{\bevot}}(t,s)\Psi=-i\bigl(\doublecolon \pmb{H}^{sol,{\elec}}_0(t)\doublecolon+\doublecolon H^{sol}_{I,g,{\beefb}}\doublecolon\bigr){{\bevot}}(t,s)\Psi\),
    for each \(\Psi\in\Dom(\doublecolon \pmb{H}^{vac}_0\doublecolon)\bigcap\Dom(\doublecolon\tilde H^{sol}_{I,g,{\beefb}}\doublecolon)\);
  \item \(\frac{d}{ds}{{\bevot}}(t,s)\Psi=+i{{\bevot}}(t,s)\bigl(\doublecolon \pmb{H}^{sol,{\elec}}_0(s)\doublecolon
    +\doublecolon H^{sol}_{I,g,{\beefb}}\doublecolon\bigr)\Psi\),
    for each \(\Psi\in\Dom(\doublecolon \pmb{H}^{vac}_0\doublecolon)\bigcap
    \Dom(\doublecolon \tilde H^{sol}_{I,g,{\beefb}}\doublecolon)\)\,.
  \end{itemize}
    \end{theorem}
\proof
This follows by means of Kato's notion of a
stable family of generators on
a Banach space \(X\) with norm \(\|\;\|\): this means a collection  \(\{A(t)\}_{t\geq 0}\)
of closed and densely defined linear operators with domain \(\hbox{Dom}A(t)\subset X\)
such that each generates a strongly continuous semi-group \(\{e^{-sA(t)}\}_{s\geq 0}\}\) on \(X\),
and subject to the following assumptions.

  \begin{enumerate}[label=(\alph*)]
  \item Either of the following equivalent conditions hold for some positive \(M\) and \(\beta \):
  \begin{itemize}
  \item
    \(\Bigl\|\prod_{j=1}^N\bigl(A(t_j)+{z}\bigr)^{-1}\Bigr\|\leq M({z}-\beta)^{-N}\)
    where here and below \(\{t_j\}\) is any nondecreasing finite set of
    non-negative times
    and \(z\) is a real number satisfying \({z}>\beta\,;\)
    \smallskip
  \item
\(\Bigl\|\prod_{j=1}^N\Exp[-s_jA(t_j)]\Bigr\|\leq M\exp[+\beta\sum_{j=1}^N s_j]\,,\) for any collection of positive numbers \(\{s_j\}\).
    \end{itemize}
\item There is a Banach space \((Y,\|\cdot\|_{Y})\) and a continuous embedding \(Y\subset X\) with dense range such
  that
  \[Y\subset\hbox{Dom}A(t)\, \forall t\geq 0\,,\]
  and with the property that for each \(s\) the semi-group \(\{e^{-tA(s)}\}_{t\geq 0}\) leaves \(Y\) invariant and restricts to a strongly continuous semi-group.
\item The map \(t\mapsto A(t)\) is continuous into \(B(Y,X)\), the space of
  bounded linear maps \(Y\to X\) given the operator
  norm, and there exists a collection \(\{S(t)\}_{t\geq 0}\)
  of linear homeomorphisms \(S(t):Y\to X\) which are continuously differentiable
  as functions of \(t\) into the space \(B(Y,X)\) in the strong pointwise sense\footnote{This means that if \(y\in Y\) the map \(t\to S(t)y\in X\) is continuously differentiable
    with respect to the norm \(\|\;\|\) of \(X\).}, and
  are such that \(S(t)A(t)S(t)^{-1}-A(t)\) is continuous as a
  function of \(t\) into \(B(X,X)\), also given the strong topology.
   \end{enumerate}
  Under these conditions it is proved in \cite{MR0407477} that
  the \(\{A(t)\}\) generate a unique strongly continuous evolution
  operator \({{\evot}}(t,s)\) such that if \(y\in Y\) then \({{\evot}}(t,s)y\) is jointly
  continuous as a function of \(s,t\) into \(Y\) and for fixed \(s\)
  is differentiable as a function
  of \(t\) into \(X\) with derivative \(-A(t){{\evot}}(t,s)y\), and also
  satisfying \[\|{{\evot}}(t,s)\|\leq M\exp[\beta|t-s|]\,,\quad\hbox{and}\quad
  {{\evot}}(t,r)={{\evot}}(t,s){{\evot}}(s,r)\,.\] for
  nonnegative \(t\geq s\geq r\).
  This applies here with \[A(t)=i\Bigl(\doublecolon \pmb{H}^{sol}_{g,{\beefb}}\doublecolon\,-\,
    \lambda\varphi({\elec}(t))\Bigr)\quad
\hbox{and}\quad
  Y=
  \Dom(\doublecolon \pmb{H}^{vac}_0\doublecolon)\bigcap\Dom(\doublecolon \tilde H^{sol}_{I,g,{\beefb}}\doublecolon)\,,\]
  with norm
  \(\|\Psi\|_{Y}=\|\Psi\|+\|\doublecolon \pmb{H}^{vac}_0\doublecolon\Psi\|
  +\|\doublecolon\tilde H^{sol}_{I,g,{\beefb}}\doublecolon\Psi\|\,.\)
To check the continuity condition, notice that for \(\Psi\in Y\)
\[
\bigl\|\bigl(A(t+h)-A(t)\bigr)\Psi\bigr\|=
|\lambda|\,\bigl\|\varphi({\elec}({t+h})-{\elec}(t))\Psi\bigr\|
\leq const.|\lambda|\,\|({\elec}({t+h})-{\elec}(t))\|_{H^{-\frac{1}{2}}}\,\|(\id+\doublecolon \pmb{H}^{vac}_0\doublecolon)\Psi\|
\]
which gives the result by the definition of the norm on \(Y\) just given. For \(S(t)\)
it then suffices to choose
\(S(t)=\doublecolon \pmb{H}^{sol,{\elec}}_{g,{\beefb}}(t)\doublecolon+z\) for any sufficiently large positive \(z\in\R\), so that
\(S(t)A(t)S(t)^{-1}-A(t)=0\), so the final condition is immediately verified.
Replacing \(t\) by \(-t\) the conditions remain
  valid, and hence the existence of the solution operator can be extended to
  all real \(t,s\), and uniqueness can be used to deduce the composition property
  holds for all real \(t,s,r\), just
  as in the derivation of Stone's theorem from the Hille-Yosida theorem.
\qed

\subsubsection{Solitonic Representations and the definition of \(\bbS^\theta(\xi)\)}\label{diffrep}
Having  established existence of the evolution operator in the preceding section,
we now introduce some representations which are needed 
in order to describe explicitly
the semiclassical dynamics, including the soliton motion. We already introduced, in
Proposition \ref{sch}, the unitary equivalences \(\I:{\fock}\to L^2(\mu_0) \) relating the vacuum representation to the vacuum Schr\"odinger representation, and
\(\mbS^\theta \,:\,L^2(\cs'(\R),\mu(\theta))\to\, L^2(\mu_0)\) relating the latter to the solitonic Schr\"odinger representation from Theorem \ref{uniteq}. The latter equivalence
generalizes when the solitonic representation is defined with respect to a soliton with centre
\(\xi\in\R\), replacing \({u}(x)=-6m^2\ssmx\) by \({u}_\xi(x)=-6m^2\ssmxi\).
For fixed \(\xi\) the equivalence is written as
\[ \mbS^\theta(\xi) \,:\,L^2(\cs'(\R),\mu(\theta,\xi))\to\, L^2(\mu_0)\,.\]
The formulae for the eigenfunctions, spectral projections etc generalize, and will be labelled with an additional sub-index
\(\xi\), thus for example \( {\tteox}\) is the zero mode around the soliton centred at \(\xi\), and the corresponding projection operator
is \(\P_{0\xi}\), see also Section \ref{eigen} for the generalized eigenfunctions \(e_{k\xi}\).
Similarly the covariant quantization map in \eqref{uniteqsol} generalizes to
\begin{equation}\label{uniteqsolxi}
  \J^{\theta}(\xi):{\mathfrak H}(\theta)
  \to L^2(\cs'(\R),\mu(\theta,\xi))\,,
\end{equation}
and composition gives a unitary map
\[
\bbS^\theta(\xi)=\mbS^\theta(\xi)\circ\J^\theta(\xi):{\mathfrak H}(\theta)\to L^2(\mu_0)\,.
\]
Concretely, the covariant quantization in \eqref{uniteqsolxi}
is obtained by using the following form for the fields in place of
\eqref{cfsgext}:
\begin{align}\la{cfsgextxi}\begin{split}
  \upphi(x;\xi)\,          & =\,-{\sqrt{\scm_{cl}}}Q{}
   {\tteox}\bigl({}x{}\bigr)
  +\frac{1}{\sqrt{2\omega^{}_d}}
(a^{}_{d}+a_{d}^\dagger) {\tteonex}\bigl({}x{}\bigr)                                          \\
                       & \phantom{=\,\Phi_Sx{}-X {\tteo}x{}}
+\frac{1}{\sqrt{2\pi}}\,\int\,\frac{1}{\sqrt{2\omega^{}_k}}\,
\bigl(a^{}_{k}e_{k\xi}\bigl({}x{}\bigr)
+a_{k}^\dagger {{e_{-k\xi}}}\bigl({}x{}\bigr)\bigr)\,dk\,,                     \\
\uppi(x;\xi)\,             & =\,-\frac{P}{\sqrt{\scm_{cl}}}
 {\tteox}\bigl({}x{}\bigr)-
i\sqrt{\frac{\omega^{}_d}{2}}
(a^{}_{d}-a_{d}^\dagger) {\tteonex}\bigl({}x{}\bigr)                              \\
                       & \phantom{===\,\Phi_Sx{}-pppp}+
\frac{1}{\sqrt{2\pi}}\,\int\,-i\sqrt{\frac{\omega^{}_k}{2}}\,
\bigl(a^{}_{k}e_{k\xi}\bigl({}x{}\bigr)
-a_{k}^\dagger {{e_{-k\xi}}}
\bigl({}x{}\bigr)\bigr)\,dk
\,.\end{split}
\end{align}
These solitonic representations are all unitarily equivalent. This is most easily seen by considering the
Schr\"odinger representations on \(L^2(\cs'(\R),\mu(\theta,\xi))\) which are unitarily equivalent because the 
measures \(\{\mu(\theta,\xi)\}_{\xi\in\R}\) are all equivalent (mutually absolutely continuous), and in particular all are
equivalent to
\(\mu(\theta,0)=\mu(\theta)\); the equivalence
is given explicitly by the Radon-Nikodym derivative:
\begin{equation}\label{defyy}
L^2(\cs'(\R),\mu(\theta,\xi))\ni\Psi\mapsto \bfhatOm(\theta,\xi,0)\Psi\in L^2(\cs'(\R),\mu(\theta))\,,\qquad
\bfhatOm(\theta,\xi,\xi')=\sqrt{\frac{d\mu(\theta,\xi)}{d\mu(\theta,\xi')}}\,.
\end{equation}
Observe that the operator \(Q\) on \(\mfrH(\theta)\) corresponds to
\(\mbQ=\phivs(-{\scm_{cl}}^{-1/2} {\tteox})\) on \(L^2(\cs'(\R),\mu(\theta,\xi))\).
Equivalence of the measures means we
can treat all the functions \(\bfhatOm,\mbQ\) as functions on the same measure space, which can conveniently be taken to
be \(\cs'\) with measure \(\mu_0\), in which case the field is written as \(\phisch\).
Composition with \(\J^{\theta}(\xi)\)  on the right gives a unitary map
\begin{align}\label{sxi}
       {\mathfrak H}(\theta) & \to L^2(\cs'(\R),\mu(\theta)) \\\notag
  f(Q)  (a_d^\dagger)^m
\prod_{j=1}^M {a}(f_j)^\dagger\Omega_0       & \mapsto
  f\bigl(\phisch(-{\scm_{cl}}^{-1/2} {\tteox})\bigr)
  \bfhatOm(\theta,\xi,0)\triplecolon\bigl(\sqrt{2\omega_d}\phisch( {\tteonex})\bigr)^m
  \prod_{j=1}^M \sqrt{2}\phisch\bigl(K^{1/4}\breve f_j(\cdot;\xi)\bigr)\triplecolon
\end{align}
where the formulae \eqref{dft} for the distorted Fourier transform generalize (for
potentials \(u_\xi(x)=-6\ssmxi\)
for arbitrary \(\xi\in\R\))  to
\begin{align}\label{dftxi}\begin{split}
  \tilde U(k;\xi)\,&=\,
\mcF_{u_\xi}(U)(k;\xi)
  \,=\,(2\pi)^{-1/2}\int\,{e_{-k\xi}}(x{})U(x)\,dx \,,\;\;\hbox{with inverse}\\
       {\breve f}(x;\xi)\,&=\,
\mcF_{u_\xi}^{-1}(f)(x;\xi)
       \,=\,(2\pi)^{-1/2}\int\,e_{k\xi}(x{})f(k)\,dk\,.
 \end{split} \end{align}
  Finally, composing on the left with \(\bbS^\theta\) from \eqref{sth} we
  define \(\bbS^{\theta}(\xi):{\mathfrak H}(\theta)\to L^2(\mu_0)\)
which maps e.g.
\begin{equation}\label{defmbs} f(Q) 
  \Omega_0\mapsto
  f\bigl(\phisch(-{\scm_{cl}}^{-1/2} {\tteox})\bigr)
  \bfOm^\theta(\xi)
  \end{equation}
  with
  \(\bfOm^\theta(\xi)=\bfhatOm(\theta,\xi,0)
\sqrt{\frac{d\mu(\theta)}{d\mu_0}}=
  \sqrt{\frac{d\mu(\theta,\xi)}{d\mu_0}}\in L^{p_*}(d\mu_0)\,,\)
  by the associativity rule for Radon-Nikodym derivatives, \cite{MR2018901}*{\S V.5.4}. Here
  the Radon-Nikodym derivative formula \eqref{rnd} generalizes for \(\xi\in\R\) to
   \begin{equation}\label{rndxi}
  \frac{d\mu(\theta,\xi)}{d\mu_0}\,=\,\det(\id+\mbO(\theta,\xi))^{\frac{1}{2}}\exp\bigl[
    -\bigl(\phisch,(C_0^{\frac{1}{2}}{K^\theta(\xi)}^{\frac{1}{2}}-\id)\phisch\bigr)_{\frac{1}{2}}\,\,
    \bigr]\,,
  \end{equation}
  where \(\mbO(\theta,\xi)=\,C_0^{\frac{1}{4}}({K^\theta(\xi)}^{\frac{1}{2}}-K_0^{\frac{1}{2}})C_0^{\frac{1}{4}}\,,\)
so that  \(\frac{d\mu(\theta,0)}{d\mu_0}=\frac{d\mu(\theta)}{d\mu_0}\) is as in \eqref{rnd}. The integrability exponent
  \(p_*\) is as in Theorem \ref{mac} (and consideration of the action of translations
  shows that the integrability exponent \(p_*\) is independent of \(\xi\)).
  In what
  follows we will need the following integrability and differentiability properties of
  \(\bfhatOm\) and \(\bfOm\), considered, as mentioned above, as functions on the measure space
\((\cs',\mu_0)\).
  \begin{lemma}\label{diffom} For all \(q\in [1,\infty)\) there exists \(\delta(q)>0\) such that
    \((-\delta(q),\delta(q))\ni h\mapsto\bfhatOm(\theta,\xi+h,\xi)\in L^q(\mu_0)\) is a \(C^2\) function, with
    bounds independent of \(\xi\). For \(p<p_*\)
    the map \(\R\ni \xi\mapsto \bfOm^\theta(\xi)\in L^{p}(\mu_0)\) is \(C^2\) with
    bounds independent of \(\xi\).
    \end{lemma}
  \begin{proof}The eigenfunctions for the compact operators \(\mbO(\theta,\xi)\) for different
    \(\xi\) are related by spatial translation, and the corresponding eigenvalues are the
    same. Therefore the determinant \(\det(\id+\mbO(\theta,\xi))\) is independent of \(\xi\).
   \begin{equation}\label{rndxih}
{\frac{d\mu(\theta,\xi+h)}{d\mu(\theta,\xi)}}
\,=\,
\det(\id+\delta_h\mbO(\theta,\xi))^{\frac{1}{2}}\exp\bigl[
    -\bigl(\phisch,(K^\theta(\xi)^{-\frac{1}{2}}{K^\theta(\xi+h)}^{\frac{1}{2}}-\id)\phisch\bigr)_{K^\theta(\xi)^{\frac{1}{2}}}\,\,
    \bigr]\,,
   \end{equation}
     where \(\delta_h\mbO(\theta,\xi)=\,{K^\theta(\xi)}^{-\frac{1}{4}}({K^\theta(\xi+h)}^{\frac{1}{2}}-{K^\theta(\xi)}^{\frac{1}{2}}){K^\theta(\xi)}^{-\frac{1}{4}}\,.\)
     The multiplicativity of the Gohberg-Krein determinant (see \cite{MR0246142}) means that
     \(\det(\id+\mbO(\theta,\xi))\times\det(\id+\delta_h\mbO(\theta,\xi))
     =\det(\id+\mbO(\theta,\xi+h))\), which is independent of \(h\) by the action of translations
     just remarked upon, and so to prove the \(L^p\) properties stated
     we only need consider the exponential factors:
\[
\exp\bigl[
    -\bigl(\phisch,(C_0^{\frac{1}{2}}{K^\theta(\xi)}^{\frac{1}{2}}-\id)\phisch\bigr)_{\frac{1}{2}}\,\,
    \bigr]\,\times\,
     \exp\bigl[
    -\bigl(\phisch,({K^\theta(\xi+h)}^{\frac{1}{2}}-K^\theta(\xi)^{\frac{1}{2}})\phisch\bigr)\,\,
    \bigr]\,.
     \]
     The left factor is covered by Theorem \ref{mac}, and is in \(L^{p_*/2}\) for some \(p_*>2\)
     independent of \(\xi\). The right factor lies in any \(L^p\) for \(h\) sufficiently small
     either from \cite{MR1642391}*{Lemma 6.4.4} or from
     the proof of \cite{MR887102}*{Proposition 9.3.1}, on account of the following\\
     \noindent
     {\em Claim:} \(\sqrt{K^\theta(\xi+h)}-\sqrt{K^\theta(\xi)}\) is Hilbert-Schmidt,
     with a norm which is \(O(h)\) as \(h\to 0\).\\
     \noindent
    To prove this:
    referring to \eqref{fc} we can write down a formula for
    the kernel of \(\sqrt{K^\theta(\xi+h)}-\sqrt{K^\theta(\xi)}\)
    as a sum of contributions arising from the discrete spectrum, namely
    \begin{align}\label{fcdis}
      \sqrt{\theta}\bigl( \texttt{e}_{0\xi+h}(x) \texttt{e}_{0\xi+h}(y)- {\tteox}(x) {\tteox}(y)\bigr)+
      \sqrt{3m^2}\bigl( \texttt{e}_{1\xi+h}(x) \texttt{e}_{1\xi+h}(y)- {\tteonex}(x) {\tteonex}(y)\bigr)\,,
      \end{align}
    which are handled very easily,
    and an oscillatory integral, on which we will concentrate. In allowing the soliton
    to have arbitrary centre \(\xi\) the quantity \({\ttF}\) in \eqref{fc1} generalizes to
    \begin{equation}\label{genf}
{{\ttF}}(k,x;\xi)={\ttF}(k,x-\xi)=(k^2+3imk\tmxi-2m^2+3m^2\ssmxi)\,,
    \end{equation}
    in terms of which
    the relevant oscillatory integral is given by:
\begin{align}\label{fcnew}
\frac{1}{2\pi}\, & \int_{\smr}\,
\frac{\bigl[{{{\ttF}}(-k,y;\xi+h)}{{\ttF}}(k,x;\xi+h)-
    {{{\ttF}}(-k,y;\xi)}{{\ttF}}(k,x;\xi)
    \bigr]e^{ik(x-y)}}
{(k^2+m^2)(k^2+4m^2)^{\frac{1}{2}}}
\;dk                                                  \\
                 & =\,N'''(x-y)\bigl(c_3(x,y,\xi)-c_3(x,y,\xi+h)\bigr)
+
N''(x-y)\bigl(c_2(x,y,\xi)-c_2(x,y,\xi+h)\bigr)\notag \\
                 & \qquad+
N'(x-y)\bigl(c_1(x,y,\xi)-c_1(x,y,\xi+h)\bigr)+
N(x-y)\bigl(c_0(x,y,\xi)-c_0(x,y,\xi+h)\bigr)\,,\notag
\end{align}
where \(N\) is defined in \eqref{defC02} and the coefficient functions
are given by
\begin{align*}
c_3(x,y,\xi)     & =\,+3m(\tmxi-\tmxiy)\notag         \\
c_2(x,y,\xi)     & =\bigl(
9m^2\tmxi\tmxiy)+3m^2(\ssmxi+\ssmxiy)
\bigr)                                                \\
c_1(x,y,\xi)     & =3m^3\bigl(2(\tmxi-\tmxiy)         \\ & \qquad-3(\tmxiy\ssmxi-\tmxi\ssmxiy)\bigr)\\
c_0(x,y,\xi)     & =-
3m^4(2\ssmxi+2\ssmxiy-3\ssmxi\ssmxiy)\,.\notag
\end{align*}
Bearing in mind the properties of the \(C^1\) function \(N\) in \eqref{defC01}-\eqref{defC03},
the important points  which imply the claim are:
\begin{itemize}
\item \(N'''\) has a singularity \(\sim -1/(\pi(x-y))\) and \(N''\) has a logarithmic singularity
  as \(x-y\to 0\);
\item \(c_3(x,y,\xi)/(x-y)\) and its derivatives with respect to \(\xi\) extend continuously to \(x=y\), and
  the same is true of \((c_3(x,y,\xi)-c_3(x,y,\xi+h)) N'''(x-y)\);
\item All \(\bigl(c_j(x,y,\xi)-c_j(x,y,\xi+h)\bigr)\) and their
 derivatives with respect to \(\xi\) are smooth functions which
  decay as \(\leq const. e^{-m|x|\wedge|y|}\), which, given the exponential
  decay of \(N(x-y),\dots N'''(x-y)\) as \(\leq const. e^{-m|x-y|}\)
  as \(|x-y|\to\infty\), ensures
  square integrability of \eqref{fcnew} with \(L^2(dxdy)\) norm \(O(h)\).
\end{itemize}

In detail, the kernel of \(\sqrt{K^\theta(\xi+h)}-\sqrt{K^\theta(\xi)}\) has a Taylor expansion
up to second order
\[
\bigl(\sqrt{K^\theta(\xi+h)}-\sqrt{K^\theta(\xi)}\bigr)(x,y)=
h{\k}_1(x,y;\xi)+\frac{1}{2}h^2{\k}_2(x,y;\xi)+h^3\delta\k(x,y;\xi,h)
\]
with all terms bounded in the space \(L^2(dxdy)\) of Hilbert-Schmidt kernels,
and the mappings \(\xi\to\k_j(x,y;\xi)\in L^2(dxdy)\) are continuous.
Indeed, by the items just listed, the \(\k_j\) and \(\delta\k\) are smooth except
for possible logarithmic singularities \(\sim\ln|x-y|\) on the diagonal, and have exponential decay
\[
\sum_{j=1}^2|\k_j(x,y;\xi)|+|\delta\k(x,y;\xi,h)|\leq const. \exp[-m(|x|+|y|)/10]
\]
uniformly for \(\xi\) on bounded intervals and \(|h|<1\), from which validity of the expansion in
\(L^2(dxdy)\) follows, and hence the claim.

Now for the differentiability properties, the Taylor expansion suggests formal expressions for the
derivatives with respect to \(h\). That
\(h\mapsto\bfhatOm(\theta,\xi+h,\xi)\in L^{q}(\mu_0)\)
really is \(C^2\) with derivatives given by these formal expressions can be proved by
approximation, using the field regularizations \(\phisch_\kappa\) defined as in \eqref{regf} and taking limits.
In detail, for any \(\phisch\in\cs'\), the mapping
\begin{equation}\label{tlimofme}
h\mapsto \left(\phisch_\kappa,\bigl(\sqrt{K^\theta(\xi+h)}-\sqrt{K^\theta(\xi)}\bigr)\phisch_\kappa\right)
=h\left(\phisch_\kappa,\k_1(\xi)\phisch_\kappa\right)
+\frac{1}{2}h^2\left(\phisch_\kappa,\k_2(\xi)\phisch_\kappa\right)
+h^3\left(\phisch_\kappa,\delta\k(\xi,h)\phisch_\kappa\right)
\end{equation}
is smooth because the regularization \(\phisch_\kappa\) is a smooth function of polynomial growth. In order to
control this in the limit \(\kappa\to+\infty\) via \cite{MR0674511}*{Theorem 5.7}
we will normal order: this just introduces the (\(\phisch\)-independent) term
\begin{align}\label{findt}
\int \sqrt{C_0}(x,y)\bigl(\sqrt{K^\theta(\xi+h)}-\sqrt{K^\theta(\xi)}\bigr)_{\kappa}(x,y)dxdy&=
h\int\sqrt{C_0}(x,y){\k}_{1\kappa}(x,y;\xi)dxdy\\&\quad+\frac{1}{2}h^2\int\sqrt{C_0}(x,y){\k}_{2\kappa}(x,y;\xi)dxdy+h^3
\int\sqrt{C_0}(x,y)\delta\k_{\kappa}(x,y;\xi,h)dxdy
\notag\end{align}
where the subscript \(\kappa\) means the kernels are regularized by convolution as in \eqref{regfc}. By the previous discussion, the kernels are all in \(L^p(dxdy)\) for \(p<\infty\), and so convergence of their regularizations
in these norms as \(\kappa\to+\infty\) follows from convolution approximation. It follows from the
converse Taylor theorem
in \cite{MR0240836}*{\S2} that \eqref{findt} converges to
define a \(C^2\) function, and so for the purposes of proving the \(C^2\)-regularity assertion
we can as well replace the quadratic expressions in
\eqref{tlimofme} by their normal ordered forms:
\[
h\doublecolon\left(\phisch_\kappa,\k_1(\xi)\phisch_\kappa\right)\doublecolon
+\frac{1}{2}h^2\doublecolon\left(\phisch_\kappa,\k_2(\xi)\phisch_\kappa\right)\doublecolon
+h^3\doublecolon\left(\phisch_\kappa,\delta\k(\xi,h)\phisch_\kappa\right)\doublecolon\,.
\]
Using \cite{MR0674511}*{Theorem 5.7} to take the limit \(\kappa\to+\infty\) in \(L^p(\mu_0)\) yields
a putative expansion up to second order in \(h\) valid in this norm, and because the coefficients are continuous in \(\xi\) the converse Taylor theorem again implies that
\(h\mapsto \bigl(\phisch,({K^\theta(\xi+h)}^{\frac{1}{2}}-{K^\theta(\xi)^{\frac{1}{2}}})\phisch\bigr)\)
is \(C^2\) into any \(L^q(\mu_0)\) (\(q<\infty\)).

This argument extends on taking exponentials to prove
\(h\mapsto\bfhatOm(\theta,\xi+h,\xi)\in L^{q}(\mu_0)\) is also \(C^2\): the expressions suggested
formally by the chain rule for the first and second derivatives of this function
are also continuous into \(L^{q}(\mu_0)\)  with \(\xi\)-independent bounds by the H\"older inequality
(always for \(h\) in a sufficiently small interval around \(0\)).  Finally the
assertion in the theorem about the map \(\xi\mapsto \bfOm^\theta(\xi)\) is a consequence of
the product rule and H\"older inequalty (recalling that the definitions of \(\bfhatOm\) and of \(\bfOm^\theta\)in
\eqref{defyy}-\eqref{rndxi},
involve a square root thus shifting the integrability index by a factor of 2.)
  \end{proof}

  The following expression for the logarithmic derivative of \eqref{rndxi}
  will be needed.
  \begin{align}\label{drndxi}
    \partial_\xi\ln \sqrt{\frac{d\mu(\theta,\xi)}{d\mu_0}}\,&=\,
    -\frac{1}{2}\,\bigl(\phisch,\partial_\xi K^\theta(\xi)^{\frac{1}{2}})\phisch\bigr)
    \quad\hbox{where, using \((\partial_x+\partial_\xi) \texttt{e}_j(x-\xi)=0\), the operator kernel is}\\\notag
\partial_\xi K^\theta(\xi)^{\frac{1}{2}}(x,y)\,&\,=
-\sqrt{\theta}\bigl({{\tteox}'}(x) {\tteox}(y)+ {\tteox}(x){{\tteox}'}(y)\bigr)-
\sqrt{3m^2}\bigl({\tteonex}'(x) {\tteonex}(y)+ {\tteonex}(x){\tteonex}'(y)\bigr)
\\\notag
&\quad+\,
\frac{1}{2\pi}\,  \int_{\smr}\,
\frac{\bigl[{\partial_\xi{{\ttF}}(-k,y;\xi)}{{\ttF}}(k,x;\xi)+
    {{{\ttF}}(-k,y;\xi)}\partial_\xi{{\ttF}}(k,x;\xi)
    \bigr]e^{ik(x-y)}}
{(k^2+m^2)(k^2+4m^2)^{\frac{1}{2}}}
\;dk                                                 
      \end{align}
  and defines a Hilbert-Schmidt operator. Let \(\phischperp\) be the transverse part of the field describing the discrete mode and the bosons,
\(\phischperp(f)=\phisch(\P^\perp_{0\xi}f)\), as in \eqref{defperp} and the discussion preceding \eqref{cfseg1}-\eqref{cfseg2},
so that \(\phisch(f)=-{\sqrt{\scm_{cl}}}( {\tteox},f\bigr)_{L^2}\,\mbQ+\phischperp(f)\).
Using \((\partial_x+\partial_\xi){\ttF}(k,x;\xi)=0\) and
  \((\partial_x+\partial_y)e^{ik(x-y)}=0\), and referring to \eqref{ekdef} and the final paragraph of \S\ref{eigen} for notation, integration by parts yields
  \begin{align}\notag
    \partial_\xi\ln \sqrt{\frac{d\mu(\theta,\xi)}{d\mu_0}}\,&=\,
   -\, \sqrt{\theta}{\sqrt{\scm_{cl}}}\mbQ\phischperp({{\tteox}'})+
    \sqrt{3m^2}{\sqrt{\scm_{cl}}}\,\mbQ\,({{\tteox}'}, {\tteonex})_{L^2}\phischperp( {\tteonex})\\
    &\qquad+{\sqrt{\scm_{cl}}}\,\mbQ\,\frac{1}{4\pi}\,  \int_{\smr}\,
         {(k^2+4m^2)^{\frac{1}{2}}}\,\bigl[{{\tteox}'}(y){{e}_{-k\xi}(y)}{e}_{k\xi}(x)\phischperp(x)\notag\\\notag
  &\qquad\qquad\qquad\qquad\qquad\qquad\qquad\qquad\qquad\qquad\qquad         +
    {{e}_{-k\xi}(y)}{e}_{k\xi}(x){{\tteox}'}(x)\phischperp(y)
    \bigr]
\;dkdxdy                                                
    \\&\qquad\qquad-\frac{1}{2}\bigl(\phischperp,\partial_\xi K^\theta(\xi)^{\frac{1}{2}}\phischperp\bigr)\notag\\
&=\,
\,    \sqrt{\theta}{\sqrt{\scm_{cl}}}\mbQ\phischperp(\partial_\xi {\tteox})
    -\,{\sqrt{\scm_{cl}}}\,\mbQ\,\phischperp\bigl(K^\theta(\xi)^{\frac{1}{2}}\partial_\xi {\tteox}\bigr)
    -\frac{1}{2}\bigl(\phischperp,\partial_\xi K^\theta(\xi)^{\frac{1}{2}}\phischperp\bigr)\,.
  \label{defrnfinal}\end{align}
\begin{remark}Observe that \(\mbQ\propto\phisch( {\tteox})\) - the operator describing quantum fluctuations \(Q\)
in the kink's location - appears only linearly, while the fluctuations corresponding to the discrete mode and the bosons appear quadratically.
The expression \eqref{defrnfinal} is in every space \(L^p(d\mu_0),\,p<\infty\) and the fact that it really is
the derivative can be established by truncating as in the preceding proof. Observe also that in the final two terms in \eqref{defrnfinal} one can
remove the \(\theta\) since only the projection of the operators onto the subspace \(\langle\tteox\rangle^\perp\) contributes. 
\end{remark}

Next let \(\chi=\chi(t,Q)\) be (say) smooth and bounded with bounded derivatives,
then \(t\mapsto \bbS^\theta(\xi(t))\chi(t,Q)\Omega'\in L^2(d\mu_0)\) is differentiable with
  \begin{align}  \label{defrnfinal15}\begin{split}
    \partial_t\Bigl(\bbS^\theta(\xi)\chi(t,Q)\Omega'\Bigr)\,=\,&\bbS^\theta(\xi)\Bigl[
\,-\dot\xi\,{\sqrt{\scm_{cl}}}\Bigl(\sqrt{\theta}\upphiperp({{\tteox}'})-\upphiperp(K^\theta(\xi)^{\frac{1}{2}}{\tteox}')\Bigr)Q\chi(t,Q)\Omega'\\
&\qquad
-\frac{1}{2}\,\dot\xi\,\bigl(\upphiperp,\partial_\xi K^\theta(\xi)^{\frac{1}{2}}\upphiperp\bigr)\chi(t,Q)\Omega'
+{\scm_{cl}}^{-1/2}\dot\xi\upphiperp(\tteox')\partial_Q\chi(t,Q)\bigr)\Omega'
\\
&
\qquad\qquad+\partial_t\chi(t,Q)\Omega'\Bigr]\,.\end{split}
    \end{align}
  Looking at the first two lines it is natural to define, given continuously differentiable
  \(t\mapsto \Lambda(t)\in\mfrF\) and \(t\mapsto\xi(t)\in\R\) and \(\chi\) as above,
  an operator \(\bbdotS^\theta(t)\) by
  \begin{equation}\label{defrnfinal2}
    \bbdotS^\theta(t)\chi\Lambda\,=\,
    \dot\xi\,\bbS^\theta(\xi)\Bigl(i{\scm_{cl}}^{-1/2}\upphiperp(\tteox')P\chi
    +\sqrt{\scm_{cl}}\upphiperp(K^\theta(\xi)^{\frac{1}{2}}{\tteox}')Q\chi
-\frac{1}{2}\,\bigl(\upphiperp,\partial_\xi K^\theta(\xi)^{\frac{1}{2}}\upphiperp\bigr)\chi
    \Bigr)\Lambda\,,
  \end{equation}
  where \(P=-i\partial_Q+i{\scm_{cl}}\sqrt{\theta}Q\). Composition with the inverse
  gives the following operator on \(L^2(\mu_0)\):
  \begin{equation}\label{defrnfinal225}
    \bbdotS^\theta(t)\circ \bbS^\theta(\xi)^{-1}\,=\,
        i\dot\xi{\scm_{cl}}^{-1/2}\phischperp(\tteox')\mbP
    +\dot\xi\sqrt{\scm_{cl}}\phischperp(K^\theta(\xi)^{\frac{1}{2}}{\tteox}')\mbQ
-\frac{1}{2}\dot\xi\,\bigl(\phischperp,\partial_\xi K^\theta(\xi)^{\frac{1}{2}}\phischperp\bigr)
    \,,
  \end{equation}
  where \(\mbQ,\mbP\) are defined by \(\mbQ=-{\scm_{cl}}^{-1/2}\phisch( {\tteox})\) and
  \(\mbP=-{\scm_{cl}}^{1/2}\pisch({\tteox})\); observe that this not quite a multiplcation operator due to the presence of
  \(\mbP\).
  The operator
  \(\bbdotS^\theta(t)\) can be thought of (heuristically at least) as the
  derivative of \(t\mapsto \bbS^\theta(\xi(t))\) at time \(t\),
  although strictly speaking this is really only the case applied to states
  \(\chi(t,Q)\Omega'\). For the case that \(\Lambda\) is determined from a multiplication operator
  \(M(t)\) on \(L^2(\mu_0)\) such that \(M(t)\bbS^\theta(\xi(t))\chi(t,Q)\Omega'=\bbS^\theta(\xi(t))\chi(t,Q)\Lambda(t)\)
  then formally the above formula would have an additional term:
    \begin{equation*} 
        \partial_t\Bigl(\bbS^\theta(\xi)\chi(t,Q)\Lambda(t)\Bigr)\,=\,
M(t)\bbdotS^\theta(t)\chi(t,Q)\Omega'+\dot M(t)\bbS^\theta(\xi(t))\chi(t,Q)\Omega'\,.
    \end{equation*}
    For example, if 
    \(t\mapsto M(t)\) is continuously differentiable into \(L^p(\mu_0)\) for all \(p<\infty\) then this formula can be justified via dominated convergence in the usual way
    as long as \(t\mapsto \bbS^\theta(\xi(t))\chi(t,Q)\Omega'\) is continuously differentiable
    into \(L^r(d\mu_0)\) for some \(r>2\). However, note that the order must be maintained as written in the above
    formula, on account of the fact that \eqref{defrnfinal225} is not a multiplication operator - moving
    \(M\) back to the domain introduces a commutator:
    \begin{align} \label{defrnfinal17}\begin{split} 
        \partial_t\Bigl(\bbS^\theta(\xi)\Lambda(t)\chi(t,Q)\Bigr)\,&=\,
\dot M(t)\bbS^\theta(\xi(t))\chi(t,Q)\Omega'+
        \bbdotS^\theta(t)\circ \bbS^\theta(\xi)^{-1}\circ \bbS^\theta(\xi)\Lambda(t)\chi(t,Q)\Omega'
        \\
\,&\qquad+[M(t),\bbdotS^\theta(t)\circ \bbS^\theta(\xi)^{-1}]\circ\bbS^\theta(t)\chi(t,Q)\Omega'
        \,.
    \end{split}\end{align}
    \noindent
    However in the case that \(M\) is an operator of multiplication by a polynomial in the {\em transverse} field,
    e.g. of the form \(\phisch(f)\) with \((f,\tteox)=0\), then because only \(\pisch(\tteox)\) appears in \eqref{defrnfinal225} the commutator vanishes, and the polynomial can be transferred back through \(\bbS^\theta\)
    to act on \(\mfrH(\theta)\). This happens for example if \(f=\tteox'\), to take an example which appears below in the proof of Lemma \ref{esterrtd}; but note in this case although \(\dot M\) is no longer a polynomial in the transverse field
    (since it is the operator of multiplication by \(-\dot\xi\phisch(\tteox'')\), and \((\tteox,\tteox'')\neq 0\),) it can be passed through to \(\mfrH(\theta)\) because there is no \(\bbdotS^\theta(t)\circ \bbS^\theta(\xi)^{-1}\) in the way.
         
  Results corresponding to the forgoing for the case \(\theta=0\) can be obtained directly by putting \(\theta=0\) in the above, so that the first term on the right is to be excluded. To see this, recall that we are just composing with the
unitary map \(g\mapsto g/\mbChi_\theta\) from \(L^2(dQ)\) to \(L^2(\mbChi_\theta^2\, dQ)\) in the \(\R\) factor. This gives
\(\bbS(\xi)\define \bbS^\theta(\xi)\bigr|_{\theta=0}:{\mathfrak H}\to L^2(\mu_0)\) 
which in particular maps
\[ g(Q) 
  \Omega_0\mapsto
  g(\mbQ)\mbChi_\theta(\mbQ)^{-1}\,
\sqrt{\frac{d\mu(\theta,\xi)}{d\mu_0}}\in L^{p_*}(d\mu_0)
  \]
  with the understanding that the right hand side is actually independent of \(\theta\), as explained in the
  discussion and remarks following \eqref{defrn}. Referring to the formula for \(\mbChi_\theta\), differentiation produces
  an extra term which precisely cancels the term proportional to \(\sqrt{\theta}\) in \eqref{defrnfinal15}.
  (It is often more convenient to leave the \(\mbChi_\theta\) factor explicit in order to express the integrability properties in Lemma \ref{diffom}.)


\begin{remark}\label{trans0}
For any real \(\xi\), mapping \(a^{}_k\) and \(a_k^\dagger\) to \(e^{-ik\xi}a^{}_k\) and \(e^{+ik\xi}a_k^\dagger\), respectively,
gives a unitarily equivalent representation related by the action of translation
\(x\mapsto x+\xi\) which induces the map
\(\bftau_\xi f=f(\cdot+\xi)\) on functions and hence
\(\bftau_\xi^*\Phi(f)=\Phi(f(\cdot+\xi))\) on
quantum fields. Because the
vacuum configuration \(\Phi_0\) is translation invariant this is not particularly significant, but does become so
for the soliton sector quantization, as \(\Phi_S\) is not translation invariant.
Note also, in this connection, that a physical state is an element of the Fock space
\(\oplus_{n=0}^\infty\Symn (L^2(\R,dx))\), which is mapped into the
Fourier Fock space \eqref{fock} by the Fourier transform. However the
Fourier transform of a function translated by \(\xi\) is \(e^{ik\xi}\) times the
Fourier transform of the function, and similarly
an n-particle wave function \(\Psi_n(k_1,\dots ,k_n)\)
is multiplied by \(\exp[i(\sum k_j)\xi]\).
These two descriptions are really active/passive descriptions of the
same mapping, related by
\[
f(k_1,\dots,k_n)\exp[i(\sum_1^n k_j)\xi]\prod_1^n a_{k_j}^\dagger|\,0\,\rangle
=f(k_1,\dots,k_n)\prod_1^n\bigl(\exp[ik_j\xi]a_{k_j}^\dagger\bigr)|\,0\,\rangle\,.
\]
We will use the notation \(\Gamma(\bftau_\xi)\) for the family of unitary operators
on Fock space so defined which implement the translation operation, i.e. 
\(\Gamma(\bftau_\xi)^*\circ\varphi(f)\circ \Gamma(\bftau_\xi)=\varphi(f(\cdot+\xi))\).
Combining this with the Weyl operators in \eqref{displace},
we obtain the action of the translations on the complete solitonic
  quantum field \eqref{cfxi1}
  \[
  \Gamma(\bftau_\xi)^*\circ \U(\delta_\xi\Phi_S)^*\circ\Phi(f)\circ \U(\delta_\xi\Phi_S)\circ \Gamma(\bftau_\xi)=\Phi(f(\cdot+\xi))\,.\]
\end{remark}

\subsubsection{Limiting dynamics - definition}\label{limdyn}

In this section we introduce an effective Hamiltonian \({\rm H}^{eff}_0\) acting on the space Hilbert space
\(\mfrH(\theta)\) which generates 
limiting semiclassical solution operator which, after mapping back to \(L^2(\mu_0)\) via \(\bbS^\theta(\xi(t))\),
is a good approximation to the true solution operator \(\bevot\). A naive first guess might be as follows: expand
the Hamiltonian about a soliton centred at \(\xi\):
\[
\Phi(x{})=\Phi_S(x-\xi)+\upphi(x{})\qquad
\Pi(x{})=\uppi(x{})\,,
\]
and extract the terms up to quadratic, namely
\[
{\rm H}^{naive}_{0}\define 
\triplecolon  \frac{1}{2}\,\int\,\Bigl[\,
    \uppi^2\,+\,\upphi\,K(\xi)\,\upphi\,\Bigr]\,dx\,\triplecolon
-\lambda\upphi(\elec)\,=
\,\frac{P^2}{2\scm_{cl}}+h_d+\h-\lambda\upphi(\elec)
\,.
\]
This is a linear perturbation of the Hamiltonian \eqref{simh} by
\(-\lambda\upphi(\elec)\). Substituting \eqref{cfsgextxi} and expanding this perturbation out further,
reveals a term
which is linear in \(Q\), to be precise:
  \[
\lambda{\sqrt{\scm_{cl}}}Q{}
   \int_{\smr}{\tteox}(x)\E(t,x)\,dx\,.
  \]
  Such a term in the Schr\"odinger equation induces an acceleration of a quantum particle,
  and indicates the necessity of allowing the (classical) parameters
  \(\xi,\eta\) to depend on time, which is one reason the above naive guess for the effective Hamiltonian is not quite correct:
  the time evolution of \(\xi\) is enacted via the \(\Delta(t)\) operators in \eqref{defdis}, and working through the details turns out to imply that
the transverse degrees of freedom are acted on by the
effective electric field \({\elec}^{eff}\) in \eqref{defeff}: for a derivation see \eqref{dds} and the subsequent discussion, while a heuristic discussion is given in \S\ref{heuristics}.
Further complications arise from the fact that in order to bound perturbatively the remainder it is necessary to extract certain ``averaged terms'' see \eqref{defv1xi}, \eqref{defv1eta} and \eqref{defv2}- after these subtractions the remainder is indeed perturbative. The first two can be cancelled out by appropriate choice of \((\dot\xi,\dot\eta)\), leading to \eqref{newtpert}, while the third has to be included in the one-particle Hamiltonian for the soliton:
\[
h_{\mbox{\tiny 1P}}\,=\,\frac{P^2}{2\scm_{cl}}+V_2Q^2\,.
\]
See \eqref{defv2},\eqref{defv22} and \eqref{defv23}  for the definition of \(V_2\). (Expressed as a Schr\"odinger operator the expression for 
\(h_{\mbox{\tiny 1P}}\) is different for \(\mfrH\) and \(\mfrH(\theta)\), but in going between the two it is only necessary to transform in
  the usual way by conjugating with the unitary transformation between \(L^2(dQ)\) and
  \(L^2(\gamma_\theta(dQ))\).)

We can now define the effective limiting dynamics: it consists 
of a classical evolution
\(t\mapsto (\xi(t),\eta(t))\) for the soliton and quantum fluctuations governed by the one-particle Hamiltonian
\(
h_{\mbox{\tiny 1P}}\), and this soliton evolution is coupled to the dynamics of the transverse modes
via the effective Hamiltonian on \(\mfrH(\theta)\) of \eqref{aost} which is given explicitly as
\begin{align}
  \label{simhE}
{\rm H}_0^{eff}\,=
        \,h_{\mbox{\tiny 1P}}+{h}^{}_d+
{\beed}(a^{}_d+a_d^\dagger)+
{{\h}}+\int\Bigl({{\beekm}}a^{}_k+{\beek}a_k^\dagger\Bigr)dk
\end{align}
where the creation/annihlation operators were introduced in the discussion surrounding
\eqref{l2g}, and \({\beed},{\beek}\) are as in \eqref{defeeze1} , or using notation \eqref{dftxi}
\begin{align}{\beed}\,&=\,\lambda\,({2\omega_d})^{-\frac{1}{2}}( {\tteonex},{\elec}^{eff})\,,\qquad\hbox{and}\;\;
  {\beek}\,=\,\frac{\lambda}{\sqrt{2\pi}}\,\int\,\frac{1}{\sqrt{2\omega_k}}\,e_{-k\xi}\,{\elec}^{eff}\,dx\,
  =\frac{\lambda}{\sqrt{2}}\,\mcF_{u_\xi}({K^{-\frac{1}{4}}{\elec}^{eff}})(k,\xi)
  \,,\label{defeeze15}
\end{align}
with \({\elec}^{eff}={\elec}-\frac{g\dot\xi\eta}{\lambda\sqrt{\scm_{cl}}}{{\tteox}'}\) ,
the effective electric field acting on the transverse modes (bosons) of the scalar field.
The effective Hamiltonian
is decomposed into three pieces: evolution of the classical soliton parameters \(\xi,\eta\) by \eqref{newtpert}, evolution of the
transverse modes (quantum fluctuations superimposed on the displacement \eqref{defeeze1}) and thirdly the quantum evolution of the
soliton wave function according to the Hamiltonian \(h_{\mbox{\tiny 1P}}\). This latter evolution does depend on the first two pieces, but does not affect them
in return. The transverse quantum fluctuations are described below, in terms of \(\cee,\eff\), but we first comment on the existence
of solutions of the coupled system of equations \eqref{newtpert} and \eqref{defeeze1}: integration of the latter system leads to the formulae
\eqref{theresult} for \(\cee,\eff\) as nonlocal functionals of \(\xi,\dot\xi,\eta\) which allows the question of existence to be reduced to a
nonlocal ODE problem, see Theorem \ref{lse} below.
  
Given \(t\mapsto (\cee(t),\eff(t,k))\) the transverse dynamics
can be analyzed in two ways:
\begin{itemize}
\item[(i)] using the vacuum sector
representation, and the duality pairing with the classical evolution as in e.g. \cite{MR529429}*{Section XI.15}, the solution operator in \(\mfrH_0\cong L^2(\mu_0)\) can be obtained;
\item[(ii)] by using the diagonalization in the soliton sector representation, and thence displaying the evolution
  in the space \(\mfrH\) generated by the time-dependent operator
  \( {\rm H}^{eff}_0\).
\end{itemize}
The latter approach gives a very explicit description of the dynamics, see \eqref{eqdd}-\eqref{4lp} below, but the former approach can also be useful and is summarized in Appendix \ref{timevo}.
To start with we will
consider the equation for the discrete mode
\begin{equation}\label{eqdd}
  i\partial_t\chi_d
  =\Bigl({h}^{}_d+
{\beed}(t)(a^{}_d+a_d^\dagger)\Bigr)\chi_d\,.
\end{equation}
The solution to \eqref{eqdd} can be obtained from a solution to the free equation as follows:
\begin{itemize}
\item Let \(\tilde\chi_d\) solve the free equation up to a phase \(\Theta_d(t)=-\int_0^t{\cee}_1(s'){\beed}(s')\,ds'\):
    \[
i\partial_t\tilde\chi_d=\bigl({h}^{}_d+{\cee}_1(t)
{\beed}(t)\bigr)\,\tilde\chi_d\,;
\]
\item Define \(\chi_d=D^{}_d({\cee}(t))\tilde\chi_d\) where the displacement operator \(D^{}_d({\cee})\), defined
for \({\cee}={\cee}_1+i{\cee}_2\in\C\),
acts on \(L^2(\R,\gamma_d(dq^{}_d))\) as a group of unitary operators:
\[
D^{}_d({\cee})\chi_d(q_d)=\exp\bigl[{\cee}\sqrt{2\omega^{}_d}\bigl(q_d-\frac{{\cee}_1}{\sqrt{2\omega^{}_d}}\bigr)\bigr]
\chi_d\bigl(q_d-\frac{2{\cee}_1}{\sqrt{2\omega^{}_d}}\bigr)\,=\,
\exp[{\cee}a_d^\dagger-\overline{{\cee}}a^{}_d]\,\chi_d\,(q_d)\,=\,
e^{-\frac{1}{2}|{\cee}|^2}e^{{\cee}a_d^\dagger}e^{-\overline{{\cee}}a^{}_d}\chi_d\,;
\]
\item Finally \(t\mapsto {\cee}(t)\) solves the classical equation of motion
    \(i\dot {\cee}-\omega_d {\cee}-{\beed}=0
    \).
  \end{itemize}
This can be verified explicitly of course, but for purposes of generalization
can usefully be derived from the commutators
\begin{align}\label{crel0}\begin{split}
&
D^{}_d({\cee})a_d^\dagger=(a_d^\dagger-\overline{{\cee}})D^{}_d({\cee})\quad\hbox{and}\quad   D^{}_d({\cee})a^{}_d=(a^{}_d-{\cee})D^{}_d({\cee})\,;
\\    
&  [i\partial_t,D^{}_d({\cee})]
  \,=\,i\dot {\cee}a_d^\dagger D^{}_d({\cee})-i\dot{\overline{{\cee}}}D^{}_d({\cee})a^{}_d-
  \frac{i}{2}\frac{d}{dt}({\cee}\overline{{\cee}})
  \,=\,iD^{}_d({\cee})\bigl(\dot {\cee}a_d^\dagger -\dot{\overline{{\cee}}}a^{}_d+\overline{{\cee}}\dot {\cee}-
  \frac{1}{2}\frac{d}{dt}({\cee}\overline{{\cee}})\bigr)\,;\\
&    [{h}^{}_d,D^{}_d({\cee})]\,=\,D^{}_d({\cee})(\omega\overline{{\cee}}a^{}_d+\omega {\cee} a^\dagger_d
    +\omega|{\cee}|^2)\,.
    \end{split}\end{align}
To summarize:
    \begin{enumerate}
    \item[(i)] The quadratic Hamiltonian \({h}^{}_d+
      {\beed}(t)(a^{}_d+a_d^\dagger)\) has lowest eigenvalue \(-{\beed}^2/\omega_d\) with corresponding
      eigenfunction \(D^{}_d(\cee_0)\id_{\smr}\), where \(\cee_0=-{\beed}/\omega_d\) and
      \(\id_{\smr}\in L^2(\R,\gamma^{}_d (dq_d))\)
      is the function identically equal to \(1\).
    \item[(ii)]  The solution operator for \eqref{eqdd} is given by
      \[
\Pexp\Bigl[-i\int_s^t\bigl({h}^{}_d+
  {\beed}(\sigma)(a^{}_d+a_d^\dagger)\bigr)d\sigma\Bigr]\chi_d
=
e^{i\Theta_d(t,s)}D^{}_d({\cee}(t))\Exp[-i(t-s){h}^{}_d]D^{}_d({\cee}(s))^*\chi_d\,.
      \]
      \end{enumerate}
The method and formulae generalize to the equation in \(\mfrH_0\):
\begin{equation}\label{solveme}
i\partial_t\psi\,=\, \h^{{\elec}}\psi\,=\,\Bigl({{\h}+\h_1}\Bigr)\psi\,,
\qquad\h_1(t)=\int\Bigl( {{\beekm}(t)}a^{}_k+{\beek}(t)a_k^\dagger\Bigr)dk\,.
\end{equation}
It is necessary to generalize the definition of the displacement operator \(D_{\cee}\) to the present
infinite dimensional setting: in the Schr\"odinger setting this is the well-known Cameron-Martin problem,
see \cite{MR887102}*{9.1.27} or \cite{MR1474726}*{Chapter 14}, and the
corresponding quantum field formalism is developed in 
\cite{MR0180856}.
In the present case the displacement operator is determined by a complex-valued Schwartz function \(k\mapsto f(k)\in\C\), and is
given by
\begin{align}\label{crel00}\begin{split}
&\D_0(f)=
\Exp[a^\dagger(f)-a({f})]=
\exp[-\frac{1}{2}\|f\|^2]\Exp[ a^\dagger(f)]\Exp[-a({f})]\,,\\&\hbox{where}\;\; a^\dagger(f)=\int\,f(k)a^{\dagger}_k\,dk\;\;\hbox{and}\;\;
a(g)=\,\int\,\overline{g(k)}a^{}_k\,dk\,.
\end{split}\end{align}
Here \(i^{-1}\bigl(a^\dagger(f)-a({f})\bigr)\) (with domain the finite particle subspace \({\Poly}\))
is essentially self-adjoint,
and generates a unitary group.
This operator \(\D_0(f)\) so defined is therefore unitary and verifies
\[[a(g),\D_0(f)]=(g,f)\D_0(f)\qquad\hbox{and}\quad [a^\dagger(g),\D_0(f)]=(f,g)\D_0(f)\,.\]
The commutation relations needed are (at time \(t\))
\begin{align}\label{crel}\begin{split}
  [i\partial_t,\D_0(f)]\,&=\,i\,\D_0(f)\Bigl(
  a^\dagger(\dot f)-a(\dot f^*)+(f,\dot f)-\frac{1}{2}\frac{d}{dt}\|f\|^2\Bigr)\\
  [\h,\D_0(f)]\,&=\,\D_0(f)\Bigl(a^\dagger(\omega^{}_\bullet f)+a(\omega^{}_\bullet f^*)
  +(f,\omega^{}_\bullet f)
  \Bigr)\\
       [{\h_1},\D_0(f)]\,&=\int {{\beekm}(t)}f(t,k)+{\beek}(t)\overline{f(t,k)}\,dk\,
=\int \overline{{{\beek}}}(t)f(t,k)+{\beek}(t)\overline{f(t,k)}\,dk
\,.
\end{split}
\end{align}
(These are valid on the finite particle subspace, in particular when applied to the vacuum.)
This leads to two observations analogous to those above:
\begin{enumerate}
\item[(i)] The lowest eigenvalue of \({{\h}}+{\h_1}\)
  (at fixed \(t\))  is \(-\int |{\beek}|^2/\omega_k\,dk\) with eigenfunction \(\D_0(\eff_0)\Omega_0\) where
  \(\eff_0(k)=-{\beek}/\omega_k\).
  \item[(ii)]
The solution to \eqref{solveme} satisfies
\[
\Psi(t)\,=\,e^{i\Theta_e(t)}\D_0(\eff(t))\Exp[-i(t-s)\h](\D_0(\eff(s)))^*e^{-i\Theta_e(s)}\Psi(s)
\]
where the phase factor satisfies
\(\Theta_e(t,0)=-\int_0^t\frac{1}{2}\int (\overline{{\beek}}(s')\eff(s',k)+{\beek}(s')\overline{\eff}(s',k))\,dkds'\) and
\(t\mapsto \eff(t,k)\) evolves according to the second equation of \eqref{defeeze1}, namely,
\(
i\dot \eff-\omega^{}_k \eff-{\beek}\,=\,0\,.
  \)
This gives the analogous formula for the solution operator \(\Pexp[-i\int_s^t({{\h}+{\h_1}(\sigma)})
  d\sigma]\) for \eqref{solveme}, following from the preceding commutation relations on a dense subspace,
and with a unique extension to the whole Hilbert space.
\end{enumerate}
Taking tensor products leads to the definition of the limiting dynamics
\begin{align*}\notag  
  \Pexp\bigl[-i\int_s^t{\rm H}^{eff}_0(\sigma)\, d\sigma\bigr]
  \,=\,\Pexp\bigl[-i\int_s^t h_{\mbox{\tiny 1P}}(\sigma)d\sigma\bigr]&\otimes
  \Pexp\bigl[-i\int_s^t\bigl({h}^{}_d+
{\beed}(\sigma)(a^{}_d+a_d^\dagger)\bigr)d\sigma\bigr]\\&\qquad\qquad\qquad\otimes
  \Pexp\bigl[-i\int_s^t ({{\h}}+{\h_1}(\sigma))d\sigma\bigr]\,,
  \end{align*}
in the space \({\mathfrak H}\) or  \({\mathfrak H}(\theta)\). We transfer this
evolution to the space
\(L^2(\mu_0)\) via the operators \(\bbS(\xi)\) or  \(\bbS^\theta(\xi)\), defining
the limiting semiclassical solution operator by
\begin{equation}\label{defld}
  \bevot_{scl}(t,s)\circ\bbS^\theta(\xi(s))\,\define\,
\bbS^\theta(\xi(t))\circ 
\Pexp\Bigl[-i\int_s^t\,{\rm H}^{eff}_0(\sigma)\,d\sigma\Bigr]\,.
\end{equation}

We will specify the quadratic transverse Hamiltonian by giving the parameters
\(\cee_0,\eff_0\) which determine the displacement of the vacuum. So for the Hamiltonian
\begin{align}\label{disvac}\begin{split}
h_{\cee_0,\eff_0}&\define
{h}^{}_d+{\beed}(a^{}_d+a^\dagger_d)+\h+\h_1\qquad\hbox{ the vacuum is}\\
  \Omega_{\cee_0,\eff_0}&\define D_d(\cee_0)\id_{\smr}\otimes\D_0(\eff_0)\Omega_0\,
  =\D_{\cee_0,\eff_0}\id_{\smr}\otimes\Omega_0\in\mfrF\,,
\end{split}\end{align}
where \(\D_{\cee,\eff}=D_d(\cee)\otimes \D_0(\eff)\).
Now the time dependence induces a fluctuation of {\em complex} values about these real values,
so it is convenient to use the same notation 
for complex values \(\cee\in \C,\; \eff\in\cs_{\smc}(\R)\) of the parameters so that (retaining the same
relation between these parameters and the linear part of the Hamiltonian:
\begin{align}
&h_{\cee,\eff}\Omega_{\cee,\eff}=e_{\cee,\eff}\Omega_{\cee,\eff}\,,\qquad e_{\cee,\eff}=
-\omega_d|\cee|^2-(\eff,\omega_\bullet \eff)_{L^2}\,,\\
&\hbox{where}\qquad
h_{\cee,\eff}\define
{h}^{}_d-\omega_d\cee_d(a^{}_d+a^\dagger_d)+\h-
\int\omega_k\Bigl( {\overline{\eff(t,k)}}a^{}_k+\eff(t,k)a_k^\dagger\Bigr)dk
\end{align}
(This Hamiltonian is still self-adjoint at fixed times).

With these notations fixed, the Hamiltonian in \eqref{simhE} can be written
\({\rm H}^{eff}_0(t)=\,h_{\mbox{\tiny 1P}}+{h}_{{\cav},{\fav}}\), with the transverse Hamiltonian labelled by
\begin{equation}\label{defcfav}
{\cav}(t)=-{\beed}(t)/\omega_d\,,\qquad {\fav}(k,t)=-{\beek}(t)/\omega_k\,;
\end{equation}
these values are significant in that they give mean values around which the bosonic modes oscillate, see below.
In the simplest case, take the transverse modes to lie in a displaced vacuum state
\(\Omega_{\cee(0),\eff(0)}\) initially, then the transverse evolution just displaces this vacuum:
given a smooth curve \(s\mapsto (\cee(s),\eff(s))\in\C\times\cs_{\smc}(\R)\) satisfying
\eqref{defeeze1},
\begin{equation}\label{evo}
  \frac{d}{ds}\Omega_{\cee,\eff}=
  \bigl(i\dot\Theta_3
  -i{h}_{{\cav},{\fav}}
  \bigr)
  \Omega_{\cee,\eff}\,,\quad\hbox{and}\quad
  \frac{d}{ds}\chi(s,\cdot)\Omega_{\cee,\eff}=
  \bigl(
  i\dot\Theta_3
  -i{\rm H}_0^{eff}
  \bigr)
  \chi(s,\cdot)\Omega_{\cee,\eff}\,,\end{equation}
or in integrated form
\begin{equation}\label{evoo}
  \Pexp\Bigl[-i\int_s^t{\rm H}^{eff}_0(\sigma)\, d\sigma\Bigr]\,\chi(s,\cdot)\Omega_{\cee(s),\eff(s)}=
  e^{-i\Theta_3(t)}
  \chi(t,\cdot)\Omega_{\cee(t),\eff(t)}
  \end{equation}
where \(t\mapsto\chi(t,\cdot)\) solves \(i\partial_t\chi =h_{\mbox{\tiny 1P}}\chi\)
and \(\Theta_3+\Theta_d+\Theta_e=0\).

Referring to
\eqref{uniteqsolxi}, we note the  transfer of the expressions for the displaced vacuum
into the Schr\"odinger representation:
\begin{equation}\label{4lp}
 \J^{\theta}(\xi)\,\id_{\smr}(Q)\otimes\Omega_{\cee,\eff}\,=\,\exp\Bigl[-\cee \cee_1
    +\cee\sqrt{2\omega_d}\bfupphi( {\tteonex})-(\eff_1,\eff)
    +\bfupphi\bigl(\mcF_{u_\xi}^{-1}(\sqrt{2\omega_\bullet}\eff)(\cdot;\xi)\bigr)\Bigr]\,\in L^2({\mu(\theta,\xi)})
\end{equation}
and thence
\begin{align}\begin{split}\label{4lp2}
  &\bfOm^\theta_{\cee,\eff}(\xi)\define   \bbS^\theta(\xi) \id_{\smr}(Q)\otimes\Omega_{\cee,\eff}
  \,=\,\rho(\cee,\eff,\xi)\,
\bbS^\theta(\xi)\id_{\smr}(Q)\otimes\Omega'
\,\in L^2(\mu_0)\,,\\
&\hbox{where}\quad\rho(c,f,\xi)\,=\,\exp\Bigl[-c c_1
    +c\sqrt{2\omega_d}\phisch( {\tteonex})-(f_1,f)
    +\phisch\bigl(\mcF_{u_\xi}^{-1}(\sqrt{2\omega_\bullet}f)(\cdot;\xi)\bigr)\Bigr]
\end{split}\end{align}
The factor \( \id_{\smr}(Q)\) which is just the function of \(Q\) identically equal to one will often be omitted.
\begin{remark}\label{hosol} Recall that the measures \(\mu(\theta,\xi)\) and \(\mu_0\) are equivalent on \(\cs'\).
  For \(p<\infty\) the map \(\xi\mapsto \bfOm^\theta_{\cee,\eff}(\xi)\in L^p(\mu_0)\) is differentiable with derivative
  \begin{align}\label{diffvac}\begin{split}
    \partial_\xi\bfOm^\theta_{\cee,\eff}(\xi)&=\Bigl(
    \cee\sqrt{2\omega_d}\phisch\bigl(\partial_\xi {\tteonex}\bigr)
    +\phisch\bigl(\partial_\xi\mcF_{u_\xi}^{-1}(\sqrt{2\omega_\bullet}\eff)
    \bigr)\Bigr)
    \bfOm^\theta_{\cee,\eff}(\xi)\\
    &\qquad+\exp\Bigl[-\cee \cee_1
    +\cee\sqrt{2\omega_d}\phisch( {\tteonex})-(\eff_1,\eff)
    +\phisch\bigl(\mcF_{u_\xi}^{-1}(\sqrt{2\omega_\bullet}\eff)
    \bigr)\Bigr]
    \partial_\xi\Bigl(
    \bbS^\theta(\xi)\Omega'
    \Bigr)\,,
    \end{split}
  \end{align}
  where the prefactor in the first line is just \(\partial_\xi\ln\rho\) and
  the last term is worked out in \eqref{defrnfinal15}.
  Direct substitution of the series for \(\Exp[ a^\dagger(\eff)]\) in the definition of
  \(\Omega_{\cee,\eff}\) and estimation gives a proof that
  \(\Omega_{\cee,\eff}\in\Dom(h(\omega_d)\oplus \h(\omega^{}_\bullet))\); consequently, if
  \(\chi\in \cs(\R)\,,\) then \(\chi(Q)\Omega_{\cee,\eff}\)
  lies in \(\Dom (\triplecolon H^{sol}_0\triplecolon)\) and so
  \(\bbS(\xi)\,\psi\,\Omega_{\cee,\eff}\in \Dom (\doublecolon H^{sol}_0\doublecolon)\), and similarly for positive \(\theta\).
  It is useful to write the formula \eqref{diffvac} making use of the decomposition into real and imaginary components. For the complex number we just write \(\cee=\cee_1+i\cee_2\), and \(\cee_1,\cee_2\) real, but for \(\eff\) the decomposition is determined by the conjugation in which the conjugate of \(\eff\) is the  function \(\eff^\flat(k)=
  \overline{\eff(-k)}\) induced by the distorted Fourier transform \eqref{dft}. We write \(\eff=\eff_1+i\eff_2\)
  with \(\eff_j^\flat=\eff_j\), so \(\eff^\flat=\eff_1^\flat-i\eff_2^\flat\)
  then \(\eff_2\) drops out of \(\upphi_{scl}\) and the first line of \eqref{diffvac} equals
  \begin{equation}\label{diffvacconj}
    \partial_\xi\ln\rho(c,f,\xi)\bfOm^\theta_{\cee,\eff}(\xi)=
    \Bigl[\phisch\bigl(\partial_\xi(K(\xi)^{\frac{1}{2}}\upphi_{scl})\bigr)+
    i\Bigl(
    \cee_2\sqrt{2\omega_d}\phisch\bigl(\partial_\xi {\tteonex}\bigr)
    +\phisch\bigl(\partial_\xi\mcF_{u_\xi}^{-1}(\sqrt{2\omega_\bullet}\eff_2)
    \bigr)\Bigr)
    \Bigr]\,\bfOm^\theta_{\cee,\eff}(\xi)\,.
  \end{equation}
  The first term in the square brackets
  can be written \(-\sqrt{\scm_{cl}}\mbQ\,\upphi_{scl}\bigl(K(\xi)^{\frac{1}{2}}{{\tteox}'}\bigr)
  +\phischperp\bigl(\partial_\xi(K(\xi)^{\frac{1}{2}}\upphi_{scl})\bigr)\), and \(K(\xi)\) can be replaced by
  \(K^\theta(\xi)\) since it is applied to elements of \(\langle\tteox\rangle^\perp\). The remaining two terms
  in the brackets can be written
  \begin{equation}\label{diffvacconj2}
-i\sqrt{\scm_{cl}}\mbQ  \Bigl(  \cee_2\sqrt{2\omega_d}(\tteox,\partial_\xi {\tteonex})
    +\bigl(\tteox,\partial_\xi\mcF_{u_\xi}^{-1}(\sqrt{2\omega_\bullet}\eff_2\bigr)
    \Bigr)
    +i\Bigl(\cee_2\sqrt{2\omega_d}\phischperp\bigl(\partial_\xi {\tteonex}\bigr)
    +\phischperp\bigl(\partial_\xi\mcF_{u_\xi}^{-1}(\sqrt{2\omega_\bullet}\eff_2)
    \bigr)\Bigr)\,.
  \end{equation}
  The second term in \eqref{diffvac} can be read off from \eqref{defrnfinal2}, and combining with \eqref{evo} (in which
  \(\xi\) is independent of time)
  we have for the total derivative
  \begin{align}\begin{split}\label{imuseful}
  \frac{d}{dt}\bfOm^\theta_{\cee,\eff}(\xi)=&
  \Bigl(\frac{d}{dt}\bbS^\theta(\xi)\Bigr)\Omega_{\cee,\eff}
  +\dot\xi\partial_\xi\ln\rho(\cee,\eff,\xi)\bfOm^\theta_{\cee,\eff}(\xi)
  +i\bbS^\theta(\xi)(\dot\Theta_3-h_{\cee_0,\eff_0})
  \Omega_{\cee,\eff}\\
  &\Bigl(\frac{d}{dt}\bbS^\theta(\xi)\circ \bbS^\theta(\xi)^{-1}
  +\dot\xi\partial_\xi\ln\rho(\cee,\eff,\xi)\Bigr)\bfOm^\theta_{\cee,\eff}(\xi)
  +i\bbS^\theta(\xi)(\dot\Theta_3-h_{\cee_0,\eff_0})\Omega_{\cee,\eff}\,.
  \end{split}\end{align}
  \end{remark}

\subsubsection{Limiting dynamics - existence theorem}\label{limdynex}
\begin{theorem}\label{lse}
Under the hypotheses (H1)-(H2) there exist continuously differentiable functions 
  \(t\mapsto (\xi(t),\eta(t))\in\R^2\)
and  \( t\mapsto (\cee(t),\eff(t,\cdot))\in \C\times \cs(\R;\C)\)     
which satisfy \eqref{newtpert} and \eqref{defeeze1}, on a time interval \(0\leq t\leq \tau_{loc}/\sqrt{g}\)
where \(\tau_{loc}>0\) is independent of \(g<1\).
     \end{theorem}
\begin{proof}
  {\em Throughout this proof only we'll write \(a\) for \(\frac{1}{2}\) to avoid fractions in indices.}
  This is proved by a reduction to the contraction mapping theorem, which is achieved by first
  ``integrating out'' \(\cee\) and \(\eff\) as follows.
  Multiplication of both equations in \eqref{defeeze1} by appropriate integrating factors  leads, for initial
data \(\eff(0),\cee(0)\), to
\begin{align}\label{theresult}\begin{split}
  \eff(t,k)&=\eff(0,k)e^{-i\omega_k t}
  \,-i\,\int_0^te^{-i\omega_k (t-\sigma)}\,{\beek}(\sigma)d\sigma=\tilde \eff(t,\xi,\eta,\eff(0,k))\\
  \cee(t)  &=e^{-i\omega_d t}\cee(0)\,-i\,\int_0^te^{-i\omega_d (t-\sigma)}\,{\beed}(\sigma)d\sigma=
             \tilde \cee(t,\xi,\eta,\cee(0))\,.
\end{split}\end{align}
Recall that \({\beed},{\beek}\) are determined from \(\xi,\eta\) by \eqref{defeeze15}, so 
we have thus defined functionals \(\tilde\cee,\tilde\eff\) of the soliton parameters \(\xi,\eta\)
which are nonlocal in that
at time \(t\) they depend upon the entire trajectories \(\{(\xi(\sigma),\eta(\sigma)\}_{0\leq\sigma\leq t}\). These
functionals can now be substituted into \eqref{newtpert} to obtain a self-contained nonlocal system of equations for
the soliton parameters:
\[
\dot\xi=g^2{\scm_{cl}}^{-1}\eta+g{{V_{-1}}}(\xi,\dot\xi,\tilde \cee,\tilde \eff)\quad \hbox{and} \quad
\dot\eta=-\frac{1}{g}\sqrt{\scm_{cl}}\lambda({\elec}, {\tteox})_{L^2}+\frac{1}{g}{V_1}(\xi,\dot\xi,\tilde \cee,\tilde \eff)\,,
\]
(The first equation determines \(\dot\xi\) uniquely for small \(g\) once \(\xi,\eta\) are known, in particular for the initial values.)
These equations can be solved via the contraction
mapping principle on the space of continuous functions \(t\mapsto Z(t)=(\xi(t),\dot\xi(t),\eta(t))\in\R^3\), where
we use a weighted norm \(|Z|_g=|\xi|+g^{-a}|\dot\xi|+g^{2-a}|\eta|\) to clarify the relevant scales. With this
scaling the equations can be written 
\begin{align}\label{cbw}\begin{split}
g^{-a}\dot\xi&={\scm_{cl}}^{-1}g^{2-a}\eta+g\Gamma_1\,,\quad\Gamma_1(\xi,\dot\xi,\eta)\define\scm_{cl}^{-a}g^{-a}\dot\xi\bigl(\tilde\upphi_{scl},{{\tteox}'}\bigr)\\
  g^{2-a}\dot\eta&= g^{a}\Gamma_2\,,\quad
                   \Gamma_2(\xi,\dot\xi,\eta)\define-\lambda \sqrt{\scm_{cl}}({\elec}, {\tteox})+\dot\xi\,\scm_{cl}^{a}
\Bigl(  \tilde\cee_2\sqrt{2\omega_d}(\tteox,\partial_\xi {\tteonex})
    +\bigl(\tteox,\partial_\xi\mcF_{u_\xi}^{-1}\sqrt{2\omega_\bullet}\tilde\eff_2\bigr)
    \Bigr)
\end{split}\end{align}
where \(\tilde\upphi_{scl}=\tilde\upphi_{scl}(t,Z)\in\cs\) means \eqref{defiscl} with \(\cee,\eff\) replaced by \(\tilde \cee,\tilde \eff\).
We now set this up to use the contraction mapping theorem on spaces of continuous functions with initial
values \(\xi_0,\eta_0\) which determine \(Z_0=(\xi_0,\dot\xi_0,\eta_0)\) with \(|Z_0|_g\) bounded uniformly in \(g\)
for all \(g<1\). (The dependence of the solutions on \(g\) is suppressed for clarity.)
Define the map \(Z\mapsto SZ=(\xi_*,\dot{\xi}_*,\eta_*)=Z_*\) obtained by solving, with \((\xi_*(0),\eta_*(0))=(\xi_0,\eta_0)\),
\[g^{2-a}\dot\eta_*=g^{a}\Gamma_2(Z)
\quad \hbox{ and {\em then}}\quad
g^{-a}\dot{\xi}_*={\scm_{cl}}^{-1}g^{2-a}\eta_*+g\Gamma_1(Z)
\]
where the arguments of \(\tilde\cee /\tilde \eff\) are, respectively,
understood to be \((t,\xi,\eta,\cee(0) /\eff(0))\).) The point of this particular
choice of \(S\) is that the estimate
\(
|g^{2-a}({\eta}_*-{\eta}_{*1})]_0^t|\leq\int_0^tg^{2-a}|(\dot{\eta}_*-\dot{\eta}_{*1})|
    \)
    can be substituted into the
    \(\xi\) equation to yield 
    \begin{align}\label{getc}\begin{split}
    |g^{-a}(\dot\xi_*-\dot\xi_{*1})(t)|&\leq {\scm_{cl}}^{-1}|g^{2-a}({\eta}_*-{\eta}_{*1})(t)|+
                                      g\,|\Gamma_1(Z(t))-\Gamma_1(Z_1(t))|\\
      &\leq {\scm_{cl}}^{-1} g^{a}\int_0^t|\Gamma_2({Z(\sigma)})-\Gamma_2({Z}_{1}(\sigma))|d\sigma+
 g\,|\Gamma_1(Z(t))-\Gamma_1(Z_1(t))|\end{split}\,.
    \end{align}
    All together we have
    \begin{align}\label{lippp}
      \max_{[0,T]}|Z_*|_{g}&\leq const.\, \Bigl(|Z(0)|_g+g^a|T|(1+g^a|T|)\max_{[0,T]}|\Gamma_2(Z)|+g(1+g^a|T|)\max_{[0,T]}|\Gamma_1(Z)|\Bigr)
      \\
      \max_{[0,T]}|(Z_*-Z_{1*})|_{g}&\leq g\max_{[0,T]}|\Gamma_1(Z)-\Gamma_1(Z_1)|+const.\, |T|\Bigl(g^{a}(1+g^{a}|T|)\max_{[0,T]}|\Gamma_2(Z)-\Gamma_2(Z_1)|+g^{1+a}\max_{[0,T]}|\Gamma_1(Z)-\Gamma_1(Z_1)|
      \Bigr)
\notag
      \end{align}
    which will now be seen to imply that \(S\) is a contraction in norm
    \(|\cdot|_g\) on short \(O(g^{-a})\) time-scales.
    The quantities
\({\beed},{\beek}\) defined in \eqref{defeeze1} depend on \(t\) and \(Z\):
\begin{align}
  {\beed}\,&=\,b_{d1}(t,\xi)+g b_{d2}\dot\xi\,\eta=b_{d1}(t,\xi)+ b_{d2}g^{-a}\dot\xi\,g^{2-a}\eta
  \,,\quad\hbox{and}\\
  {\beek}\,&=\,b_{k1}(t,\xi)+g b_{k2}(t,\xi)\dot\xi\,\eta=
  b_{k1}(t,\xi)+b_{k2}(t,\xi)g^{-a}\dot\xi\,g^{2-a}\eta
\label{defeeze2}
\end{align}
with \(b_{d2}\) a number (\(-(\tteo',\tteone)/\sqrt{\scm_{cl}}\),which evaluates to \(3m\pi/8\sqrt{2\scm_{cl}}\)), \(b_{d1}\) a \(C^1\) scalar valued function of \((t,\xi)\), while
\(b_{k1},b_{k2}\) are \(C^1\) functions of \((t,\xi)\) taking values in Schwartz space (as a function of \(k\)).
(The factors of \(g\) are chosen so that these quantities so defined can be bounded independently of
\(g\) under the scaling implicit in the definition of the norm \(|Z|_g\).) We have:
\begin{align}
  b_{d1}(t,\xi)&=\frac{\lambda}{\sqrt{2\omega_d}}\int \tteone(x-\xi){\elec}(t,x)\,dx\\
  b_{k1}(t,\xi)&=\frac{\lambda}{\sqrt{4\pi\omega_k(k-im)(k-2im)}}\,\int {\elec}(t,x){\ttF}(-k,x-\xi)e^{-ikx}\,dx
  =s_1(k,t,\xi)\\
  \partial_\xi b_{k1}(t,\xi)&=\frac{-\lambda}{\sqrt{4\pi\omega_k(k-im)(k-2im)}}\,\int {\elec}(t,x)\partial_x{\ttF}(-k,x-\xi)e^{-ikx}\,dx
=s_{11}(k,t,\xi)\\
b_{k2}(t,\xi)&=\frac{-\lambda}{\sqrt{4\pi\omega_k(k-im)(k-2im)\scm_{cl}}}\,\int {\tteo}'(x-\xi){\ttF}(-k,x-\xi)e^{-ikx}\,dx
=e^{-ik\xi}s_2(k)
\end{align}
and, with reference to \eqref{genf} it follows that \(s_1,s_{11},s_2\in\cs\) as functions of \(k\), and can be simply expressed in terms of
the distorted Fourier transform: \(s_1(k)\) is proportional to \(\omega_k^{-1/2}\mcF_{u_\xi}\E(k)\)
and similarly \(s_2(k)\)is proportional to \(e^{-ik\xi}\omega_k^{-1/2}\mcF_{u}\tteo'(k)\). This allows us to read off \(L^2\) bounds
by the unitarity of the distorted Fourier transform, while \(s_{11}\) can be directly controlled by inspection of the form of \(\ttF\)
in \eqref{genf}. This is all that is needed to read off the bounds which follow.
\begin{enumerate}
\item[(i)] By inspection of \eqref{defeeze2}, if \(\max_{[0,T]}|\xi|=R_0<\infty\) then
\[
\max_{[0,T]}|{\beed}|\leq const.(M_1,R_0)\,\times(1+\max_{[0,T]}|Z|_g^2)\,.
\]
\item[(ii)]
Similarly
\begin{align*}
\sum_{n_0,n_1=0}^1\max_{[0,T]}\,\|\partial_t^{n_0}\partial_\xi^{n_1}b_{kj}(t,\xi)\|_{L^2(dk)}&\leq
const.(R_0)\,\times \bigl(\max_{t\in [0,T]}\|{\elec}(t,x)\|_{L^2(dx)}+\max_{t\in [0,T]}\|\partial_t{\elec}(t,x)\|_{L^2(dx)}+
                                                                                               \max_{[0,T]}|Z|_g^2\bigr)\\
  &\leq
const.(M_1,R_0)\,\times \bigl(1+\max_{[0,T]}|Z|_g^2\bigr)\,,
\end{align*}
with the constant determined by the norms of \({\elec},\partial_t{\elec}\), which are subject to the assumption (H1).
\item[(iii)] Also there are corresponding bounds for differences related to two continuous functions \(Z,Z_1\) into \(\R^3\):
assume
there exist positive \(R,T\) such that
\(\max_{[0,T]}|Z|_g\leq R\) and \(\max_{[0,T]}|Z_1|_g\leq R\),
then
\begin{align*}\begin{split}
\max_{[0,T]}\Bigl(\sum_{n_0,n_1=0}^1|\partial_t^{n_0}\partial_\xi^{n_1}({\beed}(t,\xi)-b_{d}(t,\xi_1))|+\sum_{n_0,n_1=0}^1\,
&\|\partial_t^{n_0}\partial_\xi^{n_1}(b_{kj}(t,\xi)-b_{kj }(t,\xi_1))\|_{L^2(dk)}\Bigr)\\&\leq
const.(M_1,R)\,\times 
\max_{[0,T]}|Z-Z_1|_g\,.
\end{split}
\end{align*}
\end{enumerate}
Inserting these into the integrals in \eqref{theresult} gives (using \(\|\cdot\|\) for the \(L^2\) norm)
\begin{align*}
&\max_{0\leq t\leq T}\Bigl(|\tilde\cee(t,\xi,\eta,\cee(0))|
+\|\tilde\eff(t,\xi,\eta,\eff(0))\|\Bigr)\leq |\cee(0)|+\|\eff(0)\|+const.(M_1,R_0) \times T\times\bigl(1+
\max_{[0,T]}|Z|_g^2\bigr)\qquad\hbox{and}
 \\ &\max_{0\leq t\leq T}\Bigl(|\tilde\cee(t,\xi,\eta,\cee(0))-\tilde\cee(t,\xi_1,\eta_1,\cee(0))|
+\|\tilde\eff(t,\xi,\eta,\eff(0))-\tilde\eff(t,\xi_1,\eta_1,\eff(0))\|\Bigr)\leq const.(M_1,R) \times T\times
\max_{[0,T]}|Z-Z_1|_g
\end{align*}
and these in turn allow control of 
the semiclassical field \(\tilde\upphi_{scl}\) by \eqref{bdscl} and:
\begin{equation}\label{lipscl}
\|\tilde\upphi_{scl}(t,Z)-\tilde\upphi_{scl}(t,Z_1)\|_{L^2}\leq const.(M_1,R) \times T\times
\max_{[0,T]}|Z-Z_1|_g\,,
  \end{equation}
with constants having the dependencies shown, and (importantly) independent of \(g<1\). 
Combining with \eqref{lippp} gives
\[
|\Gamma_1(Z)|\leq const.(R)(|\tilde\cee|+\|\tilde\eff\|)
\leq
const. (M_1,R,|\cee(0)|,\|\eff(0)\|)(1+T)
\]
and
\[
  \Gamma_2(Z)|\leq const. (M_1,R)\bigl(1+g^a(|\tilde\cee|+\|\tilde\eff\|)\bigr)
  \leq
  const. (M_1,R,|\cee(0)|,\|\eff(0)\|)(1+g^aT)\,.
\]
The growth with \(T\) needs to be kept in mind. Similarly for the differences:
\begin{align*}
\max_{[0,T]}|\Gamma_1(Z)-\Gamma_1(Z_1)|&\leq const. (1+T)\max_{[0,T]}|Z-Z_1|_g\qquad
\max_{[0,T]}|\Gamma_2(Z)-\Gamma_2(Z_1)|\leq const. (1+g^aT)\max_{[0,T]}|Z-Z_1|_g
                                         \end{align*}
with the constants independent of \(g<1\) but still dependent upon
\(R,M_1,|\cee(0)|,\|\eff(0)\|\)). From this we deduce - and here Remark \eqref{getc} is used - that
there exist \(\Lambda_0,\Lambda_1\) with the same dependencies and
{\em in addition} depending upon a number \(\tau_*\) such that \(|T|\leq \frac{\tau_*}{g^{a}}\), then
\begin{align}
  \max_{[0,T]}|SZ(t)|_g&\leq\Lambda_0\Bigl(1+|Z(0)|_g\Bigr)\\
  \max_{[0,T]}|SZ(t)-SZ_1(t)|_g&\leq\Lambda_1\, (g^a+g^{a}T)\max_{[0,T]}|Z(\sigma)-Z_1(\sigma)|_g\,.
\end{align}
If we now consider the problem at hand, that of constructing solutions to \eqref{cbw}, we work in a
Banach space of functions
\[\Xi_{R,T}=\{Z\in C([0,T];\R^3):\max_{[0,T]}|Z|_g\leq R\;\hbox{and}\;Z(0)=Z_0\}\,,\]
and, choosing \(R\) sufficiently large (depending upon
\(|Z(0)|_g\) and \(\Lambda_0\)) there exists \(T=\tau_{loc}/g^{a}\) such that
\(S:\Xi_{R,T}\to\Xi_{R,T}\)
is a contraction. It follows that there exists a fixed point for \(S\) in
\(\Xi_{R,T}\), and hence a \(C^1\) solution of \eqref{cbw},
or equivalently \eqref{newtpert}, and by inspection this ensures the existence of a \(C^1\) solution
of \eqref{defeeze1} (taking values in \(\C\times\cs\).)
\end{proof}
This establishes local existence. The local existence argument can of course be repeated in the usual way, allowing the
solution to be extended. The more detailed estimates for the solution
which we now obtain provide potentially much longer intervals over which the solution will exist.
In particular, averaging gives asymptotic expansions which improve the bounds on \((\cee,\eff)\)
to show that the growth of \(\cee,\eff\) is actually much slower than the above proof worked with
- compare \eqref{iwasclose} and \eqref{imclose} - and so the solution can be expected to exist on long time intervals.
\begin{corollary}\label{cproof}
  Given \(R\) sufficiently large, there exists a solution to \eqref{newtpert} and \eqref{defeeze1} satisfying
  \[
\max_{[0,\frac{\tau_3}{\sqrt{g}}]}|Z|_g\leq R
  \]
  on some time interval \([0,\frac{\tau_3}{\sqrt{g}}]\), with \(\tau_3\) independent of \(g<1\), for sufficiently small \(g\).
  Furthermore for \(0\leq t\leq \frac{\tau_3}{\sqrt{g}}\) there holds
  \[
  |\xi(t)|+g^{-\frac{1}{2}}|\dot\xi(t)|+g^{-1}|\ddot\xi(t)|+g^{-\frac{3}{2}}|\dddot\xi(t)|+
  g^{\frac{3}{2}}|\eta(t)|+g|\dot\eta(t)|+g^{\frac{1}{2}}|\ddot\eta(t)|\leq const.
      \]
      with constant independent of \(g\).
  \end{corollary}
\paragraph{Asymptotic Expansion of Solution of Transverse mode dynamics}

The displaced vacuum \(\Omega_{\cee,\eff}\) which is formed dynamically is
the vacuum vector of \({h}_{\cee,\eff}\) on \(\mfrF\), but 
not of the restriction of \({\rm H}^{eff}_0\) to \(\mfrF\), which is \({h}_{{\cav},{\fav}}\).
This will lead to the need to estimate
\[
\delta h=h_{\cee,\eff}-{h}_{{\cav},{\fav}}\,,
\]
via the following asymptotic expansions.
Integration by parts of \eqref{theresult} yields
\begin{align}\notag
  \eff(t,k)&=-\frac{{\beek}(t)}{\omega_k}+\Bigl(\eff(0,k)+\frac{{\beek}(0)}{\omega_k}\Bigr)e^{-i\omega_k t}
  +\frac{1}{\omega_k}\int_0^te^{-i\omega_k (t-\sigma)}\dot {\beek}(\sigma)d\sigma\\\label{asex}
  &=\fav(t,k)+\frac{\dot {\beek}(t)}{i\omega_k^2}+
  \Bigl(\eff(0,k)+\frac{{\beek}(0)}{\omega_k}-\frac{\dot {\beek}(0)}{i\omega_k^2}\Bigr)e^{-i\omega_k t}
  -\frac{1}{i\omega_k^2}\int_0^te^{-i\omega_k (t-\sigma)}\ddot {\beek}(\sigma)d\sigma\,,
  \qquad\hbox{and similarly}\\\notag
\cee(t)  &=\cav(t)+\frac{\dot {\beed}(t)}{i\omega_d^2}+
  \Bigl(\cee(0)+\frac{{\beed}(0)}{\omega_d}-\frac{\dot {\beed}(0)}{i\omega_d^2}\Bigr)e^{-i\omega_d t}
  -\frac{1}{i\omega_d^2}\int_0^te^{-i\omega_d (t-\sigma)}\ddot {\beed}(\sigma)d\sigma\,.
\end{align}
The first line gives boundedness on suitable time-scales:
\begin{equation}\label{iwasclose}
\|\omega_k^r\eff(t,k)\|_{L^p(dk)}\leq\|\omega_k^r\eff(0,k)\|_{L^p(dk)}
+2\sup_{0\leq \sigma\leq t}\|\omega_k^{r-1}{\beek}(\sigma)\|_{L^p(dk)}+
\int_0^t\|\omega_k^{r-1}\dot {\beek}(\sigma)\|_{L^p(dk)}d\sigma
\end{equation}
for any weighting exponent \(r\in\R\), 
and the analogous bound for the absolute value of \(\cee(t)\) also holds.
The second line of the expansions indicates further the significance of the average values \({\cav},{\fav}\) defined above, as the values about
which the solutions oscillate, since their time dependence is on a faster time scale than the soliton dynamics. For
appropriate initial data these expansions indicate that \(\cee-\cav,\eff-\fav,\Im\cee,\Im\eff\) are small, in the sense that they can be bounded by the time derivatives of the inhomogeneous terms
(which are small in the present application). For example
the first formula implies for \(t>0\)
\begin{equation}\label{imclose}
\|\omega_k^r(\eff(t,k)-\fav(t,k))\|_{L^p(dk)}\leq \|\omega_k^r(\eff(0,k)-\fav(0,k))\|_{L^p(dk)}
+\bigl(2\sup_{0\leq \sigma\leq t}\|\omega_k^{r-2}\dot {\beek}(\sigma)\|_{L^p(dk)}
+|t|\sup_{0\leq \sigma\leq t}\|\omega_k^{r-2}\ddot {\beek}(\sigma)\|_{L^p(dk)}\bigr)\,,
\end{equation}
where \(r\in\R\) and \(1\leq p<\infty\), or also with a supremum over \(k\in\R\);
since \(\fav,\cav\) are real, the imaginary values are controlled automatically in the same norms
as an immediate consequence of:
\begin{equation}\label{imcloser}
|\Im\eff(t,k)|\leq |\eff(t,k)-\eff_0(t,k)|\qquad\hbox{and}\quad
|\Im\cee(t,k)|\leq|\cee(t,k)-\cee_0(t,k)|\,.
\end{equation}
(Recall from Remark \ref{reality} that the reality condition is \(\eff(t,-k)=\overline{\eff(t,k)}\).) A consequence
which will be useful is the following comparison of the two Hamiltonians defined in \eqref{disvac}-\eqref{defcfav}:
\begin{lemma}\label{compham} Let \(\delta h=h_{\cee,\eff}-{h}_{{\cav},{\fav}}\),
where \eqref{asex} hold, and \(\sigma\to {\beed}(\sigma)\) (resp. \(\sigma\to b_\bullet(\sigma)\)) are \(C^2\) into
\(\R\) (resp. \(\cs(\R)\)).  Then 
the following bounds for the Fock space \(\mfrF\) operator norm hold at time \(t\):
  \begin{align*}
  \|(\id+{\N})^{-1/2}\delta h\|
  \leq const. \Bigl[&|\cee(0)-\cav(0)|+\|\omega_k(\eff(0,k)-\fav(0,k))\|_{L^2(dk)}\\
  &+ \sup_{0\leq \sigma\leq t}|\dot {\beed}(\sigma)|+\sup_{0\leq \sigma\leq t}\|\omega_k^{-1}\dot {\beek}(\sigma)\|_{L^2(dk)}\\
  &\qquad+|t|\bigl(\sup_{0\leq \sigma\leq t}|\ddot {\beed}(\sigma)|+
  \sup_{0\leq \sigma\leq t}\|\omega_k^{-1}\ddot {\beek}(\sigma)\|_{L^2(dk)}\bigr)\Bigr]\,.
  \end{align*}
\end{lemma}
\begin{corollary}\label{imclosercorl} Assume 
  \(\sigma\to\eta(\sigma)\) is \(C^2\), \(\sigma\to\xi(\sigma)\) is \(C^3\) and
  \(\sigma\to\elec(\sigma,\cdot)\) is \(C^2\) into \(\cs(\R)\). Define \({\beed},{\beek},\cee,\eff\) by \eqref{defeeze1}, then
  the bounds in the previous lemma imply
  \begin{align*}
  \|(\id+{\N})^{-1/2}\delta h\|
  \leq const. \Bigl[&|\cee(0)-\cav(0)|+\|\omega_k(\eff(0,k)-\fav(0,k))\|_{L^2(dk)}\\
    &+ \sup_{0\leq \sigma\leq t}\|\partial_\sigma\elec(\sigma,x)\|_{L^2(dx)}
    +\sup_{0\leq \sigma\leq t}|\dot\xi|\|\elec(\sigma,x)\|_{L^2(dx)}\\
    &\phantom{ + \sup_{0\leq \sigma\leq t}\|\partial_\sigma\elec(\sigma,x)\|_{L^2(dx)}+}
    +\,g\sup_{[0,t]}\Bigl(
|\eta|(|\ddot\xi|+|\dot\xi|^2)+|\dot\eta||\dot\xi|\Bigr)\\
&\qquad
+|t|\sup_{0\leq \sigma\leq t}\|\partial_\sigma^2\elec(\sigma,x)\|_{L^2(dx)}+
|t|\sup_{0\leq \sigma\leq t}|\dot\xi|\|\partial_\sigma\elec(\sigma,x)\|_{L^2(dx)}\\
    &\phantom{|t|\sup_{0\leq \sigma\leq t}\|\partial_\sigma^2\elec(\sigma,x)\|_{L^2(dx)}+
|t|\sup_{0\leq \sigma\leq t}|\dot\xi|}+|t|\sup_{0\leq \sigma\leq t}(|\dot\xi|^2+|\ddot\xi|)\|\elec(\sigma,x)\|_{L^2(dx)}
\\
    &\qquad\qquad\qquad\qquad+\,g|t|\sup_{[0,t]}
    \Bigl(
  |\eta|(|\dddot\xi|+|\dot\xi||\ddot\xi|+|\dot\xi|^3)+|\dot\eta|
(|\ddot\xi|+|\dot\xi|^2)+|\ddot\eta|
  |\dot\xi|\Bigr)\,
\Bigr]\,.
  \end{align*}
\end{corollary}
\proof
This follows by referring to the formulae \eqref{defeeze1}, differentiating \eqref{defeff} and bounding the
resulting formulae thus:
\begin{align*}
  \|\partial_t\elec^{eff}\|&\leq\|\partial_t\elec\|+const.\,g\Bigl(
|\eta|(|\ddot\xi|+|\dot\xi|^2)+|\dot\eta||\dot\xi|\Bigr)\\
  \|\partial_t^2\elec^{eff}\|&\leq\|\partial_t^2\elec\|+const.\,g\Bigl(
  |\eta|(|\dddot\xi|+|\dot\xi||\ddot\xi|+|\dot\xi|^3)+|\dot\eta|
(|\ddot\xi|+|\dot\xi|^2)+|\ddot\eta|
  |\dot\xi|\Bigr)\,.
\end{align*}
(These hold in \(L^2(dx)\), but also in weighted \(L^2\) or Schwartz norms with corresponding assumptions
on \(\elec\) of course.)
The assertion made in the corollary is then a consequence of the
Cauchy-Schwarz inequality.
\qed


\subsubsection{Computations of error terms for the proof of Theorem \ref{main}.}
\la{repcalc}

\paragraph{Derivation of \eqref{tottdi}}
In this paragraph we work in the standard Schr\"odinger representation of the fields on \(L^2(\mu_0)\) with notation
as in Theorem \ref{sadj2} and \S\ref{dissint}.
We start with the final conclusion in Theorem \ref{extexist2}:
\begin{equation}\la{klast}\frac{d}{ds}{{\bevot}}(t,s)\Psi=+i{{\bevot}}(t,s)\bigl(\doublecolon {\pmb H}^{sol,{\elec}}_0(s)\doublecolon
    +\doublecolon H^{sol}_{I,g,{\beefb}}(\phisch)\doublecolon\bigr)\Psi\,,\quad\hbox{
    for each }\;\Psi\in\Dom(\doublecolon {\pmb H}^{vac}_0\doublecolon)\bigcap
    \Dom(\doublecolon \tilde H^{sol}_{I,g,{\beefb}}(\phisch)\doublecolon)\,,
    \end{equation}
    and consider the effect of the Weyl operators \(\mbU,\mbV\) from which
    \(\{s\mapsto\Delta(s)\}\) is built. The interaction term is a polynomial in
    the field, so by Appendix \ref{arh}
\[\mbU(f)\,\doublecolon H^{sol}_{I,g,{\beefb}}(\phisch)\doublecolon=
\doublecolon H^{sol}_{I,g,{\beefb}}(\phisch+f)\doublecolon\,\mbU(f)\,.\]
Combining the corresponding formula for
\(v(\phisch)=-3m^2\doublecolon\,\int\ssmx \phisch(x)^2\,dx\doublecolon
=\doublecolon {\pmb H}^{sol}_{0}\doublecolon-\doublecolon {\pmb H}^{vac}_0\doublecolon\) with \eqref{araki} we deduce
the  fixed time commutation relations (valid for real Schwartz functions \(f\))
\begin{align}\label{arakisol}\begin{split}
  &[\doublecolon {\pmb H}^{sol}_0\doublecolon\,,\,\mbU(f)]=\mbU(f)\Bigl(
  -\phisch(K(0)f)+\frac{1}{2}\langle f,K(0)f\rangle\Bigr)\\
    &[\doublecolon {\pmb H}^{sol}_0\doublecolon\,,\,\mbV(h)]=\mbV(h)\,\Bigl(
\pisch(h)+\frac{1}{2}\langle h,h\rangle\Bigr)\,.\end{split}
\end{align}
Next, putting \(f=\delta_\xi\Phi_S\) and \(h=-g\frac{\eta}{\sqrt{\scm_{cl}}} {\tteox}\) as in \eqref{defdis},
we obtain (for appropriate \(F\), specified below)
\begin{align}
\frac{d}{ds}\tilde\bevot(t,s)F\,&=\,
\tilde\bevot(t,s)\Bigl[i \doublecolon {\pmb H}^{sol}_0\doublecolon\,
  -i\bigl\langle\phisch-\delta_\xi\Phi_S,\lambda{\elec}(s)+g{\scm_{cl}}^{-1/2}({\dot\eta} {\tteox}
-\dot\xi\eta {{\tteox}'})
  \bigr\rangle-ig{\scm_{cl}}^{-1/2}({\eta}-\frac{{\scm_{cl}}}{g^2}{\dot\xi})\pisch( {\tteox})+i\frac{g^2\eta^2}{2\scm_{cl}}
\notag\\
&\qquad\qquad\qquad
+\doublecolon\,\int\,\frac{{\beefb}(x)}{3!}\cU^{(iii)}(\Phi_S)(\phisch-\delta_\xi\Phi_S)^3
+\frac{{\beefb}(x)}{4!}\cU^{(iv)}(\Phi_S)(\phisch-\delta_\xi\Phi_S)^4\,dx\,\doublecolon\label{dds}\\\notag
&\qquad\qquad\qquad\qquad-\phisch(K(0)\delta_\xi\Phi_S)+\frac{1}{2}(\delta_\xi\Phi_S,K(0)\delta_\xi\Phi_S)_{L^2}
\Bigr]\,F\,.
\end{align}
This holds for  \(F\in\Dom(\doublecolon {\pmb H}^{sol,{\elec}}_{g,{\beefb}}\doublecolon)\); recall that
\(\Dom(\doublecolon {\pmb H}^{sol,{\elec}}_{g,{\beefb}}(s)\doublecolon)\) is independent of \(s\), see Theorem \ref{fixt}.
In order to put the right side in a more tractable form we make use of the identity in \S\ref{dissint} to
rewrite the terms above which involve the displaced field
\(\phisch-\delta_\xi\Phi_S\). As in (iv) of Theorem \ref{sadj2} we write \(\doublecolon\pmb{H}^{sol}_{0\xi}\doublecolon\) for the quadratic Hamiltonian obtained by
expanding around \(\Phi_S(\cdot-\xi)\), and generally add \(\xi\) as a suffix to indicate this; corresponding formulae for the interaction Hamiltonian are below in \eqref{intixi}.
The second line above is just the spatially cut-off interaction evaluated on the shifted field, i.e., \(\doublecolon H^{sol}_{I,g,0,{\beefb}}(\phisch-\delta_\xi\Phi_S)\doublecolon\). The identity \eqref{trif}
   allows this to be combined with the third line to give the Hamiltonian evaluated on the unshifted field, but now defined with respect to a soliton located at \(\xi\), i.e.,
   \(\doublecolon H^{sol}_{I,g,\xi,{\beefb}}(\phisch)\doublecolon\), although at the expense of an infrared error
equal to:
\begin{equation}\label{irequal}
Err^0_{\mbox{\tiny IR}}(\phisch,\xi)=\doublecolon\int\,(1-b)\Bigl(\frac{1}{3!}\cU^{(iii)}(\Phi_{S\xi})\phisch^3+\frac{1}{4!}\cU^{(iv)}(\Phi_{S\xi})\phisch^4
-\frac{1}{3!}\cU^{(iii)}(\Phi_S)(\phisch-\delta_\xi\Phi_S)^3
-\frac{1}{4!}\cU^{(iv)}(\Phi_S)(\phisch-\delta_\xi\Phi_S)^4\,
\Bigr)\,dx\,\doublecolon\ ,.
\end{equation}
With this, the formula \eqref{dds} is equivalent to
\begin{align}\label{ddsni}\begin{split}
\frac{d}{ds}\tilde\bevot(t,s)F\,=\,
i\tilde\bevot(t,s)\Bigl[ \doublecolon \pmb{H}^{sol}_{0\xi}\doublecolon\,
  -\langle\phisch-\delta_\xi\Phi_S,\lambda{\elec}^{eff}(s)&\rangle
+\doublecolon H^{sol}_{I,g,\xi,{\beefb}}(\phisch)\doublecolon
+Err^0_{\mbox{\tiny IR}}(\phisch,\xi)+i\frac{g^2\eta^2}{2\scm_{cl}}
\\
&-\frac{g}{\sqrt{\scm_{cl}}}{\dot\eta}\langle\phisch-\delta_\xi\Phi_S, {\tteox}\rangle
 +\Bigl({\dot\xi}-\frac{g^2\eta}{\scm_{cl}}\Bigr)\frac{\sqrt{\scm_{cl}}}{g}\pisch( {\tteox})
\Bigr]\,F\,.
  \end{split}
\end{align}
Observe that we have introduced the effective electric field
\({\elec}^{eff}={\elec}-g\lambda^{-1}{\scm_{cl}}^{-1/2}\dot\xi{\eta}{{\tteox}'}\), as displayed in \eqref{defeff},
which includes a term arising from the movement of the soliton. The second term on the right hand side of the first line determines the interaction with the electric field, and should be thought of as consisting of a part aligned with
the zero mode \(\tteox\), which acts on the soliton itself, while the remainder contributes to the action on the
transverse modes, see \eqref{defeeze1} - it is here that the additional term in the effective electric field is
relevant. The precise determination of the soliton dynamics is determined by the need to control the terms in the second line of \eqref{ddsni}, but this does involve additional effects worked out in the next paragraph. As previously we include the linear term in the Hamiltonian with the following notation:
\begin{equation}\label{includelec}
{\pmb H}^{sol,{\elec}^{eff}}_{0\xi}= \doublecolon {\pmb H}^{sol}_{0\xi}\doublecolon -\lambda\phisch({\elec}^{eff})\,.
\end{equation}
\paragraph{Derivation of \eqref{dds2}} We now work out the form of the right hand side of \eqref{tottdi} under the unitary transformation \eqref{wdw} and hence obtain \eqref{dds2}. Theorem \ref{chrep}, combined with a simple translation
of the soliton by \(\xi\in\R\) implies
\begin{equation}\label{useful}
(\bbS^\theta(\xi))^*\doublecolon {\pmb H}^{sol,{\elec}^{eff}}_{0\xi}\doublecolon\,\bbS^\theta(\xi)\,=\,\frac{P^2}{2\scm_{cl}}+{h}_{{\cav},{\fav}}+\Delta\M_{scl}
\,=\,{\rm H}_0^{eff}-V_2Q^2+\Delta\M_{scl}\,,
\end{equation}
which gives the transformation of the linear and quadratic parts of the Hamiltonian. (Recall from above though that the effective Hamiltonian
\({\rm H}_0^{eff}\) also includes the correction term \(V_2Q^2\) which accounts for quantum dispersion effects.)
To complete the derivation of 
\eqref{dds2} we need the interaction terms.
  The interaction Hamiltonian is made up precisely from the terms
involving third and fourth derivatives of \(\cU\), we deduce that
  under the unitary map \(\bbS^\theta(\xi)\), the right hand side of \eqref{tri} becomes
  \[
  \triplecolon H^{sol}_{0\xi}(\upphi,\uppi)\triplecolon+{\rm H}^{sol}_{I,g,\xi,{\beefb}}(Q,\upphiperp)
  =\frac{P^2}{2\scm_{cl}}+h_d+\h+ {\rm H}^{sol}_{I,g,\xi,\xi,{\beefb}}(Q,\upphiperp)
  \]
  (up to the infrared regularization error written in the
  previous item); here \({\rm H}^{sol}_{I,g,\xi,{\beefb}}\) is obtained
  in the same way as \eqref{inti}, except the soliton is centered at \(\xi\) and the
  representation \eqref{cfsgextxi} is used in place of
  \eqref{cfsgext}; note that there are thus two origins for the \(\xi\)-dependence, which it turns out
  to be necessary to keep track of for technical reasons, by means of the following:
  \begin{align}\label{intixi}\begin{split}
      &{\rm H}^{sol}_{I,g,\xi',\xi,{\beefb}} (Q,\upphiperp)\define\int\,
2mg{\beefb}(x)\tmxip\,\mcn_{I,\xi,\delta\gamma_\xi}^3(Y,\upphiperp)
+\frac{1}{2}g^2{\beefb}(x)\mcn_{I,\xi,\delta\gamma_\xi}^4(Y,\upphiperp)
\,dx\,,\\
      &{\rm H}^{sol}_{I,g,\xi,{\beefb}}(Q,\upphiperp)\,=\,{\rm H}^{sol}_{I,g,\xi,\xi,{\beefb}}(\upphi)
  \end{split}\end{align}
in which \(\xi'\) labels the location of the soliton and \(\xi\) the representation of the fields used.
This latter issue introduces an implicit \(\xi\)-dependence in the definition of the interaction
  polynomials from \eqref{iii} which are now defined as:
\begin{align}\label{iii2} \begin{split}
&\mcn_{I,\xi,\delta\gamma_\xi}^3(Y,\upphi)\,\define\,Y^3+3Y^2\upphi+3Y
\triplecolon\upphi^2\triplecolon+
\triplecolon\upphi^3\triplecolon+
3{\delta\gamma_\xi}\upphi+3Y{\delta\gamma_\xi}\,,\qquad\hbox{
and}\\
&\mcn_{I,{\xi},\delta\gamma_\xi}^4(Y,\upphi)\,\define\,
Y^4+4Y^3\upphi+6Y^2
\triplecolon\upphi^2\triplecolon+
4Y\triplecolon\upphi^3\triplecolon+
\triplecolon\upphi^4\triplecolon+
6Y^2{\delta\gamma_\xi}+12Y{\delta\gamma_\xi}\upphi+
6{\delta\gamma_\xi}\triplecolon\upphi^2\triplecolon
+3{\delta\gamma_\xi}^2\,.
\end{split}\end{align}
with \(\delta\gamma_\xi(x)=\delta\gamma(x-\xi)\) and 
  and  \(Y=-{\sqrt{\scm_{cl}}}{Q}
   {\tteox}\).
Now under our assumptions for \({\beefb}\), the coefficient of the cubic term \(\int 2mg{\beefb}(x)\tmxi  {\tteo}(x-\xi)^3dx\,(-{\sqrt{\scm_{cl}}}{Q})^3\)
is exponentially small in the infrared cut-off \(R\) (by parity), and is better included with the infrared error term
as will be indicated with an upright font as in the following definition:
\begin{equation}\label{deferrir}
\bbS^\theta(\xi)  \eirup\bigl(\bbS^\theta(\xi)\bigr)^*\,=\, Err^0_{\mbox{\tiny IR}}(\phisch,\xi)
  -g\,Q^3\,\int 2m\scm_{cl}^{3/2}{\beefb}(x)\tmxi  {\tteo}(x-\xi)^3dx\,,
  \end{equation}
and the (modified) interaction Hamiltonian is now (by definition, with \(\xi\)-dependence suppressed)
  \begin{equation}\label{intixit}{\hat{\rm H}}^{sol}_{I}(Q,\upphiperp)\,\define\,\int\,
2mgb(x)\tmxi\,\hat\mcn_{I,\xi,\delta\gamma_\xi}^3(Y,\upphiperp(x;\xi))
+\frac{1}{2}g^2b(x)\mcn_{I,\xi,\delta\gamma_\xi}^4(Y,\upphiperp(x;\xi))
\,dx\,,\end{equation}
with 
\(
\hat\mcn_{I,\xi,\delta\gamma_\xi}^3(Y,\upphiperp)
=\mcn_{I,\xi,\delta\gamma_\xi}^3(Y,\upphiperp)-Y^3\),
i.e. compared with \eqref{iii} the \(Q^3\) piece is transferred to the infra-red error term, which can be
written
\begin{equation}\label{ffir}
\eirup(\upphi)={\rm H}^{sol}_{I,g,\xi,\xi,1-b}(\upphi)-{\rm H}^{sol}_{I,g,0,\xi,1-b}(\upphi-\delta_\xi\Phi_S)
-g\,Q^3\,\int 2m\scm_{cl}^{3/2}b(x)\tmxi  {\tteo}(x-\xi)^3dx\,.
\end{equation}
The
dependence on \(g,{\beefb},\xi\) of the interaction Hamiltonian etc will be left implied in what follows where possible.
(Note that in \eqref{ffir} the second \(\xi\) index is not zero as this is determined by the representation via
\(\bbS^\theta(\xi)\).)
The ``c-number'' contribution to the term in brackets in \eqref{dds} works out to be as in \eqref{pf1}
and just combines with other such terms to give a phase factor.

Defining \({{V_{-1}}}\) and \({V_1}\) by
  \(\dot\xi=g^2{\scm_{cl}}^{-1}\eta+{{V_{-1}}}\) and \(\dot\eta=-\frac{1}{g}\sqrt{\M_{cl}}\lambda({\elec}, {\tteox})_{L^2}+{V_1}\)
  the remaining terms can be written \[-i\lambda\phisch({\elec}^\perp(s))
  -ig{\scm_{cl}}^{-1/2}({V_1}\phisch( {\tteox})-\dot\xi{\eta}\phisch({{\tteox}'})
+\frac{i}{g}{\scm_{cl}}^{1/2}{{V_{-1}}}\pisch( {\tteox})
  \,,\]
  where \({\elec}^{\perp}={\elec}-({\elec}, {\tteox})_{L^2} {\tteox}\). Under the change of representation \(\bbS^\theta(\xi)\) this
  expression becomes
  \begin{equation}\label{extra}
-i\frac{1}{g}{{V_{-1}}}P+igQ{V_1}-i\lambda\upphi^\perp({\elec}^\perp-g\lambda^{-1}{\scm_{cl}}^{-1/2}\dot\xi{\eta}{{\tteox}'})
  \,,    \end{equation}
  as an operator on \(\mfrH\). 
 The quantity against which the field \(\upphi^\perp\) is paired in the last term of \eqref{extra} defines the
 effective electric field \({\elec}^{eff}\) defined above, which is now seen to be the field
 sensed by the bosons and the discrete mode, and appears
in the effective Hamiltonian \({\rm H}_0^{eff}\).

  \paragraph{Errors induced by time dependence - derivation of \eqref{tottd}} We have calculated the generator
  of the evolution using the representation \eqref{cfsgextxi} by applying the transformation at time \(s\)
  \begin{equation}\label{tats}\bbS_s\define \bbS^\theta(\xi(s))
  \end{equation}
  but have so far not computed the effects of this transformation
  itself being dependent on time - this produces additional error terms to be computed and controlled. This
  control is achieved by averaging, but with a subtlety in that
  the final two terms in \eqref{dds2} need to be balanced against (averages of)
  terms of the same form which arise in the time derivative of the operator \(\bbS_s\), in order to
  take advantage of the averaging effects which lie behind the adiabatic approximation. It will become clear in
  the following calculations that this requirement is behind the choices to be made for \({V_{-1}}\) and \(V_1\)
  in \eqref{defv1xi} and \eqref{defv1eta}.
  The contribution from  the time derivative of \(\bbS_s\) can be read off from \eqref{defrnfinal15}-\eqref{defrnfinal225} and
  \eqref{diffvac}-\eqref{imuseful}, giving eventually:
\begin{align}\begin{split}
\frac{d}{ds}\Bigl(\bbS_s
\chi(s,Q)\Omega_{\cee(s),\eff(s)}\Bigr)
\,=\,\dot\xi(s)\bbS_s\,\Bigl[&
  \frac{i{P}}{\sqrt{\scm_{cl}}}\upphiperp(\tteox')\,
  -i\sqrt{\scm_{cl}}{Q}  \Bigl(  \cee_2\sqrt{2\omega_d}(\tteox,\partial_\xi {\tteonex})
    +\bigl(\tteox,\partial_\xi\mcF_{u_\xi}^{-1}\sqrt{2\omega_\bullet}\eff_2\bigr)
    \Bigr)
    \\&\quad+i\cee_2\sqrt{2\omega_d}(\upphiperp-\upphi_{scl})\bigl(\partial_\xi {\tteonex}\bigr)
+i(\upphiperp-\upphi_{scl})\bigl(\partial_\xi\mcF_{u_\xi}^{-1}(\sqrt{2\omega_\bullet}\eff_2)(\cdot;\xi)\bigr)
\\&\qquad\quad +i\cee_2\sqrt{2\omega_d}(\upphi_{scl},\partial_\xi {\tteonex})
+i(\upphi_{scl},\partial_\xi\mcF_{u_\xi}^{-1}(\sqrt{2\omega_\bullet}\eff_2)(\cdot;\xi))
\\&\qquad\qquad+\,
  {\sqrt{\scm_{cl}}}\,Q\,(\upphiperp-\upphi_{scl})\bigl(K^\theta(\xi)^{\frac{1}{2}}{{\tteox}'}\bigr)\,
+\,\bigl(\upphiperp-\upphi_{scl}\bigr)\bigl(K^\theta(\xi)^{\frac{1}{2}}\partial_{\xi}\upphi_{scl}\bigr)
  \\
&\qquad\qquad\qquad-\,\frac{1}{2}\,\bigl(\upphiperp-\upphi_{scl},\partial_\xi K^\theta(\xi)^{\frac{1}{2}}(\upphiperp-\upphi_{scl})\bigr)\Bigr]\,\chi(s,Q)
\,\Omega_{\cee(s),\eff(s)}
\\&
\qquad\qquad\qquad\qquad+i\,\bbS_s\bigl((\dot\Theta_3-h_{\mbox{\tiny 1P}}-h_{\cav,\fav})
\,\chi(s,Q)
\,\Omega_{\cee(s),\eff(s)}\bigr)\,.
\end{split}
\label{who}
\end{align}
where, as usual, \({\sqrt{\scm_{cl}}}
Q=-\upphi( {\tteo})\) and \({P}=-{\sqrt{\scm_{cl}}}\uppi( {\tteo})\) determine the
position and momentum operators for the soliton on \(\mfrH(\theta)\).
The first term of \eqref{defrnfinal225} gives the first term on the right of \eqref{who}, while the second term of
\eqref{defrnfinal225} appears in the fourth line of \eqref{who}.
In deriving the remaining terms we use the formulae \eqref{evo} and  \eqref{diffvacconj}. In particular
the second and third terms on the right of \eqref{diffvacconj}, in the form
\eqref{diffvacconj2}, appear in the first and second lines
of \eqref{who}. (Note also that the third line is the (transverse) expectation of the second line in the state
\(\Omega_{c,f}\),  separated for reasons which will become clear.)
The first term on the right of \eqref{diffvacconj} expands to give
\(-\sqrt{\scm_{cl}}\mbQ\,\upphi_{scl}\bigl(K(\xi)^{\frac{1}{2}}{{\tteox}'}\bigr)
\) and \(\upphiperp\bigl(\partial_\xi(K(\xi)^{\frac{1}{2}}\upphi_{scl})\bigr)\). The first of these appears in
the fourth line of \eqref{who}; as regards the second,
observe that
\[
\upphiperp\bigl(\partial_\xi (K^\theta(\xi)^{\frac{1}{2}}\upphi_{scl})\bigr)=
\upphiperp\bigl(K^\theta(\xi)^{\frac{1}{2}}\partial_{\xi}\upphi_{scl}\bigr)+
\upphiperp\bigl(\partial_\xi K^\theta(\xi)^{\frac{1}{2}}\upphi_{scl}\bigr)
\]
then combine with the final term of \eqref{defrnfinal225}, complete the square and finally,
to get the form given in the 4th and 5th lines, add additional ``c-number'' terms which
in combination vanish identically:
\[
\Bigl(\frac{1}{2}\bigl(\upphi_{scl},\partial_\xi K^\theta(\xi)^{\frac{1}{2}}\upphi_{scl}\bigr)
+\upphi_{scl}\bigl(K^\theta(\xi)^{\frac{1}{2}}\partial_{\xi}\upphi_{scl}\bigr)\Bigr)\,=\,
\frac{1}{2}\partial_\xi\bigl(\upphi_{scl},K^\theta(\xi)^{\frac{1}{2}}\upphi_{scl}\bigr)
\,=\,\frac{1}{2}\partial_\xi\bigl(2c_1^2+2\int|f_1(k)|^2dk\bigr)\,=\,0\,.
\]
%

Now \eqref{who} will be combined with \eqref{dds2} to obtain \eqref{tottd}, with an appropriate choice of
the modulation equations \eqref{newtpert}
to allow estimation of the effects of the remaining error terms, via an averaging type argument generalizing that
in \cite{katoad}. 
%
This will be based on the expansions \eqref{asex} which allow us to separate the mean, and the
fluctuation around the mean, of \(\upphi\)
as will now be explained, initially focusing on the first line of \eqref{who}.
We use the fact that (using the representation \eqref{cfsgextxi})
if \((g, {\tteox})_{L^2}=0\) then since
\[\Bigl(\upphi(g)\id\otimes\Omega', 
\id\otimes\Omega'\Bigr)_{{\mathfrak H}(\theta)}=0\,,\] unitarity of the displacement operators
implies that, with the definition \(\Omega_{\cee,\eff}=\D_{\cee,\eff}\Omega'\),
\[\Bigl(\D_{\cee,\eff}\circ\upphi(g)\circ \D_{\cee,\eff}^*\,\id\otimes\Omega_{\cee,\eff},
\id\otimes\Omega_{\cee,\eff}\Bigr)_{{\mathfrak H}(\theta)}=0\,.\]
But the definition of the displacement operators gives
\(
 \D_{\cee,\eff}\circ\upphi\circ \D_{\cee,\eff}^*=\upphi-
  \upphi_{scl}
  \)
  where \(\upphi_{scl}\) is as in \eqref{defiscl}.
 %
  Using \( {\tteox}'=-\partial_\xi {\tteox}\), the mean (with respect to transverse fluctuations) of
  the first term on the right of  \eqref{who} is \(
{iP}{\scm_{cl}}^{-1/2}(\upphi_{scl},{\tteox}').
  \)
  This is to be cancelled with the corresponding term in \eqref{dds2}, by defining \({{V_{-1}}}\) as in 
\eqref{defv1xi}
and imposing the modulation equation \(\dot\xi=g^2{\scm_{cl}}^{-1}\eta+gV_{-1}\), as in \eqref{newtpert}.
This leaves the transverse fluctuation term
\begin{equation}
    i{\scm_{cl}}^{-1/2}\,\bbS^\theta(\xi(s))
    \Bigl[\dot\xi\,\langle\upphi-\upphi_{scl}\,,\,{\tteox}'\rangle{P\chi}\Omega_{\cee,\eff}
    \Bigr]\,;
  \end{equation}
  it is a function of \(Q\) taking values in
  \(\Omega_{\cee,\eff}^\perp\), the subspace of
  \(\mfrF\) orthogonal to \(\Omega_{\cee,\eff}\); this will be estimated in a fashion which generalizes the argument in
  \cite{katoad}.
  Similarly the remaining terms in the first line are to be balanced with the corresponding term in \eqref{dds2},
  by the definition \eqref{defv1eta},
  and imposing as modulation equation \(\dot\eta=-\frac{1}{g}\sqrt{\scm_{cl}}\lambda({\elec},{\tteox})_{L^2}+
  \frac{1}{g}V_{1}(\xi,\dot\xi,c,f)\), see \eqref{newtpert}. The term \(\lambda\phisch(\E)\)
  in the Hamiltonian involving the electric field distributed between \(V_{1}\) in \eqref{newtpert}
  and \eqref{defeeze1}, so all together this leaves only (i) terms for transverse oscillations, and (ii) certain ``c-number'' terms which only affect the phase, as will now be discussed in turn.
  
   
  The transverse oscillations arising from the time dependence of the unitary map \eqref{tats}
  are generated by the operator
  \begin{align}\label{errtd}\begin{split}
\eirtd&\define -i\dot\xi\Bigl[
  \scm_{cl}^{1/2}Q\bigl\langle\upphiperp-\upphi_{scl},K^\theta(\xi)^{\frac{1}{2}}{{\tteox}'}\bigr\rangle
+\bigl\langle\upphiperp-\upphi_{scl},K^\theta(\xi)^{\frac{1}{2}}\partial_{\xi}\upphi_{scl}
-
    ic_2\sqrt{2\omega_d}\tteonex'
+i\partial_\xi\mcF_{u_\xi}^{-1}(\sqrt{2\omega_\bullet}f_2)\bigr\rangle
\\&\qquad\qquad\qquad\qquad
+  i\scm_{cl}^{-1/2}\bigl\langle\upphiperp-\upphi_{scl}, {\tteox}'\bigr\rangle\,P-\frac{1}{2}
\triplecolon \bigl(\upphiperp-\upphi_{scl},\partial_\xi K^\theta(\xi)^{\frac{1}{2}}(\upphiperp
-\upphi_{scl})\bigr)\triplecolon
\Bigr]\,,\end{split}
  \end{align}
  which takes on the slightly cleaner form given in \eqref{cleaner} after conjugation by \(\D_{\cee,\eff}\).
  Applied to \(\psi(Q)\,\Omega_{\cee,\eff}\) this operator generates an \(\Omega_{\cee,\eff}^\perp\)-valued contribution to the wave function, whose control in \S\ref{controlf} is a central part of the proof.
  \begin{remark}
The ``c-number'' terms arise from the third line of \eqref{who}, which is pure imaginary and
contributes \(\Theta_4\) to the phase of the solution, according to
  \begin{equation}\label{phasedet3}
  i\dot\Theta_4\,=\,   ic_2\sqrt{2\omega_d}\bigl(\upphi_{scl},\partial_\xi {\tteonex}\bigr)
+i\bigl(\upphi_{scl},\partial_\xi\mcF_{u_\xi}^{-1}(\sqrt{2\omega_\bullet}f_2)(\cdot;\xi)\bigr)
  \,,
  \end{equation}
  and as with other phases we take \(\Theta_4(0)=0\);
  this can be cancelled together with the other \(\Theta_j\)
  by appropriate choice of \(\Theta_0\) in \eqref{defdis}, see \eqref{tottd}-\eqref{opp}.
  (In addition there is the field independent normal-ordering term
\(\frac{\dot\xi}{2}\bigl((C^{\theta,\perp}(\xi))^{\frac{1}{2}},\partial_\xi K^\theta(\xi)^{\frac{1}{2}}\bigr)\,,\) but this is
actually zero:
the inner product is the \(L^2(dxdy)\) inner product on the kernels of the corresponding operators,
which can be checked to be zero by parity using the formulae in \S\ref{eigen}.)
    \end{remark}
  We also record that the product rule extends the formula \eqref{tottd} applied to more general functions: let \(\psi=\psi(s,Q)\) be a smooth
  function, and \(p\) be a (generally time-dependent) polynomial {\em in the transverse field} \(\upphiperp\). Then
  \begin{align}\begin{split}\label{tottdgen}
    \frac{d}{ds}\Bigl[\tilde\bevot(t,s)\bbS_s\,\psi(s,\cdot)\D_{\cee(s),\eff(s)}\,p(s,\upphi)\Omega'\Bigr]
    \,=\,i\tilde\bevot(t,s)\bbS_s\,&\Bigl[\Bigl(\,
h_{\cav,\fav}+\frac{P^2}{2\scm_{cl}}
      +{\hat{\rm H}}^{sol}_{I}+
\eirup
+\eirtd\,\Bigr)\psi(s,Q)\D_{\cee(s),\eff(s)}\,p(s,\upphi)\Omega'\\
&\qquad-i\partial_s\psi(s,Q)
\D_{\cee(s),\eff(s)}\,p(s,\upphi)\Omega'\\
&\qquad\qquad-\psi(s,Q)\,p(s,\upphi-\upphi_{scl})\bigl(e_{c,f}-\delta h)\D_{\cee(s),\eff(s)}\Omega'\,
\Bigr]\\
&\qquad\qquad\qquad+
    \tilde\bevot(t,s)\Bigl(\frac{d}{ds}\Bigr)p(s,\phisch-\upphi_{scl})\bbS_s\psi(s,Q)\D_{\cee(s),\eff(s)}\,\Omega'
\,,
\end{split}  \end{align}
  where \(\delta h=h_{\cee,\eff}-{h}_{{\cav},{\fav}}\), and we used
  \(\frac{d}{ds}|_{\xi}\D_{\cee,\eff}\,\Omega'=(i\dot\Theta_3-ih_{\cav,\fav})\D_{\cee,\eff}\,\Omega'=i(\dot\Theta_3
  -e_{\cee,\eff}+\delta h)\D_{\cee,\eff}\Omega'\) since \((h_{\cee,\eff}-e_{\cee,\eff})\D_{\cee,\eff}\Omega'=0\), and we again used the
  choice of \(\Theta_0\) given with \eqref{tottd} to remove all the phases \(\Theta_j,\,j=1,\dots 4\). Observe that when
  \(p\) is a constant polynomial and \(\psi=\chi\),
the solution of \(i\partial_t\chi(t,\cdot)=h_{\mbox{\tiny 1P}}\chi(t,\cdot)\),
there is a cancellation between the \({\rm H}_0^{eff}-h_{\mbox{\tiny 1P}}\) and
  \((e_{\cee,\eff}-\delta h)\) terms, leading to \eqref{tottd}. The use of this formula is to derive the following
  integration by parts formula, in which we consider states which, at each time \(s\), are of the form
\(\chi(s,Q)\D_{\cee,\eff}F(s)\), with \((F(s),\Omega')=0\),  so that \(\D_{\cee,\eff}F\) is orthogonal to the
kernel of \(h_{\cee,\eff}-e_{\cee,\eff}\) and so can be written \((h_{\cee,\eff}-e_{\cee,\eff})\D_{\cee,\eff}G\) for some
\(G\). In what follows we will take  
\(G(t)=p(t,\upphi)\Omega'\) for  linear or quadratic Wick polynomial \(p\) as in \eqref{xi2p},\eqref{xi1p} and \eqref{xi0p} - these are
all seen to be polynomial in the transverse field \(\upphi^\perp\) since the test functions the field is paired with are all orthogonal
to \(\tteox\). The significance of this is, as noted following \eqref{defrnfinal17}, that \(p\) can be interchanged with the operator \(\eirtd\), and
thence moved back though
\(\bbS_s\psi(s,Q)\D_{\cee(s),\eff(s)}\) to act on \(\mfrH(\theta)\), which makes for cleaner estimates.

The next lemma employs the preceding identity to integrate by parts.
\begin{lemma}\label{5g0} Assume \(\psi(s,Q)\) is smooth and
  \(s\mapsto p(s,\upphi)\Omega'\in\Fin(L^2)\subset \mfrF\) is a \(C^1\)
  curve of finite particle vectors in the transverse Fock space determined by \(p\), and 
  \[\D_{\cee,\eff}F=
(h_{\cee,\eff}-e_{\cee,\eff})p(t,\upphi-\upphi_{scl})\D_{\cee,\eff}\Omega'
=(h_{\cee,\eff}-e_{\cee,\eff})\D_{\cee,\eff}p(t,\upphi)\Omega'  \]
at each time \(s\).  Then
  \begin{align}\label{5guys0}\begin{split}
    &
    \int_0^t\tilde\bevot(t,s)\bbS_s\psi(s,Q)\D_{\cee,\eff}F(s)\,ds=
    -i\bbS_t\psi(t,Q)\D_{\cee(t),\eff(t)}p(t,\upphi)\Omega'+i\tilde\bevot(t,0)\bbS_0\psi(0,Q)\D_{\cee(0),\eff(0)}
    p(0,\upphi)\Omega'\\
    &\quad+\,\int_0^t\tilde\bevot(t,s)\Bigl[i\Bigl(\frac{d}{ds}p(s,\phisch-\upphi_{scl})\Bigr)\bbS_s\bigl(\psi(s,Q)
    \D_{\cee,\eff}\Omega'\bigr)
  - \bbS_s\bigl(({\hat{\rm H}}^{sol}_{I}-V_2Q^2+\eirup+\eirtd-\delta h\,)\psi(s,Q)\D_{\cee,\eff}p(s,\upphi)\Omega' \bigr)
  \,\Bigr]\,ds\\
    &\qquad+\,\int_0^t\tilde\bevot(t,s)\bbS_s\Bigl(
-\psi
p(s,\upphi-\upphi_{scl})\delta h\,\D_{\cee,\eff}\Omega'+
    \bigl(i\partial_s\psi-h_{\mbox{\tiny 1P}}\psi\bigr)\D_{\cee,\eff}p(s,\upphi)\Omega'\Bigr)
  \,ds\,.
  \end{split}\end{align}
As above \(\delta h=h_{\cee,\eff}-{h}_{{\cav},{\fav}}\) at each time \(s\).
\end{lemma}
\proof
Write \(G(s)=p(s,\upphi)\Omega'\) and \(\tilde G(s)=\psi(s,Q)\D_{\cee,\eff}G(s)\).
Recalling \eqref{evo} and \eqref{tottd},
\[\frac{d}{ds}\tilde\bevot(t,s)\bbS(s)\,F=i\tilde\bevot(t,s)\bbS(s)\,{\rm H}^{tot}\,F\,,\qquad
\hbox{with}\quad
{\rm H}^{tot}\,\define\,h_{\cav,\fav}+\frac{P^2}{2\scm_{cl}}+{\hat{\rm H}}^{sol}_{I}+
\eirup
+\eirtd
\]
we use the occurence of the Hamiltonian in the left hand side of the following formula to
integrate by parts in time to get
\begin{align}\label{katoadibp}
  \int_0^t\tilde\bevot(t,s)\bbS_s\,({\rm H}^{tot}-e_{\cee,\eff}) \tilde G(s)\,ds&=-i\bbS_t\tilde G(t)
  +i\tilde\bevot(t,0)\bbS_0\tilde G(0)\\
  &\quad+\,\int_0^t\tilde\bevot(t,s)\Bigr[\bbS_s\Bigl(i\partial_s{\psi}\D_{\cee(s),\eff(s)}\,p(\upphi)\Omega'-\psi
p(\upphi-\upphi_{scl})\delta h\,\D_{\cee(s),\eff(s)}\Omega'
\Bigr)\notag\\
&\qquad\phantom{+\,\int_0^t\tilde\bevot(t,s)}
+i\Bigl(\frac{d}{ds}p(\phisch-\upphi_{scl})\Bigr)\bbS_s\psi(s,Q)\D_{\cee(s),\eff(s)}\,\Omega'
\,\Bigr]\,ds\,.\notag
\end{align}
By the formula for \({\rm H}^{tot}\)
we can write (for smooth \(\psi\))
\[
\psi(s,Q)\D_{\cee,\eff}F=
({\rm H}^{tot}-e_{\cee,\eff}) \tilde G-
\frac{P^2}{2\scm_{cl}}\tilde G
- {\hat{\rm H}}^{sol}_{I}\tilde G-\eirup\tilde G-\eirtd{\tilde G}+\delta h{\tilde G}
\,,
\]
and hence the lemma follows from \eqref{tottdgen}.
\qed

\subsubsection{Control of the error terms}\label{controlf}
  Applied to a state of the form \(\psi(Q)\D_{\cee,\eff}\Omega'\in L^2(dQ)\otimes\mfrF\) we have the formula
\begin{equation}\label{errtd15}
i\eirtd\psi\,\D_{\cee,\eff}\Omega'=\dot\xi\,
\D_{\cee,\eff}\Bigl(\psi\Xi^0 \Omega'+\,(Q\psi)\Xi^1 \Omega'+iP\psi\Xi^2 \Omega'\Bigr)\end{equation}
with
\(\;\Xi^0,\Xi^1,\Xi^2\) the Wick polynomial operators on \(\mfrF\) defined as in \eqref{errtd2}.
In particular the \(\Xi^j\Omega\in\Omega^\perp\), and so
\begin{equation}\label{errtd3}
\eirtd\psi\,\Omega_{\cee,\eff}=\dot\xi\,\psi\Omega_{\cee,\eff}^0+\dot\xi\,Q\psi\Omega_{\cee,\eff}^1
+i\dot\xi\, P\psi\Omega_{\cee,\eff}^2
\end{equation}
defines \(\Omega_{\cee,\eff}^0,\Omega_{\cee,\eff}^1,\Omega_{\cee,\eff}^2\;\hbox{in}\;\Omega_{\cee,\eff}^\perp\).
  \paragraph{Explicit formulae for the \(\Xi^a\).} Referring to the formula
  \begin{align}\label{phitrans}\upphi^\perp(x)
\,     =\,
  \frac{1}{\sqrt{2\omega^{}_d}}
(a^{}_{d}+a_{d}^\dagger) {\tteonex}\bigl({}x{}\bigr)      
\;+\,\int\,\frac{1}{\sqrt{4\pi\omega^{}_k}}\,
\bigl(a^{}_{k}e_{k\xi}(x{})+a_{k}^\dagger {{e_{-k\xi}}}(x{})\bigr)\,dk\,,
  \end{align}
  it follows that the \(\Xi^a\) are given as follows:
\begin{itemize}
\item For \(a=2\)
  \begin{equation}\label{xi2}
  \Xi^2(\upphi)
  =\beta_{2d}\,a_{d}^\dagger+
\int\bigl(\beta_2^0(k,\xi)\, a_{k}^\dagger +\beta_2^1(k,\xi)\,a_{k}^{}\bigr)\,dk\,\,,
\end{equation}
with \(\beta_{2d}=({2\omega^{}_d})^{-1/2}( {\tteone}, {\tteo}')\) and
\[
\beta^0_2(k,\xi)=\,\frac{1}{\sqrt{4\pi\omega^{}_k}}\,
\int\,{{e_{-k\xi}}}(x{}){{\tteox}'}(x)dx
\]
and analogously for \(\beta^1_2\).
The result of the \(x\)-integration \(\beta^0_2(k,\xi)=e^{-ik\xi}{\tilde\beta}^0_2(k)\) is \(e^{-ik\xi}\) times a Schwartz function of \(k\), while \(\beta_{2d}\) is independent of \(\xi\). The corresponding polynomial \(p_2\) is 
  \begin{equation}\label{xi2p}
  p_2(\upphi)
  =\dot\xi\upphi\bigl(C^{\perp,\frac{1}{2}}(\xi){\tteox}'\bigr)\,.
  \end{equation}
  Using the explicit covariance formulae in Appendix \ref{eigen} it follows that \(\xi\to C^{\perp,\frac{1}{2}}(\xi){\tteox}'\in\cs(\R)\) is smooth, which
  implies other useful results. In particular, smoothness as a function of \(\xi\) of the corresponding operator \((\id+\N)^{-\frac{1}{2}}p_2(\upphi)\)
  on \(\mfrF\) is implied.
\item Next for \(a=1\) we obtain similarly
  \begin{equation}\label{xi1}
  \Xi^1(\upphi)= \beta_{1d} (a_{d}^\dagger+a^{}_{d})+
\int\beta^0_1(k,\xi)\,dk\, a_{k}^\dagger\,+\int\beta^1_1(k,\xi)\,dk\, a_{k}^{}\,,
\end{equation}
with 
\(\beta_{1d}=\sqrt{2\omega_d\scm_{cl}}( {\tteox}', {\tteonex})=\sqrt{2\omega_d\scm_{cl}}( {\tteo}', {\tteone})\) independent of \(\xi\),
while
\[\beta^0_1(k,\xi)=\int \sqrt{\frac{\omega_k\scm_{cl}}{\pi}}e_{-k\xi}(x){\tteox}'(x)dx=e^{-ik\xi}{\tilde\beta}^0_1(k)\qquad
\hbox{with}\quad
       {\tilde\beta}^0_1(k)=\int \sqrt{\frac{\omega_k\scm_{cl}}{\pi}}e_{-k}(x){\tteo}'(x)dx\] a
       Schwartz function of \(k\). An analogous formula holds for \(\beta_1^1\). The corresponding polynomial \(p_1\) is 
  \begin{equation}\label{xi1p}
  p_1(\upphi)
  =\dot\xi\upphi\bigl({\tteox}'\bigr)\,.
\end{equation}
       As for \(p_1\) the formula implies smoothness properties as a function of \(\xi\).
\item         Finally consider \(\Xi^0\), which is made up of a part \(\Xi^{0,1}\) which is linear in the field
         while the part \(\Xi^{0,2}\) is quadratic  in the field . The first of these has an explicit form like the
         corresponding parts of \(\Xi^1\), but with \({{\tteox}'}\) replaced by
         \[R_0(x,\xi,c,f)=\P_{0\xi}^\perp\Bigl(K^\theta(\xi)^{\frac{1}{2}}\partial_{\xi}\upphi_{scl}
-
    ic_2\sqrt{2\omega_d}{\tteonex}'
+i\partial_\xi\mcF_{u_\xi}^{-1}(\sqrt{2\omega_\bullet}f_2)\Bigr)\,.
\]
The part
\(\Xi^{0,2}\)
can be understood by using the formula \eqref{k123} and substituting from \eqref{phitrans}, yielding 
\begin{align*}\Xi^{0,2}(\upphi)=-\frac{1}{2}\,\triplecolon \bigl(\upphiperp,\partial_\xi K^\theta(\xi)^{\frac{1}{2}}\upphiperp\bigr)\triplecolon
=
-\int\,dk\,{\beta}^0_{0d}&(k,\xi)(a_{d}^\dagger +a_{d})(a_{k}^\dagger
+a^{}_{-k})\\
&\quad-2^{-1}\iint\,dkdl\,{\beta_{0}^{0}}(k,l,\xi)\,\bigr(a_{k}^\dagger a_{l}^\dagger +a_{l}^\dagger a^{}_{-k}+a_{k}^\dagger a^{}_{-l}+a^{}_{-k}a^{}_{-l}\bigr)\,.
\end{align*}
Here referring to \eqref{k123}, and using the definition of \(\Lambda_\xi(x,y)\) in the second line of this equaiton, we have
\[{\beta}^0_{0d}(k,\xi)= -e^{-ik\xi}
\,\sqrt{\frac{\omega^{}_d}{2}}\,
\int\,{{e_{-k}}}(x{}){\tteone}'(x)dx\,dk+\frac{1}{\sqrt{8\pi\omega_d\omega_k}}\iint\Lambda_\xi(x,y)\tteonex(x)e_{-k\xi}(y)dxdy\,,
\]
and
\begin{align}{\beta_0^{0}}(l,n,\xi)= 
\,{\frac{1}{4\pi{\sqrt{\omega_n\omega_l}}}}\,
\iint\,{{{e_{-n\xi}}}(x{}){e_{-l\xi}}(y{})
\Lambda_\xi(x,y)}
   \, dxdy
\end{align}
and analogously for the remaining terms. All together this gives
\begin{equation}\label{xi0p}
p_0(\upphi)=\dot\xi\upphi(C^{\perp,\frac{1}{2}}(\xi)R_0)+\dot\xi\triplecolon\Bigl(\upphi,\B(\xi)\upphi\Bigr)\triplecolon
\end{equation}
where the kernel \[
\B(\xi)(x,y)=\int\,{\beta}^0_{0d}(k,\xi)\frac{\sqrt{8\pi\omega_d\omega_k}}{\omega_d+\omega_k}(e_{k\xi}(x)\tteone(y-\xi)+e_{k\xi}(y)\tteone(x-\xi))dk
+\iint\frac{4\pi\sqrt{\omega_l\omega_n}\beta_0^0(l,n,\xi)}{\omega_l+\omega_n}e_{l\xi}(x)e_{n\xi}(y)dldn\,.
\]
\end{itemize}
The explicit formulae and properties of \(\Lambda_\xi\) in Appendix \ref{eigen}
(smooth except for logarithmic diverergnce on the diagonal, and exponential decay
separately in \(x\) and \(y\)) imply that these integrals are defined. In particular
\(\sqrt{\omega_n\omega_l}\beta_0^{0}(l,n,\xi)\in L^2(dldn)\) and \(\sqrt{\omega_k}{\beta}^0_{0d}(k,\xi)\in L^2(dk)\)
with smooth dependence on \(\xi\). Together with the explicit expressions in Appendix \ref{eigen}, this in turn
implies that \(\xi\to\B(\xi)(x,y)\in L^2(dxdy)\) is smooth, and explicit expressions for derivatives can be written using formulae in that
appendix.

\begin{lemma}\label{best3} For \(M,l_1,l_2\in\Z_+\) let \(F\in\Ker(\N-M)\subset\mfrF\) be an \(M\)-particle vector, then
  \begin{align*}
&  \|\eirtd\psi(Q)\D_{\cee,\eff}F\|\leq const.(M) |\dot\xi|
\Bigl[(1+|c|+\|f\|_{L^2})^2\|\psi\|+ (1+|c|+\|f\|_{L^2})\bigl(\|Q\psi\|+\|P\psi\|\bigr)\Bigr]
  \|F\|\,,\\
    &\|{ \N}^{l_1} (h_d+\h)^{-l_2}\Xi^j(\upphi-\upphi_{scl})\,\Omega'\|\leq const.(l_1)\bigl(
    1+|c|+\|f\|_{L^2}\bigr) \quad \hbox{for \(j=1,2\,,\) while for \(j=0\)}\\
    &\|{ \N}^{l_1} (h_d+\h)^{-l_2}\Xi^0(\upphi-\upphi_{scl})\,\Omega'\|\leq const.(l_1)\bigl(
    1+|c|+\|f\|_{L^2}\bigr)^2\,.
  \end{align*}
\end{lemma}
\proof
Using the formulae above, the fact that \(\xi\mapsto\Lambda_\xi(x,y)\in L^2(dxdy)\) is continuous and
uniformly bounded, and Remarks \ref{lpart}-\ref{msa} and \eqref{best25} these follow once it is noted that the operator \(\Xi^0\) can increase the particle number \(\N\) by at most two, and \(\Xi^1,\Xi^2\) can increase it by at most one, see \eqref{errtd2}. (The case that
\(\upphi_{scl}\) is not present is included by just taking \(\cee,\eff\) to be zero.)
\qed

\paragraph{Bounds for the error terms in \eqref{revduh2}}
The following lemmas bound the three  error terms on the right of \eqref{revduh2}. We have already explained that the cubic (in \(Y=-\sqrt{\scm_{cl}}Q {\tteox}\))  term, which vanishes in the absence of the infrared cutoff, has been transferred to the
infrared error term. 
It is also necessary to treat separately
the term in \(\hat\mcn_{I,{\alpha}}^3\) which is quadratic in \(Q\) and linear in \(\upphiperp+\upphi_{scl}\). Precisely,
referring to \eqref{intixit} , \(V_2\) is defined as \(V_2=V_{2,1}+V_{2,2}+V_{2,3}\) where
\begin{equation}\label{defv2}
V_{2,1}=6mg\scm_{cl}\int b(x)\tmxi {\tteo}(x-\xi)^2\upphi_{scl}\,dx=\frac{1}{2}9m^2g\scm_{cl}
\int b(x)\tmxi{\sech{}^4 m(x-\xi)\,}\upphi_{scl}\,dx
\end{equation}
so as to annihilate the average, leaving
  \begin{equation}\label{annav}
{\hat{\rm H}}^{sol}_{I}(Q,\upphiperp+\upphi_{scl})-V_{2,1}\,Q^2
            =6mg\scm_{cl}\,Q^2\,\int b(x)\tmxi {\tteo}(x-\xi)^2\upphiperp\,dx+{\hat{\rm H}}^{sol,<2}_{I}
  \end{equation}
  where \({\hat{\rm H}}^{sol,<2}_{I}\) means the interaction Hamiltonian \eqref{intixit} but with only terms
  of order \(0\) or \(1\) in \(Q\) from \(\hat\mcn^3_{I,\delta\gamma}\) included. This is controlled by the next lemma,
  while the first term on the left will be controlled below in Lemma \ref{finest}. (The definition of
  \(V_{2,2}, V_{2,3}\) is below in Lemma \ref{esterrtd}, and the choice ensures that similar term quadratic in \(Q\) which arise in
  perturbation theory after integration by parts can be controlled.)
  \begin{lemma}\label{esthi}
     Let \(F\in\mfrF\) be an \(M\)-particle vector and \(\psi\in C^\infty(\R)\), then
    \begin{align}\label{esthieq}
        \|\,
\D_{\cee,\eff}
       {\hat{\rm H}}^{sol,<2}_{I}(\upphi+\upphi_{scl})
  \psi(s,Q)F
  \|\,&\leq\,
  const.(M)\,\Bigl[g\sum_{2\leq r+l\leq 3} \,\|{\beefb}\|_{L^2}(1+\|\upphi_{scl}\|_{L^\infty})^l\|Q^{3-r-l}\psi(s,Q)\|_{L^2}\\
    &\qquad\qquad\qquad\qquad +
  \,g^2\sum_{0\leq r+l\leq 4}\|{\beefb}\|_{L^2}(1+\|\upphi_{scl}\|_{L^\infty})^l\|Q^{4-r-l}\psi(s,Q)\|_{L^2}\,\Bigr]\,\|F\|
    \notag\end{align}
    while \({\hat{\rm H}}^{sol}_{I}\) obeys a bound identical except that the first sum also includes \(r=1,l=0\), and these cam be combined
    (non-optimally) as
    \begin{align}
    \label{esthieq2}
        \|\D_{\cee,\eff}
       {\hat{\rm H}}^{sol}_{I}(\upphi+\upphi_{scl})
  \psi(s,Q)F
  \|+\|V_2Q^2 \psi(s,Q)F\|\leq\,
  \,const.(M)\,\bigl(g\|{\beefb}\|_{L^2}+|\dot\xi^2|\bigr)\bigl(1&+\|\upphi_{scl}\|_{L^\infty}\bigr)^4\Bigl[\|\psi(s,Q)\|_{L^2}\\
    &+\|Q^2\psi(s,Q)\|_{L^2}
+
    \,g\|Q^{4}\psi(s,Q)\|_{L^2}\,\Bigr]\,\|F\|\,.
\notag
    \end{align}
    \end{lemma}
  \proof
  Referring to \eqref{iii},
  the interaction Hamiltonian
  contributes a finite linear combination of terms of the form
  \begin{align*}
    &g\int\,b_{3,r_1,r_2,,l}(x)Q^{3-r_1-2r_2-l}\delta\gamma_\xi(x)^{r_2}\triplecolon\upphiperp(x)^{r_1}\triplecolon\upphi_{scl}(x)^{l}\,dx\,
    \psi(s,Q)F\qquad(2\leq r_1+2r_2+l\leq 3)\qquad\hbox{and}
  \\
     &g^2\int\,b_{4,r_1,r_2,l}(x)Q^{4-r_1-2r_2-l}\delta\gamma_\xi(x)^{r_2}\triplecolon\upphiperp(x)^{r_1}\triplecolon\upphi_{scl}(x)^{l}\,dx\,  \psi(s,Q)F\qquad(r_1+2r_2+l\leq 4)\,,
  \end{align*}
where \(\delta\gamma_\xi=\delta\gamma(\cdot-\xi)\in\cs\) arises from the change of covariance, see \S\ref{intt}.
Here \(r_1,r_2,l\) are nonnegative integers restricted as shown, and the coefficient functions
\(b_{\bullet,r_1,r_2,l}\) arise by multiplying the infra-red cut off function \({\beefb}\) by the hyperbolic functions \(\tanh,\sech\dots\).
  Substituting into the first of these the expression for  \(\upphiperp\)
  (i.e., the unregularized version of \eqref{uperpk}), yields a finite linear combination of
  \begin{equation}\label{exc}
  g\,Q^{3-r-l}\psi(s,Q)\,\int\,b_{3,r_1,r_2,l}(x) \delta\gamma(x)^{r_2}{\tteonex}(x)^{r_3}(2\omega_d)^{-r_3/2}
  \triplecolon\upphiperposc(x)^{r_4}\triplecolon\,dx\,(a_{d}^\dagger)^{r_3}F
  \end{equation}
  where \(r_3+r_4=r_1\) and \(\upphiperposc(x)=\int\,\frac{1}{\sqrt{4\pi\omega^{}_k}}\,
  \bigl(a^{}_{k}e_{k\xi}(x{})+a_{k}^\dagger {{e_{-k\xi}}}(x{})\bigr)\,dk\). Substituting this and making use of
  \eqref{best}-\eqref{best2} gives
  the bound
  \[
const.\,g\|{\beefb}\|_{L^2}(1+\|\upphi_{scl}\|_{L^\infty})^l\|Q^{3-r-l}\psi(s,Q)\|_{L^2}\,,
\]
and 
and analogously for the second. Put \(r=r_1+2r_2\) for the statement given.\qed
\begin{lemma}\label{esterrir} Let \(F\in\mfrF\) be an \(M\)-particle vector, then
  \begin{align}\begin{split}
      \|
\D_{\cee,\eff}
\eirup(\upphi+\upphi_{scl},\xi)
  \psi(s,Q){\Omega'}
  \|\,&\leq\,   const.(M) g^{-2}e^{-m|R_g|/2}
        \sum_{p=0}^4 (1+\|\upphi_{scl}\|)^{4-p}\,\|Q^p\psi(s,Q)\|_{L^2}\,\|F\|\,.
    \end{split}\label{respd}\end{align}
\end{lemma}
\proof
The most important feature is the exponential decay in the cut-off length \(R_g\). To see how this arises, consider
the term
\[
-g\int 2m\scm_{cl}^{3/2}{\beefb}(x)\tmxi  {\tteo}(x-\xi)^3dx\,Q^3\,\psi(s,Q){\Omega'}\,.
\]
Observe that if \({\beefb}\equiv 1\) the integral vanishes by parity, so we can replace \(-{\beefb}\) by \(1-{\beefb}\).
  The assumptions \(0\leq {\beefb}\leq 1\) and \({\beefb}(x)=1\) for \(|x|\leq R_g\) ensure similarly a uniform bound
  \[
|\int 2mg\scm^{3/2}(1-{\beefb}(x))\tmxi  {\tteo}(x-\xi)^3dx|\leq const. e^{-m|R_g|/2}
\]
for large \(R_g\), since the odd integrand has exponential decay \(\sim e^{-3m|x-\xi|}\)
and \(\xi\) is confined to a bounded region. Therefore this term is bounded by
\(
ge^{-m|R_g|/2}\|Q^3\psi(s,Q)\|_{L^2}\,,
  \)
consistent with the theorem's assertion for \(g\) small.
The remaining terms in \(\eirup\), 
\[
{\rm H}^{sol}_{I,g,\xi,\xi,1-{\beefb}}(\upphi+\upphi_{scl})-{\rm H}^{sol}_{I,g,0,\xi,1-{\beefb}}(\upphi+\upphi_{scl}-\delta_\xi\Phi_S)
\]
come directly from the interaction Hamiltonian, and 
can be estimated in much the
same way as in Lemma \ref{esthi}, with the most important difference being that the integrands all involve factors
of the form \((1-{\beefb}(x))\zeta_{3,r,l}(x)\) or \((1-{\beefb}(x))\zeta_{4,r,s}(x)\), where the Schwartz functions
\(\zeta_{\bullet,r,s}\)
are exponentially decaying: \(|\zeta_{\bullet,r,s}(x)|\leq const. (e^{-m|x|}+e^{-m|x-\xi|})\). To see this, consider the
quartic piece
\[
\int\,
\frac{1}{2}g^2(1-{\beefb}(x))\Bigl(\mcn_{I,\xi,\delta\gamma_\xi}^4(Y,\upphiperp+\upphi_{scl})-
\mcn_{I,\xi,\delta\gamma_\xi}^4(Y-\delta_\xi\Phi_S,\upphiperp+\upphi_{scl})
\Bigr)
\,dx\,.
\]
(The central term \(\delta_\xi\Phi_S\) can be placed either with \(Y\) or \(\upphiperp\) - the choice made is
convenient, but not essential.) Now refer to \eqref{iii2} and 
observe that the terms like the \(\upphi^4\) which do not involve \(Y\) all cancel, leaving only a finite linear
combination of expressions
\[
g^2\int\,(1-{\beefb}(x))Q^{4-r_1-2r_2-l}\delta_\xi\Phi_S^{r_1}\delta\gamma(x)^{r_2}\triplecolon\upphiperp(x)^{r_3}\triplecolon\upphi_{scl}(x)^{4}\,dx\,  \psi(s,Q)F\qquad(r_1+2r_2+r_3+r_4\leq 4\,,\;r_1>0)\,,
\]
with a factor \(\delta_\xi\Phi_S\), which is a Schwartz function with a \(g^{-1}\) factor in front, hence the appearance of powers of \(g^{-1}\) in the assertion of the theorem. To obtain the bound stated, note that 
under the assumptions that (i)
\(\xi\) is confined to a bounded region, and (ii) \(1-{\beefb}(x)\) is
  supported in \(|x|\geq R_g\), there is, for every \(p\),
  a bound
  \[\|(1-{\beefb})\delta_\xi\Phi_S^{r_1}\|_{L^p}\leq g^{-r_1}const. e^{-m|R_g|/2}\]
  uniformly as \(R_g\to+\infty\),
  and the quartic piece can be bounded (non-optimally) by
  \begin{equation}\label{temp}
  const. (M)g^{-2}\,e^{-m|R_g|/2}\,
  \sum_{p=0}^4 (1+\|\upphi_{scl}\|_{L^\infty})^{4-p}\,\|Q^p\psi(s,Q)\|\,\|F\|
  \end{equation}
  in the same way as the proof of Lemma \ref{esthi}. For the cubic piece one can argue in exactly the same way except
  that there is the hyperbolic tangent appearing in \(\mcn_{I,\xi,\delta\gamma_\xi}^3\),
  which is easily seen to also lead to exponentially decreasing factors in the integrand to be estimated
  by writing \(\tmx-\tmxi=m\int^{\xi}_0\ssm (x-\theta)\,d\theta\).
This can be controlled as \({g\downarrow 0}\) by \eqref{temp}, but now with \(p<4\), completing the proof.
\qed

In order to successfully control the error terms induced by time-dependence, i.e., the final line of
\eqref{revduh2}, it is not sufficient to just use Lemma \ref{best3} to bound the integrand
directly as it was for the interaction and infrared terms - indeed \(\dot\xi\) is the only small factor in those
bounds, and integrating over relevant time-scales will not lead to an \(o(1)\) effect.
Instead we will make use of integration by parts in the form of Lemma \ref{5g0} to replace each of the three terms by the corresponding expression
on the right side of \eqref{5guys0}, and then combine with \(V_{2,j}Q^2\chi\) to obtain the required bounds. This will
be done in the next two lemmas.
\begin{lemma}\label{esterrtd} With the choices
     \begin{equation}
   V_{2,2}=-\dot\xi^2\bigl(\Omega',\Xi^1(h_d+\h)^{-1}\Xi^1\Omega'\bigr)=-\frac{2}{5}m^2\scm_{cl}\dot\xi^2
   \label{defv22}\end{equation}
   and
   \begin{equation}\label{defv23}
V_{2,3}=\frac{4}{5}m^2{\scm_{cl}}\dot\xi^2
\end{equation}
there holds
   \begin{align}\label{esterrtdeq}
    \bigl\|\int_0^t\,\tilde{\bevot}(t,s)&\bbS_s
\D_{\cee,\eff}
  \Bigl(
       \dot\xi(s)\,
(\Xi^0\,+\,\Xi^1Q\,+\,i\Xi^2P)-V_{2,2}Q^2-V_{2,3}Q^2
\Bigr)\chi(s,Q)\Omega'
ds\bigr\|\\\notag
&\,\leq\, const.\sup_{[0,t]}|\dot\xi|(\|\chi\|+\|P\chi\|+\|Q\chi\|)
\\\notag
&\qquad + const.\,\int_0^{t}\,\Bigl[\bigl(|\ddot\xi|+ |\dot\xi|^2\bigr)\bigl(1+|\cee|+\|\eff\|\bigr)+|\dot\xi|\bigl(|\Im \cee|+\|\Im \eff\|\bigr)\Bigr]\bigl(\|P\chi\|+\|Q\chi\|+(1+|\cee|+\|\eff\|)\|\chi\|\bigr)
\\\notag
&\qquad\quad+\,|\dot\xi|\,(g\|b\|_{L^2}+|\dot\xi|^2)(1+|c|+\|\eff\|)^4\sum_{r+l\leq 1}\bigl(\|Q^rP^l\chi\|
    +\|Q^{2+r}P^l\chi\|
+
    \,g\|Q^{4+r}P^l\chi\|\bigr)\\
&\qquad\qquad+\,g^{-2}e^{-m|R_g|/2}\,|\dot\xi|(1+|\cee|+\|\eff\|)^4
\sum_{\substack{a\leq 5\\b\leq 1}} 
  \|Q^aP^b\chi\|\notag
  \\\notag&\qquad\qquad\qquad +  \,|\dot\xi|^2(1+|\cee|+\|\eff\|)^2\sum_{\substack{r<2\\r+l\leq 2}}\|Q^rP^l\chi\|\,
  \\\notag&\qquad\qquad\qquad\qquad
  + |\dot\xi|\,(1+|\cee|+\|\eff\|)^3
  \|(\id+{\N})^{-1/2}\delta h\|\bigl(\|\chi\|+\|Q\chi\|+\|P\chi\|\bigr)
\\\notag&\qquad\qquad\qquad\qquad\qquad
+\,\Bigl(|\dot\xi(s)|\|P\chi(s,Q)\|_{L^2}+\bigl(|\dot\xi(s)|^2
+g(|c(s)|+\|f(s)\|)\bigr)\|Q\chi(s,Q)\|_{L^2}\Bigr)\,\Bigr]\,ds
  \\\notag
  &\qquad\qquad\qquad\qquad\qquad\qquad
  +\|\int_0^t\,\tilde{\bevot}(t,s)\bbS_s\,
Q^2\chi\dot\xi^2\D_{\cee,\eff}\,\triplecolon\, \Xi^1(h(\omega_d)+\h(\omega_\bullet))^{-1}\Xi^1 \,\triplecolon
\,\Omega'\,ds\|\,.
%
      \end{align}
  \end{lemma}
\proof
The evaluation of \(V_{2,2}\) reduces to computing
\[
\scm_{cl}\bigl(\upphi^\perp(K(\xi)^{\frac{1}{2}}\tteox'),\upphiperp(\tteox')\bigr)_{\mfrF}=
\frac{1}{2}\scm_{cl}\|\tteox'\|^2
\]
which gives the stated value using \(\int_{\smr}\ssssm x\ttmx dx=4/(15m)\). The reason for this choice
is to normal order a certain term, as now explained. The idea to prove \eqref{esterrtdeq}
is to integrate by parts by
means of Lemma \ref{5g0}, substituting into \eqref{5guys0} with
\begin{itemize}
\item \( \psi_j(s,Q)\) in place of \(\psi\), where for 
\(j=0,1,2\) respectively \(\psi_j\) equals \(\chi(s,Q),Q\chi(s,Q),iP\chi(s,Q)\).
\item \(\D_{\cee,\eff}F\) taken as \(\dot\xi\D_{\cee,\eff}\Xi^j\Omega'\).
  \end{itemize}

However, there is a proviso, essentially the same as in the estimation of the
interaction terms in Lemma \ref{esthi}:
care is required with the factors of
\(Q\) which arise both from \(\eirtd\) and \(\frac{d}{ds}p\) in \eqref{5guys0} when \(j=1\). Indeed from
\eqref{errtd3} we see that the former leads to a  ``\(Q^2\)'' term
\[
Q^2\chi\dot\xi^2\D_{\cee,\eff}\Xi^1(h_d+\h)^{-1}\Xi^1\,\Omega'
\]
which has to be treated separately, see Remark \ref{onehalf}.
On account of the choice \eqref{defv22}, combining with the \(V_{2,2}\) term ensures that
   the mean with respect to the transverse vacuum has been removed - or more accurately, transferred into the one-particle Hamiltonian
   \(h_{\mbox{\tiny 1P}}\) - leaving 
\[
Q^2\chi\dot\xi^2\D_{\cee,\eff}\triplecolon \Xi^1(h_d+\h)^{-1}\Xi^1 \triplecolon
\]
{\em which is now normal ordered} in the transverse variables, hence the final (seventh) line on the right of the inequality
\eqref{esterrtdeq} in the statement.
This will be bounded in Lemma \ref{a2} by using the fact that it is normal ordered (and so lies in \(\Omega_{\cee,\eff}^\perp\))
to integrate by parts a second time, again using Lemma \ref{5g0}. The choice of \(V_{2,3}\) similarly ensures that the
\(\frac{d}{ds}p_1\) term can be controlled, see item (i)(a) below in this proof.

Now to obtain the stated bound, we integrate the other terms by parts, substituting into
\eqref{5guys0} as above,
with \(p\) such that \(p_j\Omega'=\dot\xi(h_d+\h)^{-1}\Xi^j\Omega'\) (see \eqref{errtd15}-\eqref{errtd3});
in the notation used in the proof of Lemma \ref{5g0} this corresponds
  to
  \(\tilde G\) being taken, for \(j=0,1,2\), to be \(\Sigma_j\) where
\begin{align*}
  \Sigma_0(s,Q)&= \dot\xi\,\chi(s,Q)(h_{\cee,\eff}-e_{\cee,\eff})^{-1}\Omega_{\cee,\eff}^0=\dot\xi\,\chi(s,Q)\D_{\cee,\eff}\,(h_d+\h)^{-1}\Xi^0\Omega'\,,\\
  \Sigma_1(s,Q)&= \dot\xi\,Q\chi(s,Q)(h_{\cee,\eff}-e_{\cee,\eff})^{-1}\Omega_{\cee,\eff}^1=\dot\xi\,Q\chi(s,Q)\D_{\cee,\eff}\,(h_d+\h)^{-1}\Xi^1\Omega'\,,\\
  \Sigma_2(s,Q)&= i\dot\xi\,P\chi(h_{\cee,\eff}-e_{\cee,\eff})^{-1}\Omega_{\cee,\eff}^2
  =i\dot\xi\,P\chi(s,Q)\D_{\cee,\eff}\,(h_d+\h)^{-1}\Xi^2\Omega'\,,
\end{align*}
or equivalently \(G\) to be taken as \(G_j=p_j\Omega'\), with
the formulae for the Wick polynomials \(p_j\) being made completely explicit in \S\ref{controlf},
allowing their estimation by \eqref{best25} or Lemma \ref{best3}.

Now consider the various terms in \eqref{5guys0}, line by line. Summing over \(j\),
the {\em first line} can be bounded by a multiple of \(\max_{[0,t]}|\dot\xi|(\|\chi\|+\|P\chi\|+\|Q\chi\|)\),
which is the
first line of the right side of \eqref{esterrtdeq}.

Using 
\((i\partial_s-h_{\mbox{\tiny 1P}})\chi(s,Q)=0\), we deduce that the final term on {\em line three} vanishes for
  \(j=0\). For \(j=1\)  the formula
\((i\partial_s-h_{\mbox{\tiny 1P}})(Q\chi(s,Q))=i\scm_{cl}^{-1}P\chi(s,Q)\),
leads to the need to estimate also
\begin{equation}\label{wamii}
  \Bigl\|\,\int_0^t\tilde\bevot(t,s)\bbS(s)\Bigl(
  \dot\xi\,P\chi(s,Q)(h_{\cee,\eff}-e_{\cee,\eff})^{-1}\Omega_{\cee,\eff}^1\Bigr)
\,ds\,\Bigr\|\,,
  \end{equation}
the integrand of which is bounded by \(const. |\dot\xi|\|P\chi(s,Q)\|_{L^2}\), and is in turn
bounded by the seventh line in \eqref{esterrtdeq}. For \(j=2\) the relevant formula is 
\((i\partial_s-h_{\mbox{\tiny 1P}})(P\chi(s,Q))=-2iV_2Q\chi(s,Q)\), and by the explicit formulae for
\(V_{2,j}\) in \eqref{defv2} and \eqref{defv22}-\eqref{defv23},
the term analogous to \eqref{wamii} is bounded by
\[const. \int_0^t\,\bigl(|\dot\xi(s)|^2+g(|c(s)|+\|f(s)\|)\bigr)\|Q\chi(s,Q)\|_{L^2}\,ds\,,\]
which can be absorbed by the seventh line on the right of \eqref{esterrtdeq}. The first term on line three is estimated together with
the final term on {\em line two} of \eqref{5guys0}, to which line we now turn,
considering in turn the five terms for the various 
\(j=0,1,2\).

  \begin{enumerate}
  \item[(i)] Consider the first term, with \(j=2\): by
  Remark \ref{msa} and \eqref{xi2} we write
  \begin{equation}\label{xi25}
(h_d+\h)^{-1}  \Xi^2\Omega'
  =\omega_d^{-1}\beta_{d}^0\,a_{d}^\dagger\Omega'+
\int\omega_k^{-1}\beta^0(k,\xi)\,dk\, a_{k}^\dagger \Omega'\,.
\end{equation}
Referring to \eqref{xi2}, this implies
\[
p_2(s,\phisch-\upphi_{scl})=
\dot\xi\Bigl\langle\phisch-\upphi_{scl},\frac{(\tteo',\tteone)}{\omega_d}\tteonex
+\mcF_{u_\xi}^{-1}\Bigl(\frac{1}{\omega_k}\mcF_{u_\xi}\tteox'\Bigr)\Bigr\rangle\,.
\]
To control the time derivative of this, an important observation is that only the
imaginary part of \((\cee,\eff)\), which is small, contributes. To be precise, 
\[
\frac{d}{ds}\upphi_{scl}(x;\xi,\cee,\eff)=
\dot\xi\frac{\partial}{\partial\xi}\upphi_{scl}(x;\xi,\cee,\eff)+
\upphi_{scl}(x;\xi,\omega_d\cee_2,\omega_{\bullet}\eff_2)\,,
\]
i.e., only the imaginary parts \(\Im \cee,\Im \eff\) appear after substituting for
\(\dot\cee,\dot\eff\) from  \eqref{defeeze1}; as a consequence we can estimate, for example in the \(L^2\) norm at time \(s\),
\begin{equation}\label{estl2}
\|\frac{d}{ds}\upphi_{scl}(x;\xi,\cee,\eff)\|\leq const.\,
\Bigl(|\dot\xi|(|\cee|+\|\eff\|)+|\Im\cee|+\|\Im\eff\|\Bigr)\,.
\end{equation}
and analogously in Schwartz seminorms.
On differentiation
the \(\ddot\xi\) contribution can be bounded, using \eqref{best}-\eqref{best2} or \eqref{best3}, by
  \[
    \leq const. (1+|\cee|+\|\eff\|)|\ddot\xi|\|P\chi\|\,.
    \]
    There is also a contribution from differentiation of the argument, which has to be decomposed
    with respect to the subspace \(\langle \tteox\rangle\), i.e., at time \(s\)
    \[
\frac{d}{ds}\Bigl[\frac{(\tteo',\tteone)}{\omega_d}\tteonex
+\mcF_{u_\xi}^{-1}\Bigl(\frac{1}{\omega_k}\mcF_{u_\xi}\tteox'\Bigr)\Bigr]=\dot\xi(s)\Bigl(\nu_0(s)\tteox+\nu_1(s,x)\Bigr)\,,
\]
where \(\nu_1(s,\cdot)\) is a Schwartz function satisfying \((\nu_1,\tteox)=0\) at each \(s\);
the component \(\nu_0\tteox\) generates a \(Q\) when paired with the field.
Thus all together for the first term on second line with \(j=2\) we need to control
\begin{align*}
i\Bigl(\frac{d}{ds}p_2(s,\phisch-\upphi_{scl})\Bigr)\bbS_s P\chi(s,Q)
\D_{\cee,\eff}\Omega'&=
-i\dot\xi\Bigl\langle \frac{d}{ds}\upphi_{scl}(x;\xi,\cee,\eff),\frac{(\tteo',\tteone)}{\omega_d}\tteonex
+\mcF_{u_\xi}^{-1}\Bigl(\frac{1}{\omega_k}\mcF_{u_\xi}\tteox'\Bigr)\Bigr\rangle\bbS_s P\chi(s,Q)
\D_{\cee,\eff}\Omega'\\
&\quad+i{\ddot\xi}\bbS_s \D_{\cee,\eff}\Bigl\langle \upphi,
\frac{(\tteo',\tteone)}{\omega_d}\tteonex
+\mcF_{u_\xi}^{-1}\Bigl(\frac{1}{\omega_k}\mcF_{u_\xi}\tteox'\Bigr)
\Bigr\rangle P\chi(s,Q)\Omega'
\\
&\qquad+i\dot\xi^2\bbS_s \D_{\cee,\eff} \Bigl\langle\upphi+\upphi_{scl},
\nu_0(s)\tteox+\nu_1(s,x)\Bigr\rangle
P\chi(s,Q)\Omega'
\end{align*}
which is bounded by
  \[
  const.\,\int_0^{t}\,\Bigl[\bigl(|\ddot\xi|\|P\chi\|+ |\dot\xi|^2(\|P\chi\|+\|QP\chi\|)\bigr)\bigl(1+|\cee|+\|\eff\|\bigr)+|\dot\xi|\bigl(|\Im \cee|+\|\Im \eff\|\bigr)\|P\chi\|\Bigr]\,ds\,.
  \]
  Now consider the cases \(j=1\) and \(j=0\).
  \begin{enumerate}
  \item[(a)] For \(j=1\) the analogous computation leads to a term with ``\(\dot\xi^2Q^2\)'' which cannot be treated
    perturbatively. Calculating we find that
    \[
i\Bigl(\frac{d}{ds}p_1(s,\phisch-\upphi_{scl})\Bigr)\bbS_s Q\chi(s,Q)
    =-\frac{4m^2\dot\xi^2}{5}{\scm_{cl}}Q^2\chi+\mathcal{B}
    \]
    where the first term is cancelled by the choice of
\(V_{2,3}\)    
and the contribution of \(\mathcal{B}\) can be bounded by
    \[
      const.\,\int_0^{t}\,\Bigl[\bigl(|\ddot\xi|+ |\dot\xi|^2\bigr)\bigl(1+|\cee|+\|\eff\|\bigr)+|\dot\xi|\bigl(|\Im \cee|+\|\Im \eff\|\bigr)\Bigr]\|Q\chi\|\,ds\,.
      \]
      All together, both of these bounds are controlled by the second line on the right of \eqref{esterrtdeq}.
  \item[(b)] For \(j=0\), see  \eqref{errtd2},  the term linear in the field \(\upphiperp\) is similar to those already treated,
    except for the additional \(\upphi_{scl}\) which necessitates an additional factor linear in \(\cee,\eff\),
        \[
  const.\,\int_0^{t}\,\Bigl[\bigl(|\ddot\xi|\|\chi\|+ |\dot\xi|^2(\|\chi\|+\|Q\chi\|)\bigr)\bigl(1+|\cee|+\|\eff\|\bigr)+|\dot\xi|\bigl(|\Im \cee|+\|\Im \eff\|\bigr)\|\chi\|\Bigr]\bigl(1+|\cee|+\|\eff\|\bigr)\,ds\,.
  \]
The part of \(\Xi^0\) quadratic in \(\upphiperp\) can be bounded by arguments similar to those in the proof of Lemma \ref{diffom} (but easier due to the
fact that the kernel of the oscillatory part of the integral differs from the corresponding part of the
kernel of \(\partial_\xi K^{1/2}\) by a factor \(\omega_k^{-1}\) which reduces by one the order of the singularity in the integral over \(k\)), leading to the same bound, again controlled by second line on the right of \eqref{esterrtdeq}.
  \end{enumerate}
\item[(ii)] Summing over \(j=0,1,2\) the second term in the middle line of \eqref{5guys0} can be bounded, using
  \eqref{esthieq2} and Lemma \ref{best3}, by
  \begin{align}\label{respbmod}\begin{split}
  g\,const.(M)\,\int_0^t\|{\beefb}\|_{L^2}(1+|c|+\|\eff\|)^4\sum_{r+l\leq 1}\Bigl[&\|Q^rP^l\chi\|
    +\|Q^{2+r}P^l\chi\|
+
    \,g\|Q^{4+r}P^l\chi\|\,\Bigr]\,ds
    \end{split}
  \end{align}
    which is controlled by the third line on the right side  of \eqref{esterrtdeq};
  \item[(iii)] Similarly, summing over \(j=0,1,2\) and making use of Lemma \ref{esterrir}, the third term in the middle line of \eqref{5guys0} can be bounded
    again using \eqref{best3} by
  \begin{equation}\label{respdd}
const.\,g^{-2 }e^{-m|R_g|/2}\,\int_0^t\,|\dot\xi| (1+|\cee|+\|\eff\|)^4\sum_{\substack{r_2+l\leq 1\\0\leq r_1\leq 4}}\|Q^{r_1+r_2}P^l\chi\|\,ds\,,
  \end{equation}
which is in turn controlled for small \(g\) by the fourth line on right in the assertion \eqref{esterrtdeq} of the lemma;
\item[(iv)] referring to \eqref{errtd15}, and the formulae for the \(\Sigma_j\) above,
  observe that the various terms which arise from the fourth summand of the middle line
\(+i\,\int_0^t\tilde\bevot(t,s)\bbS(s)\eirtd{\Sigma}_j\,ds
\) involve \(Q^rP^l\psi\) with powers \(r,l\) in \(\{0,1,2\}\) with \(r+l\leq 2\).
As long as \(r<2\) it is sufficient to
bound them directly as
\[
\leq const. \int_0^t\,|\dot\xi|^2(1+|\cee|+\|\eff\|)^2\sum_{\substack{r<2\\r+l\leq 2}}\|Q^rP^l\chi\|\,ds\,,
\]
which is the fifth line on the right side of \eqref{esterrtdeq},
while the case \(r=2\) is treated by Lemma \ref{a2} below, after subtracting \(V_{2,2}\) to normal order, as already explained;
\item[(v)] For each \(j=0,1,2\) the fifth term on the middle line, and the first term on the third line of
  \eqref{5guys0} are bounded using Lemma \ref{best3} by the time integral of \(const. |\dot\xi|\,(1+|\cee|+\|\eff\|)^2
  \|(\id+{\N})^{-1/2}\delta h\|\|\psi_j\|\), which is in turn dominated by the sixth line on the right of \eqref{esterrtdeq}.
  \end{enumerate}
  This completes the proof of the bound in Lemma \ref{esterrtd}.\qed
   \begin{lemma}\label{a2} The final line of the inequality in Lemma \ref{esterrtd} is bounded as follows:
 \begin{align*}
   \|\int_0^t\,\tilde{\bevot}(t,s)\bbS^\theta(\xi)
   Q^2\chi\dot\xi^2\D_{\cee,\eff}\triplecolon \Xi^1(h_d+\h)^{-1}\Xi^1 \triplecolon\,\Omega
  \,ds\,
\|\,&\leq\,
const.\, \int_0^t \Bigl[(1+|\cee|+\|\eff\|)^2\bigl(|\dot\xi||\ddot\xi|
      +\, |\dot\xi|^2(|\Im c|+\|\Im f\|)\bigr)\|Q^2\chi(s,Q)\|\\
    &\qquad+
      g|\dot\xi|^2\,\|{\beefb}\|_{L^2}(1+|c|+\|\eff\|)^4(\sum_{2\leq r\leq 5}\|Q^r\chi\|+g\|Q^6\chi\|)
      \\
      & \qquad\qquad\qquad|\dot\xi|^2 e^{-m|R_g|/2}(1+|\cee|+\|\eff\|)^4
\sum_{a=2}^6 
  \|Q^a\chi\|)\\
      &\qquad\qquad
      \qquad\qquad
+|\dot\xi|^3(1+|\cee|+\|\eff\|)^2\bigl(\sum_{r\leq 3}\|Q^r\chi\|+\sum_{l\leq 2}\|PQ^l\chi\|\bigr)
\\
&\qquad\qquad\qquad\qquad\qquad +|\dot\xi|^2\,(|\Im\cee|+\|\Im\eff\|)\|Q^2\chi(s,Q)\|\\
&\qquad\qquad\qquad\qquad\qquad\qquad + |\dot\xi^2|(\|QP\chi(s,Q)\|+\|\chi(s,Q)\|)
\Bigr]\,ds\,.
  \end{align*}
    \end{lemma}
 \proof
 We apply Lemma \ref{5g0} with
 \(\psi\) replaced by \(\psi_3=Q^2\chi\) and the polynomial \(p\) taken as
 \[
p_3(\upphi)=|\dot\xi|^2\,\triplecolon\,\upphi(K^{\frac{1}{2}}(\xi)\tteox')\upphi(\tteox')\triplecolon
\]
so that
 \(F=\dot\xi^2\,\triplecolon \Xi^1(h_d+\h)^{-1}\Xi^1 \triplecolon\). We now
 estimate the various terms directly using Lemmas \ref{best3},\ref{esthi} and \ref{esterrir} as in
 (i)-(v) of the proof of Lemma \ref{esterrtd}, with the difference that there is no need to treat specialy
 factors of \(Q\) arising from \eqref{errtd3} (as all terms have additional small factors here.)
The main difference now is that the final term on the third line of \eqref{5guys0} leads to the
 need to bound, instead of \eqref{wamii}, the analogous expression involving
 \[(i\partial_s-h_{\mbox{\tiny 1P}})(Q^2\chi(s,Q))=
Q^2(i\partial_s-h_{\mbox{\tiny 1P}})\chi(s,Q)+
 \scm_{cl}^{-1}(2iQP\chi(s,Q)+\chi(s,Q))\,,\]
which is bounded by the final line of the statement.
 \qed
 
\noindent
 Finally it remains to bound the term in the interaction Hamiltonian which is quadratic in \(Q\), i.e., the first term on the right side of
 \eqref{annav} (which is what was left after extracting terms subquadratic in \(Q\), and then canceling the average by choice of \(V_{2,1}\)).
\begin{lemma}\label{finest}
  Define \(R_3(x;\xi)=6m\scm_{cl}{\beefb}(x)\tmxi {\tteo}(x-\xi)^2\,\)
so the first term on the right side of \eqref{annav} is
\(gQ^2\langle\upphiperp,R_3\rangle \), and
the corresponding term  in \eqref{revduh2} can be estimated as follows:
  \begin{align}\label{btg}
    \|ig\int_0^t\,\tilde{\bevot}(t,s)\bbS^\theta(\xi)
\D_{\cee,\eff}\langle\upphiperp,R_3\rangle
Q^2  \psi(s,Q)\Omega
  \,ds\,
\|\,&\leq\,
const.\,
\sup_{[0,t]} g \|Q^2\chi\|
\\\notag
&\qquad + const.\,g\,\int_0^{t}\,\Bigl[|\dot\xi|\bigl(1+|\cee|+\|\eff\|\bigr)+\bigl(|\Im \cee|+\|\Im \eff\|\bigr)\Bigr]\|Q^2\chi\|
\\\notag
&\qquad\quad+g\,\,\|{\beefb}\|_{L^2}(1+|c|+\|\eff\|)^4(\sum_{2\leq r\leq 5}\|Q^r\chi\|+g\|Q^6\chi\|)
\\\notag
&\qquad\qquad+\,g^{-1}e^{-m|R_g|/2}\,(1+|\cee|+\|\eff\|)^4
\sum_{a=2}^6 
  \|Q^a\chi\|
  \\\notag&\qquad\qquad\qquad +  \,g\,|\dot\xi|(1+|\cee|+\|\eff\|)^2\bigl(\sum_{r\leq 3}\|Q^r\chi\|+\sum_{l\leq 2}\|PQ^l\chi\|\bigr)
  \\\notag&\qquad\qquad\qquad\qquad
  + g\,(1+|\cee|+\|\eff\|)^3
  \|(\id+{\N})^{-1/2}\delta h\|\|Q^2\chi\|
\\\notag&\qquad\qquad\qquad\qquad\qquad
+\,(\|\chi(s,Q)\|+\|QP\chi(s,Q)\|)\,\Bigr]\,ds
%
      \end{align}
    \end{lemma}
\proof
  At each time \(s\)
    \[gQ^2\psi(s,Q)\Omega_{\cee,\eff}^3(s,Q)
  =gQ^2\psi(s,Q)\,\D_{\cee,\eff}\langle\upphiperp,R_3\rangle\,\Omega\in
  L^2(\R,dQ;\Omega_{\cee,\eff}^\perp)
  \]
so we can integrate by parts using Lemma \ref{5g0} just
as in the proof of Lemma \ref{esterrtd}. The term to be bounded can be written
  \((h_{\cee,\eff}-e_{\cee,\eff})\Sigma_3(s,Q)\) with
  \[\Sigma_3(s,Q)=
  gQ^2\psi(s,Q)\D_{\cee,\eff}\,(h(\omega_d)+\h(\omega_\bullet))^{-1}\Xi^3\,,\qquad
\Xi^3=\langle\upphiperp,R_3\rangle\,.
\]
 Putting \(\psi_3(s,Q)=Q^2\chi(s,Q)\)  the integral on the left side of \eqref{btg} can be bounded by means of
 \eqref{5guys0}. The  details
 are very similar to the \(j=2\) case considered in the proof of Lemma \ref{errtd},
the main difference being, as in the preceding lemma, the occurrence of \(Q^2\chi\) leading to the final line. 
 This aside, there is a factor \(g\) in place of \(\dot\xi\), and making these changes leads to the stated bound.
 \qed

\protect\renewcommand{\thesection}{\Alph{section}}
\protect\renewcommand{\thesubsection}{\Alph{section}.\arabic{subsection}}
\protect\renewcommand{\theequation}{\Alph{section}.\arabic{equation}}
\renewcommand*{\theHsection}{\Alph{section}}
\renewcommand*{\theHsubsection}{\Alph{section}.\arabic{subsection}}
\renewcommand*{\theHequation}{\Alph{section}.\arabic{equation}}

\setcounter{section}{0}
\setcounter{subsubsection}{0}
\setcounter{equation}{0}

\section{Appendix: Quantum Mechanics in the Kink Background}

The analysis in this article is based on spectral representations 
for the linear operators which arise on linearization around the kink.
The linearized one-particle Hamiltonian is the Schr\"odinger operator
\(
K
\) in \eqref{def-k}.
This operator is one of a ladder of differential operators whose
eigenfunctions can
be written explicitly as follows. Starting with the operator 
\(-\partial_x^2+m^2\), where \(m^2>0\), we notice the factorization
\begin{equation}\la{l0}
-\partial_x^2+m^2\,=\,A\,A^\dagger\,,\quad\hbox{where}
\end{equation}
\[
A=\partial_x+m\tmx\quad \hbox{and}\quad A^\dagger=-\partial_x+m\tmx\,.
\]
Paired with \(A\, A^\dagger\) is the operator
\begin{equation}\la{l1}
-\partial_x^2+m^2-2m^2\ssmx\,=\,A^\dagger\,A\,.
\end{equation}

This process repeats: define
\[
B=\partial_x+2m\tmx\quad \hbox{and}\quad B^\dagger=-\partial_x+2m\tmx\,,
\]
then compute that 
\begin{equation}\label{p3}
B^\dagger\,B\,=\,-\partial_x^2+4m^2-6m^2\ssmx\,\quad\hbox{and}\quad
B\,B^\dagger\,=\,A^\dagger\,A+3m^2\,.
\end{equation}

\subsection{Free covariance kernel and Bessel functions}\label{bessel}
The kernel \eqref{defC0} can be expressed either in terms of the
McDonald function on the positive real axis, or in terms of the Hankel functions
on the imaginary axis; the latter representation reads
(\cite{MR0167642}*{\S 9.6}) as
\begin{align}\label{defC01}
C_0^{\frac{1}{2}}(x,y)\,&=\,
\frac{1}{2\pi}\,\int_{\smr}\,
\frac{e^{ik(x-y)}}{(k^2+4m^2)^{\frac{1}{2}}}
\,dk
=\frac{i}{2}H_0^{(1)}(2im|x-y|)\,\bigl(=-\frac{i}{2}H_0^{(2)}(-2im|x-y|)\bigr)\\
&=\,
-\frac{1}{\pi}\ln m|x-y|\Bigl(1+m^2|x-y|^2+\frac{1}{4}m^4|x-y|^4+\dots\Bigr)+c_1|x-y|^2+c_2|x-y|^4+\dots\,.
\notag\end{align}
(It is real valued, smooth away from the origin, and exponentially decaying at infinity as are its derivatives.)
In what follows it will also be useful to consider also
\begin{align}\label{defC02}
N(x-y)\,&=\,
\frac{1}{2\pi}\,\int_{\smr}\,
\frac{e^{ik(x-y)}}{(k^2+m^2)(k^2+4m^2)^{\frac{1}{2}}}
\,dk
=(-\partial^2+m^2)^{-1}\frac{i}{2}H_0^{(1)}(2im|x-y|)\\
&=\,
-\frac{1}{2\pi}\ln m|x-y|\Bigl(|x-y|^2+a_2|x-y|^4+\dots\Bigr)+\tilde c_1|x-y|^2+\tilde c_2|x-y|^4+\dots\,.
\notag\end{align}
\(N\) solves the equation \(-N''+m^2N=\frac{i}{2}H_0^{(1)}(2im|x-y|)\), and so is
a \(C^1\) function all of whose derivatives decay exponentially as \(|x-y|\to\infty\).
The second derivative of \(N\) has a logarithmic singularity and expansion
for \(x\approx y\) of the same form as \eqref{defC01}, while in this regime its third derivative admits
an expansion of the form
\begin{align}\label{defC03}
N'''(x-y)\,&=\,
\frac{1}{2\pi}\,\,\int_{\smr}\,
\frac{(ik)^3e^{ik(x-y)}}{(k^2+m^2)(k^2+4m^2)^{\frac{1}{2}}}
\,dk\\
&=\,
-\frac{1}{\pi}\frac{1}{x-y}+\ln m|x-y|\Bigl(a_1'(x-y)+a_2'(x-y)^3+\dots\Bigr)+c_1'(x-y)+c_2'(x-y)^3+\dots\,.
\notag\end{align}
\subsection{Spectral Resolution and Covariance Operators}\label{eigen}
It follows from the ladder structure just introduced
that if \(A\,A^\dagger\phi=\epsilon\phi\), then
\(A^\dagger\, A\,A^\dagger\phi=(\epsilon+1)A^\dagger\,\phi\), and
hence that
\[
A^\dagger\,e^{ikx}\,=\,(\tmx-ik)e^{ikx}
\]
is a generalized eigenfunction of \(A^\dagger\,A\). In addition, there
is a normalizable eigenfunction, \(\smx\), which lies in the kernel
of \(A^\dagger\,A\).

In the same way, it follows that, for any \(k\in\R\), the function
\[
B^\dagger\,A^\dagger\, e^{ikx}\,=\,
(-k^2-3imk\tmx+2m^2-3m^2\ssmx)\,e^{ikx}
\]
is an eigenfunction for
\(B^\dagger\,B\).
It is  a consequence of \eqref{l0},\eqref{l1} and \eqref{p3} that 
\[A\,B\,B^\dagger\,A^\dagger=(-\partial_x^2+m^2)(-\partial_x^2+4m^2)\,.\]
It is
convenient to introduce phase factors
\[
e^{\pm i\delta_k}
\,=\,\frac{[-k^2\mp 3imk+2m^2]}
          {\sqrt{(k^2+m^2)(k^2+4m^2)}}\,,
\]
and then to normalize the generalized eigenfunctions as follows:
\begin{align}
  e^-_k(x)\,&=\,
\frac{[-k^2-3imk\tmx+2m^2-3m^2\ssmx]}
     {\sqrt{(k^2+m^2)(k^2+4m^2)}}\,e^{i\delta_k}\,e^{ikx}\\
&=\,\frac{[k^2+3imk\tmx-2m^2+3m^2\ssmx]}{(k-im)(k-2im)}
e^{ikx}     ,\\
e_k(x)\,&=\,\frac{[-k^2-3imk\tmx+2m^2-3m^2\ssmx]}
     {\sqrt{(k^2+m^2)(k^2+4m^2)}}\,e^{-i\delta_k}\,e^{ikx}\\
&=\,\frac{[k^2+3imk\tmx-2m^2+3m^2\ssmx]}{(k+im)(k+2im)}
e^{ikx}\,=\,\frac{{\ttF}(k,x)e^{ikx}}{(k+im)(k+2im)} 
\,.
  \label{ekdef}
\end{align}
These obey \((-\partial_x^2+4m^2-6m^2\ssmx)e_k(x)=(k^2+4m^2)e_k(x)\,,\)
as a consequence of the above algebraic structure, and are normalized so
that \(e_k(x)=e^{ikx}+O(e^{-m|x|}) \) as
\(x\to +\infty \), and is analytic in the upper half \(k\)-plane, and
\(e^-_k(x)=e^{ikx}+O(e^{-m|x|}) \) as \(x\to -\infty \), and \(e_k^-\) is
analytic in the lower half \(k\)-plane. Observe that \(e_{-k}(x)=\overline{e_k(x)}\) for real \(k\).
The final equality in \eqref{ekdef} defines the quantity \({\ttF}\) used in the text,
after \eqref{fc1}.

In addition, there is a pair of square-integrable eigenfunctions,
given in normalized form as:
\begin{align}
 {\tteo}(x)\,&=\,\sqrt{\frac{3m}{4}}\ssmx\,,\\
 {\tteone}(x)\,&=\,\sqrt{\frac{3m}{2}}\tmx\,\smx\,.
\end{align}
We write \(\P_0,\P_1\) for the corresponding orthogonal projection
operators,
defined by the integral kernels \(\P_a(x,y)= e_a(x) e_a(y)\) for
\(a\in\{0,1\}\). The discrete eigenfunctions obey \((-\partial_x^2+4m^2-6m^2\ssmx) {\tteo}(x)=0\) (zero mode)
and \((-\partial_x^2+4m^2-6m^2\ssmx) {\tteone}=3m^2 {\tteone}=\omega_d^2 {\tteone}\,,\) (discrete oscillatory mode). {\ttF}inally we write
\(\P_c=\id-\P_0-\P_1\) for the orthogonal projector onto the continuous spectral subspace.

These definitions are chosen so that the following orthonormality 
relations hold:
\begin{align}
\int_{\smr}\,e_{-l}(x)\,e_k(x)\,dx\,&=\,{2\pi}\,\delta(k-l)\,,\;\hbox{ for all }k,l\in\R\,,\label{on1}\\
\int_{\smr}\, \ttea(x)\, \tteb(x)\,dx\,&=\,\delta_{ab}\,,\;\hbox{ for all }a,b\in\{0,1\}\,,\\
\int_{\smr}\, \ttea(x)\,e_k(x)\,dx\,&=\,0\,,\hbox{ for all }\;a\in\{0,1\}\,\hbox{ and }k\in\R\,.
\end{align}
(In the first of these, and in related formulae, the integral is of course
to be understood as being an
\(\cs'\)-valued integral, i.e. the relation when holds when paired with
Schwartz function inside the integral:
\[
\int_{\smr}\,\langle f(k,l),e_{-l}(x)\otimes e_k(x)\rangle \,dx\,=\,{2\pi}\,\int f(k,k)dk
  \]
  for \(f\in\cs(\R^2)\) and with \(\langle,\rangle\) as the duality pairing.)
  The completeness relation takes 
the form:
\begin{equation}\label{comprel}
  \frac{1}{2\pi}\,\int\,{e_{-k}}(y)\,e_k(x)\,dk\,+\,\P_0\,+\,\P_1\,
  =\,\delta(x-y)\,.
\end{equation}
Writing \(K=B^\dagger B \), the functional calculus gives the following
formula for the integral kernel of the operator \(f(K) \):
\begin{align}\label{fc}
f(K)(x,y)\,&=
\,f(0) {\tteo}(x) {\tteo}(y)\,+\,f(3m^2) {\tteone}(x) {\tteone}(y)\,+
\,\frac{1}{2\pi}\,\int_{\smr}\,\notag
\Bigl[f(k^2+4m^2)\overline{e_k(y)}e_k(x)\Bigr]
\;dk\,
\\
&=
\,f(0) {\tteo}(x) {\tteo}(y)\,+\,f(3m^2) {\tteone}(x) {\tteone}(y)\,\\
&\quad\,+\,\frac{1}{2\pi}\,\int_{\smr}\,\notag
\Bigl[f(k^2+4m^2)(-k^2+3imk\tmy+2m^2-3m^2\ssmy)e^{ik(x-y)}\\
&\qquad\qquad\qquad\qquad\qquad\qquad\qquad\qquad\times\frac{(-k^2-3imk\tmx+2m^2-3m^2\ssmx)}
{(k^2+m^2)(k^2+4m^2)}\Bigr]
\;dk\,.\notag
\end{align}

\begin{prop}\label{ksb}
  For any \(s\in\R,r\geq 0\,,\) the operator \(K^{\frac{r}{2}}\) is bounded as an operator \(H^s\to H^{s-r}.\) For any \(s,r\in\R\), and \(\theta>0\), the operator \({(K^\theta)}^{\frac{r}{2}}\),
where \({{K^\theta}}\) was defined just prior to Theorem \ref{mac}, is bounded as an operator \(H^s\to H^{s-r}.\)
\end{prop}
\proof  Consider the second statement of the proposition. We remark that if
  \(f\in C^\infty(\R)\) is a smooth function all of whose derivatives are bounded then
  the operator \(u\mapsto fu\) is bounded on every Sobolev space \(H^s\), i.e.
  \(\|fu\|_{H^s}\leq const. \|u\|_{H^s}\). [This is an immediate consequence of the
    product rule for the case \(s\in {\{1,2,3\dots\}}\), and follows in the negative integral case by duality and in the general case by interpolation.] Making use of the formula for the kernel
\begin{align}\label{fccc}
{(K^\theta)}^{\frac{r}{2}}(x,y)\,&=\,\theta^{\frac{r}{2}} {\tteo}(x) {\tteo}(y)\,+(3m^2)^{\frac{r}{2}} {\tteone}(x) {\tteone}(y)\,\\
&\quad\,+\,\frac{1}{2\pi}\,\int_{\smr}\,\notag
\Bigl[(-k^2+3imk\tmy+2m^2-3m^2\ssmy)e^{ik(x-y)}\\
&\qquad\qquad\qquad\qquad\qquad\qquad\qquad\qquad\times\frac{(-k^2-3imk\tmx+2m^2-3m^2\ssmx)}
{(k^2+m^2)(k^2+4m^2)^{1-\frac{r}{2}}}\Bigr]
\;dk\,,\notag
\end{align}
it is only necessary to consider the final integral, by the preceding remark.
By observation, this integral can be put in
the form
\(\sum_{j=0}^4\sum_{\alpha_j=1}^{N_j}\,
f^{\alpha_j}_j(x)g^{\alpha_j}_j(y)I_{j,1-\frac{r}{2}}(x-y)\),
where each \(N_j\in{\{1,2,3\dots\}}\), the functions \(\{f_j^{\alpha_j}, g_j^{\alpha_j}\}\) are all
smooth bounded functions, whose derivatives are in fact Schwartz functions
and the \(I_{j,1-\frac{r}{2}}(z)\) are as defined in \eqref{ijgen} with
\((a,b)=({j,1-\frac{r}{2}})\). Again making
use of the remark above, the result is  consequence of the fact that
 for each \(j\in\{0,1,2,3,4\}\),
the pseudo-differential operator \((-i\partial)^j(m^2-\partial^2)^{-1}
(4m^2-\partial^2)^{-1+\frac{r}{2}}\), whose integral kernel is \(I_{j,1-\frac{r}{2}}\)
is bounded \(H^s\to H^{s-r}\). The first statement of the
proposition is proved similarly, but requires \(r\geq 0\) because \(K\) has a kernel.\qed

The regularization induced by smoothing of the field operators
as in \eqref{regf} leads to the following regularization of functions \(f\) of the operator \(K\),
under the assumption of regularity of \(f\) at zero:
\begin{align}
  \label{regfc}
  f(K)_\kappa(x,y)\,&=
  \,f(0) {\tteokap}(x) {\tteokap}(y)
  \,+\,f(3m^2) {\tteonekap}(x) {\tteonekap}(y)\\
  &\qquad+
\,\frac{1}{2\pi}\,\iiint_{\smr\times\smr\times\smr}\,
f(k^2+4m^2){\delta^{[\kappa]}}(x-x')e_k(x'{}){e_{-k}}(y'{})
{\delta^{[\kappa]}}(y-y')
\;dk\,dx'\,dy'\,,
\notag\end{align}
with \(\tteokap={\tteo}*{\delta^{[\kappa]}}\) and \(\tteonekap={\tteone}*{\delta^{[\kappa]}}\).

Finally, to handle the soliton centred at \(\xi\in\R\) we replace the operator
\(K=K(0)=(-\partial_x^2+4m^2-6m^2\ssmx)\) by \(K(\xi)=(-\partial_x^2+4m^2-6m^2\ssmxi)\) in these considerations, and
everything generalizes in the obvious way to produce discrete eigenfunctions \(\{ e_{j\xi}= e_j(\cdot-\xi)\}_{j=1,2}\) and Jost eigenfunctions \(e_{k\xi}={\ttF}(k,x;\xi)e^{ikx}/{(k+im)(k+2im)} \), with \({\ttF}\) as in \eqref{genf}.
The Jost eigenfunctions are fixed by the requirement that they all have the same
asymptotic behaviour as \(e_k\), namely: \(e_{k\xi}(x)=e^{ikx}+O(e^{-m|x|}) \) as \(x\to\infty\).
In the notation \(e_{-k\xi}\) the
\(-\) applies to \(k\), i.e., \(e_{-k\xi}(x)={\ttF}(-k,x;\xi)e^{-ikx}/{(k-im)(k-2im)} \). 
For arbitrary \(\xi\) the case \(r=1\) in \eqref{fccc} is of particular utility, in which case referring to \eqref{defC02} we have
\begin{align}\notag
  \sqrt{K^\theta(\xi)}(x,y)\,&=\,\sqrt{\theta}\tteox(x)\tteox(y)+
  \sqrt{3m^2}\tteonex(x)\tteonex(y)+
  N^{(iv)}(x-y)\\
  &\;-N'''(x-y)\Bigl(3m\tmxi-3m\tmxiy\Bigr)\notag\\&\quad+N''(x-y)
  \Bigl(4m^2-3m^2\ssmxi-3m^2\ssmxiy\Bigr)\label{k12}\\
  &\quad\;-N'(x-y)\Bigl(6m^3(\tmxi-\tmxiy)\notag\\
  &\phantom{-N'(x-y)\Bigl(6m^3}-9m^3\bigl(\tmxi\ssmxiy-\tmxiy\ssmxi\bigr)\Bigr)\notag\\
&\qquad\;+N(x-y)\Bigl((2m^2-3m^2\ssmxi)(2m^2-3m^2\ssmxiy)
  \Bigr)\,.
  \notag\end{align}
  Referring to the discussion of the function \(N\) in \eqref{defC01}-\eqref{defC03} above,
  we see that the most singular term is
  \(N^{(iv)}(x-y)\), which is independent of \(\xi\), and this aside the strongest singularities are actually
  only logarithmic because the limit
  \[
\lim_{y\to x}\frac{\tmxi-\tmxiy}{x-y}=m\ssmxi
\]
exists. On differentiation with respect to \(\xi\) we obtain an expression of the form
\begin{align}\begin{split}
  &\partial_\xi\sqrt{K^\theta(\xi)}(x,y)\,=\,-\sqrt{\theta}(\tteox'(x)\tteox(y)+\tteox(x)\tteox'(y))
  -\sqrt{3m^2}(\tteonex'(x)\tteonex(y)+\tteonex(x)\tteonex'(y))+\Lambda_\xi(x,y)\\
  &\hbox{where}\quad\Lambda_\xi(x,y)=-3m^2N'''(x-y)\Bigl(\ssmxi-\ssmxiy\Bigr)
  +\sum_{j=0}^2N^{(j)}(x-y){\tilde f}_j(x-\xi,y-\xi)
   \end{split} \label{k123}
    \end{align}
where the \({\tilde f}_j\in C^\infty(\R^2)\) can be read off directly, and verify
\[
|{\tilde f}_j(x,y)|\leq const. (e^{-m|x|}+e^{-m|y|})\,,
\]
as do all their partial derivatives. From this we can deduce (using properties of \(N\) above)  that
    \(\partial_\xi\sqrt{K^\theta(\xi)}\) is Hilbert-Schmidt, and it varies with \(\xi\) continuously in the
    Hilbert-Schmidt norm.
\subsection{Wave Operators}\label{wavop}
We now summarize the scattering theory for the
operator \(K=-\partial_x^2+4m^2-6m^2\ssmx\). Theoretically this
falls under the framework for short range scattering developed in
\cite{MR751959}*{\S XI.4, and problem 44} or
\cite{MR2108588}*{Chapter XIV}. The differential operators \(K \) and
\(K_0=(-\partial_x^2+4m^2)\) extend to define unbounded
self-adjoint operators on \(L^2(\R) \), and there exist partial isometries
\({\mathfrak W}_\pm \) such that
\[
{\mathfrak W}_\pm\;u\,=\,\lim_{t\to\pm\infty}\,e^{itK}\,e^{-itK_0}\,u
\]
and
\[
{\mathfrak W}_\pm\;e^{isK_0}\,u\,=\,e^{isK}\,{\mathfrak W}_\pm\;u\,,
\]
for all \(u\in L^2(\R)\,. \) These are the wave operators and are
isometric from \(L^2(\R) \) onto the absolutely continuous subspace of
\(K \), which is the orthogonal complement of the linear span of the
two discrete eigenfunctions \( {\tteo} \) and \( {\tteone} \). The wave operators
can be represented explicitly using the distorted Fourier transform, a
representation which  we will now derive.

Introduce \({u}(x)=-6m^2\ssmx \) and
\(R_0(z)=(-\partial_x^2-z)^{-1} \) as, respectively,
notation for the potential induced by the kink, and for the free
resolvent. The resolvent is well-defined on the complement of
the non-negative real axis, and has the integral kernel
\begin{equation}\label{fres}
R_0(z)(x,y)\,=\,\frac{i}{2\sqrt{z}}\exp[i\sqrt{z}|x-y|]
  \end{equation}
where \(\sqrt{z} \) means the square root with \(\hbox{Im\,} \sqrt{z}>0 \),
so that in a neighbourhood of the positive real axis \(z=k^2>0 \)
there holds
\[
\sqrt{k^2\pm i\epsilon}\,=\,\pm\bigl(|k|\pm
\frac{i\epsilon}{2|k|}+O(\epsilon^2)\bigr)\,.
\]

Consider now the formulae for the generalized eigenfunctions
\(e_k(x) \) which were derived in the preceding section. The fact
that only \(e^{ikx} \) appears is a consequence of the fact that
the potential \({u}(x) \) is a reflectionless potential.
The wave operators are completely determined by the phase factors
\(e^{\pm i\delta_k}\,. \)

Taking the limit \(\epsilon\downarrow 0 \) leads to the introduction
of boundary values \(R_0^\pm(k^2) \) of the free resolvent on the
upper and lower sides of the positive axis and thence,
by a calculation similar to that in
\cite{MR2108588}*{Example 14.6.10}, we obtain the following result.
\begin{lemma}\label{lap}
  For \(k\in \R \) there holds
  \[
\bigl(1+R_0^+(k^2)V\bigr)e_k\,(x)\,=\,e^{ikx}\,,
\]
while for \(k\gtrless 0 \) there holds, respectively,
\[
\bigl(1+R_0^-(k^2)V\bigr)e_k\,(x)\,=\,\pm\,e^{\pm 2i\delta_k}\,e^{ikx}\,.
\]
\end{lemma}
As in the same reference we can now read off
the following formulae for the adjoints of the wave
operators:
\begin{align}\label{wop}
  {{\mathfrak W}}_+^*\;\Bigl(
\int_{\smr} f(k)e_k(x)\,dk
\Bigr)\,&=\,\int_{-\infty}^{+\infty} f(k)e^{ikx}\,dk\,\\
   {{\mathfrak W}}_-^*\;\Bigl(
\int_{\smr} f(k)e_k(x)\,dk
  \Bigr)\,&=\,\int_0^\infty f(k)e^{+2i\delta_k+ikx}\,dk\,+\,
  \int_{-\infty}^0 f(k)e^{-2i\delta_k+ikx}\,dk\,.
\end{align}

The scattering operator \(\hat{\mathfrak{S}}\define{\mathfrak W}_-^*\circ {\mathfrak W}_+ \) has the
effect:
\[
\int_0^\infty f(k)e^{-i\delta_k+ikx}\,dk\,+\,
\int_{-\infty}^0 f(k)e^{+i\delta_k+ikx}\,dk\,
\mapsto
\int_0^\infty f(k)e^{+i\delta_k+ikx}\,dk\,+\,
  \int_{-\infty}^0 f(k)e^{-i\delta_k+ikx}\,dk\,,
  \]
  which may be written in the alternative form
  \[
\hat{\mathfrak{S}}\;\Bigl(\int_\smr g(k)e^{ikx}\,dk\Bigr)
\,
=\,
\int_0^\infty g(k)e^{+2i\delta_k+ikx}\,dk\,+\,
  \int_{-\infty}^0 g(k)e^{-2i\delta_k+ikx}\,dk\,.
  \]

  \subsection{Time Evolution}\label{timevo}
  The classical equation \(\ddot y+Ky=0\) can be written in first order form
  as
  \begin{equation}
\frac{\partial}{\partial t}
\begin{pmatrix} y\\\dot y\end{pmatrix}\,=\,\begin{pmatrix}0&1\\-K&0\end{pmatrix}
\begin{pmatrix} y\\\dot y\end{pmatrix}\,.
  \label{fof}
    \end{equation}
  This generates a one parameter group of operators
  which are continuous on \(H^{s+1}(\R)\times H^s(\R)\), for
  any \(s\in\R\), and also on \(\cs(\R)\times \cs(\R)\).

  In terms of the
  eigenfunction expansion \eqref{gfe}, the general solution of this equation
  is
  \begin{align*}
  y(t,x)\,&=\,
  y_0(t) {\tteo}(x)+\frac{1}{\sqrt{2\omega^{}_d}}\bigl(y_d(t)
  +\overline{y}_d(t)\bigr) {\tteone}(x)+
  \frac{1}{\sqrt{2\pi}}\,\int\,\frac{1}{\sqrt{2\omega^{}_k}}\,
\bigl(y^{}_{k}(t)e_k(x)
+\overline{y}^{}_{k}(t){{e^{}_{-k}}}(x)\bigr)\,dk\\
  \dot y(t,x)\,&=\,
  \dot y_0(t) {\tteo}(x)-i\sqrt{\frac{\omega^{}_d}{2}}\bigl(y_d(t)
  -\overline{y}_d(t)\bigr) {\tteone}(x)+
  \frac{1}{\sqrt{2\pi}}\,\int\,-i\sqrt{\frac{\omega^{}_k}{2}}\,
\bigl(y^{}_{k}(t)e_k(x)
-\overline{y}^{}_{k}(t){{e^{}_{-k}}}(x)\bigr)\,dk\,,
  \end{align*}
  with \(\ddot y_0=0\) and \(\dot y_k=-i\omega_ky_k\) and similarly
  for \(y_d\). This transfers to give the time-dependent Heisenberg field in the
  soliton representation:
  \begin{align}
  {\upphi}^H(t,x)\,& =\,-\sqrt{\scm_{cl}}(X+vt) {\tteo}\bigl({}x{}\bigr)
  +\frac{1}{\sqrt{2\omega^{}_d}}
(\mathbf{a}^{}_{d}(t)+\mathbf{a}_{d}^\dagger(t)) {\tteone}\bigl({}x{}\bigr)                        \\
                   & \phantom{=\,\Phi_Sx-\xi-X {\tteo}x-\xi}
+\frac{1}{\sqrt{2\pi}}\,\int\,\frac{1}{\sqrt{2\omega^{}_k}}\,
\bigl(\mathbf{a}^{}_{k}(t)e_k\bigl({}x{}\bigr)
+\mathbf{a}_{k}^\dagger(t) {{e_{-k}}}\bigl({}x{}\bigr)\bigr)\,dk\,.\notag
  \end{align}
  Thus the time evolution of the creation and annihilation operators in the
  Heisenberg picture, which are written in boldface,
  is \(\mathbf{a}^{}_k(t)=a^{}_ke^{-it\omega^{}_k}\),
  \(\mathbf{a}^{\dagger}_k(t)=a^{\dagger}_ke^{+it\omega^{}_k}\) and similarly for
  the discrete mode. We can now give a description of the evolution determined by the semiclassical Hamiltonian
  \(\triplecolon H^{sol}_{0}\triplecolon\) on states of the form
  \(f(q)a_d^m\prod_j a(\chi_j)\Omega\)
  where \(f\) and \(\tilde\chi_j\) are Schwartz. (The \(\tilde\chi_j\)
  are distorted Fourier transforms,
  see \eqref{dft}, of Schwartz functions \(\chi_j\in\langle\{ {\tteo}, {\tteone}\}\rangle^\perp\).)
  Then if \(\Omega'\) is the vacuum in the Fock space \(\mfrF\), see \eqref{l2g}, we have
  \[
  \Exp[-it\triplecolon H^{sol}_{0}\triplecolon]f(q)(a_d^\dagger)^m\prod_j a(\chi_j)\Omega'
  \,=\,
  \psi(t,q)e^{imt\omega_d}(a_d^\dagger)^m\prod_j a^\dagger(e^{it\omega_\bullet^{}}\chi_j)\Omega'\,,
  \]
  where \(i\partial_t\psi(t,q)=-\frac{g^2}{2\scm_{cl}}\partial_q^2\psi(t,q)\) and
  \(\psi(0,q)=f(q)\). Explicitly
  \(
a(e^{it\omega_\bullet^{}}\chi_j)=\int a^\dagger_k(e^{i\omega_kt}\tilde\chi_j(k))\,dk\,.
\) (For purposes of comparison, the coordinate \(q\) is rescaled according to \(q=gQ\) in the main body of the paper.)

  For the purposes of
  quantization in the vacuum representation, it is useful to
  introduce \(\alpha=2^{-\frac{1}{2}}(K_0^\frac{1}{4}y+iK_0^{-\frac{1}{4}}\dot y)\) and
  its complex conjugate \(\overline{\alpha}\), in terms of which the
  evolution equation can be written
  \begin{equation}\label{timevoeq}
  \frac{d\eta}{dt}\,=\,\begin{pmatrix}-iK_0^\frac{1}{2}&0\\0&iK_0^\frac{1}{2}\end{pmatrix}\eta
  -\frac{i}{2}K_0^{-\frac{1}{4}}VK_0^{-\frac{1}{4}}\begin{pmatrix}{\; }1^{ }_{ }&{\;  }1^{ }_{ }\\-1^{ }_{ }&-1^{ }_{ }\end{pmatrix}\,\eta\,,
    \qquad \eta(t)=\begin{pmatrix}\alpha(t)\\\overline{\alpha}(t)\end{pmatrix}\,.
    \end{equation}
    (Here \(K_0\) and \({u}(x)=-6m^2\ssmx \) are as in the preceding appendix.)
    The solution of this can be written \(\eta(t)={\mathfrak u}(t-t_0)\eta(t_0)\) in terms of a
 \(2\times 2\) matrix of operators \({\mathfrak u}(t)\) whose entries
satisfy \({\mathfrak u}_{22}=\overline{\mathfrak u}_{11}\) and
\({\mathfrak u}_{12}=\overline{\mathfrak u}_{21}\), and which is pseudo-unitary in the sense that
\[
  {\mathfrak u}^*\begin{pmatrix}\id&0\\0&-\id\end{pmatrix}{\mathfrak u}=\begin{pmatrix}\id&0\\0&-\id\end{pmatrix}=
  {\mathfrak u}\begin{pmatrix}\id&0\\0&-\id\end{pmatrix}{\mathfrak u}^*\,.
    \]
    The \(\{{\mathfrak u}(t)\}_{t\in\smr}\) constitute a strongly continuous one parameter group
    of operators on \(L^2(\R;\C^2)\) which have operator norm
    \(\|{\mathfrak u}(t)\|_{L^2\to L^2}\in [e^{-L|t|},e^{+L|t|}]\) for some \(L>0\), and satisfy the usual differentiability properties for quite general \(V \)
    (see \cite{MR529429}*{Theorem XI.104}).
    For the case at hand, the presence of the zero mode \(K {\tteo}=0 \)
    shows that the bounds cannot be time-independent: there is a solution
    \(\eta_Z(t)={\mathfrak u}(t)iK_0^{-\frac{1}{4}} {\tteo}
    =tK_0^{\frac{1}{4}} {\tteo}+iK_0^{-\frac{1}{4}} {\tteo}\) growing in
    time (as well as the constant solution \(K_0^{\frac{1}{4}} {\tteo} \)); see
    remark \ref{zeromod}.      

    The group
    of operators
    \(\{{\mathfrak u}(t)\}\) induces an evolution of the quantum fields in the
    Heisenberg picture: the Heisenberg field at time \(t\) can be written
    \[\phisch^H(f,t)\,=\,\frac{1}{\sqrt{2}}\Bigl(
\mathbf{a}(K_0^{-\frac{1}{4}}f,t)+\mathbf{a}^\dagger(K_0^{-\frac{1}{4}}f,t)\Bigr)\,,
\quad
\pisch^H(f,t)\,=\,\frac{-i}{\sqrt{2}}\Bigl(
\mathbf{a}(K_0^{\frac{1}{4}}f,t)-\mathbf{a}^\dagger(K_0^{\frac{1}{4}}f,t)\Bigr)
\]
where \(f\in\cs(\R)\) and the evolution of the 
creation and annihilation operators is given 
in terms of \({\mathfrak u}(t)\):
\begin{align}
  \mathbf{a}(f,t)\,&=\,{a}({\mathfrak u}_{11}(t)^Tf)+
  {a}^\dagger({\mathfrak u}_{12}(t)^Tf)\,,\label{ha1}\\
    \mathbf{a}^\dagger(f,t)\,&=\,{a}(\overline{\mathfrak u}_{12}(t)^Tf)+
    {a}^\dagger(\overline{\mathfrak u}_{11}(t)^Tf)\,.\label{ha2}
  \end{align}
(Here the \(\overline{\,\cdot\,}\) means complex conjugate and \({}^T\) means
the transpose with respect to the bilinear form \((f,g)\mapsto (\overline{f},g)_{L^2}=\int fg\),
i.e. \(\int A^Tfg=\int f Ag\).)

In the presence of an electric field, i.e. for the classical evolution
\[
\frac{\partial}{\partial t}
\begin{pmatrix} y\\y_t\end{pmatrix}\,=\,\begin{pmatrix}0&1\\-K&0\end{pmatrix}
\begin{pmatrix} y\\y_t\end{pmatrix}\,+\,\begin{pmatrix}0\\\lambda{\elec}\end{pmatrix}\,,
  \]
  the corresponding formulae for the evolution of the creation annihilation operators in
  the Heisenberg picture is
\begin{align}
  \mathbf{a}(f,t)\,&=\,\mathbf{a}({\mathfrak u}_{11}(t,t_0)^Tf,t_0)+
  \mathbf{a}^\dagger({\mathfrak u}_{12}(t,t_0)^Tf,t_0)+g\,,\label{ha3}\\
    \mathbf{a}^\dagger(f,t)\,&=\,\mathbf{a}(\overline{\mathfrak u}_{12}(t,t_0)^Tf,t_0)+
    \mathbf{a}^\dagger(\overline{\mathfrak u}_{11}(t,t_0)^Tf,t_0)+\overline g\,,\label{ha4}\\
    g(t)\,&=\,\frac{i\lambda}{\sqrt{2}}\int_{t_0}^t\bigl({\mathfrak u}_{11}(t,s)-{\mathfrak u}_{12}(t,s)\bigr)
    \bigl((-\partial_x^2+4m^2)^{-\frac{1}{4}}{\elec}\bigr)\,ds\,.\label{ha5}
  \end{align}

    \begin{theorem}\label{segsha}
      The evolution of the field operators determined by \eqref{ha1}-\eqref{ha2}
      (resp. \eqref{ha3}-\eqref{ha5})
      is unitarily implementable on Fock space, i.e. there exists a family
      of evolution operators \(\{exp[-i(t-s)\doublecolon H^{sol}_0\doublecolon]\}\)
     (resp. \(\{{{\evot}_{scl}(t,s)}\}\))
      defined on \({\fock}\),
      which induce the above actions
      and map the
      finite particle space, in particular the Fock vacuum \(\Omega^{}_0\),
      into the subspace  \(\bigcap\limits_{s=1,2\dots}\Dom(\N_0^s)\subset{\fock}\)
      of smooth vectors for the number operator \(\N_0\).
      \end{theorem}
    \proof This is essentially a consequence of basic results on unitary implementability
    explained in \cite{MR1178936}*{Chapter 4} and \cite{MR0180855},\cite{MR0180856}, and \cite{MR0218081}.
    To obtain the precise statement we need on the smoothness of the transformed vacuum with
    respect to the number operator we take as starting point the discussion in 
    \cite{MR516713}
    and \cite{MR529429}*{\S XI.15}. In
    particular, see pages 313-314 of the latter reference for
    the unitary implementability of the finite time classical evolution operator
    for the charged case and
    \cite{MR516713} for a treatment of the neutral case appropriate to
    a real scalar field.
The verification of
the Hilbert-Schmidt hypothesis on \({\mathfrak u}_{11}^{-1}{\mathfrak u}_{12}\)
amounts to checking that the integral
    \[
\int_{\smr^2}\,|\hat V(k-l)|^2(k^2+4m^2)^{-\frac{1}{2}}(l^2+4m^2)^{-\frac{1}{2}}dkdl
\]
is finite, which certainly holds since \(\hat V\in\cs(\R)\). The statement that the
quantum flow maps the Fock vacuum into a smooth vector of \(\N_0\) is proved by consideration of the formula
\begin{equation}\label{putvac}
\Exp\bigl[-\frac{1}{2}\Lambda(a^\dagger,a^\dagger)\bigr]\Omega^{}_0\,.
\end{equation}
Here \(\Lambda\) is the bilinear form associated with the operator
\[
-\bigl({\mathfrak u}_{22}^{-1}{\mathfrak u}_{21}\bigr)=\bigl({\mathfrak u}_{22}^{-1}{\mathfrak u}_{21}\bigr)^T\,.
\]
The fact that this operator is equal to its transpose is a consequence of the pseudo-unitarity property
above, see \cite{MR516713}. It is shown on p. 122 of this reference that if
\(c_n=2^{-4n}(n!)^{-2}\|(\Lambda(a^\dagger,a^\dagger)^n\Omega^{}_0\|^2\)
then the radius of convergence of \(\sum c_nz^n\) is greater than one, so that
\(c_n\leq C^2e^{-2rn}\) for some positive \(C,r\). This implies that the formula \eqref{putvac}
defines a smooth vector for \(\N_0\) since \(\N_0^s(\Lambda(a^\dagger,a^\dagger))^n\Omega^{}_0
=(2n)^s(\Lambda(a^\dagger,a^\dagger))^n\Omega^{}_0\) whose square norm is bounded by
\(C^2|2n|^{2s}e^{-2rn}\) which is summable.

To extend this to the inhomogeneous case,
note firstly that since
\[
\sum_{j=0}^N (-1)^j\frac{a^\dagger(g)^j}{j!}
\sum_{n=0}^\infty\frac{(-1)^n}{2^nn!}\Lambda(a^\dagger,a^\dagger)^n\Omega^{}_0
\]
is a well-defined smooth vector for \(\N\) by the preceding discussion,
in order to prove that \eqref{putvac} also defines a
a smooth vector for \(\N\) it suffices to prove the convergence of
\[
\sum_{j=0}^\infty (-1)^j\frac{a^\dagger(g)^j}{j!}
\sum_{n=0}^\infty\frac{(-1)^n}{2^nn!}\Lambda(a^\dagger,a^\dagger)^n\Omega^{}_0
\,=\,\lim_{N\to\infty}
\sum_{j=0}^N (-1)^j\frac{a^\dagger(g)^j}{j!}
\sum_{n=0}^\infty\frac{(-1)^n}{2^nn!}\Lambda(a^\dagger,a^\dagger)^n\Omega^{}_0
\]
in the topology defined by the seminorms \(\|\N^s(\,\cdot\,)\|\).
Recalling that on an \(n\)-particle state the bound
\(\|a^\dagger(f)^j\|\leq\prod_{i=1}^j\sqrt{n+i}\|f\|^j\)
holds in operator norm, we deduce the convergence in the seminorm
\(\|\N^s(\,\cdot\,)\|\)
for each \(s\) from the fact that, by the Stirling approximation,
\[
\frac{1}{2}\sqrt{2\pi j}\bigl(\frac{j}{e}\bigr)^j\leq j!\leq{2}\sqrt{2\pi j}\bigl(\frac{j}{e}\bigr)^j
\,,\quad\hbox{for all}\;j\geq j_0\geq 1
\]
we obtain, using \((a+b)^p\leq 2^{p-1}(a^p+b^p)\) with \(p=s+j/2\),
\begin{align}
\sum_{j=j_0}^\infty\sum_{n=0}^\infty(2n+j)^{s}\prod_{i=0}^j\sqrt{2n+i}\frac{\sqrt{c_n}\|f\|^j}{j!}
&\leq\notag
const.\sum_{j=j_0}^\infty\sum_{n=0}^\infty e^{-rn}(2n+j)^{s+\frac{j}{2}}\|f\|^j(\frac{j}{e})^{-j}(2\pi j)^{-\frac{1}{2}}\\
&\leq
const.\sum_{j=j_0}^\infty\sum_{n=0}^\infty e^{-rn}2^{s+\frac{j}{2}-1}(2n)^{s+\frac{j}{2}}\|f\|^j(\frac{j}{e})^{-j}(2\pi j)^{-\frac{1}{2}}
\la{sum1}\\
&\quad+
const.\sum_{j=j_0}^\infty\sum_{n=0}^\infty e^{-rn}2^{s+\frac{j}{2}-1}j^{s+\frac{j}{2}}\|f\|^j(\frac{j}{e})^{-j}(2\pi j)^{-\frac{1}{2}}\,.
\la{sum2}\end{align}
\eqref{sum2} is convergent since it is dominated by \(\sum_j\sum_n e^{-rn}j^{-\frac{1}{2}}(const.(s))^je^{-\frac{1}{2}jln j}<\infty\).
To prove convergence of \eqref{sum1}, note that
\[\sum_{n=0}^\infty e^{-rn}n^{s+\frac{j}{2}}\leq\sum_{n=0}^\infty\int_{n}^{n+1} e^{-r(t-1)}t^{s+\frac{j}{2}}dt
=\int_0^\infty t^{s+\frac{j}{2}}e^{-r(t-1)}dt=e^rr^{s+\frac{j}{2}+1}\Gamma\bigl(s+\frac{j}{2}+1\bigr)\]
so that again using the Stirling approximation, it is bounded by
\begin{align*}
&\sum_{j\geq j_0}\bigl(\frac{s+\frac{j}{2}}{j}\Bigr)^{\frac{1}{2}}(const.(s))^j\bigl(\frac{s+\frac{j}{2}}{e}\bigr)^{s+\frac{j}{2}}\bigl(\frac{j}{e}\bigr)^{-j}\\
&  \leq
\sum_{j\geq j_0}\bigl(\frac{s+\frac{j}{2}}{j}\Bigr)^{\frac{1}{2}}(const.(s))^j2^{s+\frac{j}{2}-1}\bigl(s^{s+\frac{j}{2}}+(\frac{j}{2})^{s+\frac{j}{2}}\bigr)e^{-s-\frac{j}{2}}\bigl(\frac{j}{e}\bigr)^{-j}
\\
&\leq\sum_{j\geq j_0}(const.'(s))^jj^{s}{e^{\frac{j}{2}\ln(j)}}e^{-j\ln j}<\infty\,.
\end{align*}
This establishes that the formula \eqref{putvac} defines a smooth vector for the number operator.

\begin{remark}\label{zeromod}
  The presence of the zero mode shows itself in the possibility of growth in the norm of operators such as \(  \mathbf{a}^\dagger(f,t)(\N_0+1)^{-\frac{1}{2}} \), where
  \[
    \mathbf{a}^\dagger(f,t)\,=\,
  \Exp[+it\doublecolon H^{sol}_{0}\doublecolon ]
  \mathbf{a}^\dagger(f,0)
  \Exp[-it\doublecolon H^{sol}_{0}\doublecolon ]\,,
  \]
  in contrast to the boundedness in the vacuum case which is an immediate
  consequence of \eqref{useme} and \eqref{best}. Indeed taking the inner
  product of \eqref{fof} with \((0, {\tteo})^T\) and
  \(( {\tteo},(t_0-t) {\tteo})^T\) leads to the identities
  \[
  \uppi^H( {\tteo},t_0)=\uppi^H( {\tteo},0)\quad\hbox{and}\quad
  \upphi^H( {\tteo},t_0)=  \upphi^H( {\tteo},0)+t_0\uppi^H( {\tteo},0)
  \]
  and hence
  \[
    \mathbf{a}^\dagger(K_0^{-\frac{1}{4}} {\tteo},t_0)\Omega_0 \,=\,
    \mathbf{a}^\dagger(K_0^{-\frac{1}{4}} {\tteo},0)\Omega_0 
 +it_0\mathbf{a}^\dagger(K_0^{+\frac{1}{4}} {\tteo},0)\Omega_0 
\,.
\]
From this it follows that
\[
\lim_{t_0\to\infty}\frac{\|    \mathbf{a}^\dagger(K_0^{-\frac{1}{4}} {\tteo},t_0)\Omega_0\|}{t_0}
\,=\,
\|    \mathbf{a}^\dagger(K_0^{\frac{1}{4}} {\tteo},0)\Omega_0\|\,.
\]
  \end{remark}

\section{Appendix: proofs of some technical results}\label{arh}
\subsection{Cameron-Martin shift on Wick polynomials}Working in the Schr\"odinger representation
the Cameron-Martin shift \(\bfdel_g:\phisch\to\phisch+g\) induces an equivalent measure and is unitarily
implementable  by an operator \(\mbU(g)\) on \(L^2(\mu_0)\), see\eqref{rndis} and \eqref{displace}. In terms
of the collection of all measurable functions, the Cameron-Martin shift gives rise to an automorphism
which preserves the Wick product, see \cite{MR1474726}*{Theorem 14.1} for this perspective. Now the formula
\eqref{rndis} makes it clear that \(\mbU(g)\) preserves the subspace \(L^{2+}=\cup_{q>2}L^q\).
This latter subspace serves as
a domain for the Wick monomials \(\doublecolon\int {\beefb}(x)\phisch(x)^n\,dx\doublecolon\), regarded as unbounded operators
on \(L^2(\mu_0)\), and this is the meaning to be attached to the formula
\begin{equation}\label{wwu}
\mbU(g)\circ\,\doublecolon\int b(x)\phisch(x)^n\,dx\doublecolon\,\circ\mbU(g)^*=
\doublecolon\int b(x)(\phisch(x)+g(x))^n\,dx\doublecolon
=\sum \binom{n}{j}\doublecolon\int b(x)g(x)^{n-j}\,\phisch(x)^j\,dx\doublecolon\,,
\end{equation}
i.e., it holds as an operator identity on \(L^{2+}\).
To prove this formula, use a succession of approximations to reduce to the case of a linear combination of
Wick powers \(\doublecolon \phisch(f)^n\doublecolon\) for which the result holds by \cite{MR1474726}*{Theorem 14.1}.
Firstly, using the regularized Schr\"odinger field
\(\phisch_\kappa\), the continuity of \(x\mapsto \phisch_\kappa(x)\in L^p(\mu_0)\,,p<\infty\) allows one to approximate
\(\int_\alpha^\beta\doublecolon\phisch_\kappa(x)^j\doublecolon\,dx\) by a finite Riemann sum
\(\sum b_i\doublecolon\phisch_\kappa(x_i)^j\doublecolon\), and hence to
deduce from the foregoing reference that
\begin{equation}
\mbU(g)\circ\,\doublecolon\int_\alpha^\beta\phisch_\kappa(x)^n\,dx\doublecolon\,\circ\mbU(g)^*
=\sum \binom{n}{j}\doublecolon\int_\alpha^\beta g(x)^{n-j}\,\phisch_\kappa(x)^j\,dx\doublecolon\,.
\end{equation}
Approximating \({\beefb}\) in \(L^1(\R,dx)\) by a finite linear combination of indicator functions of intervals, and using the
Minkowski inequality for \(L^p(\mu_0)\) valued functions of \(x\in\R\), this gives
\eqref{wwu} with \(\phisch\) replaced by \(\phisch_\kappa\).
Finally, using results from \cite{MR0674511}*{\S5}, one can take the limit \(\kappa\to+\infty\) in
any \(L^p(\mu_0)\,,p<\infty\) to obtain \eqref{wwu}. 
\subsection{The Hamiltonian under displacement}\label{dissint}
Comparison of the expansion of classical Hamiltonian functional \eqref{ham} about either
\(\Phi_S\) or \(\Phi_{S\xi}=\Phi_S(\cdot-\xi)\) implies the identity
\begin{align}\notag
&\int \Bigl(\frac{1}{2}\varphi K(0)\varphi-\varphi K(0)\delta_\xi\Phi_S
+\frac{1}{2}\delta_\xi\Phi_SK(0)\delta_\xi\Phi_S+\frac{1}{3!}\cU^{(iii)}(\Phi_S)(\varphi-\delta_\xi\Phi_S)^3
+\frac{1}{4!}\cU^{(iv)}(\Phi_S)(\varphi-\delta_\xi\Phi_S)^4\Bigr)\,dx\\
&\qquad=\int\Bigl(
\frac{1}{2}\varphi K(\xi)\varphi+\frac{1}{3!}\cU^{(iii)}(\Phi_{S\xi})\varphi^3+\frac{1}{4!}\cU^{(iv)}(\Phi_{S\xi})\varphi^4\Bigr)\,dx
\,,\qquad\qquad \delta_\xi\Phi_{S}=\Phi_{S}-\Phi_S(\cdot-\xi)\,.
\notag
\end{align}
This identity holds for all \(C^1\) functions \(\varphi\) for which the relevant integration by parts is allowed,
but since this is always against exponentially decreasing functions, the identity holds under a condition of polynomial growth for \(\varphi\).
In particular, this is true for the smoothed Schr\"odinger fields \(\phisch_\kappa\),
obtained as in \eqref{regf}, since the measure is defined on tempered distributions and the
regularizations \eqref{regf} of tempered distributions have polynomial growth.
Recalling the definition \eqref{hintxi}
and rearranging, the above identity for \(\phisch_\kappa\) implies
\begin{align}\notag
&\int \Bigl(\frac{1}{2}\phisch_\kappa K(0)\phisch_\kappa-\phisch_\kappa(K(0)\delta_\xi\Phi_S)
  +\frac{1}{2}\delta_\xi\Phi_SK(0)\delta_\xi\Phi_S\Bigr)\,dx\,+\,
H^{sol}_{I,g,0,{\beefb}}(\phisch_\kappa-\delta_\xi\Phi_S)\\
 &\qquad=\int\,
\frac{1}{2}\phisch_\kappa K(\xi)\phisch_\kappa\,dx\,+
H^{sol}_{I,g,\xi,{\beefb}}(\phisch_\kappa)+H^{sol}_{I,g,\xi,1-{\beefb}}(\phisch_\kappa)-H^{sol}_{I,g,0,1-{\beefb}}(\phisch_\kappa-\delta_\xi\Phi_S)\,.
\label{tri}
\end{align}
Think of this as an identity between operators acting on the Hilbert space \(L^2(\mu_0)\). 
Substituting \(K(0)=K_0+u\) and \(K(\xi)=K_0+u_\xi\) we observe that the \(K_0\) pieces on
both sides are identical so can be replaced by \(  \doublecolon {\pmb H}^{vac}_{0}\doublecolon\).
Adding and subtracting appropriately we obtain the operator identity
\begin{align}\notag
  \doublecolon {\pmb H}^{vac}_{0}\doublecolon-
  &\frac{1}{2}\doublecolon \int\,6m^2\,\ssmx\,\phisch(x)^2\,dx\doublecolon
    +H^{sol}_{I,g,0,{\beefb}}(\phisch_\kappa-\delta_\xi\Phi_S)
+  \int \Bigl(-\phisch_\kappa K(0)\delta_\xi\Phi_S
  +\frac{1}{2}\delta_\xi\Phi_SK(0)\delta_\xi\Phi_S\Bigr)\,dx
  \\\notag
  &\qquad=
  \doublecolon \pmb{H}^{vac}_{0}\doublecolon
-
  \frac{1}{2}\doublecolon \int\,6m^2\,\ssmxi\,\phisch(x)^2\,dx\doublecolon
  +
 H^{sol}_{I,g,\xi,{\beefb}}(\phisch_\kappa)
\\
&\qquad\qquad+\frac{1}{2} \int\,6m^2\,(\ssmxi-\ssmx)\,(\doublecolon\phisch(x)^2\doublecolon-\phisch_\kappa(x)^2)\,dx
\\
&\qquad\qquad\qquad+H^{sol}_{I,g,\xi,1-{\beefb}}(\phisch_\kappa)-H^{sol}_{I,g,0,1-{\beefb}}(\phisch_\kappa-\delta_\xi\Phi_S)\,.
\notag
\end{align}
It is now possible to normal order this identity (still valid)
and take the limit \(\kappa\to+\infty\), using the
results on convergence of Wick powers in \cite{MR0674511}*{\S5}, as above.
The final two terms can be worked out explicitly to reveal that 
the coefficients of the various powers of \(\phisch_\kappa\) are of the form of \(1-{\beefb}(x)\) multiplied by a Schwartz function of
exponential decrease, allowing us to take the limit \(\kappa\to+\infty\) to obtain the formula \eqref{deferrir} for the error term
induced by infra-red cutoff in terms of the Schr\"odinger fields \(\phisch\)
in the vacuum representation. The penultimate line has limit zero. To express
the overall conclusion, recall from Theorem \ref{sadj2} the operator \(\doublecolon {\pmb H}^{sol}_{0}\doublecolon\)
obtained by quantization of
the quadratic part of the Hamiltonian using the Schr\"odinger representation; similarly linearising about
a soliton located at \(\xi\in\R\) as in \eqref{arbloc}, yields the corresponding operator
\(\doublecolon {\pmb H}^{sol}_{0\xi}\doublecolon\). With these definitions the limit of the
preceding calculations leads to
\begin{equation}
  \doublecolon {\pmb H}^{sol}_{0}\doublecolon
-
\phisch(K(0)\delta_\xi\Phi_S)
+\frac{1}{2}\int\delta_\xi\Phi_SK(0)\delta_\xi\Phi_S\,dx
  +\doublecolon H^{sol}_{I,g,0,{\beefb}}(\phisch-\delta_\xi\Phi_S)\doublecolon
  =\doublecolon {\pmb H}^{sol}_{0\xi}\doublecolon+
\doublecolon H^{sol}_{I,g,\xi,{\beefb}}(\phisch)\doublecolon
+Err^0_{\mbox{\tiny IR}}\,,
\label{trif}
\end{equation}
which, with reference to \eqref{araki},  reads as
\begin{equation}
\mbU(\delta_\xi\Phi_S)^*\Bigl(  \doublecolon {\pmb H}^{sol}_{0}\doublecolon
  +\doublecolon H^{sol}_{I,g,0,{\beefb}}(\phisch)\doublecolon\Bigr)\mbU(\delta_\xi\Phi_S)
  =\doublecolon {\pmb H}^{sol}_{0\xi}\doublecolon+
\doublecolon H^{sol}_{I,g,\xi,{\beefb}}(\phisch)\doublecolon
+Err^0_{\mbox{\tiny IR}}\,,
\label{trif2}
\end{equation}
the infrared error term being as given in \S\ref{repcalc}.
\subsection{Proof of Lemma \ref{altregham}}
We start with the formula
\begin{align*}
\bigl((K\varphi)_\kappa-K\varphi_\kappa\bigr)\,&=\,
\frac{1}{\sqrt{2\pi}}\,\int\frac{-6m^2}{\sqrt{2\omega^{}_k}}\,\int
{\delta^{[\kappa]}}(z)\bigl(
\ssm (x-z{})-\ssmx\bigr)
\bigl(a^{}_ke^{ik(x-z){}}+a_k^\dagger e^{-ik(x-z){}}\bigr)
\,dz\,dk\\
&\,=\,
\frac{-2m/\kappa}{\sqrt{2\pi}}
\int\frac{-6m^2}{\sqrt{2\omega^{}_k}}\,\int
\,z'{\delta^{[1]}}(z')\,
\int_0^1\,\ssm(x-z'\theta/\kappa{})\\
&\phantom{
\,=\,
\frac{-2m/\kappa}{\sqrt{2\pi}}
\int\frac{-6m^2}{\sqrt{2\omega^{}_k}}\,a^{}_k
}
\times\,\tanh m(x-z'\theta/\kappa{})
\bigl(a^{}_ke^{ik(x-z'/\kappa){}}+a_k^\dagger e^{-ik(x-z'/\kappa){}}\bigr)
\,d\theta\,dz'\,dk\,.
\end{align*}
(In deriving this, various terms do drop out due to
the fact that differentiation does commute with convolution, so only
the final \(-6m^2\ssmx \) in \eqref{def-k} causes error terms.)

A typical term in \(\int\,\varphi_\kappa\bigl((K\varphi)_\kappa
-K\varphi_\kappa\bigr)\,dx \)
is that involving two annihilation operators; it can be
written
\begin{align*}
\frac{1}{2\pi}\,\int\int\int\int\int
\frac{6m^3}{\kappa\sqrt{\omega^{}_k\omega^{}_l}}
a^{}_l\,a^{}_k\,z'{\delta^{[1]}}(z'){\delta^{[1]}}(w')&
\,\int_0^1\,\ssm(x-\theta z'/\kappa{})
\tanh m(x-\theta z'/\kappa{})\,d\theta\,
\\
&\times\;
\exp[ik(x-z'/\kappa{})]\exp[il(x-w'/\kappa{})]\,
dw'\,dz'\,dk\,dl\,dx\,.
\end{align*}
Now to show that has limit zero as a bilinear form on
\({\Poly(\varphi)}\times{\Poly(\varphi)} \), take a matrix element between two vectors
in \({\Poly(\varphi)} \), leading to an integral
\begin{align*}
\frac{1}{2\pi}\,\int\int\int\int\int
\frac{6m^3}{\kappa\sqrt{\omega^{}_k\omega^{}_l}}
F(k,l)\,z'{\delta^{[1]}}(z'){\delta^{[1]}}(w')&
\,\int_0^1\,\ssm(x-\theta z'/\kappa{})
\tanh m(x-\theta z'/\kappa{})\,d\theta\,
\\
&\times\;
\exp[ik(x-z'/\kappa{})]\exp[il(x-w'/\kappa{})]\,
dw'\,dz'\,dk\,dl\,dx\,,
\end{align*}
with \(F\in\cs(\R^2)\), which is \(O(\kappa^{-1}) \).

The same holds for all the terms arising except for the
one involving \(a^{}_la_k^\dagger \), which is not normal ordered. This
term can be normal ordered, and then the previous argument does apply but
at the expense of an additional c-number term, which is given by
\begin{align*}
Err_0\,=\,\frac{1}{2\pi}\,\int\int\int\int
\frac{6m^3}{\kappa\omega^{}_k}
\,z'{\delta^{[1]}}(z'){\delta^{[1]}}(w')
\,\int_0^1\,\ssm(x&-\theta z'/\kappa{})
\tanh m(x-\theta z'/\kappa{})\,d\theta\,
\\
&\times\;
\exp[-ik(x-z'/\kappa{})]\exp[ik(x-w'/\kappa{})]\,
dw'\,dz'\,dk\,dx\,.
\end{align*}
We bound the inner \(w' \) integral
as
\[
\sup_{x,\xi,k}\,\Bigl|
\int
{\delta^{[1]}}(w')
\exp[ik(x-w'/\kappa{})]\,
dw'
\Bigr|\,\leq\,const.\,\frac{\int(|{\delta^{[1]}}|+|{\delta^{[1]}}'|)dw'}{(1+|k|/\kappa)}\,,
\]
and thence bound
\[
|Err_0|\,\leq\,\frac{C}{\kappa}\,\int\frac{1}{\omega^{}_k}\frac{1}{(1+|k|/\kappa)}\,dk\,\leq\,
{C'}\,\int_0^\infty\frac{1}{(2m+k)(\kappa+k)}\,dk\,
=C'\,\frac{\ln\kappa/(2m)}{\kappa-2m}
=O\Bigl(\frac{\ln\kappa}{\kappa}\Bigr)\,.\qedhere
\]
\subsection{Proof of Lemma \ref{crf}}
The first assertion of (b) is proved in 
\cite{MR0674511}*{Section 5}. The second can be proved by a modification
of that argument as follows. After a change of variables
\(2\pi\iint
\doublecolon \phisch_\kappa(x)
{\delta^{[\kappa]}}(x-x')\ssmx'\phisch(x')
 \doublecolon \,dx'dx
 \)
 can be written as
 \[
 \int\int\int\int\int
 \frac{{\delta^{[\kappa]}}(w){\delta^{[\kappa]}}(z)}{2\sqrt{\omega_k\omega_l}}
 e^{i(k+l)x-ilw-ikz}\ssm (x-z)
 \doublecolon (a_l+a^\dagger_{-l})(a^{}_k+a^\dagger_{-k})dzdwdx\,
 \doublecolon (a_l+a^\dagger_{-l})(a^{}_k+a^\dagger_{-k})\doublecolon\, dkdl\,.
 \]
 Thus the inner \(dzwdx \) integral determines the kernel whose \(L^2 \)
 properties determine the required \(L^p(d\mu_0) \) properties according
 to \cite{MR0674511}*{Section 5}.
 The \(dw \) integral just gives the Fourier transform
 \(\widehat{\delta^{[1]}}(l/\kappa) \). The \(z \) integral is a convolution
 of the function \(\delta^{[\kappa]}(z)e^{-ikz}\) with the function
\(h(x)\define\ssmx \). Noting that the Fourier transform of the
former function is just \(\widehat{\delta^{[1]}}((\cdot+k)/\kappa) \),
the convolution theorem implies that the \(x,z \) integral
gives \(\hat h(-(k+l))\widehat{\delta^{[1]}}(-l/\kappa)\). Thus all together
we are left with
\[
\iint
\frac{\widehat{\delta^{[1]}}(l/\kappa)\hat h(-(k+l))\widehat{\delta^{[1]}}(-l/\kappa)}
     {2\sqrt{\omega_k\omega_l}}
      \doublecolon (a_l+a^\dagger_{-l})(a^{}_k+a^\dagger_{-k})dzdwdx\,
 \doublecolon (a_l+a^\dagger_{-l})(a^{}_k+a^\dagger_{-k})\doublecolon\,
\,dkdl\,.
\]
Since the function \(\frac{\hat h(-(k+l))}{{2\sqrt{\omega_k\omega_l}}}\) is
square integrable, the dominated convergence theorem implies that this
kernel converges to \(\frac{\hat h(-(k+l))}{{2\sqrt{\omega_k\omega_l}}}\) in
\(L^2 \) as \(\kappa\to+\infty \), and hence the results follows by
\cite{MR0674511}*{Theorem 5.7}.
The same calculation applied in the Fock space implies statement (a), 
via Theorem 4.2 in the same reference.
\subsection{Proof of Lemma \ref{zpd}}
We compute
\begin{align*}\int\,\Bigl[
  6m^2&\gamma_\kappa\,\ssmx\,-\,6m^2\iiint
  \frac{e^{ik(x'-y)}}{2(k^2+4m^2)^{\frac{1}{2}}}\delta^{[\kappa]}(x-y)\delta^{[\kappa]}(x'-x)\ssm x'dx'dydk
    \Bigr]\,dx\\
  &=3m^2\int\,\Bigl[\iiint
  \frac{e^{ik(x'-y)}}{(k^2+4m^2)^{\frac{1}{2}}}\delta^{[\kappa]}(x-y)\delta^{[\kappa]}(x'-x)\bigl(\ssmx-\ssm x'\bigr)dx'dydk
  \Bigr]\,dx\\
  &=3m^2\int\,\Bigl[\iiint
  \frac{e^{ik(w-z)/\kappa}}{(k^2+4m^2)^{\frac{1}{2}}}\delta^{[1]}(z)\delta^{[1]}(w)\bigl(\ssmx-\ssm (x+w/\kappa)\bigr)dzdwdk
  \Bigr]\,dx\\
  &=3m^2\iiint
  \frac{\widehat{\delta^{[1]}}(k/\kappa)e^{ikw/\kappa}}{(k^2+4m^2)^{\frac{1}{2}}}\delta^{[1]}(z)\delta^{[1]}(w)\bigl(\ssmx-\ssm (x+w/\kappa)\bigr)dwdkdx\\
  &\leq
  \frac{const.}{\kappa}\int \frac{|\widehat{\delta^{[1]}}(k/\kappa)|}{(k^2+4m^2)^{\frac{1}{2}}}
  dk\,=\,O\bigl(\frac{\ln\kappa}{\kappa}\bigr)\,.
\end{align*}

\small
\section*{Acknowledgements}
This work has been partially supported by STFC consolidated grant ST/P000681/1 and St John’s College, Cambridge. It was completed while visiting the Institute for Analysis, Leibniz University,
Hannover and the author thanks Elmar Schrohe for hospitality.
\setcounter{equation}{0}

\small
\baselineskip=13pt

\begin{bibdiv}
\begin{biblist}

\bib{MR0240836}{book}{
   author={Abraham, Ralph},
   author={Robbin, Joel},
   title={Transversal mappings and flows},
   note={An appendix by Al Kelley},
   publisher={W. A. Benjamin, Inc., New York-Amsterdam},
   date={1967},
   pages={x+161},
   review={\MR{0240836}},
}
  
  \bib{MR0167642}{book}{
   author={Abramowitz, Milton},
   author={Stegun, Irene A.},
   title={Handbook of mathematical functions with formulas, graphs, and
   mathematical tables},
   series={National Bureau of Standards Applied Mathematics Series},
   volume={55},
   publisher={For sale by the Superintendent of Documents, U.S. Government
   Printing Office, Washington, D.C.},
   date={1964},
   pages={xiv+1046},
   review={\MR{0167642}},
}

\bib{andrews}{article}{
author={Andrews,Mark },
title={The evolution of free wave packets},
journal={American Journal of Physics},
volume={76},
number={12},
pages={1102-1107},
year={2008},
doi={10.1119/1.2982628},
}

\bib{MR0127821}{article}{
   author={Araki, H.},
   title={Hamiltonian formalism and the canonical commutation relations in
   quantum field theory},
   journal={J. Mathematical Phys.},
   volume={1},
   date={1960},
   pages={492--504},
   issn={0022-2488},
   review={\MR{0127821}},
   doi={10.1063/1.1703685},
}
  
\bib{MR1178936}{book}{
      author={{Baez, John C.}},
      author={{Segal, Irving E.}},
      author={{Zhou, Zheng-Fang}},
       title={Introduction to algebraic and constructive quantum field theory},
      series={Princeton Series in Physics},
   publisher={Princeton University Press, Princeton, NJ},
        date={1992},
        ISBN={0-691-08546-3},
      review={\MR {1178936}},
}

\bib{MR0496006}{article}{
      author={B\'{e}llissard, J.},
      author={Fr\"{o}hlich, J.},
      author={Gidas, B.},
       title={Soliton mass and surface tension in the {$(\lambda \mid \phi \mid
  ^{4})_{2}$} quantum field model},
        date={1978},
        ISSN={0010-3616},
     journal={Comm. Math. Phys.},
      volume={60},
      number={1},
       pages={37\ndash 72},
         url={http://projecteuclid.org/euclid.cmp/1103904021},
      review={\MR{0496006}},
}

\bib{MR1642391}{book}{
      author={{Bogachev, Vladimir I.}},
       title={Gaussian measures},
      series={Mathematical Surveys and Monographs},
   publisher={American Mathematical Society, Providence, RI},
        date={1998},
      volume={62},
        ISBN={0-8218-1054-5},
      review={\MR {1642391}},
}

\bib{MR2098271}{book}{
   author={Bourbaki, Nicolas},
   title={Integration. II. Chapters 7--9},
   series={Elements of Mathematics (Berlin)},
   note={Translated from the 1963 and 1969 French originals by Sterling K.
   Berberian},
   publisher={Springer-Verlag, Berlin},
   date={2004},
   pages={viii+326},
   isbn={3-540-20585-3},
   review={\MR{2098271}},
}
		

\bib{MR2018901}{book}{
   author={Bourbaki, Nicolas},
   title={Integration. I. Chapters 1--6},
   series={Elements of Mathematics (Berlin)},
   note={Translated from the 1959, 1965 and 1967 French originals by
   Sterling K. Berberian},
   publisher={Springer-Verlag, Berlin},
   date={2004},
   pages={xvi+472},
   isbn={3-540-41129-1},
   review={\MR{2018901}},
}
		


\bib{cole}{book}{
      author={{S. Coleman}},
       title={Aspects of symmetry},
   publisher={Cambridge University Press, Cambridge},
        date={1985},
}


\bib{Dashen}{article}{
      author={{R.F. Dashen, B. Hasslacher and A. Neveu}},
       title={Nonperturbative methods and extended hadron models in field
  theory, parts i-iii},
        date={1974},
     journal={Phys.Rev.},
      volume={D10},
       pages={4114\ndash 4142},
}


\bib{MR0588684}{article}{
   author={Hagedorn, George A.},
   title={A time dependent Born-Oppenheimer approximation},
   journal={Comm. Math. Phys.},
   volume={77},
   date={1980},
   number={1},
   pages={1--19},
   issn={0010-3616},
   review={\MR{0588684}},
}

\bib{MR0674511}{incollection}{
      author={{Glimm, J.}},
      author={{Jaffe, A.}},
       title={Boson quantum field models},
        date={1972},
       pages={77\ndash 143},
      review={\MR {0674511}},
}

\bib{MR0247845}{article}{
      author={{Glimm, James}},
      author={{Jaffe, Arthur}},
       title={A $\lambda \phi ^{4}$ quantum field without cutoffs. I},
        date={1968},
     journal={Phys. Rev. (2)},
      volume={176},
       pages={1945\ndash 1951},
      review={\MR {0247845}},
}

\bib{MR0282243}{article}{
      author={{Glimm, James}},
      author={{Jaffe, Arthur}},
       title={Singular perturbations of selfadjoint operators},
        date={1969},
        ISSN={0010-3640},
     journal={Comm. Pure Appl. Math.},
      volume={22},
       pages={401\ndash 414},
      review={\MR {0282243}},
}

\bib{MR810217}{book}{
      author={{Glimm, James}},
      author={{Jaffe, Arthur}},
       title={Quantum field theory and statistical mechanics},
   publisher={Birkh\"auser Boston, Inc., Boston, MA},
        date={1985},
        ISBN={0-8176-3275-1},
        note={Expositions; Reprint of articles published 1969--1977},
      review={\MR {810217}},
}

\bib{MR887102}{book}{
      author={{Glimm, James}},
      author={{Jaffe, Arthur}},
       title={Quantum physics},
     edition={2},
   publisher={Springer-Verlag, New York},
        date={1987},
        ISBN={0-387-96476-2},
        note={A functional integral point of view},
      review={\MR {887102}},
}

\bib{MR0246142}{book}{
      author={{Gohberg, I. C.}},
      author={{Kre\u \i n, M. G.}},
       title={Introduction to the theory of linear nonselfadjoint operators},
      series={Translated from the Russian by A. Feinstein. Translations of
  Mathematical Monographs, Vol. 18},
   publisher={American Mathematical Society, Providence, R.I.},
        date={1969},
      review={\MR {0246142}},
}

\bib{MR2108588}{book}{
      author={{H\"ormander, Lars}},
       title={The analysis of linear partial differential operators. ii},
      series={Classics in Mathematics},
   publisher={Springer-Verlag, Berlin},
        date={2005},
        ISBN={3-540-22516-1},
        note={Differential operators with constant coefficients; Reprint of the
  1983 original},
      review={\MR {2108588}},
}

\bib{MR0503137}{article}{
   author={Jackiw, R.},
   title={Quantum meaning of classical field theory},
   journal={Rev. Modern Phys.},
   volume={49},
   date={1977},
   number={3},
   pages={681--706},
   issn={0034-6861},
   review={\MR{0503137}},
   doi={10.1103/RevModPhys.49.681},
}

\bib{MR1474726}{book}{
      author={{Janson, Svante}},
       title={Gaussian hilbert spaces},
      series={Cambridge Tracts in Mathematics},
   publisher={Cambridge University Press, Cambridge},
        date={1997},
      volume={129},
        ISBN={0-521-56128-0},
      review={\MR {1474726}},
}

\bib{katoad}{article}{
   author={Kato, Tosio},
   title={On the Adiabatic Theorem of Quantum Mechanics},
   date={1950},
   journal={J. Phys. Soc. Jpn},
   pages={435-439},
}


\bib{MR0407477}{article}{
   author={Kato, Tosio},
   title={Quasi-linear equations of evolution, with applications to partial
   differential equations},
   conference={
      title={Spectral theory and differential equations (Proc. Sympos.,
      Dundee, 1974; dedicated to Konrad J\"{o}rgens)},
   },
   book={
      publisher={Springer, Berlin},
   },
   date={1975},
   pages={25--70. Lecture Notes in Math., Vol. 448},
   review={\MR{0407477}},
}


\bib{MR0180856}{article}{
      author={{Kristensen, P.}},
      author={{Mejlbo, L.}},
      author={{Poulsen, E. Thue}},
       title={Tempered distributions in infinitely many dimensions. ii.
  displacement operators},
        date={1964},
        ISSN={0025-5521},
     journal={Math. Scand.},
      volume={14},
       pages={129\ndash 150},
      review={\MR {0180856}},
}

\bib{MR0180855}{article}{
      author={{Kristensen, P.}},
      author={{Mejlbo, L.}},
      author={{Poulsen, E. Thue}},
       title={Tempered distributions in infinitely many dimensions. i.
  canonical field operators},
        date={1965},
     journal={Commun. Math. Phys.},
      volume={1},
       pages={175\ndash 214},
      review={\MR {0180855}},
}

\bib{MR0218081}{article}{
      author={{Kristensen, P.}},
      author={{Mejlbo, L.}},
      author={{Poulsen, E. Thue}},
       title={Tempered distributions in infinitely many dimensions. iii. linear
  transformations of field operators},
        date={1967},
        ISSN={0010-3616},
     journal={Comm. Math. Phys.},
      volume={6},
       pages={29\ndash 48},
      review={\MR {0218081}},
}

\bib{MR2068924}{book}{
   author={Manton, Nicholas},
   author={Sutcliffe, Paul},
   title={Topological solitons},
   series={Cambridge Monographs on Mathematical Physics},
   publisher={Cambridge University Press, Cambridge},
   date={2004},
   pages={xii+493},
   isbn={0-521-83836-3},
   review={\MR{2068924}},
   doi={10.1017/CBO9780511617034},
}

\bib{MR0129790}{book}{
      author={{Messiah, Albert}},
       title={Quantum mechanics. vol. i},
      series={Translated from the French by G. M. Temmer},
   publisher={North-Holland Publishing Co., Amsterdam; Interscience Publishers
  Inc., New York},
        date={1961},
      review={\MR {0129790}},
}

\bib{MR0493420}{book}{
      author={{Reed, Michael}},
      author={{Simon, Barry}},
       title={Methods of modern mathematical physics. II. Fourier analysis,
  self-adjointness},
   publisher={Academic Press [Harcourt Brace Jovanovich, Publishers], New
  York-London},
        date={1975},
      review={\MR {0493420}},
}

\bib{MR529429}{book}{
      author={{Reed, Michael}},
      author={{Simon, Barry}},
       title={Methods of modern mathematical physics. III},
   publisher={Academic Press [Harcourt Brace Jovanovich, Publishers], New
  York-London},
        date={1979},
        ISBN={0-12-585003-4},
        note={Scattering theory},
      review={\MR {529429}},
}

\bib{MR751959}{book}{
      author={{Reed, Michael}},
      author={{Simon, Barry}},
       title={Methods of modern mathematical physics. I},
     edition={2},
   publisher={Academic Press, Inc. [Harcourt Brace Jovanovich, Publishers], New
  York},
        date={1980},
        ISBN={0-12-585050-6},
        note={Functional analysis},
      review={\MR {751959}},
}

\bib{MR0162118}{book}{
  AUTHOR = {Robertson, A. P.},
    AUTHOR = {Robertson, W. J.},
     TITLE = {Topological vector spaces},
    SERIES = {Cambridge Tracts in Mathematics and Mathematical Physics, No.
              53},
 PUBLISHER = {Cambridge University Press, New York},
      YEAR = {1964},
     PAGES = {viii+158}
}

\bib{MR516713}{article}{
      author={{Ruijsenaars, S. N. M.}},
       title={On Bogoliubov transformations. II. The general case},
        date={1978},
        ISSN={0003-4916},
     journal={Ann. Physics},
      volume={116},
      number={1},
       pages={105\ndash 134},
      review={\MR {516713}},
}

\bib{MR1741419}{book}{
    AUTHOR = {Schaefer, H. H.},
 AUTHOR = {Wolff, M. P.},
    TITLE = {Topological vector spaces},
    SERIES = {Graduate Texts in Mathematics},
    VOLUME = {3},
   EDITION = {Second},
 PUBLISHER = {Springer-Verlag, New York},
      YEAR = {1999},
     PAGES = {xii+346},
      ISBN = {0-387-98726-6}
}

\bib{MR0489552}{book}{
      author={{Simon, Barry}},
       title={The $P(\phi )_{2}$ euclidean (quantum) field theory},
   publisher={Princeton University Press, Princeton, N.J.},
        date={1974},
        note={Princeton Series in Physics},
      review={\MR {0489552}},
}

\bib{MR2097788}{book}{
   author={Spohn, Herbert},
   title={Dynamics of charged particles and their radiation field},
   publisher={Cambridge University Press, Cambridge},
   date={2004},
   pages={xvi+360},
   isbn={0-521-83697-2},
   review={\MR{2097788}},
   doi={10.1017/CBO9780511535178},
}

\bib{MR1186038}{article}{
   author={Stuart, David M. A.},
   title={Perturbation theory for kinks},
   journal={Comm. Math. Phys.},
   volume={149},
   date={1992},
   number={3},
   pages={433--462},
   issn={0010-3616},
   review={\MR{1186038}},
}

\bib{MR1810509}{article}{
   author={Stuart, David M. A.},
   title={Modulational approach to stability of non-topological solitons in
   semilinear wave equations},
   language={English, with English and French summaries},
   journal={J. Math. Pures Appl. (9)},
   volume={80},
   date={2001},
   number={1},
   pages={51--83},
   issn={0021-7824},
   review={\MR{1810509}},
   doi={10.1016/S0021-7824(00)01189-2},
}

\bib{MR2360179}{article}{
      author={{Stuart, David M. A.}},
       title={Analysis of the adiabatic limit for solitons in classical field
  theory},
        date={2007},
        ISSN={1364-5021},
     journal={Proc. R. Soc. Lond. Ser. A Math. Phys. Eng. Sci.},
      volume={463},
      number={2087},
       pages={2753\ndash 2781},
      review={\MR {2360179}},
}

\bib{MR2905846}{book}{
   author={Weinberg, Steven},
   title={Lectures on quantum mechanics},
   publisher={Cambridge University Press, Cambridge},
   date={2013},
   pages={xx+358},
   isbn={978-1-107-02872-2},
   review={\MR{2905846}},
}

\bib{MR0783974}{article}{
   author={Weinstein, Michael I.},
   title={Modulational stability of ground states of nonlinear Schr\"odinger
   equations},
   journal={SIAM J. Math. Anal.},
   volume={16},
   date={1985},
   number={3},
   pages={472--491},
   issn={0036-1410},
   review={\MR{0783974}},
   doi={10.1137/0516034},
}

\bib{MR0999137}{book}{
   author={Yamasaki, Y.},
   title={Measures on infinite-dimensional spaces},
   series={Series in Pure Mathematics},
   volume={5},
   publisher={World Scientific Publishing Co., Singapore},
   date={1985},
   pages={x+256},
   isbn={9971-978-52-0},
   review={\MR{0999137}},
   doi={10.1142/0162},
}

\end{biblist}
\end{bibdiv}

\end{document}